%% file: review.tex
\documentclass[rmp,onecolumn,amsfonts,nofootinbib]{revtex4}
\usepackage{epsfig}
\usepackage{amsmath}
\usepackage{amssymb}
\usepackage{amsfonts}
\usepackage[mathscr]{eucal}

\newcommand{\be}{\begin{equation}}
\newcommand{\bea}{\begin{eqnarray}}
\newcommand{\rep}[1]{{{\bf {#1}}}} 
\newcommand{\eea}{\end{eqnarray}}
\newcommand{\ba}{\begin{array}}
\newcommand{\ea}{\end{array}}
\newcommand{\ee}{\end{equation}}

\newcommand{\llra}{\longleftrightarrow}

\expandafter\ifx\csname mathbbm\endcsname\relax

\else

\fi
\makeatletter
\@addtoreset{equation}{section}

\makeatother

\def\cA{{\cal A}}
\def\cB{{\cal B}}
\def\co{{\cal O}}
\def\cd{{\cal D}}
\def\cj{{\cal J}}
\def\cn{{\cal N}}
\def\cw{{\cal W}}
\def\neqf{{\cal N}=4}
\def\gymsq{g^2_{YM}}
\def\gym{g_{YM}}
\def\ct{{\cal T}}
\def\a'{\alpha'}
\def\soff{$SO(4)\times SO(4)$ }
\def\sof{$SO(4)$ }
\def\soe{$SO(8)$ }
\def\eights{${\bf 8_s}$}
\def\eightc{${\bf 8_c}$}
\def\ztwo{\mathbb{Z}_2}
\def\sugra{supergravity}

\def\susy{supersymmetry}
\def\cf{{\it cf.}}

\def\susyc{supersymmetric}
\def\rep{representation}

\def\pl{plane-wave}
\def\bg{background}
\def\ads{$AdS_5\times S^5$ }
\def\opt{operator}
\def\lc{light-cone }
\def\lcg{light-cone gauge }
\def\Tr{{\rm Tr}}
\def\vac{|vac\rangle}

\begin{document}
\hfill
\vbox{
    \halign{#\hfil         \cr
           SU-ITP-03/21\cr
           SLAC-PUB-10202\cr
           hep-th/0310119 \cr
           } 
      }  
\vspace*{2mm}

\title{The Plane-Wave/Super Yang-Mills Duality}

\author{Darius Sadri}
\email{darius@itp.stanford.edu}
\affiliation{Department of Physics, Stanford University
382 via Pueblo Mall, Stanford CA 94305-4060, USA}
\affiliation{Stanford Linear Accelerator Center,
Stanford  CA 94309, USA}
\author{M. M. Sheikh-Jabbari}
\email{jabbari@itp.stanford.edu}
\affiliation{Department of Physics, Stanford University
382 via Pueblo Mall, Stanford CA 94305-4060, USA}

\begin{abstract}  

We present a self-contained review of the Plane-wave/super-Yang-Mills duality, 
which states that strings on a plane-wave background are dual to a particular large R-charge 
sector of $\neqf, \ D=4$ superconformal $U(N)$ gauge theory. This duality is a specification of the 
usual AdS/CFT correspondence in the ``Penrose limit''. The Penrose limit of
$AdS_5 \times S^5$ leads to the maximally supersymmetric ten dimensional plane-wave
(henceforth ``the'' plane-wave) and corresponds to restricting to the large R-charge sector,
the BMN sector, of the dual superconformal field theory.
After assembling the necessary background knowledge, we state the
duality and review some of its supporting evidence. 
We review the suggestion by 't Hooft that Yang-Mills theories with gauge groups of large rank
might be dual to string theories and the realization of this conjecture in the form of the
AdS/CFT duality.
We discuss plane-waves as exact solutions of supergravity and their appearance as
Penrose limits of other backgrounds, then present an overview of string theory on the plane-wave
background, discussing the symmetries and spectrum.
We then make precise the statement of the proposed duality, classify the BMN operators, and
mention some extensions of the proposal.
We move on to study the gauge theory side of the duality, studying both quantum
and non-planar corrections to correlation functions of BMN operators, and their
operator product expansion. The important issue of operator mixing and the resultant
need for re-diagonalization is stressed.
Finally, we study strings on the plane-wave via light-cone string field theory, and
demonstrate agreement on the one-loop correction to the string mass spectrum and the corresponding 
quantity in the gauge theory. A new presentation of the relevant superalgebra is given.
\begin{center}
{\hspace{1cm}} Extended version of the  article to be published in {\it Reviews of Modern 
Physics}.
\end{center}

\end{abstract}                                                                 

\maketitle
\tableofcontents
\section{Introduction}
\label{introduction}
\input{intro.tex}

\section{Plane-waves as solutions of supergravity}
\label{ppwave}
\input{ppgr.tex}

\section{Penrose limits and plane-waves}
\label{penroselimit}
\input{penrose.tex}


\section{Plane-waves as backgrounds for string theory}
\label{stringbg}
\input{stringbg.tex}


\section{Stating the plane-wave/SYM duality}
\label{BMNproposal}
\input{BMN.tex}


\section{Spectrum of strings on plane-waves from gauge theory
I: free strings}
\label{noninteractingstrings}
\input{noninter.tex}


\section{Strings on plane-waves from gauge theory II: interacting strings}
\label{interactingstrings}
\input{inter.tex}


\section{Plane-wave light-cone String Field Theory}
\label{SFT}
\input{SFT.tex}




\section{Concluding remarks and open questions}
\label{conclusion}
\input{conclusion.tex}


\section*{Acknowledgments}

We would like to thank I. Bars, K. Dasgupta, M. Fabinger, J. Gomes,  
S. Hellerman, 
Jonathan Hsu,C. Kristjansen, M. van Raamosdonk, R. Roiban and G. Semenoff
for discussions or useful comments on the manuscript. 
The work of M. M Sh-J is supported in part by NSF grant PHY-9870115 and in part by funds from 
Stanford Institute for Theoretical Physics.
The work of D. S. is supported by the Department of Energy, Contract DE-AC0376F00515.


\appendix

\section{Conventions for ${\cal N}=4,\ D=4$ supersymmetric gauge theory}
\label{ConventionD=4}
\input{appendix1.tex}

\section{Conventions for ten dimensional fermions}
\label{ConventionD=10}
\input{appendix2.tex}

\bibliographystyle{apsrmp}
\bibliography{rmp-review}
\newpage

\end{document}

%% file: intro.tex

In the late 1960's a theory of strings was first proposed as a
model for the strong interactions describing the dynamics of hadrons.
However, in the early 1970's, results from deep inelastic scattering experiments led to
the acceptance of the ``parton'' picture of hadrons, and this led to the development
of the theory of quarks as basic constituents carrying color quantum numbers, and
whose dynamics are described by Quantum Chromo-Dynamics (QCD), which is an $SU(N_c)$
Yang-Mills gauge theory with $N_f$ flavor of quarks.
According to the standard model of particle physics, $N_c=3,\ N_f=6$.
With the acceptance of QCD as the theory of strong 
interactions the old string theory became obsolete. However, in 1974 't Hooft 
\cite{'tHooft:1974jz,'tHooft:1974hx} observed a property of $SU(N_c)$ gauge theories
which was very suggestive of a correspondence or ``duality'' between the gauge 
dynamics and string theory.

To study any field theory we usually adopt a perturbative expansion, 
generally in powers of the coupling constant of the theory. The first 
remarkable observation of 't Hooft was that the true expansion parameter
for an $SU(N)$ gauge theory (with or without quarks) is not the
Yang-Mills coupling $\gymsq$, but rather $\gymsq$ dressed by $N$, in the
combination $\lambda$, now known as the 't Hooft coupling:
\be\label{tHooftcoupling}
\lambda=g^2_{YM} N \ .
\ee

The second remarkable observation 't Hooft made was that in addition to the 
expansion in powers of $\lambda$ one may also classify the Feynman 
graphs appearing in the correlation function of generic gauge theory 
operators in powers of ${1}/{N^2}$. This observation is based on the 
fact that the operators of this gauge theory are built from simple 
$N \times N$ matrices. One is then led to expand any correlation function in a double
expansion, in power of $\lambda$ as well as ${1}/{N^2}$.
In the ${1}/{N^2}$ expansion, which is a useful one for large 
$N$, the terms of lowest order in powers of ${1}/{N^2}$ arise from
the subclass of Feynman diagrams which can be drawn on a sphere
(a one-point compactification of the plane), once the 't Hooft double line notation
is used. These are called {\it planar} graphs. In the same spirit one can classify 
all Feynman graphs according to the lowest genus surface that they may be placed
on without any crossings.
For genus $h$ surfaces, with $h>0$, such diagrams are called {\it non-planar}.
The lowest genus non-planar surface is the torus with $h=1$.
The genus $h$ graphs are suppressed by a factor of
$\left({1}/{N^2}\right)^h$ with respect to the planar diagrams.
According to this $1/N$ expansion, at large $N$, but 
finite 't Hooft coupling $\lambda$, the correlators are dominated by 
planar graphs.

The genus expansion of Feynman diagrams in a gauge theory resembles a 
similar pattern in string theory: stringy loop diagrams are suppressed by 
$g_s^h$ where $h$ is now the genus of the string worldsheet and $g_s$ is 
the string coupling constant. The Feynman graphs in the large $N$
limit form a continuum surface which may be (loosely) interpreted as the string 
worldsheet. In section \ref{tHooft} of the introduction,
we will very briefly sketch the mechanics of the 't Hooft large $N$ expansion.

In the mid 1970's, string theory was promoted from an effective theory
of strong dynamics to a theory of fundamental strings and put forward as a 
candidate for a quantum theory of gravity \cite{Scherk:1974ca}. 
Much has been learned since then about the five different ten dimensional string theories.
In particular, by 1997, a web of various dualities relating these string theories, their 
compactifications to lower dimensions, and an as yet unknown, though more
fundamental theory known as M-theory, had been proposed and compelling pieces of evidence 
in support of these dualities uncovered \cite{Witten:1995ex,Hull:1995ys}.
We do not intend to delve into the details of these dualities, for 
such matters the reader is referred to the various books and reviews, e.g.
\cite{Polchinski:1998rq,Polchinski:1998rr,Johnson:2003gi}.

Although our understanding of string and M-theory had been much improved
through the discovery of these various dualities, before 1997, the observation of
't Hooft had not been realized in the context of string theory.
In other words the 't Hooft strings and the ``fundamental'' strings seemed to be 
different objects. Amazingly, in 1997 a study of the near  horizon geometry of D3-branes
\cite{Maldacena:1998re} led to the conjecture that
\vskip 3mm
{\it  \centering Strings of type IIB string theory on the $AdS_5 \times S^5$
background are the 't Hooft strings of an $\neqf$, $D=4$ supersymmetric 
Yang-Mills theory.}
\vskip 3mm

According to this conjecture any physical object or process in the
type IIB theory on $AdS_5 \times S^5$ background can be equivalently described by $\neqf$, $D=4$ super Yang-Mills (SYM) theory \cite{Gubser:1998bc,Witten:1998qj,Aharony:1999ti}.
In particular, the 't Hooft coupling \eqref{tHooftcoupling} is related to the $AdS$ radius $R$ as
\be\label{AdSraduis}
\left(\frac{R}{l_s}\right)^4=g^2_{YM} N 
\ee
where $l_s$ is the string scale. 
On the string theory side of the duality, ${l_s}/{R}$ appears as the worldsheet 
coupling; hence when the gauge theory is weakly coupled the two 
dimensional worldsheet theory is strongly coupled and 
non-perturbative, and vice-versa. In this sense the $AdS/CFT$ duality
\cite{Witten:1998qj,Aharony:1999ti} is a weak/strong duality. 
Due to the (mainly technical) difficulties of solving the worldsheet 
theory on the $AdS_5 \times S^5$ background\footnote{For a recent work in the
direction of solving this two dimensional theory see \cite{Bena:2003wd} and the references therein.},
our understanding of the string theory side of the duality has been mainly  
limited to the low energy supergravity limit, and in order for the supergravity 
expansion about the $AdS$ background to be trustworthy, we generally need to keep the 
$AdS$ radius large. At the same time we must also ensure the suppression of
string loops. As a result, most of the development and checks of the duality from the string theory side
have been limited to the regime of large 't Hooft coupling 
and the $N \to \infty$ limit on the gauge theory side.
A more detailed discussion of this conjecture, the $AdS/CFT$ duality, 
will be presented in section \ref{AdS/CFT}. 

One might wonder if it is possible to go beyond the supergravity
limit and perform real string theory calculations from the gauge theory side.
We would then need to have similar results from the string theory side 
to compare with, and this seems notoriously difficult, at least at the moment.

The $\sigma$-model for strings on $AdS_5 \times S^5$ is difficult to solve.
However, there is a specific limit in which $AdS_5 \times S^5$ reduces to a plane wave
\cite{Gueven:2000ru,Blau:2001ne,Blau:2002dy,Blau:2002mw,Blau:2002js,Blau:2003rt},
and in this limit the string theory $\sigma$-model becomes solvable
\cite{Metsaev:2001bj,Metsaev:2002re}.
In this special limit we then know the string spectrum,
at least for non-interacting strings, and one might ask if we can find the
same spectrum from the gauge theory side.
For that we first need to understand how this
specific limit translates to the gauge theory side.
We then need a definite proposal
for mapping the operators of the gauge theory to (single) string states. This proposal,
following the work of Berenstein-Maldacena-Nastase 
\cite{Berenstein:2002jq},
is known as the BMN conjecture. It will be introduced in section \ref{Intro:BMNconjecture} of the introduction and is discussed in more depth in section \ref{BMNproposal}.
The BMN conjecture is supported by some explicit and
detailed calculations on the gauge theory side. Spelling out different elements of this conjecture
is the main subject of this review.

In section \ref{ppwave} we review plane-waves as solutions of
supergravities which have a {\it globally defined null Killing vector field}, and emphasize
an important property of these backgrounds: they are exact solutions without $\alpha'$
corrections. Also in this section, we discuss Penrose diagrams and some general properties
of \pl s. We will focus mainly on the ten dimensional maximally
supersymmetric \pl\ background.
This maximally supersymmetric plane-wave will be referred to as ``the'' plane-wave to distinguish 
it from other plane-wave backgrounds.
We study the isometries of this \bg s
as well as the corresponding supersymmetric extension, and we show that
this \bg\ possesses a $PSU(2|2) \times PSU(2|2) \times U(1)_- \times U(1)_+$ superalgebra.
We also discuss the spectrum of type IIB \sugra\ on the \pl\ \bg .

In section \ref{penroselimit} we review the procedure for taking the 
Penrose limit of any given geometry. We then argue that this procedure
can be extended to solutions of supergravities to generate new solutions.
As examples we work out the Penrose limit of some $AdS_p \times S^q$ spaces, the $AdS$ 
orbifolds, and conifold geometry. For the case of orbifolds we argue that
one may naturally obtain ``compactified'' \pl s, where the compact 
direction is either light-like or space-like. Moreover, we discuss how
taking the Penrose limit manifests itself as a contraction at the level of the superalgebra.
In particular, we show how to obtain the superalgebra of the \pl , discussed in section \ref{ppwave}, as a (Penrose) contraction 
of $PSU(2,2|4)$ which is the superalgebra of the $AdS_5\times S^5$ \bg .

Having established the fact that  \pl\ \bg s form  $\alpha'$-exact solution 
of supergravities, they form a particularly simple  \bg s for string theory. In section
\ref{stringbg} we work out the $\sigma$-model action for type IIB strings 
on the plane-wave \bg\ in the light-cone gauge. 
Formulating a theory in the light-cone gauge has the advantage that only physical (on-shell)
degrees of freedom appear and ghosts are decoupled \cite{Polchinski:1998rq}.
For the particular case of strings on plane-waves, due to the existence of the
globally defined null Killing vector field, fixing the light-cone gauge has an
additional advantage: the energies (frequencies) are conserved in this gauge and as
a result the well known problem associated with non-flat spaces, namely
particle (string) production is absent. Adding fermions is done using the
Green-Schwarz formulation, and as usual redundant fermionic degrees freedom arise from
$\kappa$ symmetry. After fixing the 
$\kappa$-symmetry, we obtain the fully gauge fixed action from which
one can easily read off the spectrum of (free) strings on this \bg .
We also present the \rep\ of the \pl\ superalgebra in terms of stringy modes.

In sections \ref{BMNproposal},  \ref{noninteractingstrings} and \ref{interactingstrings},
we return to the 't Hooft expansion, though in the BMN sector of the 
gauge theory, and deduce the spectrum of strings on the plane-wave 
obtained in section \ref{stringbg}, from gauge theory calculations. 
In this sense these sections are the core of this review.
In section \ref{BMNproposal} we present the BMN or \pl / SYM duality conjecture, and 
in section \ref{Extensions} we discuss some variants, e.g. how one may analyze
strings on a $Z_K$-orbifold of the plane-wave geometry from ${\cal N}=2$, $D=4$ $U(N)^K$ quiver theory.

In section \ref{noninteractingstrings} we present the first piece of supporting evidence for the duality,
where we focus on the planar graphs. Reviewing 
the results of \cite{Kristjansen:2002bb, Gross:2002su, Constable:2002hw} we show that the 't Hooft 
expansion is modified for the BMN sector of the gauge theory, and we are led to a 
new type of ``'t  Hooft expansion'' with a different effective coupling. 
In section \ref{amonalous} we argue how and why anomalous dimensions of operators in 
the BMN sector correspond to the free string spectrum obtained in section \ref{stringbg}. In fact,
through a calculation to all orders in the 't Hooft coupling, but in the planar limit, we recover from a purely gauge theoretic analysis exactly the same spectrum we found in section \ref{stringbg}.
In section \ref{OPE} we discuss the operator product expansion (OPE) of two BMN operators and the fact that this OPE only involves the BMN operators. In other words, the BMN sector of the gauge theory is closed under the OPE. 

In section \ref{interactingstrings}, we move beyond the planar limit
and consider the contributions arising from non-planar graphs to the spectrum,
which correspond on the string theory side to inclusion of loops.
We will see that the genus counting parameter should also  be
modified in the BMN limit. Moreover, as we will see, the suppression of higher 
genus graphs with respect to the planar ones is not universal, and in fact
depends on the sector of the operators we are interested in.
One of the intriguing consequences of the non-vanishing higher genus contributions is 
the possible mixing between the original single trace BMN operators 
with double and in general multi trace \opt s. The 
mixing effects will force us to modify the original BMN dictionary. 
After making the appropriate modifications, we present the results of the one-loop (genus one) corrections to the string spectrum.

After discussing the string spectrum on the plane-wave at both planar and non-planar order
from the gauge theory side of the duality, we tackle the question of string interactions on 
the plane-wave \bg\ in section \ref{SFT}, with the aim of obtaining one-loop corrections to string spectrum from the string theory side.
This provides us with a non-trivial check of the plane-wave/SYM duality.
{}From the string theory point of view, the presence of the non-trivial 
background, in 
particular the RR form, makes using the usual machinery for computing 
string scattering amplitude via vertex operators cumbersome, and one is led to
to develop the string field theory formulation.
{}From the gauge theory side, as we will discuss in section
\ref{interactingstrings}, the nature of difficulties is different: it is not a trivial task to 
distinguish single, double and in general multi-string states. 
We work out \lc string field theory on this \bg\ and use this setup
to calculate one-loop corrections to the string spectrum.
We will show that the data extracted from non-planar gauge theory correlation
functions is in agreement with their string theoretic counterparts.
In section \ref{SFTgeneral} we present some basic facts and necessary
background regarding \lc string field theory. In section
\ref{cubicSFT} we work out the three-string vertex in the \lc string field theory on the \pl\ \bg . Then in section \ref{SFTcontact} we consider higher order string interactions and calculate one-loop corrections to the string mass spectrum.
%

Finally, in section \ref{conclusion}, we conclude by summarizing the main
points of the review, and mention some interesting related 
ideas and developments in the literature.
We also discuss some of the open questions in the formulation of strings on
general \pl\ \bg s and the related issues on the gauge theory side of the conjectured
duality.

\subsection{'t Hooft's large $N$ expansion}\label{tHooft}

Attempts at understanding strong dynamics in gauge theories led 't Hooft to introduce a remarkable
expansion for gauge theories with large gauge groups, with the rank of the gauge group $\sim N$
\cite{'tHooft:1974jz,'tHooft:1974hx}. He suggested treating the rank of the gauge group as a
parameter of the theory, and expanding in $1/N^2$, which turns out to correspond to the genus of the surface onto which the Feynman diagrams can be mapped without overlap, yielding a topological
expansion analogous to the genus expansion in string theory, with the gauge theory Feynman graphs
viewed as ``string theory'' worldsheets. In this correspondence, the planar (non-planar) Feynman
graphs may be thought of as tree (loop) diagrams of the corresponding ``string theory''.


Asymptotically free theories, like $SU(N)$ gauge theory with sufficiently few matter fields,
exhibit dimensional transmutation, in which the scale dependent coupling gives rise to a
fundamental scale in the theory. For QCD, this is the confinement scale $\Lambda_{QCD}$.
Since this is a scale associated with physical effects, it
is natural to keep this scale fixed in any expansion. This scale appears as a constant of
integration when solving the $\beta$ function equation, and it can be held fixed for large $N$ if
we also keep fixed the product $\gymsq N$ while taking $N \rightarrow \infty$. This defines the new
expansion parameter of the theory, the 't Hooft coupling constant $\lambda \equiv \gymsq N$.

\begin{figure}[ht]
\centering
\epsfig{figure=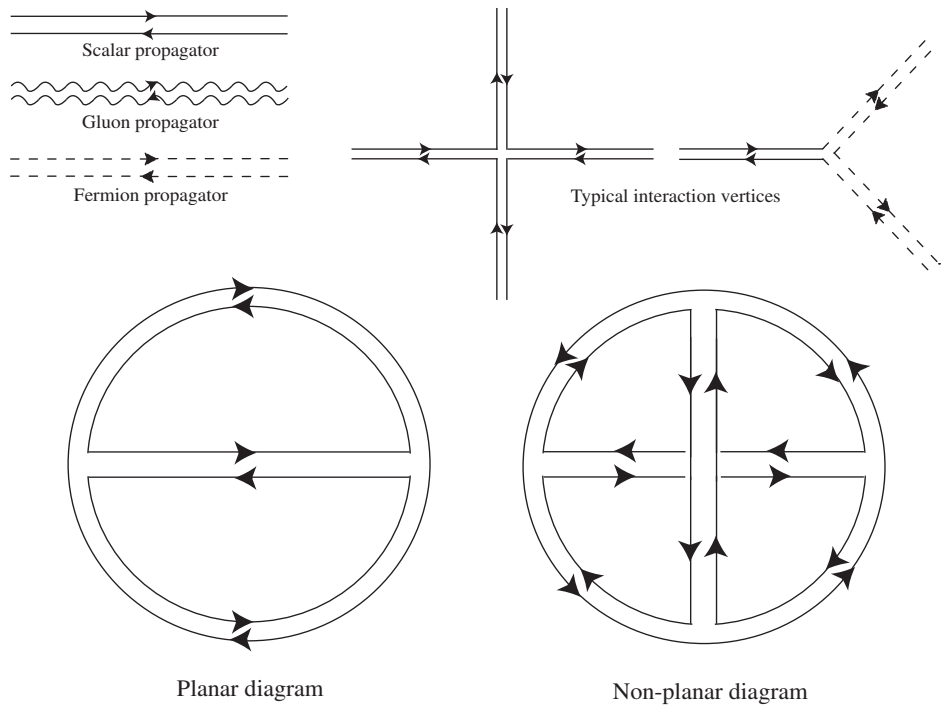,width=123.75mm,height=92.8125mm}
\begin{center}
\caption{Typical Feynman rules for adjoint fields and sample planar and non-planar
diagrams.}
\label{LargeN}
\end{center}
\end{figure}

 
To see how the expansion works in practice, we can consider the action for a gauge theory, for
example the $\neqf$ super-Yang-Mills\footnote{Supersymmetry is not consequential to this discussion
and we ignore it for now.} theory written down in component form in \eqref{neqf-action-components}.
All the fields in this action are in the adjoint representation. We have scaled our fields so that
an overall factor of $1/\gymsq$ appears in front. We write this in terms of $N$ and the 't Hooft coupling 
$\lambda$, using $1 / \gymsq = N / \lambda$. The perturbation series for this theory can be
constructed in terms of Feynman diagrams built from propagators and vertices in the usual way. With
our normalization, each propagator contributes a factor of $\lambda / N$, and each vertex a factor
of $N / \lambda$. Loops in diagrams appear with group theory factors coming from summing over the
group indices of the adjoint generators. These give rise to an extra factor of $N$ for each loop. A
typical Feynman diagram will be associated with a factor \be
  \lambda^{P-V} N^{V-P+(L+1)}
\ee
if the diagram contains $V$ vertices, $P$ propagators and $L$ loops.
These diagrams can be interpreted as simplical complexes if we choose to draw them using the 't 
Hooft double line notation. For $U(N)$, the group index structure of adjoint fields is that  
of a direct product of a fundamental and an anti-fundamental. The propagators can be drawn with 
two lines showing the flow of each index, and the arrows point in opposite directions (see FIG. 
\ref{LargeN}).

\vskip 7mm

The vertices are drawn in a similar way, with directions of arrows indicating the fundamental or anti-fundamental indices of the generators.
In this diagrammatic presentation, the propagators form the edges and the insides of loops are considered the faces. The one point compactification of the plane then means that the diagrams give rise to closed, compact and orientable surfaces, with Euler characteristic
$\chi = V - P + F = 2 - 2h$, where $h$ is the genus of the surface.
The number of faces is one more than the number of loops, since the group theory always gives rise to an extra factor of $N$ for the last trace. In the simplical decomposition, with the one-point compactification, the outside of the diagram becomes another face, and can be interpreted as the last trace.

The perturbative expansion of the vacuum persistence amplitude takes the form of a double expansion
\be \label{thooft-double-expansion}
  \sum_{h=0}^{\infty} N^{2 - 2h} P_h (\lambda)
\ee
with $h$ the genus and $P_h$ some polynomial in $\lambda$, which itself admits a power series 
expansion
\be \label{thooft-double-expansion-part2}
  P_h(\lambda)=\sum_{n=0}^{\infty} C_{h,n} \lambda^n
\ee
The simple idea is that all the diagrams generated for the vacuum correlation function can be 
grouped in classes based on their genera, and all the diagrams in each class will have varying 
dependences on the 't Hooft coupling $\lambda$. Collecting together all the diagrams in a given 
class again into groups sharing the same dependence on $\lambda$, we can extract the $h$ and $n$ 
dependent constant $C_{h,n}$.
It is clear from \eqref{thooft-double-expansion} that for large $N$, the dominant contributions 
come from diagrams of the lowest genus, the planar (or spherical) diagrams.

The double expansion \eqref{thooft-double-expansion} and \eqref{thooft-double-expansion-part2}
looks remarkably similar to the perturbative
expansion for a string theory with coupling constant $1/N$ and with the expansion in powers of $\lambda$ playing the role of the worldsheet expansion. The analogy extends to the genus expansion, with the Feynman diagrams loosely forming a sort of discretized string worldsheet. At large $N$, such a string theory would be weakly coupled. The string coupling measures the difference in the Euler character for worldsheet diagrams of different topology. This has long suggested the existence of a duality between gauge and string theory. We of course also have to account for the mapping of non-perturbative effects on the two sides of the duality.

So far we have considered only the vacuum diagrams, though the same arguments go through when
considering correlation functions with insertions of the fields. The action appearing in the
generating functional of connected diagrams must be supplemented with terms coupling the
fundamental fields to currents, and these terms will enter with a factor of $N$. The
planar\footnote{With the point at infinity identified, planar diagrams become spheres, and higher
genus diagrams spheres with handles.} (leading) contributions to such correlation functions with
$j$ insertions of the fields will be suppressed by an extra factor of $N^{-j}$ relative to the vacuum
diagrams. The one particle irreducible three and four point functions then come with factors of
$1/N$ and $1/N^2$ relative to the propagator, suggesting that $1/N$ is the correct expansion
parameter. The expansion \eqref{thooft-double-expansion} for these more general correlation
functions still holds if we account for the extra factors of $N$ coming from the insertions of the
fields. The extra factor depends on the number of fields in the correlation function, but
is fixed for the perturbative expansion of a given correlator.

The picture we have formed is of an oriented closed string theory. Adding matter in the fundamental
representation would correspond to including propagators with a single line, and these could then form the edges
of the worldsheets, and so would correspond to a dual theory with open strings (with the added
possibility of D-branes). Generalizations to other gauge groups such as $O(N)$ and $Sp(N)$ would
lead to unorientable worldsheets, since their adjoint representations (which are real) appear
like products of fundamentals with fundamentals. This viewpoint has been applied to other types of
theories, for example, non-linear sigma models with a large number of fundamental degrees of
freedom.

The new ingredient relevant to our discussion will be the following: for a conformally invariant
theory such as $\neqf$ SYM, the $\beta$ function vanishes for all values of the coupling $\gym$ (it
has a continuum of fixed points). There is no natural scale in this theory that should be held
fixed. This makes limits different from the 't Hooft limit possible, and we take advantage of such
an opening via the so called BMN limit, which we discuss at length in what
follows. Not all such limits are well-defined. In the BMN limit, we will consider operators with
large numbers of fields. If the number of fields is scaled with $N$, generically, higher genus
diagrams will dominate lower genus ones, and the genus expansion will break down. The novel feature
of the BMN limit is that the combinatorics of these large numbers of fields conspire in a way that
makes it possible for diagrams of all genera to contribute without the relative suppression typical
in the 't Hooft limit. In this sense, the BMN limit is the balancing point between two regions, one
where the diagrams of higher genus are suppressed and don't contribute in the limit, and the other
where the limit is meaningless.

A concise introduction to the basic ideas underlying the large $N$ expansion can be found in
\cite{'tHooft:2002yn}, with a more detailed review presented in \cite{'tHooft:1994gh}. Applications
to QCD are given in \cite{Manohar:1998xv}. A review of the large $N$ limit in field theories and
the relation to string theory can also be found in \cite{Aharony:1999ti}, which discusses many
issues related to $AdS$ spaces, conformal field theories and the celebrated $AdS/CFT$
correspondence.






\subsection{String/gauge theory duality}\label{AdS/CFT}

't Hooft's original demonstration that the large $N$ limit of $U(N)$ gauge theory, as we 
have already discussed, is dual to a string theory, has sparked many attempts to construct such a duality explicitly. One such attempt \cite{Gross:1993tu,Gross:1993hu} was to construct the dual to two-dimensional pure $QCD$ as a map from two-dimensional worldsheets of a given genus into a two-dimensional target space. $QCD_2$ is almost a topological theory, with the correlation functions depending only on the topology and area of the manifold on which the theory is formulated, making the theory exactly solvable.  The partition function of this string theory sums over all branched coverings of the target space, and can be evaluated by discretizing the target using a two-dimensional simplical complex with an $N \times N$ matrix placed at each link.
The partition function thus constructed can be evaluated exactly via an expansion 
in terms of group characters,
giving rise to a matrix model, whose solution has been given in
\cite{Kazakov:1996vy,Kostov:1997bs,Kostov:1998bn}.
Zero dimensional $QCD$ was considered in \cite{Brezin:1978sv}, as a toy model which retains all the diagrammatic but with trivial propagators, allowing 
the investigation of combinatorial counting in matrix models.

Another realization of 't Hooft's observation, this time via conventional string theory, is the
celebrated AdS/CFT correspondence. The duality is suggested by the two viewpoints presented by
D-branes. The low energy effective action of a stack of $N$ coincident D3-branes is given by
$\neqf$ super-Yang-Mills theory with gauge group $U(N)$. While away from the brane the theory is
type IIB closed string theory, there exists a {\it decoupling limit} where the closed strings of 
the bulk are decoupled from the gauge theory living on the brane \cite{Maldacena:1998re}.
D3-branes are $1/2$ BPS, breaking $16$ of the supercharges of
the type IIB vacuum, which in the decoupling limit will be non-linearly realized as the superconformal supercharges
in the $\neqf$ worldvolume theory of the branes which exhibits superconformal invariance. For large
$N$, the stack of D-branes will back-react, modifying the geometry seen by the type IIB strings.
In the low energy description given by supergravity, the presence of the D-brane is seen in the
form of the vacuum for the background fields like the metric and the Ramond-Ramond fields. These
are two different descriptions of the physics of the stack of D-branes, and the ability to take
these different viewpoints is the essence of the AdS/CFT duality according which 
type IIB superstring theory on the $AdS_5\times S^5$ background is dual 
to (or can be equivalently described by) ${\cal N}=4,\ D=4$ $U(N)$ 
supersymmetric Yang-Mills theory with a prescribed mapping between string theory and gauge 
theory objects.

The specific prescription for the correspondence is suggested by the matching of the global
symmetry groups and their representations on the two sides of the duality. The matching extends to
the partition function of the $\neqf$ SYM on the boundary of $AdS_5$ ($R \times S^3$) and the
partition function of IIB string theory on $AdS_5 \times S^5$ \cite{Witten:1998qj}
\be\label{WittenFormula}
\Big\langle e^{\int d^4x \; \phi_0(x){\cal O} (x)}\Big\rangle_{\rm CFT}={\cal 
Z}_{string}\left[\phi|_{boundary}=\phi_0(x)\right]\ ,
\ee
where the left-hand-side is the generating function of correlation functions of 
gauge invariant operators ${\cal O}$ in the gauge theory (such correlation functions are obtained 
by taking derivatives with respect to $\phi_0$ and setting 
$\phi_0=0$) and the right-hand-side is the full partition function of (type IIB) string theory
on the \ads \bg\ with the boundary condition that the field $\phi=\phi_0$ on the $AdS$ boundary \cite{
Aharony:1999ti}.
The dimensions of the operators ${\cal O}$ 
(i.e. the charge associated with the behavior of
the operator under rigid coordinate scalings)  correspond to the free-field
masses of the bulk excitations. Every operator in the gauge theory can be put in one to one
correspondence with a field propagating in the bulk of the $AdS$ space, e.g. the gauge invariant chiral
primary operators and their descendents on the Yang-Mills side can be put in a one to one 
correspondence with the
the supergravity modes of the type IIB theory. 
In the low energy approximation to
the string theory, we have type IIB supergravity, with higher order $\alpha^\prime$ corrections
from the massive string modes. (Note, however, that the \ads\ \bg\ itself is an exact solution to 
\sugra\ with all $\alpha'$-corrections included.) 
The relation \eqref{AdSraduis} between the radius of $AdS_5$ (and
also $S^5$) shows that when the gauge theory is weakly coupled, the radius of $AdS_5$ is small in
string units. In this regime, the supergravity approximation breaks down. 
Of course, to make the duality complete, we have to find the mapping
for the non-perturbative objects and effects on the two sides of the duality.

This conjecture has been generalized and restated for string theories on many deformations of the 
$AdS_5\times S^5$ background, such as $AdS_5\times T^{1,1}$ \cite{Klebanov:1998hh}, the orbifolds of \ads space \cite{Gukov:1998kk, Lawrence:1998ja, Kachru:1998ys}, and even non-conformal cases \cite{ Klebanov:2000hb, Polchinski:2000uf} and $AdS_3\times S^3\times 
M_4$ \cite{Giveon:1998ns, Kutasov:1999xu}. For example in the 
Klebanov-Witten case, the statement is that type IIB strings on $AdS_5\times T^{1,1}$ \bg \ are the
't Hooft strings of an $\cn =1$ super-conformal field theory and for the $AdS_3\times S^3$ case, 
't Hooft strings of the $\cn=(4,4)$ $D=2$ super-conformal field theory are dual to strings
on $AdS_3\times S^3\times T^4$. The latter have been made explicit by the Kutasov-Seiberg 
construction \cite{Giveon:1998ns, Kutasov:1999xu}. In general it is a non-trivial task to
determine the 't Hooft string picture of a given gauge theory.

\subsection{Moving away from \sugra\ limit, strings on plane-waves}
\label{Intro:BMNconjecture}
                                                                                                         
Although Witten's formula \eqref{WittenFormula} is precise, from a practical
point of view,
our calculational ability does not go beyond the large $N$ limit which
corresponds to the supergravity limit on the string theory side
(except for quantities which are protected by supersymmetry, the
calculations on both sides of the duality beyond the large $N$ limit
exhibit the same level of difficulty).
However, one may still hope to go beyond the supergravity limit
which corresponds to restricting to some particular sector of the gauge
theory.
                                                                                                         
In this section we recall some basic observations and facts which
led BMN to their conjecture \cite{Berenstein:2002jq} as well as a brief summary of the results 
obtained based
on and in support of the conjecture. These observations and results will be
discussed in some detail in the main part of this review.
                                                                                                         
\begin{itemize}
                                                                                                         
\item Although so far we have not been able to solve the string $\sigma$-model
in the \ads \bg\ and obtain the spectrum of (free) strings, the Penrose
limit \cite{Penrose:1976, Gueven:2000ru} of \ads geometry results in another
maximally supersymmetric \bg\ of type IIB which is the \pl\ geometry. The
corresponding $\sigma$-model (in the \lc gauge) is solvable,
allowing us to deduce the spectrum of (free) strings on this \pl\ \bg .

\item Taking the Penrose limit on the gravity side corresponds to
restricting the gauge theory to \opt s with
a large charge under one of its global symmetries
(more precisely the  R-symmetry charge associated with a $U(1)\subset SO(6)_R$) $J$, the BMN
sector, and simultaneously taking the large $N$ limit.
                                                                                                         
\item The BMN sector of ${\cal N}=4,\ D=4$ $U(N)$ SYM theory is comprised
of \opt s with large conformal dimension $\Delta$ and large R-charge $J$,
such that
\begin{subequations}\label{BMNlimit1}
\begin{align}
\frac{1}{\mu} p^- &\equiv \Delta-J={\rm fixed} \, , \\
\alpha'\mu p^+ &\equiv \frac{1}{2\sqrt{ g^2_{YM} N}} (\Delta+J)={\rm fixed} \, ,
\end{align}
\end{subequations}
together with
\be\label{BMNlimit2}
g_{YM}={\rm fixed} \ ,\ \ \ \ \ \frac{J^2}{N}={\rm fixed},\ \ \
N\to\infty \ , \ \ \ J\to\infty\ .
\ee
In the above $\frac{1}{\mu} p^-$ and $\alpha'\mu p^+$ are the corresponding
string \lc Hamiltonian and \lc momentum, respectively.
The parameter $\mu$ is a convenient but auxiliary parameter, the role of
which will become clear in the following sections.
                                                                                                         
\item In (\ref{BMNlimit1}a), $p^-$ should be understood as the {\it full} \pl\ 
\lc string (field) theory Hamiltonian. Explicitly, one  can interpret (\ref{BMNlimit1}a)
as an equality between two operators, the \pl\ \lc string field theory Hamiltonian, $H_{SFT}$, on 
one
side and the difference between the dilatation and the R-charge operators on the other side, i.e.
\be\label{improvedBMN}
\frac{1}{\mu}H_{SFT}={\cal D}- {\cal J}\ ,
\ee
where ${\cal D}$ is the dilatation operator and ${\cal J}$ is the
R-charge generator. Therefore
according to the identification (\ref{improvedBMN}) which is the (improved form of the original) 
BMN
conjecture, the spectrum of strings, which are the eigenvalues of the \lc Hamiltonian $p^-=H_{SFT}$,
should be equal to the spectrum of the dialtation operator, which is the Hamiltonain of the
${\cal N}=4,$ gauge theory on $\mathbb{R}\times S^3$, restricted to the operators in the BMN
sector of the gauge theory (defined through (\ref{BMNlimit1}) and (\ref{BMNlimit2})).

\item As stated above, equation (\ref{improvedBMN}) sets an equivalence between two operators.
However, the second part of the BMN conjecture is about the correspondence between the Hilbert spaces
that these operators act on; on the string theory side, it is the string (field) theory Hilbert
space which is comprised of direct sum of the zero string, single string, double string and ...
string states, quite similarly to the flat space case \cite{Polchinski:1998rq}. On the gauge
theory side it is  the so-called BMN operators, the set of $U(N)$ invariant operators of large R-charge $J$ and large dimension in the free gauge theory, subject to \eqref{BMNlimit1}
and \eqref{BMNlimit2}.
                                                                                                         
\item According to the BMN proposal \cite{Berenstein:2002jq}, single string states map to
certain single trace \opt s in the gauge theory.\footnote{The proposal as stated is only true for 
free strings. As we will see in section \ref{interactingstrings}, this proposal should be modified 
once string interactions are included.}
In particular, the single string vacuum state in the sector with \lc momentum $p^+$,
$(\alpha'\mu p^+)^2 =\frac{J^2}{g^2_{YM}N}$, is identified with the
chiral-primary BPS \opt
\be\label{BMNvacuum}
|0,p^+\rangle \longleftrightarrow {\cal N}_J \Tr (Z^J) \vac\ ,
\ee
where ${\cal N}_J$ is a normalization constant that will be fixed later in
section \ref{noninteractingstrings}. In the above, $Z=\frac{1}{\sqrt{2}}(\phi^5+i\phi^6)$,
where $\phi^5$ and $\phi^6$ are two of the six scalars of the ${\cal
N}=4,\ D=4$
gauge multiplet. The R-charge we want to consider, $J$, is the eigenvalue
for a $U(1)$ generator, $U(1)\subset SU(4)_R$, so that $Z$ carries unit charge and all the other bosonic modes in the vector multiplet, four scalars and four gauge fields, have zero R-charge.
Since all scalars have $\Delta=1$ (classically), for $Z$ and hence
$Z^J$, $\Delta-J=0$.
The advantage of identifying string vacuum
states with the chiral-primary \opt s is two-fold: {\it i)} they have
$\frac{1}{\mu} p^-=\Delta-J=0$, and {\it ii)} their anomalous dimension is zero and hence the
corresponding $p^-$ remains zero to all orders in the 't Hooft coupling
and even non-perturbatively, see for example \cite{D'Hoker:2002aw}.
                                                                                                         
\item As for stringy excitations above the vacuum, BMN conjectured that we
need to work with certain ``almost'' BPS \opt s, i.e. certain \opt s
with large $J$ charge and with $\Delta-J\neq 0$, but $\Delta-J \ll J$. In
particular, single closed string states were (originally) proposed to be dual to
single trace \opt s with $\Delta-J= 2$. The exact form of these \opt s and a more
detailed discussion regarding them will be presented in section
\ref{BMNproposal}. As we will see in sections \ref{Mixingclues} and
\ref{mixing}, however, this identification of the single string Hilbert
space with the single trace \opt s, because of the mixing between single trace
and multi trace \opt s, should be modified. Note that this mixing is present
both for chiral primaries and ``almost'' primaries.

\item As is clear from (\ref{BMNlimit2}), in the BMN limit the 't
Hooft coupling goes to infinity and naively any perturbative calculation in the
gauge theory (of course except for chiral-primary two and three
point functions) is not trustworthy. However, the fact that we are
working with ``almost'' BPS \opt s motivates the hope that, although
the anomalous dimensions for such \opt s are non-vanishing,
being close to primary, nearly saturating 
the BPS bound, some of the nice properties of primary \opt s might be inherited by the ``almost'' 
primary \opt s.

\item We will see in section \ref{interactingstrings},
as a result of explicit gauge theory calculations
with the BMN \opt s,
that the 't Hooft coupling in
the BMN sector is dressed with powers of  $1/J^2$. More explicitly, the effective
coupling in the BMN sector is $\lambda'$, rather than the 't Hooft coupling
$\lambda$, where
\be\label{lambda'}
\lambda'\equiv \frac{\lambda}{J^2}=g^2_{YM} \frac{N}{J^2}=(\alpha'\mu p^+)^{-2} \, .
\ee
The last equality is obtained using (\ref{BMNlimit1}) and
(\ref{BMNlimit2}).
                                                                                                         
\item Moreover, we will see that the ratio of non-planar to planar graphs is controlled by powers 
of the genus counting parameter
\be\label{BMNgenuscounting}
g_2\equiv \frac{J^2}{N}=4\pi g_s (\alpha'\mu p^+)^{2} \, ,
\ee
which also remains finite in the BMN limit (\ref{BMNlimit2}). Note that in
(\ref{BMNgenuscounting}) $g_s=e^{\phi}$, where $\phi$ is the value of the dilaton field\footnote{
Note that for the plane-wave background we are interested in, the dilaton is constant.},
is {\it
not} the coupling for strings on the plane-wave, although it is related to it.

\item One can do better than simply finding the free string
mass spectrum; we can study real interacting strings, their
splitting and joining amplitudes and one loop corrections to the mass spectrum.
As we will discuss in sections \ref{interactingstrings} and \ref{SFT}, the
{\it one-loop} mass corrections compared to the tree level results are suppressed by powers of
the ``effective one-loop string coupling'' (\cf\ \eqref{masscorrection-first-order-lambda'g2})
\be\label{stringcoupling}
g^{eff}_{one-loop}=
\sqrt{\lambda' g_2^2}=g_{YM}\frac{J}{\sqrt{N}}=4\pi g_s\
\alpha'\mu p^+ .
\ee
It has been argued that all higher genus (higher loop) results replicate
the same pattern, i.e.
$g_2$ always appears in the combination $\lambda'g_2^2$. This has been built into a quantum
mechanical model for strings on \pl s, the string bit model \cite{Vaman:2002ka, Verlinde:2002ig}.
However, {\it a priori} there is no reason why such a structure should exist and in principle $g_2$ 
and $\lambda'$ can appear in any combination. Using 
another quantum mechanical model constructed to capture some features of the BMN operator dynamics,
it has been argued that at $g_2^4$ level there are indeed $\lambda'g_2^4$ corrections to the mass 
spectrum \cite{Beisert:2002ff, Plefka:2003nb}.

\item The above observations revive the hope that we might be able to do a full-fledged
{\it interacting} string theory computation using {\it perturbative} gauge theory
with (modified) BMN \opt s.

\end{itemize}

We should note that the BMN proposal has, since its inception, undergone many
refinements and corrections. However,
a full and complete understanding of the dictionary of strings on \pl s
and the (modified) BMN \opt s is not yet at our disposal, and the field
is still dynamic.
Some of the open issues will be discussed in the main
text and in particular in section \ref{conclusion}.

Finally, we would like to remind the reader that in this review we have
tried to avoid many detailed and lengthy calculations, specifically
in sections \ref{noninteractingstrings}, \ref{interactingstrings}
and \ref{SFT}. In
fact, we found the original papers on these calculations quite clear and
useful and for more details the reader is encouraged to consult with the
references provided.

%% file: ppgr.tex

Plane-fronted gravitational waves with parallel rays, {\it pp-waves}, are a general class of 
spacetimes and are defined as spacetimes which support a
covariantly constant null Killing vector field $v^\mu$,
\begin{equation}\label{covariant}
\nabla_\mu v_\nu \: = \: 0\ ,\ \ \ \ \ \  v^\mu v_\mu \: = \: 0 \, .
\end{equation}
In the most general form, they have metrics which can be written as
\begin{equation}\label{ppmetric}
ds^2 \: = \:
-2 du dv -F(u,x^I) du^2 + 2A_J(u,x^I) du dx^J +g_{JK}(u,x^I) dx^J dx^K \, ,
\end{equation}
where $g_{JK}(u,x^I)$ is the metric on the space transverse to a pair of
light-cone directions given by $u,v$ and the coefficients
$F(u,x^I)$, $A_J(u,x^I)$ and $g_{JK}(u,x^I)$ are constrained by
(super-)gravity equations of motion.
The pp-wave metric (\ref{ppmetric}) has a null Killing vector
given by $\frac{\partial}{\partial v}$ which is in fact
covariantly constant by virtue of the vanishing of the $\Gamma^v_{v u}$ component of the 
Christoffel symbol.
                                                                                                         
The most useful pp-waves, and the ones generally considered in the
literature, have $A_J = 0$ and are flat in the transverse directions, i.e.
$g_{IJ}= \delta_{IJ}$, for which the metric becomes 
\be\label{pp'metric}
ds^2 \: = \:
-2 du dv -F(u,x^I) du^2 + \delta_{IJ} dx^I dx^J\ .
\ee
As we will discuss in the next subsection, existence of a covariantly constant null Killing vector 
field guarantees the  $\alpha'$-exactness of these supergravity solutions \cite{Horowitz:1990bv}.

A more restricted class of pp-waves, {\it plane-waves}, are those
admitting a {\it globally defined} covariantly constant null Killing
vector field. One can show that for \pl s
$F(u,x^I)$ is quadratic in the $x^I$ coordinates of the transverse space,
but still can depend on the coordinate $u$,
$F(u,x^I) = f_{IJ}(u) x^I x^J$, so that the metric takes the form
\begin{equation} \label{plane-wave-metric}
ds^2 \: = \:
-2 du dv -f_{IJ}(u) x^I x^J du^2 + \delta_{IJ} dx^I dx^J \ .
\end{equation}
Here $f_{IJ}$ is symmetric and by virtue of the only non-trivial
condition coming form the equations of motion, its trace is related to
the other field strengths present.
For the case of vacuum Einstein equations, it is traceless.
                                                                                                         
There is yet a more restricted class of plane-waves, homogeneous plane-waves, for which $f_{IJ}(u)$ 
is a constant, hence their metric is of the
form
\be\label{homometric}
ds^2 \: = \:
-2 du dv -\mu^2_{IJ} x^I x^J du^2 + dx^I dx^I \ ,
\end{equation}
with $\mu^2_{IJ}$ being a constant.\footnote{
This usage of the term homogeneous is not universal. For example,
the term symmetric plane-wave has been used in \cite{Blau:2002js}
for this form of the metric, reserving homogeneous for a wider subclass of plane-waves.}

\subsection{Penrose diagrams for plane-waves}\label{Penrosediagram}

Penrose diagrams \cite{Penrose:1963ij}
are useful tools which capture the 
causal structure of spacetimes. 
The idea is based on the observation that metrics which are conformally 
equivalent share the same light-like geodesics, and hence such spacetimes have the same causal structure (for a more 
detailed discussion see \cite{Townsend:1997ku}). Generically, if there is enough symmetry, it is
possible to bring a given metric into a conformally flat form or to the form conformal to 
Einstein static 
universe (a $d+1$ dimensional cylinder with the metric $ds^2=-dt^2+d\Omega^2_d$), where usually 
coordinates have a finite range.\footnote{ As a famous example in which in the conformally 
Einstein-static-universe coordinate system the range of one of the coordinates is not finite, we 
recall the global cover of \ads\ where global time is ranging over all real numbers 
\cite{Aharony:1999ti}.}
One can then use this coordinate system to draw Penrose diagrams. Therefore, Penrose diagrams are 
constructed so that the light rays always move on $45^\circ$ lines \cite{Misner:1970} 
and in which one can visualize the whole causal structure, singularities, horizons and 
boundaries of spacetimes. 
It may happen that a given  coordinate system does not cover the entire spacetime, or it may be
possible to (analytically) extend them. A Penrose diagram is a very useful way to see if there is a possibility of extending the spacetime and provides us with a specific prescription for
doing so. In the conformally flat coordinates (or coordinate system in which the metric is Einstein static universe up  to a conformal factor) all the information characterizing the spacetime is embedded in the conformal factor. It may happen that, within the range of the coordinates, this conformal factor is regular and finite. In such cases we can simply extend the range of coordinates to the largest possible range.
However, if the conformal factor blows up (or vanishes), we have a boundary or singularity. For 
example in a $d+1$ dimensional flat space, the conformal factor blows up and hence we have a ($d$ dimensional) 
light-like boundary with no possibility for further extension \cite{Misner:1970}.

Let us consider the \pl s and analyze their Penrose diagrams. Since in this review we will only
be interested in a special homogeneous \pl\ of the form \eqref{homometric} with $\mu^2_{IJ}=\mu^2\delta_{IJ},\ 
I,J=1,2,\cdots , d-1$, we will narrow our focus to this special case. 
First we note that with the coordinate transformation \cite{Berenstein:2002sa}
\be
y^0=\frac{1}{2\mu}(1+\mu^2 x^2)\tan\mu u +v\ ,\ \ 
y^d=\frac{1}{2\mu}(1-\mu^2 x^2)\tan\mu u -v\ ,\ \ 
y^I=\frac{x^I}{\cos\mu u}\ ,
\ee
the metric becomes
\be\label{midwaymetric}
ds^2=\frac{1}{1+\mu^2 (y^0+y^D)^2}
(\eta_{\check{\alpha}\check{\beta}}dy^{\check{\alpha}}dy^{\check{\beta}})\ ,
\ee
where $\check{\alpha}, \check{\beta}=0,1,\cdots, d$ and 
$\eta_{\check{\alpha}\check{\beta}}=diag(-,+,\cdots, +)$. 
Note that $y^+\equiv y^0+y^d\in (-\frac{\pi}{2\mu}, \frac{\pi}{2\mu})$. To study the 
possibility of analytic extension of the spacetime, particularly over the
 range of $y^+$, and to draw
the Penrose diagram, we make another coordinate transformation 
\begin{subequations}
\begin{align}
r^2=y_iy_i\ &, \ \ i=1,\cdots , d\ , \ \ \ \  y^d=r\cos\theta \, , \\ 
y^0\pm r &=\frac{1}{\mu}\tan \frac{\psi\pm \xi}{2}\ ,\ \ \ \  \psi, \xi, \theta\in [0,\pi]\ ,
\end{align}
\end{subequations}
in which all the coordinates have a finite range. In these coordinates \eqref{midwaymetric} is 
conformal to Einstein static universe
\be 
ds^2= 
\frac{1}{\mu^2}\frac{1}{(\cos\psi+\cos\xi)^2+ 
(\sin\psi+\sin\xi\cos\theta)^2}[-d\psi^2+d\xi^2+\sin^2\xi^2d\Omega^2_{d-1}] \, .
\ee
To simplify the conformal  factor we perform another coordinate transformation
\[
-\tan\alpha\sin\beta=\cot\theta, \ \ \ \
-\sin\alpha\cos\beta=\cos\xi, \ \ \ \ \alpha\in [0,\frac{\pi}{2}],\ \beta\in[0,2\pi]\ ,
\]
yielding
\be\label{Extendedppwave} 
ds^2= 
\frac{1}{\mu^2}\frac{1}{|e^{i\psi}-\sin\alpha e^{i\beta}|^2}\left(-d\psi^2+d\alpha^2+\sin^2\alpha 
d\beta^2+\cos^2\alpha d\Omega^2_{d-2}\right) \, .
\ee
The metric \eqref{Extendedppwave} is conformal to the Einstein space $-d\psi^2+d\Omega^2_d$ and we have already made all the possible analytic extensions (after relaxing the range of $\psi$).
This manifold, however, does not cover the whole Einstein manifold, because the conformal factor vanishes at (and only at) $\alpha=\frac{\pi}{2},\ \psi=\beta$, and this light-like direction is not part of the analytically continued plane-wave spacetime. Note that at $\alpha=\frac{\pi}{2}$ the $d-2$ sphere shrinks to zero and the metric essentially reduces to the $\psi,\beta$ plane. 

The boundary of a spacetime is a locus which does not belong to the spacetime but is in 
causal contact with the points in the bulk of the spacetime; we can send and receive
light rays from the bulk to the boundary in a finite time. It is straightforward to see that in fact the light-like direction $\alpha=\frac{\pi}{2},\ \psi=\beta$ is the boundary of the analytically continued \pl \ geometry.

\begin{figure}[ht]
\begin{center}
\epsfig{figure=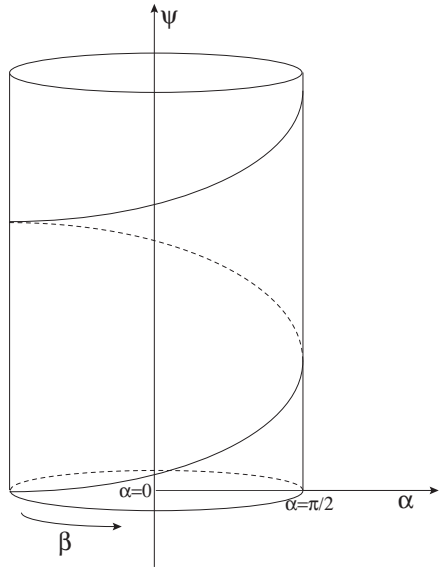, width=60mm,height=80.4mm}
\caption{Penrose diagram of the \pl . The analytically continued \pl\ fills the whole Einstein 
static universe except for the light-like direction $\alpha=\frac{\pi}{2},\ \psi=\beta$, which is its boundary. Note that at each point (except for $\alpha=\frac{\pi}{2}$) there is an $S^{d-2}$ of finite radius, which is suppressed in the figure.}
\label{penrosediagram}
\end{center}
\end{figure}

In sum,  we have shown that the \pl\ has a one dimensional light-like boundary. All the information  
about the (analytically continued) \pl\ geometry has been depicted in the Penrose diagram FIG. 
\ref{penrosediagram}.

In the metric \eqref{Extendedppwave}, we have chosen our coordinate system so that the entire
$\mu$-dependence is gathered in an overall $1/\mu^2$ factor. Although in the original coordinate
system the \pl\ metric has a smooth $\mu\to 0$ limit (which is flat space), for the analytically continued version \eqref{Extendedppwave} this is no longer the case. The reason
is that some of our analytic extensions do not have a smooth $\mu\to 0$ limit.
In particular, note that our arguments about boundary and causal structure of the \pl\ 
cannot be smoothly extended to the $\mu=0$ case.  
Finally we would like to note that the technique detailed here cannot be directly applied to a 
generic plane or pp-wave of the form \eqref{pp'metric}. For these cases one needs to use the method of ``Ideal asymptotic Points'', due to Geroch, Kronheimer and Penrose 
\cite{Geroch1972},
application of which to pp-waves can be found in 
\cite{Hubeny:2002zr, Marolf:2002ye}. 

\subsection{Plane-waves as $\a'$-exact solutions of \sugra}\label{alpha'exact}

In this subsection we discuss a property of pp-waves of the form \eqref{pp'metric} which makes them specially interesting from the string theory point of view: they are $\alpha'$-exact solutions of \sugra\
\cite{Horowitz:1990bv, Amati:1988ww}.
Supergravities arise as low energy effective 
theories of strings, and can receive $\alpha'$-corrections. Such corrections generically involve 
higher powers of curvature and form fields \cite{Green:1987se}. The basic observation made by G. 
Horowitz and A. Steif \cite{Horowitz:1990bv}
is that pp-wave metrics of the form \eqref{pp'metric} have a  covariantly constant null Killing vector, $n_\mu=\frac{\partial}{\partial v}$, and their curvature 
is null (the only non-zero components of their curvature are $R_{uIuJ}$). Higher 
$\alpha'$-corrections to the \sugra\ equations of motion are in general comprised of all
second rank tensors constructed from powers of the Riemann tensor and its derivatives. (The only possible term involving only one Riemann tensor should be of the form 
$R^{\mu\alpha\nu\beta}_{\ \ \ \ \ \ ;\mu\nu}$, which is zero by virtue of the Bianchi identity.) On the other hand, any power of the Riemann tensor and its covariant derivatives with only two free indices is also zero, because $n_\mu$ is null and $\nabla_\mu n_\nu=0$ \eqref{covariant}.
The same argument can be repeated for the form fields, noting that for pp-waves which are solutions of \sugra\ these form fields should have zero divergence and be null. As a result,
all the $\alpha'$-corrections for \sugra\ solutions with metric of the form \eqref{pp'metric} 
vanish, i.e. they also solve $\alpha'$-corrected \sugra \ equations of motion. 
This argument about $\alpha'$-exactness does not hold for a generic pp-wave of the form 
\eqref{ppmetric} with $g_{IJ}(u,x^I) \neq \delta_{IJ}$. The transverse metric, $g_{IJ}$,
may itself receive $\alpha'$-corrections, however, there are no extra
corrections due to the wave part of the metric \cite{Fabinger:2003jn}.
We would like to comment that pp-waves are generically singular solutions with no (event) 
horizons \cite{Hubeny:2002pj},
however, \pl s of the form \eqref{plane-wave-metric} for which  
$f_{IJ}(u)$ is a smooth function of $u$, are not singular.

\subsection{Maximally supersymmetric plane-wave and its symmetries}\label{maxsusy}

Hereafter in this review we shall only focus on  a very special \pl\ solution of ten dimensional type IIB supergravity which admits 32 supersymmetries and by ``the \pl '' we will mean this maximally \susyc\ solution. In fact, demanding a solution of 10 or 
eleven dimensional supergravity to be 
maximally \susyc\  is very restrictive; flat space, \ads and a special \pl\ in type IIB theory in 
ten dimensions and flat space, $AdS_{4,7}\times S^{7,4}$ and a special \pl\ in eleven dimensions 
are the only possibilities \cite{Figueroa-O'Farrill:2002ft}. Note that type IIA does 
not admit any maximally \susyc\ solutions other than flat space.

Here, we only focus on the ten dimensional \pl\ which is a special case of \eqref{homometric} 
with $\mu^2_{IJ}=\mu^2\delta_{IJ}$. This metric, however, is not a solution to source-free type IIB \sugra\ equations of motion and we need to add form fluxes. It is not hard to see that with 
$\mu^2_{IJ}=\mu^2\delta_{IJ}$ the only possibility is turning on  a constant self-dual RR five-form flux; moreover, the dilaton should also be  a constant. As we will see in section \ref{penroselimit}, this 
\pl\ is closely related to the \ads solution. The (bosonic) part of this \pl\ solution is then
\begin{subequations}\label{planewavemetric}
\begin{align}
ds^2  =  -2 dx^+ dx^- -\mu^2(x^i x^i +& x^a x^a) {(dx^+)}^2 + dx^i dx^i+ dx^a dx^a, \\
F_{+ijkl} = \frac{4}{g_s} \mu\ \epsilon_{ijkl}\ &,\ \ \ \ \ \  F_{+abcd}=\frac{4}{g_s} \mu \ 
\epsilon_{abcd}\ ,\\
e^{\phi} = g_s={\rm constant}\ ,  \ \ i,j&=1,2,3,4\ ,\ \ a,b=5,6,7,8\ ,\ \ 
\end{align}
\end{subequations}
In the above, $\mu$ is an auxiliary but convenient parameter, and can be easily 
removed by taking $x^+\to x^+/\mu$ and $x^-\to \mu x^-$ (which is in fact a light-cone boost).

Let us first check that the \bg\ \eqref{planewavemetric} is really  maximally supersymmetric. Note that this will ensure it is also a \sugra\ solution, because \sugra\ equations of motion are nothing but the commutators of the \susy\ variations.
For this we need to show that the gravitino and dilatino variations vanish for 32 independent 
(Killing) spinors, i.e.
\be\label{killingspinor}
\delta_{\epsilon}\psi^\alpha_\mu\equiv (\hat\cd_\mu)^\alpha_{\ \beta}\ \epsilon^\beta=0\ ,\ \ \ \delta_{\epsilon}\lambda^\alpha\equiv (\tilde\cd)^\alpha_{\ \beta}\ \epsilon^\beta=0\ ,\ 
\mu=0,1,\cdots 
,9,  
\  \alpha=1,2\ ,
\ee
have 32 solutions, where the dilatino $\lambda^\alpha$, gravitinos $\psi_\mu^\alpha$ and 
Killing spinors $\epsilon^\alpha$ are all 32 component ten dimensional Weyl-Majorana fermions of the same chirality (for our notations and conventions see Appendix \ref{SO(8)fermions}), and
the supercovariant derivative $\hat\cd_\mu$ {\it in string frame} is defined as (see for example
\cite{Cvetic:2002nh, Bena:2002kq, Sadri:2003ib})
\be\label{covderivative}
(\hat{\cd}_{\mu})^{\alpha}_{\ \beta}=
\delta^{\alpha}_{\ \beta}\ \nabla_\mu
+\frac{1}{8}(\sigma^3)^{\alpha}_{\ \beta}\  \Gamma^{\nu\rho} H_{\mu \nu\rho}
+\frac{ie^\phi}{8}\left[(\sigma^2)^{\alpha}_{\ \beta}\ 
\Gamma^{\nu}\partial_\nu \chi-\frac{i}{3!}(\sigma^1)^{\alpha}_{\ \beta}\ 
\Gamma^{\nu\rho\lambda}F_{\nu\rho\lambda}+\frac{1}{2\cdot 5!}
(\sigma^2)^{\alpha}_{\ \beta}\ \Gamma^{\nu\rho\lambda\sigma\delta}F_{\nu\rho\lambda\sigma\delta}
\right] \Gamma_\mu ,
\ee
\be
(\tilde\cd)^\alpha_{\ \beta}=\frac{1}{2}\delta^{\alpha}_{\ \beta}\ 
\Gamma^{\nu}\partial_\nu\phi-\frac{1}{4\cdot 3!}(\sigma^3)^{\alpha}_{\ \beta}\  \Gamma^{\mu\nu\rho} 
H_{\mu 
\nu\rho}-\frac{i}{2}e^{\phi}\left[(\sigma^2)^{\alpha}_{\ \beta}\ \Gamma^{\nu}\partial_\nu \chi
-\frac{i}{2\cdot 3!} 
(\sigma^1)^{\alpha}_{\ \beta}\ \Gamma^{\nu\rho\lambda}F_{\nu\rho\lambda}\right]\ , 
\ee
with the spin connection $\omega_\mu^{\hat{a}\hat{b}}$ appearing in the covariant derivative
$\nabla_\mu=(\partial_{\mu}+\frac{1}{4} \omega_\mu^{\hat{a}\hat{b}}\Gamma_{\hat{a}\hat{b}})$ and the hatted Latin indices used for the tangent space.
In these expressions $\phi$ is the dilaton, $\chi$ the axion, $H$ the three-form
field strength of the NSNS sector, and the $F$'s represent the appropriate RR
field strengths.

For the \bg\ \eqref{planewavemetric} 
$(\tilde\cd)^\alpha_{\ \beta}$ is identically zero and  
$(\hat\cd)^\alpha_{\ \beta}$ take a simple form
\be\label{plsupercovariant}
(\hat{\cd}_{\mu})^{\alpha}_{\ \beta} =\delta^{\alpha}_{\ \beta}\ (\partial_\mu+\frac{1}{4} 
\omega_\mu^{ab}\Gamma_{ab})+\frac{ig_s}{16\cdot 5!}(\sigma^2)^{\alpha}_{\ \beta}\ 
\Gamma^{\nu\rho\lambda\sigma\delta}F_{\nu\rho\lambda\sigma\delta}\ .
\ee
In order to work out the spin connection $\omega_\mu^{\hat{a}\hat{b}}$ we need the vierbeins 
$e_\mu^{\hat 
a}$ which are
\be\label{vierbein}
e_+^{\ +}=e_-^{\ -}=1\ ,\ e_i^{\ j}=\delta_i^{\ j}\ , e_{{a}}^{\ {b}}=\delta_a^{\ b}\ ,\ 
e_+^{\ -}=\frac{1}{2}\mu^2(x_ix_i+x_ax_a)\ ,
\ee   
and therefore
\be\label{spinconnection}
\omega_+^{\ -i}=\mu^2x_i\ ,\ 
\omega_+^{\ -a}=\mu^2x_a\ ,\ 
\ee
are the only non-vanishing components of $\omega_\mu^{\hat{a}\hat{b}}$. 

The Killing spinor equation can now be written as
\be\label{killing'}
({\bf 1}\cdot \partial_{\mu}+\Omega_{\mu})^{\alpha}_{\ \beta}\ \epsilon^\beta=0 \, ,
\ee
with
\be\label{Omega}
\Omega_-=0 \, , \ \ 
(\Omega_I)^{\alpha}_{\ \beta}=\frac{i\mu}{4} \Gamma^+ (\Pi +\Pi') \Gamma^I (\sigma^2)^{\alpha}_{\ 
\beta} \, , \ \ 
(\Omega_+)^{\alpha}_{\ \beta}=-\frac{1}{2}\mu^2 x^I\Gamma^{+I}\ \delta^{\alpha}_{\ \beta}
+\frac{i\mu}{4} (\Pi +\Pi') \Gamma^+\Gamma_+\ (\sigma^2)^{\alpha}_{\ \beta} \, .
\ee
In the above $I=\{i, a\}=1,2,\cdots 8,\ \Pi=\Gamma^{1234}$ and $\Pi'=\Gamma^{5678}$. 
The $\Omega$'s satisfy a number of useful identities such as
\bea\label{Omegaproperties}
\Gamma^+\Omega_I=\Omega_I\Gamma^+=\Gamma^+\Omega_+=0\ &,&\ \ \Omega_I\Omega_J=\Omega_I\Omega_+=0\ ,
\cr
\Omega_+\Omega_I=-\frac{\mu^2}{4} (1+\Pi\Pi')\Gamma^{+I}\cdot {\bf 1}\ &,&\ \ 
\Omega_+\Gamma^+=\frac{i\mu}{2} (\Pi+\Pi')\Gamma^{+}\cdot {(\sigma^2)^{\alpha}_{\ \beta}}\ .
\eea
We first note that the ($\mu=-$) component of \eqref{killing'} is simply $\partial_-\epsilon=0$, so all Killing spinors should be $x^-$-independent. The $\mu=I$ component can be easily solved by taking
\be\label{Ikilling}
\epsilon^{\alpha}=({\bf 1}-x^I{\Omega_I})^{\alpha}_{\ \beta}\chi^\beta \, ,
\ee
where $\chi^\beta$ is an arbitrary $x^I$-independent fermion of positive ten dimensional chirality. 
Plugging  \eqref{Ikilling} into \eqref{killing'}, using the identity $\Omega_I\Omega_+=0$ and the fact that $({\bf 1}+x^I{\Omega_I})({\bf 1}-x^I{\Omega_I})={\bf 1}$ the ($\mu=+$) component of the Killing spinor equation takes the form
\bea\label{Omega+}
\left({\bf 1}\cdot \partial_{+}+\Omega_+({\bf 1}-x^I\Omega_I)\right)^{\alpha}_{\ 
\beta}\chi^\beta=0\ . 
\eea
Equation \eqref{Omega+} has an $x^I$ independent piece and a part 
which is linear in $x^I$. These two should vanish separately. 
Using the identities given in Appendix \ref{SO(8)fermions} and after some straightforward
Dirac matrix algebra, one can show that if $\Gamma^-\chi=0$, \eqref{Omega+} simply reduces to 
$\partial_+ \chi=0$. That is, any constant $\chi$ with $\Gamma^-\chi=0$ is a Killing spinor.
These provide us with $2\times 8=16$ solutions. Now let us assume that $\Gamma^-\chi\neq 0$. 
Without loss of generality, all such spinors can be chosen to satisfy $\Gamma^+\chi=0$. For these choices of $\chi$'s the $x^I$-dependent part of \eqref{Omega+} vanishes identically and the $x^I$-independent part becomes
\[
\left({\bf 1}\cdot \partial_{+}+{i\mu}  \Pi\ (\sigma^2)^{\alpha}_{\ \beta}\right)\chi^\beta=0
\, ,
\]
where we have used the fact that $\Gamma^+\chi=0$ implies $\Pi\chi=\Pi'\chi$.
This equation can be easily solved with \cite{Blau:2001ne}
\be\label{killingsolution}
\chi^\alpha=\left( \delta^{\alpha}_{\ \beta}\ \cos {\mu x^+}-i\Pi\ (\sigma^2)^{\alpha}_{\ 
\beta}\ \sin{\mu x^+}\right)
\chi_0^\beta\ ,
\ee
where  
$\chi_0^\beta$ is an arbitrary constant spinor of positive ten dimensional chirality. 
We have shown that equations \eqref{killingspinor} have 32 linearly independent 
solutions and hence the \bg\ \eqref{planewavemetric} is maximally supersymmetric. 
We would like to note that \eqref{killingsolution} clearly shows the ``wave'' nature of our \bg\ (note the periodicity in $x^+$, the light-cone time), a fact which is not manifest in the coordinates we have chosen. This wave nature can be made explicit in the so-called Rosen coordinates (\cf\ section 
\ref{penroseguven}).

\subsubsection{Isometries of the \bg }\label{isometry}

The \bg\ \eqref{planewavemetric} has a number of isometries, some of which are manifest. In 
particular, the solution is invariant under translations in the $x^+$ and $x^-$ directions. These translations can be thought of as two (non-compact) $U(1)$'s with the generators
\be \label{U(1)s}
i\frac{\partial }{\partial x^+ } \equiv P_+=-P^- \ , \ \ \ 
i\frac{\partial }{\partial x^- } \equiv P_-=-P^+ \ .
\ee
Due to the presence of the $(dx^+)^2$ term, a boost in the ($x^+, x^-$) plane is not a 
symmetry of the metric. However, the combined boost and $\mu$ scaling:
\be\label{lightconeboost}
x^-\to \sqrt{\frac{1-v}{1+v}}x^-\ ,\ \ 
x^+\to \sqrt{\frac{1+v}{1-v}}x^+\ ,\ \ 
\mu\to \sqrt{\frac{1-v}{1+v}}\mu\ ,
\ee
is still a symmetry.

Obviously, the solution is also invariant under two $SO(4)$'s which act on the $x^i$ and $x^a$ 
directions. The generators of these $SO(4)$'s will be denoted by $J_{ij}$ and  $J_{ab}$ where
\be\label{JijJab}
J_{ij}=-i(x_i\frac{\partial}{\partial x^j}-x_j\frac{\partial}{\partial x^i})\ ,\ \ \
J_{ab}=-i(x_a\frac{\partial}{\partial x^b}-x_b\frac{\partial}{\partial x^a})\ .
\ee
Note that although the metric possesses $SO(8)$ symmetry, because of the five-form flux this symmetry is broken to $SO(4)\times SO(4)$. 
There is also a $\mathbb{Z}_2$ symmetry which exchanges these two $SO(4)$'s, acting as
\be\label{Z2}
\{x^i\} {\overset{\ztwo}{\llra}} \{x^a\}\ .
\ee

So far we have identified 14 isometries which are generators of a $U(1)\times U(1)\times SO(4)\times SO(4)\rtimes \ztwo$ symmetry group. One can easily see that translations along the $x^I=(x^i,\ x^a)$ directions are not symmetries of the metric.  However, we can show that if along with translation in $x^I$ we also shift $x^-$ appropriately, i.e.
\be
\left\{\begin{array}{cc}
x^I\to x^I+\epsilon_1^I\ \cos\mu x^+\cr
x^-\to x^--\epsilon_1^I\ \mu x^I\sin\mu x^+
\end{array}
\right. 
\ ,\ \ \ \ \  
\left\{\begin{array}{cc}
x^I\to x^I+\epsilon_2^I\ \sin\mu x^+\cr
x^-\to x^-+\epsilon_2^I\ \mu x^I\cos\mu x^+
\end{array} 
\right.
\ ,
\ee
where $\epsilon_1^I$ and $\epsilon_2^I$ are arbitrary but small parameters, the metric and the five-form remain unchanged. These 16 isometries are generated by the Killing vectors
\be\label{KILI}
L_I=-i\left(\cos\mu x^+\frac{\partial}{\partial x^I}-\mu x^I\ \sin\mu x^+
\frac{\partial}{\partial x^-}\right)\ ,\ \ \
K_I=-i\left(\sin\mu x^+\frac{\partial}{\partial x^I}+\mu x^I\ \cos\mu x^+
\frac{\partial}{\partial x^-}\right)\ ,
\ee
satisfying the following algebra
\be\label{Heisenberg}
[L_I, L_J]=0\ ,\ \ \ \ [L_I,K_J]=\mu \delta_{IJ}\frac{\partial}{\partial x^-}= i\mu P^+\ 
\delta_{IJ}\ ,\ \ \ \ [K_I, K_J]=0\ , 
\ee
\be\label{HKL}
[P^-, L_I]=i\mu\ K_I\ ,\ \ \ \
[P^-, K_I]=-i\mu\ L_I\ .
\ee
Equations \eqref{Heisenberg} are in fact an eight (or a pair of four) dimensional Heisenberg-type algebra(s) with ``$\hbar$'' being equal to $\mu P^+$ \cite{Das:2002cw}. 
Note that $P^+$ commutes with the generators of the two $SO(4)$'s as well as $K_I$ and $L_I$. In other words $P^+$ is in the center of the isometry algebra which has 30 generators $(J_{ij},\ J_{ab},\ P^+,\ P^-,\ K_i,\ K_a,\ L_i,\ L_a )$. It is also easy to check that $K_i,\ L_j$ and $K_a,\ L_b$ transform as vectors (or singlets) under the corresponding $SO(4)$ rotations. Altogether, the algebra of Killing vectors is
$[h(4)\oplus h(4)]\oplus so(4)\oplus so(4)\oplus u(1)_+\oplus u(1)_- $, where $h(4)$ is the four dimensional Heisenberg algebra.

In addition to the above 30 Killing vectors generating continuous symmetries, there are some discrete symmetries, one of which is the $\ztwo$ discussed earlier. There is also the CPT symmetry \cite{Schwarz:2002bc}
\be\label{timereversal}
x^I\to - x^I\ ,\ \ \ \ \ x^\pm\to\ -x^\pm\ ,\ \ \ \ \ \ \mu\to\ -\mu \ ,
\ee    
(note that we also need to change $\mu$). 

Finally, it is useful to compare the \pl\ isometries to that of flat space, the ten dimensional 
Poincare algebra consisting of $P^+,\ P^-,\ P^I=-i\frac{\partial}{\partial x^I}$ and
$J^{+-}$ (light-cone boost), $J^{+I},\ J^{-I}$ and $J^{IJ}$ (the \soe rotations).
Among these 55 generators, $P^+,\ P^- $ and $J^{ij},\ J^{ab}$ are also present in the set of \pl\ isometries. However, as we have discussed, $J^{+-}$ and $J^{-I}$ are absent. As for 
rotations generated by $J^{ia}$, only a particular rotation, namely the $\ztwo$ defined in 
\eqref{Z2}, is present. From \eqref{KILI} it is readily seen that $K_I$ and $L_I$ are a linear combination of $P_I$ and $J^{+I}$,
\be\label{PIJ+I}
P_I=-i\frac{\partial}{\partial x^I}\ ,\ \ \ 
J^{+I}=x^+P^I -x^IP^+ \ ,
\ee
and it is easy to show that
$[P^-, J^{+I}]=-i P^I\ ,\ [P^I, J^{+J}]=-i \delta_{IJ} P^+$.
In summary, $J^{+-},\ J^{-I}$ and $J^{ia}$ (which are altogether 25 generators) are not present 
among the Killing vectors of the \pl\ and therefore the number of isometries of the \pl\ is $55-25=30$, agreeing with our earlier results.

\subsubsection{Superalgebra of the \bg} \label{planewavesusy}

As we have shown, the \pl\ \bg\ \eqref{planewavemetric} possesses 32 Killing spinors and
in section \ref{isometry} we worked out all the isometries of the \bg . In this subsection we combine 
these two results and present the superalgebra of the \pl\ geometry \eqref{planewavemetric}. Noting the Killing spinor equations and its solutions it is straightforward to work out the  
supercharges and their superalgebra (e.g. see \cite{Green:1987se}). 

As discussed earlier, the solutions to the Killing spinor equations are all $x^-$-independent. This implies that supercharges should commute with $P^+$. However, as discussed in \ref{isometry}, $P^+$ commutes with all the bosonic isometries, and so is in the center of the whole superalgebra. Then, we noted that Killing spinors fall into two 
classes, either $\Gamma^+\chi=0$ or $\Gamma^-\chi=0$. The former lead to {\it kinematical} 
supercharges,  $Q^{+\alpha}$ with the property that $\Gamma^+Q^{+\alpha}=0$, while the latter lead 
to {\it dynamical} supercharges $Q^{-\alpha}$, which satisfy $\Gamma^-Q^{-\alpha}=0$. Since both 
sets of dynamical and kinematical supercharges have the same (positive) ten dimensional chirality, 
the $Q^{+\alpha}$ are in the \eights\ 
and $Q^{-\alpha}$ in the \eightc\ \rep \ of the \soe fermions (for details of the conventions see 
Appendix \ref{SO(8)fermions}).

For the \pl\ \bg , however, it is more convenient  to use the \soff decomposition instead of $SO(8)$. The relation between these two has been worked out and summarized in Appendix \ref{SO(4)fermions}. 
We will use 
$q_{\alpha\beta}$ and $q_{\dot\alpha\dot\beta}$ for the kinematical supercharges and
$Q_{\dot\alpha\beta}$ and $Q_{\alpha\dot\beta}$ for the dynamical ones. Note that all $q$ and
$Q$ are complex fermions.

The superalgebra in the \soe basis can be found in \cite{Blau:2002dy, Metsaev:2001bj}.
Here we present it in the \soff basis:

\begin{itemize}

\item {\it Commutators of bosonic generators with kinematical supercharges:}

\end{itemize}
\bea\label{Jq}
[J^{ij}, q_{\alpha\beta}]=\frac{1}{2}\ (i\sigma^{ij})_\alpha^{\ \rho}\ q_{\rho\beta}\ &,& \ \ 
[J^{ij}, q_{\dot\alpha\dot\beta}]=\frac{1}{2}\ (i\sigma^{ij})_{\dot\alpha}^{\ \dot\rho}\ 
q_{\dot\rho\dot\beta}\ , \cr 
[J^{ab}, q_{\alpha\beta}]=\frac{1}{2}\ (i\sigma^{ab})_\beta^{\ \rho}\ q_{\alpha\rho}\ &,& \ \ 
[J^{ab}, q_{\dot\alpha\dot\beta}]=\frac{1}{2}\ (i\sigma^{ab})_{\dot\beta}^{\ \dot\rho}\ 
q_{\dot\alpha\dot\rho}\ , \
\eea
\be\label{KLq}
[K^{I}, q_{\alpha\beta}]=[L^{I}, q_{\alpha\beta}]=0\ ,\ \ 
[K^{I}, q_{\dot\alpha\dot\beta}]=[L^{I}, q_{\dot\alpha\dot\beta}]=0\ ,\ \ 
\ee
\be\label{qcenter}
[P^{+}, q_{\alpha\beta}]=[P^{+}, q_{\dot\alpha\dot\beta}]=0\ ,\ \ 
\ee
\be\label{Hq}
[P^{-}, q_{\alpha\beta}]=+i\mu q_{\alpha\beta}\ ,\ \ 
[P^{-}, q_{\dot\alpha\dot\beta}]=-i\mu q_{\dot\alpha\dot\beta}\ .
\ee 
\begin{itemize}

\item {\it Commutators of bosonic generators with dynamical supercharges:}

\end{itemize}
\bea\label{JQ}
[J^{ij}, Q_{\alpha\dot\beta}]=\frac{1}{2}\ (i\sigma^{ij})_\alpha^{\ \rho}\ Q_{\rho\dot\beta}\ &,& \ 
\ 
[J^{ij}, Q_{\dot\alpha\beta}]=\frac{1}{2}\ (i\sigma^{ij})_{\dot\alpha}^{\ \dot\rho}\ 
Q_{\dot\rho\beta}\ , \cr 
[J^{ab}, Q_{\dot\alpha\beta}]=\frac{1}{2}\ (i\sigma^{ab})_\beta^{\ \rho}\ Q_{\dot\alpha\rho}\ &,& \ 
\ 
[J^{ab}, Q_{\alpha\dot\beta}]=\frac{1}{2}\ (i\sigma^{ab})_{\dot\beta}^{\ \dot\rho}\ 
Q_{\alpha\dot\rho}\ , \
\eea
\bea\label{KQ}
[K^{i}, Q_{\alpha\dot\beta}]=\frac{\mu}{2}\ (\sigma^i)_{\alpha}^{\ \dot\rho}q_{\dot\rho 
\dot\beta}
&, &\ \ \
[K^{a}, Q_{\alpha\dot\beta}]=-\frac{\mu}{2}\ (\sigma^a)_{\dot\beta}^{\ \rho}q_{\alpha\rho}\ ,\cr
[K^{i}, Q_{\dot\alpha\beta}]=\frac{\mu}{2}\ (\sigma^i)_{\dot\alpha}^{\ \rho}q_{\rho \beta}
\ &, &\ \ \
[K^{a}, Q_{\dot\alpha\beta}]=\frac{\mu}{2}\ (\sigma^a)_{\beta}^{\ \dot\rho}q_{\dot\alpha\dot 
\rho}\ ,
\eea
\bea\label{LQ}
[L^{i}, Q_{\alpha\dot\beta}]=-\frac{\mu}{2}\ (\sigma^i)_{\alpha}^{\ \dot\rho}q_{\dot\rho 
\dot\beta}\ 
&,& \ \ 
[L^{a}, Q_{\alpha\dot\beta}]=\frac{\mu}{2}\ (\sigma^a)_{\dot\beta}^{\ \rho}q_{\alpha\rho} ,
\cr 
[L^{i}, Q_{\dot\alpha\beta}]=\frac{\mu}{2}\ (\sigma^i)_{\dot\alpha}^{\ \rho}q_{\rho \beta}\ 
&,& \ \ 
[L^{a}, Q_{\dot\alpha\beta}]=\frac{\mu}{2}\ (\sigma^a)_{\beta}^{\ \dot\rho}q_{\dot\alpha\dot 
\rho} 
,
\eea
\be\label{Qcenter}
[P^{+}, Q_{\alpha\dot\beta}]=0\ ,\ \ [P^{+}, Q_{\dot\alpha\beta}]=0\ ,
\ee
\be\label{HQ}
[P^{-}, Q_{\alpha\dot\beta}]=0\ ,\ \ 
[P^{-}, Q_{\dot\alpha\beta}]=0 \ .
\ee 
\begin{itemize}

\item {\it Anticommutators of supercharges:}

\end{itemize}
\be\label{qq}
\{q_{\alpha \beta},q^{\dagger\rho \lambda}\}=2P^+\delta_{\alpha}^{\ \rho}
\delta_{\beta}^{\ \lambda}\  ,
\ \ \ 
\{q_{\alpha \beta},q^{\dagger\dot\alpha \dot\beta}\}=0\ ,\ \ 
\{q_{\dot\alpha \dot\beta},q^{\dagger\dot\rho \dot\lambda}\}=2P^+\delta_{\dot\alpha}^{\ 
\dot\rho}\delta_{\dot\beta}^{\ \dot\lambda}\  ,
\ee
\bea\label{qQ}
\{q_{\alpha \beta},Q^{\dagger\dot\rho \lambda}\}=i (\sigma^i)_{\alpha}^{\ \dot\rho}
\delta_{\beta}^{\ \lambda} (L^i+ K^i)\ &,& \ \ \ 
\{q_{\alpha \beta},Q^{\dagger\rho \dot\lambda}\}=i (\sigma^a)_{\beta}^{\ \dot\lambda}
\delta_{\alpha}^{\ \rho} (L^a+ K^a)\ ,
\cr
\{q_{\dot\alpha \dot\beta},Q^{\dagger\dot\rho \lambda}\}=i (\sigma^a)_{\dot\beta}^{\ \lambda}
\delta_{\dot\alpha}^{\ \dot\rho} (L^a- K^a)\ &,& \ \ \ 
\{q_{\dot\alpha \dot\beta},Q^{\dagger\rho \dot\lambda}\}=i (\sigma^i)_{\dot\alpha}^{\ \rho}
\delta_{\dot\beta}^{\ \dot\lambda} (L^i- K^i)\ ,
\eea
\bea\label{QQ}
\{Q_{\alpha \dot\beta},Q^{\dagger\rho \dot\lambda}\}&=&2\ \delta_{\alpha}^{\ \rho} 
\delta_{\dot\beta}^{\ \dot\lambda}\ P^- +
\mu (i\sigma^{ij})_{\alpha}^{\ \rho} \delta_{\dot\beta}^{\ \dot\lambda}\ J^{ij} +
\mu (i\sigma^{ab})_{\dot\beta}^{\ \dot\lambda}\delta_{\alpha}^{\ \rho} J^{ab} \ ,
\cr
\{Q_{\alpha \dot\beta},Q^{\dagger\dot\rho \lambda}\}&=& 0 \ , \\
\{Q_{\dot\alpha \beta},Q^{\dagger\dot\rho \lambda}\}&=&2\ \delta_{\dot\alpha}^{\ \dot\rho} 
\delta_{\beta}^{\ \lambda}\ P^- +\mu 
(i\sigma^{ij})_{\dot\alpha}^{\ \dot\rho} \delta_{\beta}^{\ \lambda}\ J^{ij} + \mu
(i\sigma^{ab})_{\beta}^{\ \lambda}\delta_{\dot\alpha}^{\ \dot\rho} J^{ab} \ . \nonumber
\eea

Let us now focus on the part of the superalgebra containing only dynamical supercharges and 
$SO(4)$ generators, i.e. equations \eqref{JQ}, \eqref{Qcenter}, \eqref{HQ} and \eqref{QQ}.
Adding the two $so(4)$ algebras to these, we obtain a superalgebra, which is of course a 
subalgebra of the full superalgebra discussed above. (We have another sub-superalgebra which only contains kinematical supercharges, $P^\pm$ and $J$'s, but we do not consider it here.)
The bosonic part of this sub-superalgebra is 
$U(1)_+\times U(1)_-\times SO(4)\times SO(4)\rtimes \ztwo$, where $U(1)_\pm$ is generated by 
$P^{\pm}$ and $U(1)_+$ is in the center of the algebra. 
Next we note that the algebra does not mix $Q_{\alpha\dot \beta}$ and $Q_{\dot\alpha\beta}$. This sub-superalgebra is not a simple superalgebra and it can be written as a (semi)-direct product of two simple superalgebras. Noting that $Spin(4)= SU(2)\times SU(2)$, we have four $SU(2)$ factors and $Q_{\alpha\dot \beta}$ 
and $Q_{\dot\alpha\beta}$ transform as doublets of two of the $SU(2)$'s, {\it each coming from 
different $SO(4)$ factors}. In other words the two $SO(4)$'s mix to give two $SU(2)\times SU(2)$'s.
This superalgebra falls into Kac's classification of superalgebras \cite{Kac:1977em}
and can be identified 
as $PSU(2|2)\times PSU(2|2)\times U(1)_- \times U(1)_+$. (The bosonic part of the $PSU(n|n)$ 
supergroup is $SU(n)\times SU(n)$ while that of $SU(m|n)$ for $m\neq n$ is $SU(n)\times SU(m)\times U(1)$.)
As mentioned earlier the two $PSU(2|2)$ supergroups
share the same $U(1)$, $U(1)_-$, which is generated by $P^-$. The $\ztwo$ symmetry defined 
through \eqref{Z2} is still present and at the level of superalgebra exchanges the two $PSU(2|2)$ factors.
It is interesting to compare the ten dimensional maximally \susyc\ \pl\ superalgebra with that of the eleven dimensional one which is $SU(4|2)$ \cite{Dasgupta:2002ru}.
One of the main differences is that in our case the light-cone Hamiltonian, $P^-$ commutes with the supercharges ({\it cf.} \eqref{HQ}), and as a result, as opposed to the eleven dimensional case, all states in the same $PSU(2|2)\times PSU(2|2)\times U(1)_-$ supermultiplet have the same mass. Here we do not intend to study this superalgebra and its \rep s in detail, however, this is definitely an important question which so far has not been addressed in the literature. For a more detailed discussion on $SU(m|n)$ supergroups and their unitary \rep s the reader is encouraged to look at \cite{BahaBalantekin:1981pp, 
Dasgupta:2002ru, Motl:2003rw} and for the $PSU(n|n)$ case \cite{Berkovits:1999im}.

%


\subsection{Spectrum of \sugra\  on the  \pl\ \bg}
\label{sugraspectrum}

The low energy dynamics of string theory can be understood in terms of an effective
field theory in the form of supergravity \cite{Green:1987se}. In particular, the lowest
lying states of string theory on the maximally supersymmetric \pl\ background 
\eqref{planewavemetric} should correspond to the states of (type IIB) supergravity on  this 
background. We are thus led to analyze the spectrum of modes in such a theory. 

As in the flat space, the kinematical supercharges acting on different states would generate 
different ``polarizations'' of the same state, while dynamical supercharges would lead to various 
fields in the same supermultiplet. In the \pl\ superalgebra we discussed in the previous section, 
as it is seen from \eqref{Hq} kinematical supercharges do {\it not} commute with the light-cone 
Hamiltonian and hence we expect different ``polarizations'' of the same multiplet to have different 
masses (their masses, however, should differ by an integer multiple of $\mu$), the fact that will
be explicitly shown in this section. This should be
contrasted with the flat space case, where light-cone Hamiltonian commutes with all supercharges,
kinematical and dynamical. For the same reason different states in the 
Clifford vacuum (which are related by the action of kinematical supercharges) will carry different 
energies, and so this vacuum is non-degenerate. However, chosen a Clifford vacuum, the other states 
of the same multiplet are related by the action of dynamical supercharges and hence
should have the same (light-cone) mass (\cf\ \eqref{HQ}).


As a warm up, let us first consider a scalar (or any bosonic) field $\phi$ with mass $m$ 
propagating on such a background, with classical equation of motion
\be \label{massive-scalar-eom}
  \left( \Box - m^2 \right) \: \phi \: = \: 0 \, ,
\ee
with the d'Alembertian acting on a scalar given as
\be \label{dalembertian}
  \Box \: = \: \frac{1}{\sqrt{|g|}} \partial_\mu
  \left( \sqrt{|g|} g^{\mu \nu} \partial_\nu \right) \: = \:
  -2 \partial_+ \partial_- + \mu^2 x^I x^I \partial_- \partial_-
  + \partial_I \partial_I \, .
\ee
Here, the index $I$ corresponds to the eight transverse directions, and the repeated indices
are summed. Then, \eqref{massive-scalar-eom} for the fields with 
$
\partial_{\pm} \phi=ip^{\mp}\phi
$
reduces to 
\be\label{Harmonic-Oscil-wave-equation}
  \left(2 p^{+}p^{-} - (\mu p^+)^2 x^I x^I  + \partial_I \partial_I\right)\phi=0\, ,
\ee
which is nothing but a Schrodinger equation for an eight dimensional harmonic oscillator, with 
frequency equal to $\mu p^+$. Therefore,  choosing the $x^I$ dependence of $\phi$ through Gaussians
times Hermite polynomials (the precise form 
of which can be found in \cite{Bak:2002ku})  the spectrum of the light-cone 
Hamiltonian, $p^-$, is obtained to be 
as
\be \label{sugra-lc-energy-spectrum}
  p^- \: = \: \mu \left( \sum_{i=1}^8 n_i + 4 \right) + \frac{m^2}{2 p^+}
\ee
for some set of positive or zero integers ${n_i}$.
The spectrum is discrete for massless fields ($m=0$), in which case it is also independent of the 
light-cone
momentum $p^+$. This means that for massless fields, we can not form wave-packets
with non-zero group velocity ($\sim \partial p^-/\partial p^+$), and hence scattering 
of such massless states can not take place.\footnote{The particles are confined in the transverse
space by virtue of the harmonic oscillator potential, but one can consider scattering
in a two dimensional effective theory on the $(p^-,p^+)$ subspace \cite{Bak:2002ku}.}
The discreteness of the spectrum arises from the requirement that the
wave-function be normalizable in the transverse directions, and this is translated through
a coupling of the transverse and light-cone directions in the equation of motion into
the discreteness of the light-cone energy. The flat space limit ($\mu\to 0$) is not well-defined 
for
these modes, but could be restored if we add the non-normalizable solutions to the
equation of motion, in which case the flat space limit would allow a continuum
of light-cone energies. The case of vanishing light-cone momentum is not easily
treated in light-cone frame.
For the massive case ($m\neq 0$) the light-cone energy does pick up a $p^+$ dependence allowing us
to construct proper wave-packets for scattering.


These considerations can be applied to the various bosonic fields in supergravity.
The low energy effective theory relevant here is type IIB supergravity, whose
action (in string frame) is \cite{Polchinski:1998rr}
\begin{subequations}
\begin{align}
S &=S_{NS}+S_{R}+S_{CS}\ ,\\
S_{NS}&=\frac{1}{2k_{10}^2}\int d^{10}x\ \sqrt{-\det g}\ e^{-2\phi}\left(R+
4\partial_{\mu}\phi\partial^{\mu}\phi-\frac{1}{2}|H_3|^2\right)\ ,\\
S_{R}&=-\frac{1}{4k_{10}^2}\int d^{10}x\ \sqrt{-\det g} \left(|F_1|^2+
|\tilde{F}_3|^2+\frac{1}{2}|\tilde{F}_5|^2
\right)\ ,\\
S_{CS}&=-\frac{1}{4k_{10}^2}\int d^{10}x\ C_4\wedge H_3\wedge F_3\ ,
\end{align}
\end{subequations}
where $H_3=dB^{NS}$ and $F_1=d\chi$ are the NSNS three-form and the RR scalar field strengths, 
respectively and 
\[
\tilde{F}_3=F_3-\chi\wedge H_3\ ,\ \ \ \ 
\tilde{F}_5=F_5-\frac{1}{2} B^{RR}_2\wedge H_3\ +\frac{1}{2} B_2^{NS}\wedge F_3\ ,
\]
and $F_3=dB^{RR}$ and $F_5=d f_4$. The equation of motion for $F_5$, which is nothing but the 
self-duality condition ($\tilde{F}_5=^*\!\!\! \tilde{F}_5$), should be imposed by hand.
We note here that the NSNS and RR terminology in the supergravity action is motivated by the flat 
space results of string theory, which as we point out later in section \ref{genericsinglestring}, 
does not correspond to our $SO(4) \times SO(4)$ decomposition
of states (see footnote \ref{NSNSRR}). The mapping of the fields presented in this
section and the string states will be clarified in section \ref{physicalstringspectrum}.

To study the physical on-shell spectrum of supergravity on the plane-wave background with 
non-trivial five-form flux \eqref{planewavemetric}, we linearize the supergravity equations
of motion around this background, and work in light-cone gauge,
by setting $\xi_{\mu \ldots \nu -}=0$, with $\xi_{\mu \ldots \nu -}$ generically any of the bosonic (other than scalar) tensor fields,
considered as perturbations around the background. Then, the $\xi_{\mu \ldots \nu +}$ components are not dynamical and are completely fixed in terms of the other physical modes, after imposing the constraints coming from the equations of motion for the gauge fixed components
$\xi_{\mu \ldots \nu -}$. Therefore, in this gauge we only deal with $\xi_{I \ldots J}$ modes, where $I, \cdots, J=1,2,\cdots, 8$. 
Setting light-cone gauge for fermions is
accomplished by projecting out spinor components by the action of an appropriate
combination of Dirac matrices \cite{Metsaev:2002re}. The advantage of using the \lc gauge is that in this gauge only the physical modes appear.

It will prove useful to first decompose the physical fluctuations of the
supergravity fields in terms of $SO(8) \rightarrow SO(4) \times SO(4)$
representations \cite{Metsaev:2002re,Das:2002cw}.
In the bosonic sector we have a complex scalar, combining the NSNS dilaton and RR scalar,
a complex two-form (again a combination of NSNS and RR fields), a real four-form, and a graviton.
Using the notation of section \ref{maxsusy}, we can decompose these
into $SO(4) \times SO(4)$ representations.
We label $SO(8)$ indices by $I,J,K,L$, and indices in the first $SO(4)$ by $i,j,k,l$ and those of 
the second with $a,b,c,d$. The decomposition of the bosonic fields is given in
TABLE \ref{sugratable}. 

\begin{table}[ht] 
\begin{center}
\begin{tabular}{|c|c|c|c|} \hline
Field & Components & $SO(4)\times SO(4)$ & Real degrees of freedom \\ \hline\hline
Complex scalar & $\Phi$ & $({\bf 1},{\bf 1})$ & 2\\ 
\hline
& $b_{ij}$ & $({\bf 3^+},{\bf 1}) \oplus ({\bf 3^-},{\bf 1})$ & 12 \\
Complex two-form & $b_{ab}$ & $({\bf 1},{\bf 3^+}) \oplus ({\bf 1},{\bf 3^-})$ & 12 \\
& $b_{ia}$ & $({\bf 4},{\bf 4})$ & 32\\
\hline
& $f_{ia}$ & $({\bf 4},{\bf 4})$ & 16 \\ 
Real four-form & $f_{ijab}$ & $({\bf 3^+},{\bf 3^+}) \oplus ({\bf 3^-},{\bf 3^-})$ & 18 \\
& $f$ & $({\bf 1},{\bf 1})$ & 1 \\
\hline
& $\tilde{h}_{ij}$ & $({\bf 9},{\bf 1})$ & 9 \\ 
Graviton & $\tilde{h}_{ab}$ & $({\bf 1},{\bf 9})$ & 9 \\ 
& $h_{ia}$ & $({\bf 4},{\bf 4})$ & 16 \\ 
& $h$ & $({\bf 1},{\bf 1})$ & 1 \\ 
\hline\hline
\end{tabular}
\caption{
$SO(4)\times SO(4)$ decomposition of bosonic supergravity fields.
${\bf 3^+}$ and ${\bf 3^-}$ are the self-dual and anti-self-dual
projections of the ${\bf 6}$ of $SO(4)$.
The complex scalar and two-form are defined as $\Phi=\chi+i e^\phi$ and
$b=B^{NS}+iB^{RR}$, and we have also defined
the pseudo-scalar ``trace'' piece of the four-form potential
$f=\epsilon^{ijkl} f_{ijkl}/6$, and
$f_{ia} = \frac{1}{3} \epsilon_i^{\ j k l} f_{a j k l}$.
The graviton $h_{IJ}$ and four-form $f_{IJKL}$ are fluctuations around a non-trivial plane-wave background.
$h=h_{ii}=-h_{aa}$ is the trace of the $SO(4)$ ``gravitons'', and
$\tilde{h}_{ij} = h_{ij} - \frac{1}{4} \delta_{ij} h_{kk}$.}
\label{sugratable}
\end{center}
\end{table}


The fermionic spectrum consists of a complex spin $1/2$ dilatino of negative chirality
and a complex spin $3/2$ gravitino with positive chirality.
For the dilatino, $16$ degrees of freedom survive the light-cone projection.
For the gravitino, we note that removing the spin $1/2$ component by projecting
out the $\gamma$-transverse components leaves $112$ degrees of freedom.
The details of the decomposition of $SO(8)$ fermions into representations of
$SO(4) \times SO(4)$ can be found in appendix \ref{SO(4)fermions}.
Using the notation of the appendix, the dilatino is in the \eightc\ and the gravitino in \eights.
Fermions can be decomposed along the same lines, using the
result of appendix \ref{SO(4)fermions}.

The dilaton is decoupled, in the linear regime, from the four-form, and
is the simplest field to deal with. Its equation of motion is simply that of
a complex massless scalar field \eqref{massive-scalar-eom}. Its lowest energy state
has $p^-=4\mu$, with a discretum of energies above it.

The graviton and four-form field are coupled in this background, leading to coupled
equations of motion.
The coupled Einstein and four-form potential equations of motion, after
linearizing and going to light-cone gauge, and using the self-duality of the
five-form field strength, imply the equation
\be
  \Box h_{ij} - 2 \mu \delta_{ij} \partial_- f = 0 \, ,
\ee
There is a similar expression for the other
$SO(4)$ projections of the metric and four-form.
We see that the trace (which we have yet to separate out) of the $SO(4)$ metric and four-form 
projections mix with each other. The equation of motion for the four-form, coupled to the metric 
through the covariant derivative, implies
\be
  \Box f + 8 \mu \partial_- h = 0 \, ,
\ee
These are a pair of coupled equations which can be diagonalized by the field
redefinition and using $h$ defined above
\be
  c = h_{ii} + i f \, .
\ee
The equations governing the new fields are
\be
  \Box \tilde{h}_{ij} = 0 \, , \ \ \ \ \ \ \ \ \ \
  \left( \Box + i 8 \mu \partial_- \right) c = 0 \, ,
\ee
together with the complex conjugate of the second. These are equations of motion for massive
scalar fields. Fourier transforming as before, we can compare these equations to
\eqref{massive-scalar-eom} and \eqref{sugra-lc-energy-spectrum},
to arrive at the light-cone energy spectrum, which is
\be
  p^-(\tilde{h}_{ij})=\mu (n+4) \, , \ \ \ \ \ \ \ \ \ \
  p^-(c)=\mu (n+8) \, , \ \ \ \ \ \ \ \ \ \
  p^-(c^\dagger)= 0 \ , \ \ \ \ \  n\in \mathbb{Z}^+\ ,
\ee
and obviously similar results for the components along the other $SO(4)$. Note that
$c^\dagger$ is the {\it only}  combination of fields whose light-cone 
energy is allowed to vanish.
Similar reasoning leads, for the mixed (in terms of $SO(4) \times SO(4)$) components of
the metric and four-form, to
\be
  \left( \Box + 4 i \mu \partial_- \right) \mathfrak{h}_{ia} = 0 \, ,
\ee
and its conjugate, where we have diagonalized the equations by defining
\be
  \mathfrak{h}_{ia} = h_{ia} + i f_{ia} \, ,
\ee
These lead to the light-cone energy for $\mathfrak{h}_{ia}$
\be
  p^-(\mathfrak{h}_{ia}) = \mu (n+6) \, , \ \ \ \ \ \ \ \ \ \
  p^-(\mathfrak{h}^\dagger_{ia}) = \mu (n+2) \, .  
\ee
Finally, for $f_{ijab}$, we can show that $p^-=\mu (n+4)$.

The complex two-form can be studied in the same way as the four-form and graviton, resulting in 
similar equations, but with different masses.
The two-form can be decomposed into representations that transform as two-forms
of each of the $SO(4)$'s, each of which can be further decomposed into self-dual and anti-self-dual components, with respect to the Levi-Cevita tensor of each $SO(4)$.
The self-dual part will carry opposite mass from the anti-self-dual projection.
The decomposition will also include a second rank tensor with one leg in each $SO(4)$,
which will obey a massless equation of motion (for the \soff decomposition see TABLE 
\ref{sugratable}.  The lowest
light-cone energy for the physical modes of the 
two-form take the values $p^-/\mu=2,4,6$, with the 
middle value associated with the mixed tensor and the difference of energies between the self-dual 
and anti-self-dual forms equal to four.

The analysis of the fermion spectrum follows along essentially the same lines, with minor
technical complications having to do with the spin structure of the fields (inclusion of
spin connection and some straight-forward Dirac algebra). These technicalities are not
illuminating, and we merely quote the results. The interested reader is directed to
\cite{Metsaev:2002re}. For the spin $1/2$ dilatino the lowest light-cone energies for the 
physical modes can take the values $p^-/\mu=3,5$, while for the spin $3/2$ gravitino the range is
$p^-/\mu=1,3,5,7$. It is worth noting that the lowest states of fermions/bosons are 
odd/even  integers in  $\mu$ units. This is compatible with what we expect from the superalgebra.




%% file: penrose.tex

As discussed in the previous section \pl s are particularly nice geometries with the important 
property 
of having globally defined null Killing vector field. They are also special from the \sugra\ point 
of view because they are $\alpha'$-exact (\cf\ section \ref{alpha'exact}). In this section we 
discuss a general limiting procedure, known as Penrose limit \cite{Penrose:1976}
which generates a \pl\ geometry out of any given space-time. This procedure has also been  extended to 
\sugra\ by Gueven \cite{Gueven:2000ru}, 
hence applied to \sugra\ this limit is usually 
called 
Penrose-Gueven limit e.g. see \cite{Blau:2002dy, Blau:2001ne}.
Although the Penrose limit can be applied to any space-time, if we start with solutions of  
Einstein's equations (or more generally the \sugra\ equations of motion)  we end up with
a plane-wave which is still a (super)gravity solution. In other words Penrose-Gueven  limit
is a tool to generate new \sugra\ solutions out of any given solution.
In this section first we summarize three steps of taking Penrose limits and then apply that to some
interesting examples such as $AdS$ spaces and their variations and finally in section 
\ref{susycontract} we study contraction of the \susy\ algebra corresponding to $AdS_5\times S^5$,
$PSU(2,2|4)$ \cite{Aharony:1999ti}, under the Penrose limit.

\subsection{Taking Penrose limits}\label{penroseguven}

The procedure of taking the Penrose limit can be summarized as follows:
\newline
{\it i)} Find a light-like (null) geodesic in the given space-time metric.
\newline
{\it ii)} Choose the proper coordinate system so that the metric looks like
\be\label{uvmetric}
ds^2 = R^2\left[-2 du d\tilde{v} + d\tilde{v}\left(d\tilde{v}+A_I(u,\tilde{v},\tilde{x}^I) 
d\tilde{x}^I\right) +g_{JK}(u,\tilde{v},\tilde{x}^I) d\tilde{x}^J d\tilde{x}^K\ \right].
\end{equation}
In the above $R$ is a constant introduced to facilitate the limiting procedure, the null geodesic
is parametrized by the affine parameter $u$, $\tilde{v}$ determines the distance between such null 
geodesics and $\tilde{x}^I$ parametrize the rest of coordinates. Note that any given metric can be 
brought to the form \eqref{uvmetric}.
\newline
{\it iii)} Take $R\to\infty$ limit together with the scalings
\be\label{penrosescalings}
\tilde{v}=\frac{v}{R^2}\ ,\ \ \ \tilde{x}^I=\frac{x^I}{R}\ ;\ \ \ u,v,x^I={\rm fixed}.
\ee
In this limit $A_I$ term drops out and $g_{IJ}(u,d\tilde{v},\tilde{x}^I)$ now becomes only a 
function of $u$, therefore
\be\label{plrosen}
ds^2 = -2 du d{v}+g_{IJ}(u)dx^Idx^J\ .
\ee
This metric is a \pl , though in the Rosen coordinates 
\cite{Rosen:1937}. Under the coordinate transformation
\[
x^I\to h_{IJ}(u) x^J\ , \ \ \ \ v\to v+\frac{1}{2} g_{IJ}h'_{IK}h_{JL}x^Kx^L\ ,
\]
with $h_{IK}g_{IJ}h_{JL}=\delta_{KL}$  and $h'_{IJ}=\frac{d}{du}h_{IJ}$ the metric takes the more 
standard form of 
\eqref{plane-wave-metric}, the Brinkmann coordinates \cite{Brinkmann:1923, 
Hubeny:2002vf}. 
The only non-zero component of the Riemann curvature of \pl\ \eqref{plane-wave-metric} is 
$R_{uIuJ}=f_{IJ}(u)$ and the Weyl tensor of any plane-wave is either null or vanishes.

\vskip .3cm

The above steps can be understood more intuitively. Let us start with an observer which 
boosts up to the speed of light. Typically such a limit in the (general) relativity
is singular, however, these singularities may be avoided by ``zooming'' onto a region 
infinitesimally close to the (light-like) geodesic the observer is moving on, in the particular 
way given in \eqref{penrosescalings}, so that at the end of the day from the original space-time 
point of view we remove all parts, except a very narrow strip close to the geodesic. And then 
scale up  the strip to fill the whole space-time, which is nothing but a \pl . The covariantly 
constant null Killing vector field of  \pl s correspond to the null direction of the 
original space-time along which the observer has boosted.
To demonstrate how the procedure works here we work out some explicit examples.

\subsubsection{Penrose limit of $AdS_p\times S^q$ spaces}

Let us start with a $AdS_p\times S^q$ metric in the global $AdS$ coordinate system 
\cite{Aharony:1999ti}
\be\label{AdSpSq}
ds^2=R_a^2(-\cosh^2\rho d\tau^2+d\rho^2+\sinh^2\rho\ d\Omega^2_{p-2})+
R_s^2(\cos^2\theta d\phi^2+d\theta^2+\sin^2\theta\ d\Omega^2_{q-2})\ .
\ee
We then boost along a circle of radius $R_s$ in $S^q$ directions, i.e. we choose the light-like 
geodesic along $\tau - \frac{R_s}{R_a}\phi$ direction at $\rho=\theta=0$. Next, we send $R_a,\ 
R_s\to \infty$ in the 
same rate, so that
\be\label{k2}
\frac{R^2_s}{R^2_a}=k^2={\rm fixed}
\ee
and scale the coordinates as
\begin{subequations}\label{x+-xy}
\begin{align}
x^+=\frac{1}{2}(\tau+\frac{R_s}{R_a}\phi)\ ,&\ \ \  
x^-={R^2_a}(\tau-\frac{R_s}{R_a}\phi)\ ,\\ 
\rho=\frac{x}{R_a}\ &,\ \ \ 
\theta=\frac{y}{R_s}\ ,
\end{align}
\end{subequations}
keeping $x^+, x^-, x, y$ and all the other coordinates fixed. Inserting \eqref{k2} and 
\eqref{x+-xy} into \eqref{AdSpSq} and dropping  $\co (\frac{1}{R^2_a})$ terms we obtain
\be\label{adspenrose}
ds^2  =  -2 dx^+ dx^- -(x^i x^i+k^2y^a y^a) {(dx^+)}^2 + dx^i dx^i+ dy^a dy^a, 
\ee
where $i=1,2,\cdots ,p-1$ and $a=1,2,\cdots, q-1$. For the case of $(p,q)=(5,5)$ and $(3,3)$ 
$k=\frac{R_s}{R_a}=1$, $(4,7)$ $k=\frac{R_s}{R_a}=1/2$ and $(7,4)$ $k=\frac{R_s}{R_a}=2$  
\cite{Maldacena:1998re}. 

Since $AdS_p$ and $S^q$ are not Ricci flat, $AdS_p\times S^q$ geometries can be \sugra\ solution 
only if they are accompanied with the appropriate fluxes; for the case of $AdS_5\times S^5$ that is 
a (self-dual) five-form flux of type IIB, for $AdS_4\times S^7$ and $AdS_7\times S^4$ four-form
flux of eleven dimensional \sugra\ and for $AdS_3\times S^3$ that is three-form RR or NSNS flux 
\cite{Maldacena:1998re}. 
Let us now focus on the $AdS_5\times S^5$ case and study the behaviour of the five-form flux under 
the Penrose limit. The self-dual five-form flux on $S^5$ is proportional to $N={R^4_s}/{g_s}$, 
explicitly  \cite{Aharony:1999ti}
\be\label{fiveformflux}
F_{S^5}={4} N d\Omega_5,\ \ \ F_{AdS_5}= ^*\!\!F_{S^5}\ ,
\ee
where $d\Omega_5$ is the volume form of a five-sphere of unit radius. The numeric factor  
${4}$ is just a matter of \sugra\ conventions and we have chosen our conventions so that the 
ten dimensional (super)covariant derivative is given by \eqref{covderivative}. Taking the Penrose 
limit we find that
\[
F=\frac{{4}}{g_s} dx^+\wedge(dx^1\wedge dx^2\wedge dx^3\wedge dx^4+ dy^1\wedge dy^2\wedge 
dy^3\wedge dy^4).
\]
Finally the metric can be brought to the form \eqref{planewavemetric} through the coordinate 
transformation
\[
x^+\to \mu x^+\ ,\ \ \ \ x^-\to \frac{1}{\mu} x^-.
\]
We would like to note that as we see from the analysis presented here, for the $AdS_5\times 
S^5$ case the $x^i$ come from the $AdS_5$ and $y^a$ from the $S^5$ directions. However, 
after the Penrose  limit there is no distinction between  the $x^i$ or $y^a$ 
directions. This 
leads to the $\ztwo$ symmetry of the \pl\ (\cf\ \eqref{Z2}).

Starting with a maximally supersymmetric solution, e.g. $AdS_5\times S^5$, after the Penrose limit
we end up with another maximally supersymmetric solution, the \pl . In fact that is a general 
statement that under the Penrose limit we never lose any supersymmetries, and as we will show 
in the next subsection even  we may gain some. It has been shown that all  \pl s, whether 
coming as Penrose limit or not, at least 
preserve 
half of the maximal possible supersymmetries (i.e. 16 supercharges for the type II theories)
giving rise to kinematical 
supercharges e.g. see \cite{Cvetic:2002si} and a class of them which may preserve 
more than 16  necessarily have a constant dilaton \cite{Figueroa-O'Farrill:2002ft}. 

\subsubsection{Penrose limits of $AdS_5\times S^5$ orbifolds}\label{penroseorbifold}

As the next example we work out the two Penrose limits of half supersymmetric $AdS_5\times 
S^5/\mathbb{Z}_K$ orbifold, the metric of which can be recast to 
\cite{Alishahiha:2002ev}
\be\label{AdSorbifold}
ds^2=R^2\left[-\cosh^2\rho d\tau^2+d\rho^2+\sinh^2\rho\ d\Omega^2_{3}+
\cos^2\theta d\phi^2+d\theta^2+\sin^2\theta\ d\Omega^2_{S^3/\mathbb{Z}_K}\right],
\ee
with
\be
d\Omega^2_{S^3/\mathbb{Z}_K}=\frac{1}{4}(\sin^2\gamma d\delta^2 
+d\gamma^2)+\frac{1}{K^2}[d\chi-\frac{K}{2}(1-\cos\gamma)d\delta]^2, 
\ee
where $\gamma, \delta$ and $\chi$ all range from zero to $2\pi$.
We now can take the limit \eqref{x+-xy}. Readily it is seen that  we end up with the
half supersymmetric $\mathbb{Z}_K$ orbifold of the maximally \susyc\ \pl\ \eqref{planewavemetric}.
All the above arguments can be repeated for $AdS_5/\mathbb{Z}_K\times S^5$, that is a geometry 
whose metric is \eqref{AdSorbifold} after the exchange of $d\Omega_3$ and 
$d\Omega^2_{S^3/\mathbb{Z}_K}$. It is straightforward to see that after the Penrose limit 
both $AdS_5/\mathbb{Z}_K\times S^5$ and $AdS_5\times S^5/\mathbb{Z}_K $ half \susyc\ orbifolds 
become identical (recall the $\ztwo$ symmetry \eqref{Z2}).

In the orbifold case there is another option for the geodesic to boost along, the $\chi$ direction 
in \eqref{AdSorbifold}. Let us consider the following Penrose limit: $R\to\infty$ and the scaling
\begin{subequations}\label{x+-rxy}
\begin{align}
x^+=\frac{1}{2}(\tau+\frac{1}{K}\phi)\ ,&\ \ \  
x^-={R^2}(\tau-\frac{1}{K}\phi)\ ,\\ 
\rho=\frac{x}{R}\ ,\ \ \ 
\theta=&\frac{\pi}{2}-\frac{y}{R}, \ \ \ \  \gamma=\frac{2x}{R}\ ,\ \ \ r,x,y={\rm fixed}.
\end{align}
\end{subequations}
Inserting the above into \eqref{AdSorbifold} and renaming $\delta-x^+$ as $x^+$, it is easy 
to 
observe that we again find the maximally \susyc\ \pl\ of \eqref{planewavemetric}. In other words 
the 
orbifolding is disappeared and we have enhanced supersymmetry from 16 to 32 
\cite{Alishahiha:2002ev}.
The orbifolding, however, is not completely washed away. As it is seen from (\ref{x+-rxy}a) the 
$x^-$ direction is a circle of radius $R_-=R^2/2K$. In particular,  if together with $R^2$ we also 
send $K\sim 
R^2\to \infty$ there is the possibility of keeping $R_-$ finite \cite{Mukhi:2002ck}
i.e. the 
Penrose limit of $AdS_5\times S^5/\mathbb{Z}_K$ orbifold can naturally lead to a light-like 
compactification of the \pl .
Penrose limits of more complicated $AdS$ orbifolds may be found in 
\cite{Takayanagi:2002hv, Alishahiha:2002jj, Oh:2002sv, Floratos:2002uh, Alishahiha:2002bc},
among which there are cases naturally leading to 
various 
toroidally, light-like as well as space-like,  compactified \pl s 
\cite{Bertolini:2002nr}.

\subsubsection{Penrose limit of $AdS_5\times T^{1,1}$}

As the last example we consider the case in which the Penrose-Gueven limit enhances eight 
supercharges to 32, the Penrose limit of $AdS_5\times T^{1,1}$ \cite{Itzhaki:2002kh, 
PandoZayas:2002rx, Gomis:2002km}.
$T^{1,1}$ is a five dimensional Einstein-Sasaki manifold \cite{Acharya:1998db,
Morrison:1998cs} 
whose metric is given by 
\cite{Klebanov:1998hh, Candelas:1990js}
\be\label{T11metric}
ds^2_{T^{1,1}}=\frac{R^2}{9}(d\psi+\cos\theta_1 d\phi_1+\cos\theta_2 d\phi_2)^2+\frac{R^2}{6}
(d\theta_1^2+\sin^2\theta_1d\phi_1^2+d\theta_2^2+\sin^2\theta_2d\phi_2^2)\ .
\ee
Then the $AdS_5\times T^{1,1}$ solution is obtained by replacing \eqref{T11metric} for $S^5$ term 
(i.e. the term proportional to $R^2_s$ in \eqref{AdSpSq}) together with the self-dual five-form flux
given in \eqref{fiveformflux}. Next consider the Penrose limit
\begin{subequations}\label{T11penrose}
\begin{align}
x^+=\frac{1}{2}\tau+\frac{1}{6}(\psi+\phi_1+\phi_2)\ ,&\ \ \  
x^-={R^2}\left(\tau-\frac{1}{3}(\psi+\phi_1+\phi_2)\right)\ ,\ \ \ R\to\infty \\
\rho=\frac{x}{R}\ ,\ \ \ 
\theta_1=&\frac{1}{\sqrt{6}}\frac{r_1}{R}, \ \
\theta_2=\frac{1}{\sqrt{6}}\frac{r_2}{R}, 
\end{align}
\end{subequations}
with $ x,r_1,r_2,x^+,x^-={\rm fixed}$. It is easy to see that expanding $AdS_5\times T^{1,1}$ in
$\frac{1}{R}$ and keeping the leading terms we again find the maximally \susyc\ \pl\ 
\eqref{planewavemetric}.
Finally we would like to remind the reader that in the literature Penrose limits of several other 
geometries, such as AdS Schwarzchild black-hole have been studied e.g. see 
\cite{PandoZayas:2002rx, Hubeny:2002vf, Fuji:2002vs, Brecher:2002ar, Gursoy:2002tx}.

\subsection{Contraction of the superconformal algebra $PSU(2,2|4)$ under the Penrose 
limit}\label{susycontract}

In previous subsection we showed how to obtain the \pl\ \eqref{planewavemetric} from the
$AdS_5\times S^5$ solution. In this part we continue similar line of logic and show that
under the Penrose limit the isometry  group of $AdS_5\times S^5$, $SO(4,2)\times SO(6)$ exactly  
reproduces the isometry group of the \pl\ discussed in section \ref{isometry}. As the first point 
we note that  $SO(4,2)\times SO(6)$ and the isometry group of section \ref{isometry}
both have 30 generators. In fact we will show that this correspondence goes beyond the bosonic 
isometries and extends to the whole $AdS_5\times S^5$ superalgebra, $PSU(2,2|4)$ 
\cite{Minwalla:1998ka}.
The contraction of $PSU(2,2|4)$ superalgebra under Penrose limit has been 
considered in \cite{Hatsuda:2002xp}.

\subsubsection{Penrose contraction of the bosonic isometries}\label{bosoniccontraction}

The bosonic part of the $AdS_5\times S^5$ isometries is comprised of the four dimensional   
conformal group $SO(4,2)$ times $SO(6)$, the generators of which are
\[
J_{\hat\mu\hat\nu}\ ,\ \ \ J_{\hat A \hat B}\ ,\ \ \ {\hat\mu=-1,0,1,2,3,4},\ \hat A=1,2,\cdots, 6.
\]
Being $SO(4,2)\times SO(6)$ generators they satisfy
\begin{subequations}\label{so42so6algebra}
\begin{align}
[J_{\hat\mu\hat\nu}, J_{\hat\rho\hat\lambda}] &= 
i(\hat\eta_{\hat\mu\hat\rho}J_{\hat\nu\hat\lambda}+{\rm Permutations})\\
[J_{\hat A\hat B}, J_{\hat C\hat D}] &=
i(\delta_{\hat A\hat C}J_{\hat B\hat D}+{\rm Permutations})\ ,
\end{align}
\end{subequations}
where $\hat\eta_{\hat\mu\hat\nu}=diag(-,-,+,+,+,+)$. In order to take the Penrose limit it is 
more convenient to decompose them as 
\begin{subequations}\label{so42decompose}
\begin{align}
J_{\hat\mu\hat\nu} &=\Big\{ J_{ij},\ L_i=\frac{1}{R}(J_{-1,i}+J_{0i}),\  
K_i=\frac{1}{R}(J_{-1,i}-J_{0i}),\ \cd =J_{-1,0}\Big\}\\
J_{\hat A\hat B} &=\Big\{ J_{ab},\ L_a=\frac{1}{R}(J_{5a}+J_{6a}),\  
K_a=\frac{1}{R}(J_{5a}-J_{6a}),\ \cj =J_{56}\Big\}
\end{align}
\end{subequations}
where $i,j$ and $a,b$  vary from 1 to 4 and also we redefine $\cd$ and $\cj$ as
\begin{subequations}\label{DJP+-}
\begin{align}
\cd &=\mu R^2 P^+ +\frac{1}{2\mu} P^- \\
\cj &=\mu R^2 P^+ -\frac{1}{2\mu} P^-\ . 
\end{align}
\end{subequations}
Note that in the above $R$ and $\mu$ are auxiliary parameters introduced to facilitate the 
procedure of taking the Penrose limit. In the above parametrization the Penrose limit
\eqref{x+-xy} becomes $R\to\infty$ and keeping $J_{ij},\ {J_{ab}},\ K_{i}, L_i,\ K_a, L_a$ and 
$P^+$, $P^-$ fixed. It is straightforward to show that \eqref{so42so6algebra} goes over to the
$[h(4)\oplus h(4)]\oplus so(4)\oplus so(4)\oplus u(1)_+ \oplus u(1)_-$ 
discussed in detail in section 
\ref{isometry}.
 
\subsubsection{Penrose contraction on the fermionic generators}\label{fermionicpenrosecontraction}

The supersymmetry of $AdS_5\times S^5$ fits into the Kac classifications of the superalgerbas 
\cite{Kac:1977em} and is $PSU(2,2|4)$ (e.g. see \cite{Dobrev:1985qv, Minwalla:1998ka}), 
meaning that the bosonic part of the 
algebra is $su(2,2)\oplus su(4)\simeq so(4,2)\oplus so(6)$. Usually in the literature this 
superalgebra is either written using $so(3,1)$ notations (e.g. see 
\cite{D'Hoker:2002aw}) or ten dimensional type IIB notations (e.g. see \cite{Metsaev:1998it}) for 
fermions.
For our purpose, where we merely need the simplest form of the algebra, it is more convenient to 
directly use $so(4,2)$ or $so(6)$ spinors. The supercharges carry spinorial indices of both of 
the $SO(4,2)$ and $SO(6)$ groups. First we recall that $spin(4,2)=su(2,2)$ and $spin(6)=su(4)$, 
therefore the supercharges should carry fundamental indices of $su(2,2)$ and $su(4)$ (\cf\ 
appendix \ref{sixdimfermion}), 
explicitly $Q_{\hat{I}J}$ where both of $\hat{I}$ and $J$ run from one to four and the hatted 
index is $su(2,2)$ spinorial index and the unhatted one that of $su(4)$. In fact both of these 
indices are Weyl indices of the corresponding groups. Some more details of these six dimensional
spinors are gathered in appendix \ref{sixdimfermion}. The fermionic part of $PSU(2,2|4)$ 
superalgebra in this notation reads as
\begin{subequations}\label{su224}
\begin{align}
[J_{\hat{\mu}\hat{\nu}}, Q_{\hat{I}J}] &=\frac{1}{2}(i\gamma_{\hat{\mu}\hat{\nu}})_{\hat{I}}^{\ 
\hat{K}}Q_{\hat{K}J}  
\label{su(224)MQ1}
\\
[J_{\hat{A}\hat{B}}, Q_{\hat{I}J}] &=-\frac{1}{2}(i\gamma_{\hat{A}\hat{B}})_{J}^{\ {K}}
Q_{\hat{I}K}
\label{su(224)MQ2}
\\
\{Q_{\hat{I}J}, Q^{\dagger\hat{K}L}\} & = 
2\delta_J^{\ L}(i\gamma^{\hat{\mu}\hat{\nu}})_{\hat{I}}^{\ \hat{K}}J_{\hat{\mu}\hat{\nu}}+
2\delta_{\hat{I}}^{\ \hat{K}}(i\gamma^{\hat{A}\hat{B}})_J^{\ L} J_{\hat{A}\hat{B}}
\label{su(224)QQ}
\end{align}
\end{subequations}
Having the algebra written in the above notation  and  using the decomposition \eqref{psiIJ} we can 
readily 
take the Penrose limit,  if together with \eqref{so42decompose} and \eqref{DJP+-} we scale the 
supercharges as
\be\label{Qdecompose}
Q_{\hat{I}J}\to ( {\sqrt{\mu} R}q_{\alpha\beta},\ 
{\sqrt{\mu} R}q_{\dot\alpha\dot\beta},\ 
\frac{1}{\sqrt{\mu}}Q_{\alpha\dot\beta},\ \frac{1}{\sqrt{\mu}}Q_{\dot\alpha\beta})\ ,
\ee
where we have introduced proper scalings for the kinematical and dynamical supercharges ($q$ and 
$Q$ respectively). Inserting \eqref{Qdecompose} into \eqref{su224}, sending $R\to \infty$ and 
keeping the leading terms, it is straightforward to see that 
\eqref{su224} contracts to the superalgebra of the \pl\ studied in some detail in section 
\ref{planewavesusy}.

%% file: stringbg.tex

As discussed in section \ref{alpha'exact} \pl s are $\alpha'$-exact solutions of \sugra\ and hence 
provide us with nice \bg s for string theory. In fact noting the simple form of the metric 
\eqref{plane-wave-metric} it can be seen that the bosonic part of the $\sigma$-model action in 
this \bg\ {\it in the light-cone gauge} takes a very simple form and for $f_{IJ}=constant$ 
\cite{Metsaev:2001bj, Russo:2002rq, Alishahiha:2002nf, Hyun:2002wu, Sugiyama:2002tf}
and $f_{IJ}\propto u^{-2}$ \cite{Papadopoulos:2002bg} and some more general cases  
\cite{Blau:2002js} it is even exactly solvable. In this review, however, we will only 
focus on the maximally \susyc\ \pl\ of \eqref{planewavemetric} and work out the Green-Schwarz
action for this \bg . Note that, due to the presence of the RR fluxes the RNS formulation of 
string theory can not be used.  The Green-Schwarz formulation of superstring theory on some other 
\pl\ or pp-wave \bg s has also been considered in the literature, see for example 
\cite{Hikida:2002in, Russo:2002qj, Berkovits:2002vn, Maldacena:2002fy, Cvetic:2002nh, 
Sadri:2003ib, Mizoguchi:2002qy, Fuji:2002vs, Gimon:2002nr, Kunitomo:2003ja, Walton:2003nd}.

\subsection{Bosonic sector of type IIB strings  on the plane-wave \bg }\label{bosonicstrings}

The bosonic string $\sigma$-model action  in the \bg\ \eqref{planewavemetric} 
which has metric $G_{\mu\nu}$ and a vanishing NSNS two-form, is
\cite{Polchinski:1998rq}
\bea\label{bosonicaction}
S&=&\frac{1}{4\pi\alpha'}\int d^2\sigma g^{ab}\ G_{\mu\nu}\partial_aX^\mu\partial_bX^\nu \cr
&=&\frac{1}{4\pi\alpha'}\int d^2\sigma g^{ab}\ \left(-2\partial_aX^+\partial_bX^- 
+\partial_aX^I\partial_bX^I -\mu^2 X_I^2\partial_aX^+\partial_bX^+\right)\ ,
\eea
where $g_{ab}$ is the worldsheet metric, $\sigma^a=(\tau, \sigma)$ are the worldsheet coordinates and $I=1,2,\cdots ,8$.
Note that the RR \bg\ fluxes do not appear in the bosonic action. We first need to fix the two 
dimensional gauge symmetry, a part of which is done by choosing
\be\label{gauge1}
\sqrt{-g}g^{ab}=\eta^{ab}\ ,\ \ \ \ -\eta_{\tau\tau}=\eta_{\sigma\sigma}=1\ .
\ee
To fix the residual worldsheet diffeomorphism invariance, we note that the equation of motion for $X^+$, $ (\partial_\tau^2-\partial_\sigma^2)X^+=0$, has a general solution of the form 
$f(\tau+\sigma)+g(\tau-\sigma)$. We choose $f(x)=g(x)=\frac{1}{2}\alpha' p^+ x$, i.e.
\be\label{lightconeX+}  
X^+=\alpha' p^+\tau\ ,\ \ \ \  p^+>0\ .
\ee
The choices \eqref{gauge1} and \eqref{lightconeX+} completely fix the gauge symmetry. This is the {\it light-cone gauge}. In this gauge $X^+$ and $X^-$ are not dynamical variables 
anymore and are completely determined by $X^I$'s through the constraints resulting from
\eqref{gauge1} 
\cite{Green:1987fi}
\[
\frac{\delta {\cal L}}{\delta g_{\tau\sigma}}=0 \ , \ \ \
\frac{\delta {\cal L}}{\delta g_{\tau\tau}}=\frac{\delta {\cal L}}{\delta g_{\sigma\sigma}}=0\ .
\]
Using the solution \eqref{lightconeX+} 
for  $X^+$ and setting $-g_{\tau\tau}=g_{\sigma\sigma}=1$, these constraints become 
\be\label{Virasoro}
\partial_{\sigma}X^-=\frac{1}{\alpha' p^+} \ \partial_{\sigma}X^I\partial_{\tau}X^I \ , 
\ee
\be\label{lightconeX-}
\partial_{\tau}X^-=\frac{1}{2\alpha' p^+} \ \biggl(
\partial_{\tau}X^I\partial_{\tau}X^I+\partial_{\sigma}X^I\partial_{\sigma}X^I-
(\mu\alpha' p^+)^2 X^IX^I\biggr) \, .
\ee
We can now drop the first term in \eqref{bosonicaction} and replace $X^+$ with its \lc solution. After rescaling  $\tau$ and $\sigma$ by $\alpha' p^+$, we obtain the \lc action
\be\label{LCbosonicaction}
S^{bos.}_{l.c.}=\frac{1}{4\pi\alpha'}\int d\tau\int_0^{2\pi\alpha' p^+} d\sigma  \left[
\partial_\tau X^I\partial_\tau X^I -\partial_\sigma X^I\partial_\sigma X^I 
-\mu^2 X_I^2\right] \, .
\ee
This action is quadratic in $X^I$'s and hence it is solvable. The equations of motion for 
$X^I$, 
\be\label{bosoniceom}
\left(\partial_\tau^2-\partial_\sigma^2-\mu^2\right) X^I=0\ ,
\ee
should be solved together with the closed string boundary conditions
\be\label{bosboundary}
X^I(\sigma+2\pi\alpha' p^+)=X^I(\sigma)\ .
\ee
In fact $X^\pm$ should also satisfy the same boundary condition. {}From \eqref{lightconeX+} it is 
evident that $X^+$ satisfies this boundary condition. We will come back to the boundary condition 
on $X^-$ at the end of this subsection. The solutions to these equations are
\bea\label{bosonicmode}
X^I=x_0^I\cos\mu\tau+\frac{p^I_0}{\mu p^+}\sin\mu\tau\ +\sqrt{\frac{\alpha'}{2}}\sum_{n=1}^\infty
\frac{1}{\sqrt{\omega_n}}&\biggl[&
\alpha^I_n\ e^{\frac{-i}{\alpha' p^+}(\omega_n\tau+n\sigma)}+\tilde{\alpha}^I_n\ 
e^{\frac{-i}{\alpha' 
p^+}(\omega_n\tau-n\sigma)}+\cr
&&\alpha^{I\dagger}_n\ e^{\frac{+i}{\alpha' p^+}(\omega_n\tau+n\sigma)}+
\tilde{\alpha}^{I\dagger}_n\ e^{\frac{+i}{\alpha' p^+}(\omega_n\tau-n\sigma)}\biggr] \, ,
\eea
where
\be\label{omega}
\omega_n=\sqrt{n^2+(\alpha' \mu p^+)^2}\ ,\ \ \ \ n\geq 0\ ,
\ee
and $\alpha$ and $\tilde\alpha$ correspond to the right and left moving modes. The
case of $n=0$ has been included for later convenience.
The canonical quantization conditions
\be\label{[XP]}
[X^I(\sigma,\tau), P^J(\sigma',\tau)]=i\delta^{IJ}\delta(\sigma-\sigma') \, ,
\ee
where $P^I=\frac{1}{2\pi\alpha'}\partial_\tau X^I$, yield
\be\label{bosoniccommutations}
[x_0^I, p_0^J]=i\delta^{IJ}\ ,\ \ \ 
[\alpha^I_n, \alpha^{J\dagger}_m]= [\tilde{\alpha}^I_n, \tilde{\alpha}^{J\dagger}_m]=
\delta^{IJ}\delta_{mn}\ .
\ee
Next, using the \lc action we work out the \lc Hamiltonian
\be\label{bosoniclcH}
H^{bos.}_{l.c.}=\frac{1}{4\pi\alpha'}\int_0^{2\pi\alpha' p^+}d\sigma \ 
\big[(2\pi\alpha')^2 P_I^2 +(\partial_\sigma X^I)^2+\mu^2 X^2_I \big] \, .
\ee
As we expect, the light-cone Hamiltonian density is the momentum conjugate to 
light-cone time $X^+$, $P^-=\frac{2}{\alpha' p^+}(\partial_\tau X^- +\mu^2 X^2_I)$. Plugging the mode expansion \eqref{bosonicmode}
into \eqref{bosoniclcH} we obtain
\be\label{bosonicH2}
H^{bos.}_{l.c.}=\frac{1}{\alpha' p^+}\left[\alpha'\mu p^+ \alpha_0^{I\dagger}\alpha_0^I+
\sum_{n=1}^\infty \omega_n(\alpha_n^{I\dagger}\alpha_n^{I}+
\tilde{\alpha}^{I\dagger}_n\tilde{\alpha}^I_n)\right] +\frac{8}{\alpha' p^+}
\left(\frac{1}{2}\alpha' \mu p^++\sum_{n=1}^\infty \omega_n \right)\ ,
\ee
where the last term is the zero point energies of bosonic oscillators (after normal ordering) and we have defined
\be
\tilde{\alpha}_0^I \equiv
\alpha_0^I=\frac{1}{\sqrt{2\mu p^+}}p_0^I-i \sqrt{\frac{\mu p^+}{2}} x^I_0 \, .
\ee
It is easy to check that $[\alpha_0^I,\alpha_0^{J\dagger}]=\delta^{IJ}$. We will see in the 
next subsection that this zero point energy is canceled against the zero point energy of the fermionic modes, a sign of supersymmetry.

Now let us check whether $X^-$ also satisfies the closed string boundary condition $
X^-(\sigma+2\pi \alpha' p^+)=X^-(\sigma)$. From \eqref{Virasoro} we learn that
\bea\label{levelmatching}
X^-(\sigma+2\pi \alpha' p^+)-X^-(\sigma) &=&\int_0^{2\pi \alpha' p^+} d\sigma \partial_\sigma 
X^I\partial_\tau X^I\cr
&=& \sum_{n=1}^\infty n(\alpha_n^{I\dagger}\alpha_n^{I}- 
\tilde{\alpha}^{I\dagger}_n\tilde{\alpha}^I_n)=0 \, ,
\eea
where we have used the mode expansion \eqref{bosonicmode}. Equation \eqref{levelmatching} is the {\it level matching} condition, which is in fact a constraint on the physical excitations 
of a closed string \cite{Polchinski:1998rq}.

The vacuum of the light-cone string theory, $|0,p^+\rangle$ is defined as a state satisfying
\be\label{bosvacuum}
\tilde\alpha_n^I|0,p^+\rangle = \alpha_n^I|0,p^+\rangle = 0 \ ,\ \ \    n\geq 0\ .
\ee
Note that this vacuum is specified with the light-cone momentum $p^+$, i.e. for different values of $p^+$ we have a different string theory vacuum state and hence a different Fock space built from it. As we see from \eqref{bosonicH2}, in the \pl\ \bg\ all the string modes, including the zero modes, are massive. In other words all the \sugra\ modes 
(created by $\alpha_0^\dagger$ 
) are also massive, in agreement with the discussion in section 
\ref{sugraspectrum}.

Before moving on to the fermionic modes, we would like to briefly discuss strings on  
compactified \pl s. Such compactified \pl s 
may naturally arise in the Penrose limit of particular \ads orbifolds (\cf\ dicussions of section 
\ref{penroseorbifold}).
Let us consider the compactification of $X^-$ on a circle of radius $R_-$:
\be\label{R-}
X^-\equiv X^- +2\pi R_-\ .
\ee
As a result of this compactification the light-cone momentum $p^+$, which is the momentum conjugate to the $X^-$ direction, should be quantized
\be\label{momentumX-}
p^+=\frac{m}{R_-}\ , \ \ \ m\in \mathbb{Z}-\{0\}\ .
\ee
For fixed $m$, we are in fact studying the discrete light-cone quantization (DLCQ) of 
strings on \pl s \cite{Mukhi:2002ck, Alishahiha:2002jj}. After compactification, we might also have winding modes along the $X^-$ direction. The $X^-$ winding number $w$ is related to $X^I$ excitation modes through the constraint \eqref{Virasoro}:
\[ w = \frac{1}{2\pi R_-}\int_0^{2\pi\alpha' p^+} d\sigma\ \partial_\sigma X^-=
\frac{\alpha'}{R_- }\int_0^{2\pi\alpha' p^+} d\sigma\ \partial_\sigma X^I P_I\ , \ \ w\in 
\mathbb{Z}\ .
\]
This equation together with \eqref{momentumX-} gives the ``improved'' level matching condition 
for strings which is $m w=\sum_{n>0} n(\alpha_n^{I\dagger}\alpha_n^{I}-
\tilde{\alpha}^{I\dagger}_n\tilde{\alpha}^I_n)$. 
The string theory vacuum state is now identified by two integers $m$ and $w$. 
As for toroidal compactifications in the transverse directions and T-duality for strings on \pl s, we will not discuss them here and the interested reader is referred to the available literature, see for exmaple \cite{Michelson:2002wa, Ideguchi:2003rk, Mizoguchi:2003be}.

\subsection{Fermionic sector of type IIB strings  on the plane-wave \bg }

The fermionic sector of the Green-Schwarz superstring action for type IIB strings is 
\cite{Green:1987fi, Cvetic:1999zs}
\be\label{fermionaction}
S_F=\frac{i}{4\pi\alpha'}\int d^2\sigma\ (\theta^\alpha)^\top\ 
(\beta^{ab})_{\alpha\rho}\partial_a X^\mu \Gamma_\mu\  
({\hat D}_b)^{\rho}_{\ \beta}\theta^{\beta}+ {\cal O}(\theta^3)\ .
\ee
In the above $\theta^\alpha,\  \alpha=1,2$ are two fermionic worldsheet fields
giving embedding
coordinates of ${\cal N}=2$ type IIB superspace, i.e. they are 32 component ten dimensional Weyl-Majorana fermions of the same chirality,
\be\label{beta}
(\beta^{ab})_{\alpha\rho}=\sqrt{-g}g^{ab}\delta_{\alpha\rho}-\epsilon^{ab}(\sigma^3)_{\alpha\rho}\ 
,
\ee
and $({\hat D}_b)^{\rho}_{\ \beta}$ is the pull-back of the supercovariant derivative
\eqref{covderivative} to the worldsheet, which for our background becomes
\be\label{Db}
({\hat D}_b)^{\rho}_{\ \beta}=
\delta^{\rho}_{\ \beta}\partial_b+\partial_b X^\nu\ (\Omega_\nu)^{\rho}_{\ \beta} \, ,
\ee
and $\Omega_\nu$ is given in \eqref{Omega}.\footnote{
The ${\cal O}(\theta^3)$ terms come from the higher order $\theta$ contributions to the 
supervielbein. Explicitly, the Green-Schwarz Lagrangian for a general \bg\ is 
\[
{\cal L}=g^{ab}\Pi_a^\mu\Pi_b^\nu G_{\mu\nu}+{\cal L}_{WZ}\ ,
\]
with $\Pi_a^\mu=\partial_a Z^N E_N^\mu$, and where $Z^M=(X^\mu,\ \theta^{A\alpha})$ are the type IIB superspace coordinates and $E_N^M$ are the supervierbeins (see \cite{Metsaev:2001bj}). 
One can then show that after fixing the \lcg for the \pl\ \bg\, all ${\cal O}(\theta^3)$ 
corrections to $E_N^M$ vanish \cite{Metsaev:2001bj}
and the action reduces to \eqref{fermionaction} without ${\cal O}(\theta^3)$ 
terms. We do not enter into 
these complications and the interested reader is 
referred to e.g. \cite{Metsaev:2001bj}. A similar procedure for the M2-brane action in the eleven dimensional \pl\ \bg\ has been carried out in \cite{Dasgupta:2002hx}.} Our notations for ten dimensional type IIB fermions is summarized in Appendix \ref{SO(8)fermions}; and by 
$(\theta^\alpha)^\top$ we mean the transposition in the fermionic indices.
The $\epsilon_{ab}$ term in \eqref{beta} is in fact coming from the Wess-Zumino term in the 
Green-Schwarz action.

\subsubsection{Fixing $\kappa$-symmetry and  fermionic spectrum}

$\kappa$-symmetry is a necessary fermionic symmetry in order to have spacetime \susy\ 
for the on-shell string modes. In fact by fixing the $\kappa$-symmetry we remove half of the 
fermionic gauge (unphysical) degrees of freedom so that after gauge-fixing we are left with 16 
physical fermions, describing on-shell spacetime fermionic modes. This number of fermionic degrees of freedom is exactly equal to the number of physical bosonic degrees of freedom coming from the $X^I$ modes after fixing the \lc gauge (note that there are left and right modes).

It has been shown that the action \eqref{fermionaction} for the \pl\ \bg\ possesses the necessary $k$-symmetry \cite{Metsaev:2001bj}, and to obtain the physical fermionic 
modes we need to gauge fix it, which can be achieved by choosing
\be\label{kappasymmetry}
\Gamma^+\theta^\alpha=0, \ \ \ \alpha=1,2.
\ee
Similar to the flat space case \cite{Green:1987fi}, the above suffices to fix the full 
$\kappa$-symmetry of the \pl\ \bg\ \cite{Metsaev:2001bj}.
As is shown in the Appendix \ref{SO(8)fermions}, by imposing \eqref{kappasymmetry} we can 
reduce the ten dimensional fermions to $SO(8)$ \rep s, and since the two $\theta^\alpha$ have the same ten dimensional chiralities, both of them end up to be in the same $SO(8)$ fermionic \rep, which we have chosen to be ${\bf 8_s}$. 

To simplify the action we note that \eqref{kappasymmetry} implies
\[
(\theta^\alpha)^\top \Gamma^I\theta^\beta=0\ \ \forall \alpha, \beta \, , \ \ \ \ \ 
(\Omega_I)^{\alpha}_{\ \beta}\theta^\beta=0 \, .
\]
{}From the $\partial_aX^\mu\Gamma_\mu$ term in the action only $\partial_aX^+\Gamma_+$, and
from the $\Omega_\mu$ terms only $\Omega_+$ survive and hence
\[
S^{fer}_{l.c.}=\frac{i}{4\pi\alpha'}\int d\tau\int_0^{2\pi\alpha' p^+} d\sigma\ 
\left[(\theta^\alpha)^\top\ 
(\beta^{ab})_{\alpha\rho}(\partial_a X^+ \Gamma_+)\  
(\delta^{\rho}_{\ \beta}\partial_b+\partial_b X^+\ (\Omega_+)^{\rho}_{\ \beta})
\theta^{\beta}\right] .
\]   
Next, we use  \eqref{Omega} and \eqref{lightconeX+} to further simplify the action; after
some straightforward algebra we obtain
\be\label{Sferlc}
S^{fer}_{l.c.}=\frac{-i}{4\pi\alpha'}\int d\tau\int_0^{2\pi\alpha' p^+} d\sigma\ 
\left[
\theta^\dagger \partial_\tau\theta+
\theta \partial_\tau\theta^\dagger+
\theta\partial_\sigma\theta+
\theta^\dagger \partial_\sigma\theta^\dagger-2i \mu 
\theta^\dagger\Pi \theta\right] .
\ee
Note that in the above we have replaced $\theta^1$ and $\theta^2$ which are now eight component 
\eights\ fermions with their complexified version ({\it cf.} Appendix \ref{SO(8)fermions}, 
eq.~\eqref{complexfermions}).
The last term in the action is a mass term resulting from the RR five-form flux of the \bg . As we 
see the spin connection does not contribute to the action after fixing the $\kappa$-symmetry.

The above action takes a particularly nice and simple form if we adopt \soff \rep s for fermions
({\it cf.} Appendix \eqref{SO(4)fermions}). In that case $\theta$ and $\theta^\dagger$ are replaced 
with $\theta_{\alpha\beta},\ \theta_{\dot\alpha\dot\beta}$ and their complex conjugates, where
$\alpha$ and $\dot\alpha$ are Weyl indices of either of the $SO(4)$'s.
In this notation the action reads
\bea\label{Sferlcso4}
S^{fer}_{l.c.}=\frac{-i}{4\pi\alpha'}\int d\tau\int_0^{2\pi\alpha' p^+} d\sigma && 
\biggl[
\theta_{\alpha\beta}^\dagger\partial_\tau\theta^{\alpha\beta}+
\theta^{\alpha\beta}\partial_\tau\theta^\dagger_{\alpha\beta}+
\theta_{\alpha\beta}\partial_\sigma\theta^{\alpha\beta}+
\theta^{\dagger\alpha\beta}\partial_\sigma\theta^\dagger_{\alpha\beta}-2i\mu
\theta_{\alpha\beta}^\dagger\theta^{\alpha\beta}+\cr
&&
\theta_{\dot\alpha\dot\beta}^\dagger\partial_\tau\theta^{\dot\alpha\dot\beta}+
\theta^{\dot\alpha\dot\beta}\partial_\tau\theta^\dagger_{\dot\alpha\dot\beta}+
\theta_{\dot\alpha\dot\beta}\partial_\sigma\theta^{\dot\alpha\dot\beta}+
\theta^{\dagger\dot\alpha\dot\beta}\partial_\sigma\theta^\dagger_{\dot\alpha\dot\beta}-2i\mu
\theta_{\dot\alpha\dot\beta}^\dagger\theta^{\dot\alpha\dot\beta}\biggr]\ .
\eea
As we see $\theta_{\alpha\beta}$ and $\theta^{\dot\alpha\dot\beta}$ decouple from each other.
The coupled equations of motion for the fermions are
\bea\label{fermioneom}
(\partial_\tau+\partial_\sigma)(\theta_{\alpha\beta}+\theta^\dagger_{\alpha\beta})-
i\mu(\theta_{\alpha\beta}-\theta^\dagger_{\alpha\beta})&=& 0\ , \cr
(\partial_\tau-\partial_\sigma)(\theta_{\alpha\beta}-\theta^\dagger_{\alpha\beta})-
i\mu(\theta_{\alpha\beta}+\theta^\dagger_{\alpha\beta})&=& 0\ .
\eea
The solution to the above is
\bea
\label{fermionicmodes}
\theta=\frac{1}{\sqrt{p^+}} \beta_0 e^{i\mu\tau}
+\frac{1}{\sqrt{2p^+}}\sum_{n=1}^\infty &&
c_{-n} \Big[(1-\rho_{-n})\beta_n\ e^{\frac{-i}{\alpha' 
p^+}(\omega_n\tau+n\sigma)}+
(1+\rho_{-n}){\beta}^\dagger_n\  e^{\frac{+i}{\alpha' p^+}(\omega_n\tau+n\sigma)}\Big]
\cr &+&
c_{n} \Big[(1-\rho_{n})
\tilde{\beta}_n\ 
e^{\frac{-i}{\alpha' p^+}(\omega_n\tau-n\sigma)}+(1+\rho_{n})\tilde{\beta}^{\dagger}_n\ 
e^{\frac{+i}{\alpha' p^+}(\omega_n\tau-n\sigma)}\Big]
\eea
where $\omega_n$ is defined in \eqref{omega} and 
\be\label{rhoc}
\rho_{\pm n}=\frac{\omega_n\pm n}{\alpha'\mu p^+}\ ,\ \ \
c_{\pm n}=\frac{1}{\sqrt{1+\rho_{\pm n}^2}}\ .
\ee
In the above, since there was no confusion, we have dropped the fermionic indices. 
$\theta^{\dot\alpha\dot\beta}$'s also satisfy a similar equation, with similar solutions. 

Imposing the canonical quantization conditions
\be\label{thetalambda}
\{\theta^{\alpha\beta}(\sigma,\tau), 
\theta_{\rho\lambda}^\dagger(\sigma',\tau)\}=2\pi\alpha'
\delta^{\alpha}_{\ \rho}\delta^{\beta}_{\ \lambda}\delta(\sigma-\sigma') \, ,
\ee
leads to
\be\label{fermioniccommutations}
\{\beta_0, \beta^\dagger_0\}= 1,\ \ \ 
\{\beta_n, \beta^{\dagger}_m\}= \{\tilde{\beta}_n, \tilde{\beta}^{\dagger}_m\}=\delta_{mn}\ ,
\ee
where again we have suppressed the fermionic indices.

Using the \lc action and the mode expansion \eqref{fermionicmodes}, we work out the \lc 
Hamiltonian:
\be\label{fermionicH2}
H^{fer.}_{l.c.}=\frac{1}{\alpha' p^+}\left[\alpha'\mu p^+ \beta_0^{\dagger}\beta_0+
\sum_{n=1}^\infty \omega_n(\beta_n^{\dagger}\beta_n+
\tilde{\beta}^{\dagger}_n\tilde{\beta}_n)\right] -\frac{8}{\alpha' p^+}
\left(\frac{1}{2}\alpha' \mu p^++\sum_{n=1}^\infty \omega_n \right)\ ,
\ee
where in the above we have used $\beta_n^{\dagger}\beta_n$ as a shorthand for
$\beta_{n\alpha\beta}^{\dagger}\beta_n^{\alpha\beta}+ 
\beta_{n\dot\alpha\dot\beta}^{\dagger}\beta_n^{\dot\alpha\dot\beta}$ for $n\geq 0$.

In the full \lc Hamiltonian, which is a sum of bosonic and fermionic contributions, the zero 
point energies cancel and
\be\label{H2}
{\cal H}^{(2)}_{l.c.}=\frac{1}{\alpha' p^+}\left[\alpha'\mu p^+ (\alpha_0^{I\dagger}\alpha_0^I+
\beta_0^{\dagger}\beta_0)+\sum_{n=1}^\infty \omega_n(\alpha_n^{I\dagger}\alpha_n^{I}+
\tilde{\alpha}^{I\dagger}_n\tilde{\alpha}^I_n+\beta_n^{\dagger}\beta_n+
\tilde{\beta}^{\dagger}_n\tilde{\beta}_n)\right]\ .
\ee 

\subsection{Physical spectrum of closed strings on the \pl\ \bg }\label{physicalstringspectrum}

Having worked out the Hamiltonian and the mode expansions we are now ready to summarize and list 
the low lying string states in the \pl\ \bg . First, we note that the level matching condition 
\eqref{levelmatching} also receives contributions  from fermionic modes. Again using the
fact that $\frac{\delta (L_b+L_f)}{\delta g_{\tau\sigma}}=0$ we find that a term 
like $\theta^\dagger\theta$ should be added to the right-hand-side of \eqref{Virasoro} and hence 
the improved level matching condition in which the fermionic modes have been taken into account is
\be\label{totallevelmatching}
\sum_{n=1}^\infty n(\alpha_n^{I\dagger}\alpha_n^{I}
+\beta_n^{\dagger}\beta_n-\tilde{\alpha}^{I\dagger}_n\tilde{\alpha}^I_n
-\tilde{\beta}^{\dagger}_n\tilde{\beta}_n)|\Psi\rangle=0 \, ,
\ee
with $|\Psi\rangle$ a generic physical closed string state.

As usual the {\it free} string theory Fock space, ${\mathbb{H}}$, is \cite{Polchinski:1998rq}
\be\label{Fock}
{\mathbb{H}}
=|{\rm vacuum}\rangle \overset{\infty}{\underset{m=1}{\oplus}}\ 
{\mathbb{H}}_m \, ,
\ee
where ${\mathbb{H}}_m$, the $m$-string Hilbert space, is nothing but $m$-copies of (or the
direct product of $m$) single-string Hilbert spaces ${\mathbb{H}}_1$. The string theory vacuum 
state in the sector with \lc momentum $p^+$, which will be denoted by $|{\rm v}\rangle$, is the 
state that is annihilated by all $\alpha_n$ and $\beta_n$:
\be\label{rmv}
\alpha_n|{\rm v}\rangle=\tilde\alpha_n|{\rm v}\rangle=0\ , \ \ \ 
\beta_n|{\rm v}\rangle=\tilde\beta_n|{\rm v}\rangle=0\ , \ \ \ \forall n\geq 0\ .
\ee
{\it Convention:} Hereafter we will suppress the \lc momentum in the vacuum state and the \lc 
momentum $p^+$ is implicit in $|{\rm v}\rangle$. Again, we have defined
$\tilde\beta_0=\beta_0$ for later convenience.

This state is clearly invariant under \soff symmetry and has zero energy. However, it is possible 
to define some other ``vacuum'' states which are invariant under the full $SO(8)$. These states 
all necessarily have higher energies. Two such vacua which have been considered in the 
literature are 
\cite{Metsaev:2002re, Spradlin:2002ar}
\be\label{vacuums}
|0\rangle \equiv 
{\beta_0^{\dagger}}_{11}{\beta_0^{\dagger}}_{12}{\beta_0^{\dagger}}_{21}{\beta_0^{\dagger}}_{22}| 
{\rm v}\rangle\ ,\  \ {\rm or}
\ \ \ | \dot{0}\rangle \equiv 
{\beta_0^{\dagger}}_{\dot{1}\dot{1}}{\beta_0^{\dagger}}_{\dot{1}\dot{2}}
{\beta_0^{\dagger}}_{\dot{2}\dot{1}}{\beta_0^{\dagger}}_{\dot{2}\dot{2}}|{\rm  v}\rangle\ .
\ee
It is evident that both $|0\rangle$ and $|\dot{0}\rangle$ have energy equal to $4\mu$.
The interesting and important property of $| 0\rangle$ and $|\dot{0}\rangle$  is that they are 
\soe 
invariant and hence it is natural to assign them with  positive $\ztwo$ eigenvalues. (Note 
that as discussed in section \ref{isometry} $\ztwo$ is a specific \soe rotation). On the other 
hand it is 
not hard to check that under $\ztwo$ 
\[ 
\beta_{0\ 12}\llra \beta_{0\ 21}\ \ {\rm and}\ \  \beta_{0\ \dot{1}\dot{2}}\llra 
\beta_{0\ \dot{2}\dot{1}}\ .
\] 
Therefore $|{\rm v}\rangle$ and $|0\rangle$ should have opposite $\ztwo$ 
charges \cite{Chu:2002eu}; with the positive assignment for $|0\rangle$, $|{\rm v}\rangle$ 
should have negative $\ztwo$ eigenvalue. 
Giving negative $\ztwo$ charge to $|{\rm 
v}\rangle$ at first sight may look strange, however, this charge assignment is the more natural one 
noting the arguments of section \ref{sugraspectrum}. The $|{\rm v}\rangle$
vacuum state, which has zero energy (mass), in fact arises from a combination of   metric and the 
five-form field excitations. On the other hand since the full transverse metric is traceless, the traces of the \sof parts of the metric should have opposite signs and hence we expect $|{\rm v}\rangle$ to be odd under $\ztwo$.                                                                                  $|0\rangle$ and $|\dot{0}\rangle$, are coming from the excitations the of axion-dilaton field 
which is an \soe scalar and therefore the natural assignment is to choose them to be even under 
$\ztwo$ \cite{Pankiewicz:2003kj}.

Based on the vacuum state $|{\rm v}\rangle$, we can build the single string Hilbert space 
${\mathbb{H}}_1$ by the action of pairs of right and left-mover (bosonic or fermionic) modes on the vacuum. This would guarantee that the level matching condition \eqref{totallevelmatching} is satisfied.  Note that the above does {\it not} exhaust all
the possibilities when we have zero mode excitations. In fact if we only excite $n=0$ modes
the level matching condition \eqref{totallevelmatching} is fulfilled for any number of excitations. Therefore, we consider generic $n$ and $n=0$ cases separately.

\subsubsection{Generic single string states}\label{genericsinglestring}

These states are generically of the form

\begin{subequations}\label{bosonicsinglestrings}
\begin{align}
{\rm Bosonic\ modes:\ \ } & 
\alpha^{i\dagger}_n \tilde\alpha^{j\dagger}_n|{\rm v}\rangle ,\ \  
\alpha^{a\dagger}_n \tilde\alpha^{b\dagger}_n|{\rm v}\rangle ,\ \  
\alpha^{i\dagger}_n \tilde\alpha^{a\dagger}_n|{\rm v}\rangle ,\ \  
\alpha^{a\dagger}_n \tilde\alpha^{i\dagger}_n|{\rm v}\rangle ,\ 
\\ & 
\beta^{\dagger}_{n\alpha\beta} \tilde\beta^{\dagger}_{n\rho\lambda}|{\rm v}\rangle ,\ \  
\beta^{\dagger}_{n\dot\alpha\dot\beta} \tilde\beta^{\dagger}_{n\dot\rho\dot\lambda}|{\rm v}\rangle 
, \ \  
\beta^{\dagger}_{n\alpha\beta} \tilde\beta^{\dagger}_{n\dot\rho\dot\lambda}|{\rm v}\rangle ,\ \
\beta^{\dagger}_{n\dot\alpha\dot\beta} \tilde\beta^{\dagger}_{n\rho\lambda}|{\rm v}\rangle ,
\end{align}
\end{subequations}
\begin{subequations}\label{fermionicsinglestrings}
\begin{align}
{\rm Fermionic\ modes:\ \ } &
\alpha^{i\dagger}_n \tilde\beta^{\dagger}_{n\alpha\beta}|{\rm v}\rangle,\ \  
\alpha^{a\dagger}_n \tilde\beta^{\dagger}_{n\alpha\beta}|{\rm v}\rangle ,\ \  
\beta^{\dagger}_{n\alpha\beta} \tilde\alpha^{i\dagger}_n|{\rm v}\rangle ,\ \  
\beta^{\dagger}_{n\alpha\beta} \tilde\alpha^{a\dagger}_n|{\rm v}\rangle ,\ 
\\ & 
\alpha^{i\dagger}_n \tilde\beta^{\dagger}_{n\dot\alpha\dot\beta}|{\rm v}\rangle ,\ \  
\alpha^{a\dagger}_n \tilde\beta^{\dagger}_{n\dot\alpha\dot\beta}|{\rm v}\rangle ,\ \  
\beta^{\dagger}_{n\dot\alpha\dot\beta} \tilde\alpha^{i\dagger}_n|{\rm v}\rangle ,\ \  
\beta^{\dagger}_{n\dot\alpha\dot\beta} \tilde\alpha^{a\dagger}_n|{\rm v}\rangle ,\ 
\end{align}
\end{subequations}
with $n\neq 0$. All the above states have mass equal to $2\omega_n$,  though they are in different 
\soff \rep s. The first line of \eqref{bosonicsinglestrings} for which both of left and 
right-movers are coming from bosonic modes, in the usual conventions, comprise the ``NSNS'' sector 
and the second line of \eqref{bosonicsinglestrings} the ``RR'' modes.\footnote
{In the usual (flat space) conventions NSNS and RR modes come from the decomposition 
of two bosonic 
and two fermionic modes of $SO(8)$, respectively \cite{Green:1987fi}. It is worth noting that this 
classification does 
{\it not} hold in our case in the sense that two bosonic modes (or equivalently two bosonic 
insertions)
give rise to a combination of metric and the self-dual five-form, while two fermionic insertions
give rise to two-forms, NSNS and RR. So, as we see there is a mixture of the usual NSNS and RR
modes which appear from two bosonic or fermionic stringy modes. This is not surprising recalling
that in our case we are dealing with the \pl\ \bg\ and $SO(4)\times SO(4)$ \rep s instead of
flat space and $SO(8)$. Therefore, in our notations we reserve ``NSNS'' and ``RR'' (instead of 
NSNS and RR) to distinguish this difference with  flat space.\label{NSNSRR}}

It is instructive to work out the \soff \rep s of these modes. Here we will only study the bosonic 
modes and the fermionic modes are left to the reader.
First we note that 
{\it i)} $|{\rm v}\rangle$ is \soff singlet, 
{\it ii)} $\alpha^{i\dagger}_n$ and  $\alpha^{a\dagger}_n$ are respectively in 
$({\bf 4},{\bf 1})$ and $({\bf 1},{\bf 4})$ of \soff and 
{\it iii)} as discussed in Appendix \ref{SO(4)fermions},  
$\beta^{\dagger}_{n\alpha\beta}$ and $\beta^{\dagger}_{n\dot\alpha\dot\beta}$
are in $({\bf (2,1)},{\bf (2,1)})$ and $({\bf (1,2)},{\bf (1,2)})$ respectively.
Therefore,  $\alpha^{i\dagger}_n \tilde\alpha^{j\dagger}_n|{\rm v}\rangle$ 
is in the \soff \rep
\be\label{4141}
({\bf 4},{\bf 1})\otimes ({\bf 4},{\bf 1})=
({\bf 1},{\bf 1})\oplus ({\bf 9},{\bf 1})\oplus({\bf 3^+},{\bf 1})\oplus({\bf 3^-},{\bf 1}), 
\ee
where by ${\bf 3^\pm}$ we mean the self-dual (or antiself-dual) part of ${\bf 6}$ of $SO(4)$.
Likewise $\alpha^{a\dagger}_n \tilde\alpha^{b\dagger}_n|{\rm v}\rangle$ can be decomposed into
$({\bf 1},{\bf 1})\oplus ({\bf 1},{\bf 9})\oplus({\bf 1},{\bf 3^+})\oplus({\bf 1},{\bf 3^-})$.
$\alpha^{i\dagger}_n \tilde\alpha^{a\dagger}_n|{\rm v}\rangle$ and 
$\alpha^{a\dagger}_n \tilde\alpha^{i\dagger}_n|{\rm v}\rangle$ are both in $({\bf 4},{\bf 4})$
because
\be\label{4114}
({\bf 4},{\bf 1})\otimes ({\bf 1},{\bf 4})=({\bf 4},{\bf 4}) .
\ee
Now let us consider the ``RR'' modes; for two 
$\beta^{\dagger}_{n\alpha\beta}$ or $\beta^{\dagger}_{n\dot\alpha\dot\beta}$ excitations we note 
that
\begin{subequations}\label{2121}
\begin{align}
\left(({\bf {2},1}),({\bf  {2}, 1})\right)
\otimes \left(({\bf {2},1}),({\bf {2}, 1})\right) & =
({\bf 1},{\bf 1})\oplus ({\bf 3^+},{\bf 3^+})\oplus({\bf 3^+},{\bf 1})\oplus({\bf 1},{\bf 3^+}) ,\ 
\\
\left(({\bf 1,{2})},({\bf 1, {2}})\right)
\otimes \left(({\bf 1,{2})},({\bf 1,{2}})\right) &=
({\bf 1},{\bf 1})\oplus ({\bf 3^-},{\bf 3^-})\oplus({\bf 3^-},{\bf 1})\oplus({\bf 1},{\bf  3^-}) ,
\\
{\rm and\ for\ one\ } \beta^{\dagger}_{n\alpha\beta\ }  {\rm and\ one\ } 
\beta^{\dagger}_{n\dot\alpha\dot\beta\ }  {\rm type}\  {\rm excitations} 
\ \ \ \ \ \ \ \ \ \ \ \ \ &
{\hspace{9cm}}
\cr
\left(({\bf {2}, 1)},({\bf {2}, 1})\right)
\otimes \left(({\bf 1,{2})},({\bf 1,{2}})\right) & = ({\bf 4},{\bf 4})\ .
\end{align}
\end{subequations}

\subsubsection{Zero mode excitations}\label{zeromodes}

Now let us restrict ourselves to the excitations which only involve $\alpha^\dagger_0$ and 
$\beta^\dagger_0$ modes. Compared to the previous case, there are two specific 
features to note. One is that the left and right-movers are essentially the same (e.g. 
there is no independent $\tilde\alpha_0^\dagger$ or $\tilde\beta_0^\dagger$) and second, any number of excitations are physically allowed (there are no restrictions imposed by the level matching condition 
\eqref{totallevelmatching}).

Here we only consider strings with only two  excitations, i.e. those with mass equal to $2\mu$.
These modes are very similar to \eqref{bosonicsinglestrings}  and \eqref{fermionicsinglestrings}  after setting $n=0$.
This means that
the modes of the form $\alpha^{i\dagger}_0\alpha^{j\dagger}_0|{\rm v}\rangle$,  
are symmetric in  $i$ and $j$ indices. In other words, in the decomposition
\eqref{4141} only $({\bf 1},{\bf 1})\oplus ({\bf 9},{\bf 1})$ survive. Similarly, 
$\alpha^{a\dagger}_0\alpha^{b\dagger}_0|{\rm v}\rangle$ type states are in 
$({\bf 1},{\bf 1})\oplus ({\bf 1},{\bf 9})$ \rep . The $\alpha^{i\dagger}_0\alpha^{b\dagger}_0|{\rm 
v}\rangle$ states, however, would lead to a single $({\bf 4},{\bf 4})$ \rep . In sum the 36 
``NSNS'' zero modes are in 
$({\bf 1},{\bf 1})\oplus ({\bf 9},{\bf 1})\oplus ({\bf 1},{\bf 1})\oplus ({\bf 1},{\bf 9})\oplus 
({\bf 4},{\bf 4})$. 

In the decomposition of ``RR'' modes among (\ref{2121}a) and (\ref{2121}b) we should keep modes 
which
are antisymmetric. Explicitly they are
$\epsilon^{\alpha\rho}\beta^{\dagger}_{0\alpha\beta}\beta^{\dagger}_{0\rho\lambda}|{\rm 
v}\rangle$ in $({\bf 1},{\bf 3^+})$, 
$\epsilon^{\beta\lambda}\beta^{\dagger}_{0\alpha\beta}\beta^{\dagger}_{0\rho\lambda}|{\rm 
v}\rangle$ in $({\bf 3^+},{\bf 1})$,
$\epsilon^{\dot\alpha\dot\rho}\beta^{\dagger}_{0\dot\alpha\dot\beta}
\beta^{\dagger}_{0\dot\rho\dot\lambda}|{\rm v}\rangle$  in $({\bf 1},{\bf 3^-})$, and  
$\epsilon^{\dot\beta\dot\lambda}\beta^{\dagger}_{0\dot\alpha\dot\beta}
\beta^{\dagger}_{0\dot\rho\dot\lambda}|{\rm v}\rangle$  in $({\bf 3^-},{\bf 1})$ of 
$SO(4)\times SO(4)$.
Therefore altogether, the 28 ``RR'' modes are in  
$({\bf 3^+},{\bf 1})\oplus ({\bf 3^-},{\bf 1})\oplus ({\bf 1},{\bf 3^+})\oplus ({\bf 1},{\bf 3^-})\oplus 
({\bf 4},{\bf 4})$ \rep s. 
 
The above may be compared with the \sugra\ modes discussed in section \ref{sugraspectrum}. As we 
see there is a perfect matching.  This, basically indicates that there 
exists a {\it low energy} limit in the \pl\ \bg\ so that the effective dynamics of strings is 
governed by the \sugra\
modes; in such a limit, the lowest modes of strings created by 
$\alpha^\dagger_0$ and $\beta^\dagger_0$ would decouple from the rest of string spectrum. 
For such a decoupling to happen two necessary conditions should be met; 
first  $\omega_n\gg \alpha' \mu p^+$ for any $n\geq 1$, and second, strings should be 
``weakly coupled'', i.e. $g^{eff}_s\ll 1$. The former is satisfied if $\alpha'\mu p^+ \ll 1$. 

\subsection{Representation of the \pl\ superalgebra in terms of string modes}\label{stringsusy}

String theory on the \pl\ \bg\ in the \lcg that we discussed earlier has the same \susy\ as the 
\bg\ 
whose algebra was introduced in section \ref{maxsusy}. In this section we will explicitly construct 
the \rep s of that algebra in terms of string modes.

\subsubsection{Bosonic generators}\label{bosonicgenerators}

As in the flat space case \cite{Green:1987fi}, to find the \rep\ of 30 bosonic isometries of 
the \pl\ \bg\ in terms of string modes we start with their \rep s in terms of coordinates and their derivatives and then replace them with string worldsheet fields and their momenta respectively. Noting \eqref{Jq}, \eqref{JQ} and \eqref{HQ}, we learn that some of these bosonic generators should also have a part which is quadratic in stringy fermionic modes. Putting this all together we have
\be\label{PHstring}
{P}^+=p^+ {1\!\! 1}\ , \ \ \ \  
{P}^-={\cal H}^{(2)}_{l.c.}\ , 
\ee
\bea\label{JJstring}
{J}^{ij} &=& \int_{0}^{2\pi\alpha' p^+} d\sigma\left[(X^i 
P^j-X^jP^i)
-\frac{i}{4\pi \alpha'}\bigl(\theta^\dagger_{\alpha\beta}(\sigma^{ij})^\alpha_{\ 
\rho}\theta^{\rho\beta}
+\theta^\dagger_{\dot\alpha\dot\beta}(\sigma^{ij})^{\dot\alpha}_{\ 
\dot\rho}\theta^{\dot\rho\dot\beta}\bigr)\right]\cr 
{J}^{ab} &=& \int_{0}^{2\pi\alpha' p^+} d\sigma\left[(X^a 
P^b-X^bP^a)-\frac{i}{4\pi \alpha'}\bigl(\theta^\dagger_{\alpha\beta}(\sigma^{ab})^\beta_{\ 
\rho}\theta^{\alpha\rho}
+\theta^\dagger_{\dot\alpha\dot\beta}(\sigma^{ab})^{\dot\beta}_{\ 
\dot\rho}\theta^{\dot\alpha\dot\rho}\bigr)\right]\ ,
\eea
\bea\label{KLstring}
{K}^{I} &=& \int_{0}^{2\pi\alpha' p^+} d\sigma\left[
\sin\mu\tau P^I+\frac{\mu}{2\pi\alpha'}  X^I\cos\mu\tau\right],\cr
{L}^{I} &=& \int_{0}^{2\pi\alpha' p^+} d\sigma\left[
\cos\mu\tau P^I-\frac{\mu}{2\pi\alpha'}  X^I\sin\mu\tau\right]\ ,
\eea
Note that in the above ${P}^+$ is proportional to the identity \opt\  which is compatible with the discussions of section \ref{planewavesusy} that the $U(1)$ generated by ${P}^+$ is in the center of the superalgebra. It is straightforward to check that these generators really satisfy the desired algebras.

\subsubsection{Fermionic generators}\label{fermionicgenerators}

As discussed in section \ref{planewavesusy} there are two classes of supercharges, the kinematical and dynamical ones. Let us first focus on the kinematical supercharges. {}From 
\eqref{Jq}-\eqref{Hq} one can see that $q_{\alpha\beta}$ should be proportional to 
$\theta_{\alpha\beta}$, explicitly
\be\label{qstring}
q_{\alpha\beta}=\frac{\sqrt{2}}{2\pi\alpha'} \int_{0}^{2\pi\alpha' p^+} d\sigma\ 
\theta_{\alpha\beta}\ ,\ \ 
q_{\dot\alpha\dot\beta}=\frac{\sqrt{2}}{2\pi\alpha'} \int_{0}^{2\pi\alpha' p^+} d\sigma\ 
\theta_{\dot\alpha\dot\beta}\ ,
\ee

As for the dynamical supercharges, we note that unlike $q$'s which are in the {\it complex} 
\eights\ of $SO(8)$, they are in the {\it complex} ${\bf 8_c}$. Next we note that if $\theta$ is in 
\eights\ then $\gamma^I\theta$ is in \eightc\ (it has opposite \soe chirality). We also expect 
$Q$'s 
to contain first order $X$'s and $P$'s, so that their anticommutator would generate the 
Hamiltonian, which is quadratic in $X$'s and $P$'s. Putting these together and demanding  $Q$'s to \eqref{JQ}-\eqref{HQ} fixes them to be
\begin{subequations}\label{Qstring}
\begin{align}
Q^{(0)}_{\alpha\dot\beta}=\frac{1}{2\pi\alpha'}\int_{0}^{2\pi\alpha' p^+}\!\!\! d\sigma
\left[
(2\pi\alpha' P^i-i\mu X^i)(\sigma_i)_{\alpha}^{\ \dot\rho}\theta^\dagger_{\dot\rho\dot\beta}+
(2\pi\alpha' P^a+i\mu X^a)(\sigma_a)_{\dot\beta}^{\ \rho}\theta^\dagger_{\alpha\rho}+
i\partial_\sigma X^i (\sigma_i)_{\alpha}^{\ \dot\rho}\theta_{\dot\rho\dot\beta}+
i\partial_\sigma X^a (\sigma_a)_{\dot\beta}^{\ \rho}\theta_{\alpha\rho}
\right]\\
Q^{(0)}_{\dot\alpha\beta}=\frac{1}{2\pi\alpha'}\int_{0}^{2\pi\alpha' p^+}\!\!\! d\sigma
\left[
(2\pi\alpha' P^i-i\mu X^i)(\sigma_i)_{\dot\alpha}^{\ \rho}\theta^\dagger_{\rho\beta}+
(2\pi\alpha' P^a+i\mu X^a)(\sigma_a)_{\beta}^{\ \dot\rho}\theta^\dagger_{\dot\alpha\dot\rho}+
i\partial_\sigma X^i (\sigma_i)_{\dot\alpha}^{\ \rho}\theta_{\rho\beta}+
i\partial_\sigma X^a (\sigma_a)_{\beta}^{\ \dot\rho}\theta_{\dot\alpha\dot\rho}\right]
\end{align}
\end{subequations}
The  superscript $(0)$ on $Q$'s emphasizes that they are only linear in $X$ and $P$'s. 
As we will argue in section \ref{SFT}, however, when we consider interacting strings there are 
corrections to the Hamiltonian as well as the dynamical supercharges, and in fact both $H$ and $Q$ should be viewed as a power series expansion in the string coupling, and at zeroth order they match with $Q^{(0)}$ and ${\cal H}^{(2)}$ presented here. 

One may also try to insert the mode expansions and express the generators of the superalgebra in terms of string creation-annihilation \opt s. Doing so, it is easy to see that the ``kinematical'' generators, ${K}^I,\ {L}^I$, $q_{\alpha\beta}$ and $q_{\dot\alpha\dot\beta}$, which have a linear dependence on the string worldsheet fields, only depend on the zero modes. The ``dynamical'' generators, ${J}_{ij}$,  ${J}_{ab}$, $Q^{(0)}_{\alpha\dot\beta}$, 
$Q^{(0)}_{\dot\alpha\beta}$, and  ${\cal H}^{(2)}$, however, are quadratic and hence they depend on all the stringy operators.

%% file: BMN.tex
In section \ref{penroselimit} we demonstrated the fact that \pl s may generically arise as Penrose limits of given geometries and in particular the maximally supersymmetric  \pl\ appears as the Penrose limit of $AdS_5\times S^5$ geometry. On the other hand, as briefly discussed in the introduction \cite{Gubser:1998bc, Witten:1998qj, Aharony:1999ti}, 
type IIB string theory on $AdS_5\times S^5$ \bg\ is dual to the 
${\cal N}=4,\ D=4$ (super-conformal) gauge theory. In this section we show the latter duality can be revived for type IIB strings on the maximally supersymmetric \pl .

The basic idea of the BMN proposal \cite{Berenstein:2002jq}   is to start with the usual AdS/CFT duality and find what parallels the procedure of taking the Penrose limit in the dual gauge theory side. As we argued in section \ref{penroselimit} the process of taking the Penrose limit consists of finding a light-like geodesic and rescaling the other light-like direction, as well as all the other transverse directions, in the appropriate way given in \eqref{penrosescalings}. For the case  of \ads the geodesic was chosen as a combination of a direction in $S^5$ and the global time \eqref{x+-xy}. The generator of translation along this light-like geodesic, $P^-$, is then a combination of translation along the global time and rotation along the $S^1$ inside $S^5$ (\ref{x+-xy}a). According to the AdS/CFT duality, however, translation along global time corresponds to the dilatation operator (or equivalently Hamiltonian \opt\ in the radial quantization) of the ${\cal N}=4$ gauge theory on $\mathbb{R}^4$ while the rotation in the $S^1$ direction corresponds to a $U(1)$ of the R-symmetry. Explicitly the dilatation 
\opt\ ${\cal D}$ is the generator of $U(1)_D\in SU(2,2)\simeq SO(4,2)$ (the conformal group in 
four dimensions) and ${\cal J}$ is the generator of $U(1)_J\in SU(4)\simeq SO(6)$ R-symmetry (\cf\ \eqref{so42decompose}).

As an initial step towards building the \pl /SYM duality we state the proposal in this section.
As mentioned in the introduction, section \ref{Intro:BMNconjecture}, this duality can be stated as the operator equality (\ref{improvedBMN}) supplemented with a correspondence between the Hilbert spaces on both sides, where the operators act. In the first part of this section we show how the ${\cal N}=4$ gauge theory fields fall into the $SO(4)\times SO(4)$ representations, which is the first step in making the correspondence with the string theory. Then in the later parts of this section we state the duality and introduce the BMN operators. Our conventions for the ${\cal N}=4$ gauge theory fields and the action of the theory is summarized in Appendix 
\ref{ConventionD=4}. In section \ref{Extensions}, we discuss some generalizations and extensions
of the BMN proposal to orbifolds of the plane-wave and compactified plane-waves, and the BMN sector of the ${\cal N}=1$ Klebanov-Witten theory.

\subsection{Decomposition of ${\cal N}=4$ fields into ${\cal D},\ {\cal J}$ 
eigenstates}\label{DJdecompose}

The matter content of the ${\cal N}=4$ gauge multiplet naturally falls into the \rep s of
$SO(4,2)\times SO(6)$ (for more details see for example \cite{D'Hoker:2002aw}). 
However, in order to trace the Penrose limit in the gauge theory and state the BMN proposal we need to study their \rep s in the $SO(4)\times SO(4)\times U(1)\times U(1)$ subgroup of $SO(4,2)\times SO(6)$. The ${\cal N}=4$ gauge multiplet contains six real scalars, $\phi_I,\ I=1,\cdots, 6$, four gauge fields $A_a,\ a=1,2,3,4$, and eight complex Weyl fermions, $\psi_{\alpha}^A,\ \alpha=1,2$ and $A=1,2,3,4$ \cite{Wess:1992cp} (also see Appendix \ref{ConventionD=4}). Here we are only interested in $U(N)$ gauge theories where scalars and fermions are both in the adjoint \rep\ of the $U(N)$, so they are $N\times N$ hermitian matrices. $A_a$ are not in the adjoint \rep\ however (but they do transform in the adjoint for global transformations), and as in any gauge theory one might consider the covariant derivative of the gauge theory
\be\label{gaugecovarderivative}   
D_a=\partial_a+ iA_a
\ee
which is in the adjoint of the local $U(N)$. In all our arguments we will consider Euclidean gauge theory on $\mathbb{R}^4$ so the $a$ index of $D_a$ is an $O(4)$ index. We might, however, switch between field theories on $\mathbb{R}^4$ and its conformal map, $\mathbb{R}\times S^3$.

The eigenvalues of ${\cal J}$ will be denoted by $J$. Since $\cj$ is the generator of a $U(1)$ 
subgroup of $U(4)$ R-symmetry group, the gauge fields are trivial under it. That is,
\be
[{\cal J}, D_a]=0 \ ;
\ee 
in other words $D_a$ has charge $J=0$. The scalars, however, decompose into two sets. We choose ${\cal J}$ to make rotations in the $\phi^5$ and $\phi^6$ plane, i.e.
\be \label{define-Z}
Z=\frac{1}{\sqrt{2}}(\phi^5+i\phi^6)\ ;\ \ \  [{\cal J}, Z]=+Z\ , 
\ee
and hence $[{\cal J}, Z^\dagger]=-Z^\dagger$. Therefore $Z$ has $J=1$ (and $Z^\dagger,\ J=-1$).
The other four scalars, which will be denoted by $\phi_i,\ i=1,2,3,4$ commute with ${\cal J}$ and 
have $J=0$.
The 16 fermionic fields also decompose into two sets of eight with $J=\pm \frac{1}{2}$.

The eigenvalue of ${\cal D}$ will be denoted by $\Delta$.  For fields in the $\cn =4$ gauge 
multiplet at {\it 
free}  field theory level,  $\Delta=1$ for scalars and 
$D_a$ and $\Delta=\frac{3}{2}$ for fermions. Hereafter we will use $\Delta_0$ to denote the 
dimension of 
operators at free field level (the engineering dimensions) and $\Delta$ for the full interacting 
theory. More explicitly,
\bea\label{freedimensions}
[\cd , Z(0)]=(1+ O(g^2_{YM})) Z(0)\ &,&\ \ [\cd , Z^\dagger(0)]=(1+ O(g^2_{YM})) Z^\dagger(0) \cr
[\cd , \phi_i(0)]=(1+ O(g^2_{YM})) \phi_i(0)\ &, &\ \ 
[\cd , D_a(0)]=(1+ O(g^2_{YM})) D_a(0) \\ 
{}[\cd , \psi^A_{\ \alpha}(0) ]= (\frac{3}{2} + O(g^2_{YM}))\psi^A_{\ \alpha}(0)\ &,&\ \
[\cd , \psi^A_{\ \dot\alpha}(0) ]=(\frac{3}{2} + O(g^2_{YM}))\psi^A_{\ \dot\alpha}(0)\ .
\nonumber
\eea  

After taking out the two $U(1)$ factors (${\cal D},\ {\cal J}$) of the $SO(4,2)\times SO(6)$ 
(or $SU(2,2)\times SU(4)$), the bosonic part of four dimensional superconformal group, we remain with an $SO(4)\times SO(4)$ (one $SO(4)\in SO(4,2)$ and the other $SO(4) \in SO(6)$) subgroup. 
We also need to find the $SO(4)\times SO(4)$ \rep\ of the fields. Obviously $Z$ and 
$Z^\dagger$ are singlets of both $SO(4)$'s, the $({\bf 1},{\bf 1})$ \rep , $\phi_i$ are in 
$({\bf 1}, {\bf 4})$ and $D_a$ are in $({\bf 4}, {\bf 1})$.
The $SO(4)\times SO(4)$ \rep\ of fermions can be worked out noting the arguments of section
\ref{susycontract} and Appendix \ref{sixdimfermion}. Explicitly, we first note that $SO(4)\simeq
SU(2)\times SU(2)$ and 
as for the usual four dimensional Euclidean Weyl fermions, they are in $({\bf {2},1})$ 
or $({\bf 1, {2}})$ of each $SO(4)$'s (\cf\ Appendix \ref{SO(4)fermions}). The 
$SO(4)\times SO(4)\times U(1)\times U(1) $ \rep s of all fields of the ${\cal N}=4$ gauge multiplet 
have been summarized in TABLE \ref{tableone}. Note that $\Delta_0-J$ for all the fields in 
TABLE \ref{tableone}, bosonic and fermionic, is integer-valued.

\begin{table}[ht] 
\begin{center}
\begin{tabular}{|c|c|c|c|} \hline
Field & $\Delta_0-J$ & $\Delta_0+J$ & $SO(4)\times SO(4)$ \\ \hline\hline
$Z$ & 0 & 2& $({\bf 1},{\bf 1})$ \\ 
$Z^\dagger$ & 2 & 0& $({\bf 1},{\bf 1})$\\ 
$\phi_i$ & 1 & 1& $({\bf 1},{\bf 4})$\\ 
$D_a$ & 1 & 1& $({\bf 4},{\bf 1})$\\ \hline
$\psi_{\alpha\beta}$ & 1 & 2& $\left(({\bf{2},1}),({\bf {2},1})\right)$ \\ 
$\psi_{\dot\alpha\dot\beta}$ & 1 & 2& $\left(({\bf 1,{2}}),({\bf 1,{2}})\right)$ \\ 
$\psi_{\alpha\dot\beta}$ & 2 & 1& $\left(({\bf {2},1}),({\bf 1, {2}})\right)$ \\ 
$\psi_{\dot\alpha\beta}$ & 2 & 1& $\left(({\bf 1,{2}}),({\bf 
{2}, 1})\right)$ \\ 
\hline
\end{tabular}
\caption{
$SO(4)\times SO(4)\times U(1)\times U(1) $ \rep s of all fields of the ${\cal N}=4$ gauge 
multiplet. The dimensions are those of the free theory. For the $J$ charge 
of fermions note that $\psi_{\alpha\beta}$ and $\psi_{\dot\alpha\beta}$ are related 
by CPT and hence have opposite $J$ charge; similarly for the other two fermionic modes.} 
\label{tableone}
\end{center}
\end{table}
 
\subsection{Stating the BMN proposal}\label{BMNconjecture}

Having worked out the $SO(4)\times SO(4)\times U(1)_D\times U(1)_J $ \rep\ of the ${\cal N}=4$ 
fields, we are ready to take the BMN limit, restricting to the \opt s with 
parameterically large R-charge $J$, but finite $\Delta_0-J$. In fact, starting with the AdS/CFT 
correspondence, the BMN limit on the gauge theory side parallels the Penrose limit on the gravity side, according which 
\begin{subequations}\label{cdcj}
\begin{align}
-i\frac{\partial}{\partial\phi} & \longleftrightarrow\ {\cal J} \\
i\frac{\partial}{\partial\tau} & \longleftrightarrow\ {\cal D}
\end{align}
\end{subequations}
Then, \eqref{x+-xy} or \eqref{DJP+-} imply that 
\begin{subequations}\label{partialxpm}
\begin{align}
i\mu\frac{\partial}{\partial x^-}  =\frac{i\alpha'}{2R^2}(
\frac{\partial}{\partial\tau}-\frac{\partial}{\partial\phi}) 
& \longleftrightarrow\ \frac{1}{2\sqrt{g^2_{YM}N}}({\cal D}+{\cal J}) \, , \\
\frac{i}{\mu}\frac{\partial}{\partial x^+}  
=i(\frac{\partial}{\partial\tau}+\frac{\partial}{\partial\phi}) 
& \longleftrightarrow\ {\cal D}-{\cal J} \, ,
\end{align}
\end{subequations}
where in (\ref{partialxpm}a) we have used (\ref{AdSraduis}). On the gravity (string 
theory) side $i\frac{\partial}{\partial x^-}$ and $i\frac{\partial}{\partial x^+}$ are the \lc 
momentum and the \lc Hamiltonian, respectively. Taking the Penrose limit \eqref{x+-xy} is then 
equivalent to  taking $g^2_{YM}N$ and $J$ to infinity while keeping $\frac{J^2}{g^2_{YM}N}$ fixed 
(see \eqref{BMNlimit1} and \eqref{BMNlimit2}).
According to (\ref{partialxpm}a) the value of $\frac{J^2}{g^2_{YM}N}$ is equal to the string \lc momentum (squared) on the string theory side (see (\ref{BMNlimit1}b)).

In summary, part one of the  \pl /SYM duality can be stated as
\vskip .5cm

{\it The \lc string field theory Hamiltonian in the \pl\ \bg\ is equal to the difference between 
the dilatation operator ${\cal D}$ and the R-charge operator ${\cal J}$:}
\be\label{BMNconjecture1}
\frac{1}{\mu} H_{SFT}= {\cal D}-{\cal J}\ ,
\ee
{\it in the sector of the gauge 
theory consisting of gauge invariant operators with parametrically large R-charge, the BMN sector.}
\vskip .5cm

The more detailed discussion about the construction and form of the BMN operators and also 
correspondence between the Hilbert spaces on string and gauge theory sides, i.e., part two of the \pl /SYM duality, will be presented in the next subsection.

\subsection{The BMN operators}\label{BMNoperators} 

As mentioned earlier in the \pl /SYM duality the relevant \opt s in the gauge theory side are those with large R-charge $J$; these are the so-called BMN \opt s where $\cd -\cj $ acts.
In this section we present such gauge invariant \opt s. The BMN \opt s can be classified
by the number of traces (over the $N\times N$ gauge theory indices) involved, and also the value of $\Delta_0-J$. In fact, because of the BPS bound \cite{D'Hoker:2002aw}
$\Delta \geq J$ and when $\Delta=J$ the BPS bound is saturated.
This can be seen from TABLE \ref{tableone} and the fact that the value of $\Delta_0-J$ for 
composite operators is just the sum of $\Delta_0-J$ of the basic fields present in the composite 
operator. Besides the value of $\Delta_0-J$ and number of traces to completely specify the \opt\ 
we need to identify its $SO(4)\times SO(4)$ \rep .

\subsubsection{BMN \opt s with $\Delta_0-J=0$}\label{deltaj0}

The first class of the BMN \opt s we consider are those with $\Delta_0-J=0$, in the usual $\cn 
=4$ conventions  these are 
chiral-primary operators \cite{D'Hoker:2002aw}. According to 
TABLE \ref{tableone}, such \opt s can only be 
composed of $Z$ 
fields. Therefore they are necessarily $SO(4)\times SO(4)$ singlets and hence this class of BMN 
\opt s 
is completely specified with the number of traces, the simplest of which is of course the single 
trace \opt   
\begin{equation}\label{singlevac} 
\co^J (x) = \frac1{\sqrt{JN_0^J}}\Tr Z^J (x)\ ,\ \ \ N_0=\frac{1}{8\pi^2}g^2_{YM}N\ .
\end{equation}
The normalization is fixed so that the planar two-point function of $\co^J(x)$ and 
$\co^{J\dagger}(0)$ is equal to $\frac{1}{|x|^{2J}}$; we will come back to this point in 
section \ref{noninteractingstrings}.A. We would like to stress that the point  $x$  where the above 
operator is defined is in $\mathbb{R}^4$. One can then define a state by acting \eqref{singlevac} 
on the vacuum of the gauge theory on $\mathbb{R}^4$, which will be denoted by $|vac\rangle$. In 
this way there is a natural  
one-to-one correspondence between BMN states and BMN operators. Hence, in this review we will not 
distinguish between BMN operators and BMN states and they will be used interchangeably.   
According to the second part of BMN proposal the above single-trace \opt\ (or state) corresponds to a single string state on the string theory side:
\be
|{\rm v}\rangle \longleftrightarrow \ \co^J (0)|vac\rangle \, ,
\ee
where $|{\rm v}\rangle$ is the single-string vacuum with the \lc momentum $p^+$ \eqref{rmv} . 

The next state belonging to this class is the double-trace \opt
\begin{equation}\label{doublevac} 
\ct^{J,r} = (\co^{r\cdot J} \co^{(1-r)\cdot J}) (x) = \frac1{J\sqrt{r(1-r)N^J}}
:\Tr Z^{J_1} (x)\Tr Z^{J-J_1} (x): 
\end{equation}
where ${J_1}/{J}=r$ and $J_1$ ranges between one and $J-1$. Of course the above operator is a 
BMN operator if $J_1$ is of the
order of $J$. In a similar way \eqref{doublevac} was proposed to
correspond to the double-string state with the total \lc momentum $p^+$, with the partition
$r\cdot p^+$ and $(1-r)\cdot p^+$. One can then  straightforwardly generalize the above to 
multi-trace \opt s. 

We would like to point out that each of the $\co^J$ or $\ct^{J,r}$ operators are chiral-primaries. In 
other words they are  half BPS states of the four dimensional superconformal algebra $PSU(2,2|4)$.  
Being chiral-primary these operators (states) are eigenstates of the dilatation operator and have $\Delta-J=0$ exactly \cite{D'Hoker:2002aw}. We should stress that from  the $PSU(2|2)\times PSU(2|2)\times U(1)_-$ 
superalgebra discussed in section \ref{planewavesusy}, however, these \opt s 
form a complete supermultiplet, which in this case is in fact a singlet, and are still half BPS
in the sense that all the dynamical supercharges $Q_{\alpha\dot\beta}$ and $Q_{\dot\alpha\beta}$ annihilate them.

\subsubsection{BMN \opt s with $\Delta_0-J=1$ \label{deltaj1}}

The next level of states are those with $\Delta_0-J=1$. In order to obtain such BMN states we should insert one of the fields in TABLE \ref{tableone} which have $\Delta_0-J=1$ into \eqref{singlevac} 
or \eqref{doublevac}. Therefore, there are eight bosonic states (corresponding to insertions of 
$\phi_i$ 
or $D_a$) and eight fermionic states (corresponding to insertions of $\psi_{\alpha\beta}$ or 
$\psi_{\dot\alpha\dot\beta}$). Each of these insertions may be viewed as {\it impurities} 
in 
the line of $Z$'s. Due to cyclicity of the trace it does not matter where in the sequence of $Z$'s 
these impurities are inserted. These $8+8$ states 
complete a supermultiplet of $PSU(2|2)\times PSU(2|2)\times U(1)_-$ superalgebra.
Here, we should  emphasize that in the full superconformal $PSU(2,2|4)$ algebra \rep s, however,
these states are {\it descendents} of chiral primaries and are in the same 
short supermultiplet as chiral primaries. From the $PSU(2|2)\times PSU(2|2)\times U(1)_-$ 
superalgebra point of view
they are in different multiplets than chiral primaries with $\Delta-J=0$.

As examples we present two such single-trace \opt s

\begin{equation}\label{oneimpurity}
\co^J_i = \frac1{\sqrt{N_0^{J+1}}}\Tr\left(\phi_i Z^J\right)\ ,\ \ \ 
\co^J_a = \frac1{\sqrt{N_0^{J+1}}}\Tr\left(D_aZ Z^{J-1}\right)\ .
\end{equation}
These \opt s correspond to $\alpha^{i\dagger}_0$ or $\alpha^{a\dagger}_0$ on the string theory 
side. Note that
in the closed string theory a physical state should satisfy the level matching condition 
\eqref{totallevelmatching} and is generically
composed of (equal energy excitations) of left and right modes. 
The operators \eqref{oneimpurity}, however, correspond to ``zero momentum'' string states and 
satisfy the level matching condition.

In the same spirit as \eqref{doublevac} the double-trace $\Delta_0-J=1$ BMN operators can be 
obtained by combining $\co^J$ with \eqref{oneimpurity}, e.g.
\begin{equation}\label{oneimpuritytwotrace} 
\ct^{J,r}_i = (\co^{r\cdot J}_i \co^{(1-r)\cdot J}) (x) = \frac1{\sqrt{(1-r)\cdot J N_0^J}}
:\Tr \phi_i Z^{J_1} (x)\Tr Z^{J-J_1} (x): 
\end{equation}
where, as in \eqref{doublevac}, $r$ is the ratio $J_1/J$; we will use this notation
throughout the rest of this paper.

We would like to note that all the \opt s of this class, e.g. those presented in 
\eqref{oneimpurity} and \eqref{oneimpuritytwotrace}, are descendents of chiral-primaries and are 
exact eigenstates of $\cd-\cj$, with $\Delta-J=1$.

\subsubsection{BMN \opt s with $\Delta_0-J=2$}\label{deltaj2}

To obtain BMN \opt s with $\Delta_0-J=2$  we can either have two insertions of fields with 
$\Delta_0-J=1$ or a single insertion of a $\Delta_0-J=2$ field from TABLE \ref{tableone} into the sequence of $Z$'s.\footnote{Most of the papers which have appeared so far have only considered insertions of bosonic fields, and even among the bosonic insertions the focus has mainly been on the 
$\phi_i$ fields. The $D_a$ insertions have been considered in \cite{Klose:2003tw, Gursoy:2002yy}. 
Two fermionic insertions has been briefly discussed in \cite{Eden:2003sj}.}  For the 
case of two $\Delta_0-J=1$ 
insertions, the position of the insertions
is important, however, due to the cyclicity of the trace only the relative positions of the 
insertions is relevant. We fix our conventions so that one of the impurity fields always appears at the beginning of the sequence. In the single $\Delta_0-J=2$ insertion, similar to the case of \ref{deltaj1}, the insertion position is immaterial.
For the single-trace \opt s with two $\Delta_0-J=1$ insertions, there are $J+1$ choices, depending on the relative positions of the insertions, which we may use as their ``discrete'' Fourier modes. To begin, let us consider the case where both of the insertions are of the $\phi_i$ form:
\be\label{ijBMNstringmode}
\co^J_{ij,\ n} =\frac1{\sqrt{JN_0^{J+2}}}\left[\sum_{p=0}^J\ e^{2\pi i pn/J}\ 
\Tr(\phi_i Z^p\phi_j Z^{J-p})-\delta_{ij}\Tr (Z^\dagger Z^{J+1}) \right]\ .
\ee
As we will show in section \ref{noninteractingstrings}, once we turn on the gauge theory coupling, individual \opt s of the form 
\be\label{Op}
\widetilde{\co}_p\equiv \Tr(\phi_i Z^p\phi_j Z^{J-p})
\ee 
are no longer eigenvectors of the dilatation \opt\ $\cd$. However, the 
$\co^J_{ij,\ n}$ \opt s, at {\it planar level} (and of course in the large $J$ limit) 
have definite $\cd$ eigenvalue (scaling dimension).   
Using \eqref{ijBMNstringmode} it is easy to check that 
\be\label{ij,ji}
\co^J_{ij,\ n}=\co^J_{ji,\ -n}\ .
\ee

One may then consider two $D_a$ or one $\phi_i$ and one $D_a$ insertions:
\be\label{abBMNstringmode}
\co^J_{ab,\ n} =\frac{1}{2}\cdot\frac1{\sqrt{JN_0^{J+2}}}\left[\sum_{p=0}^J\ e^{2\pi i pn/J}\ 
\Tr((D_aZ) Z^p (D_b Z) Z^{J-p})+\Tr ((D_a D_b Z) Z^{J+1}) \right]\ .
\ee
\be\label{aiBMNstringmode}
\co^J_{ia,\ n} =\frac{1}{\sqrt 2}\cdot\frac1{\sqrt{JN_0^{J+2}}}\left[\sum_{p=0}^J\ e^{2\pi i pn/J}\ 
\Tr(\phi_i Z^p(D_aZ) Z^{J-p})+\Tr ((D_a\phi_i) Z^{J+1}) \right]\ .
\ee
Note that in the above equation of motion, $D_aD_a Z=0$ should also be imposed on the fields.
The normalization of $\co^J_{ij,\ n}$
\opt s have been fixed 
so that the two point function of these operators, in the planar free gauge theory limit, is of the 
form $\langle vac| \co^{\dagger J}_{ij,\ n}(x) \co^J_{i'j',\ n}(0)\vac 
=\delta_{ii'}\delta_{jj'}\frac{1}{|x|^{2(J+1)}}$, and similarly for   
$\co^J_{ia,\ n}$ and $\co^J_{ab,\ n}$ \opt s. The difference in factors of $1$, $\frac{1}{2}$ 
and $\frac{1}{\sqrt{2}}$ in the normalization is a consequence of our conventions in 
which  $\langle vac| Z^\dagger(\frac{x^\mu}{|x|})Z(0)\vac =1$ and $\langle vac| 
\phi_i^\dagger(\frac{x^\mu}{|x|})\phi_j(0)\vac =\delta_{ij}$, while   
$\langle vac| (D_a Z)^\dagger(\frac{x^\mu}{|x|})(D_bZ(0))\vac =2\delta_{ab}$.

The second part of the \pl /SYM duality which is a map between the string theory Hilbert space and BMN \opt s can then be stated as
\vskip .5 cm
{\it The \opt s $\co^J_{ij,\ n}$, $\co^J_{ab,\ n}$ and  $\co^J_{ai,\ n}$ correspond to 
the ``NSNS'' modes of the single-string sector of free closed string theory on the \pl\ \bg\ (\cf\ section \ref{genericsinglestring} and \ref{zeromodes}). Explicitly,} 
\bea\label{BMNconjecture2}
\co^J_{ij,\ n} &\llra & \alpha^\dagger_{i, n}{\tilde\alpha}^\dagger_{j, n} \, , \cr
\co^J_{ab,\ n} &\llra & \alpha^\dagger_{a, n}{\tilde\alpha}^\dagger_{b, n}\, ,
\ \ \ \ \ \ \ \forall n\geq 0 \\
\co^J_{ia,\ n} &\llra & \alpha^\dagger_{i, n}{\tilde\alpha}^\dagger_{a, n} \, ,\nonumber
\eea
{\it where $\alpha^\dagger_{i, n}$ and ${\tilde\alpha}^\dagger_{i, n}$ are the left and
right-moving string modes 
defined in \eqref{bosonicmode}. The ``RR'' and ``NSR'' or ``RNS'' modes (note the comment in 
footnote \ref{NSNSRR}) and 
all of the fermionic modes, can be obtained in a 
similar way through insertions of fermionic $\psi^A$-fields, two $\psi^A$-fields for the bosonic
modes and one $\psi^A$ and 
one $\phi_i$ or $D_a$ for fermionic  modes. On the string theory side the inner product on the 
Hilbert space is the usual one in which $m$ and $n$ string states are orthogonal to each other 
unless $m=n$. On the gauge theory side, however, the inner product corresponds to the two-point 
function of the corresponding BMN \opt s.}  
\vskip .5 cm

We should warn the reader that identifying the inner product on the Hilbert space with the two-point functions on the gauge theory side already suggests that the correspondence
\eqref{BMNconjecture2} should be modified because the two-point functions of the single and double trace \opt s  generically do not vanish. This will bring some more complications into our dictionary which will be discussed in detail in section \ref{mixing}.
\vskip .5cm

The operators $\co^J_{ij,\ n}$, $\co^J_{ab,\ n}$ and  $\co^J_{ai,\ n}$ form the bosonic 
states of a $PSU(2|2)\times PSU(2|2)\times U(1)_-$ supermultiplet. Note that from the 
superconformal
$PSU(2,2|4)$ algebra point of view they are only a part of the bosonic states of a supermultiplet.\footnote{The \opt s we 
have presented here are given in the BMN ($J\to\infty$) limit. However, as discussed in 
\cite{Beisert:2002tn}, there is a generalization of such \opt s for finite $J$, 
based on \susy .
The form of such \opt s is slightly different than the BMN ones, differing in the Fourier phase factor, where in $2\pi i np/J$, $J$ should be replaced with $J+3$. These \opt s are in fact ``generalized'' Konishi \opt s, interpolating between the usual Konishi operators \cite{Konishi:1984hf} (at $J=0\ {\rm or}\ 1$) and $J=\infty$, the BMN \opt s.} In general since the supercharges of the $PSU(2|2)\times PSU(2|2)\times U(1)_-$ commute with the Hamiltonian, $P^-$, (\cf\ \eqref{HQ}) all the states in the same supermultiplet must have the same energy or mass. This should be contrasted with the $PSU(2,2|4)$ superconformal algebra where states with different $\Delta-J$ appear in the same multiplet, e.g. chiral-primaries and their descendents \eqref{singlevac}, \eqref{oneimpurity} and $\Delta_0-J=2$ with $n=0$ BMN \opt s discussed earlier, fall  into the same $PSU(2,2|4)$ supermultiplet \cite{Beisert:2002tn}.

These \opt s, as they are written, 
are not in irreducible \rep s of $SO(4)\times SO(4)$. Following  the discussions of section 
\ref{physicalstringspectrum} (\cf\ \eqref{4141}) one can decompose
$\co^J_{ij,\ n}$ into
$\frac{1}{2}\sum_{i=1}^4 \co^J_{ii, n}$ in $({\bf 1},{\bf 1})$,  
$\co^J_{(ij), n}=\frac{1}{2}(\co^J_{ij, n}+\co^J_{ji, n})$ in
$({\bf 9},{\bf 1})$,  and $\frac{1}{2}(\co^J_{[ij], 
n}\pm \frac{1}{2}\epsilon_{ijkl}\co^J_{[kl], n})$ (where $\co^J_{[ij], n}=\frac{1}{2}(\co^J_{ij, 
n}-\co^J_{ji, n})$) in $({\bf 3^{\pm}},{\bf 1})$ \rep s of $SO(4)\times SO(4)$.
Similar decompositions can be made for $\co^J_{ab,\ n}$ states. Noting \eqref{4114}, 
the $\co^J_{ai,\ n}$  states form a $({\bf 4},{\bf 4})$ of $SO(4)\times SO(4)$. For the cases where we have two fermionic  $\psi$-field insertions
the decomposition can be carried out using 
(\ref{2121}a)
if we have two $\psi_{\alpha\beta}$ insertions and (\ref{2121}b)
if we have two $\psi_{\dot\alpha\dot\beta}$ insertions. We might also have one 
$\psi_{\alpha\beta}$ and one $\psi_{\dot\alpha\dot\beta}$ insertion, whose decomposition can 
be read from (\ref{2121}c).

\vskip .3cm

The $n=0$ case, i.e. $\co^J_{ij,\ 0}$, $\co^J_{ab,\ 0}$ and  $\co^J_{ai,\ 0}$, correspond 
to supergravity modes of the strings in the \pl\ \bg . 
At first sight it may seem that we should not expect to find \sugra\ modes and the 
results of \ref{sugraspectrum} from gauge theory, because the truncation of stringy excitations to 
the \sugra\ modes only makes sense when all the other excitations are much heavier than the lowest 
modes, which noting \eqref{omega} is when $\alpha' \mu p^+\ll 1 $. As we will see in the next 
section
this is the limit where the ``improved'' 't Hooft coupling \eqref{lambda'} is very large and one 
cannot trust the gauge theory analysis. However, one should note that
from superalgebra point of view these states are a part of short (BPS) multiplets of 
the $PSU(2,2|4)$ superconformal algebra \cite{Beisert:2002tn}  as well as the \pl\ 
superalgebra $PSU(2|2)\times PSU(2|2)\times U(1)_-$,
and hence it is natural to expect them to be protected by \susy .
Noting \eqref{ij,ji} we see that  
$({\bf 3^+},{\bf 1})$, $({\bf 3^-},{\bf 1})$, $({\bf 1},{\bf 3^+})$ and  $({\bf  1},{\bf  3^-})$ 
\rep s are absent in these supergravity modes. These 
\rep s which correspond  the fluctuations of  type IIB NSNS or RR two-form fields 
(see section \ref{sugraspectrum}), can arise from  two fermionic insertions. Note that for 
supergravity modes ($n=0$ case), due to the fact that fermions anticommute, we only remain 
with the totally antisymmetric \rep s of (\ref{2121}a) and (\ref{2121}b) which are
$({\bf 3^+},{\bf 1})$, $({\bf 3^-},{\bf 1})$ and $({\bf 1},{\bf 3^+})$,  $({\bf  1},{\bf 3^-})$. 
Then the two $({\bf 4},{\bf 4})$ \rep s arising from $\co^J_{ai,\ 0}$ and 
$\psi_{\alpha\beta}$, $\psi_{\dot\alpha\dot\beta}$ insertions form the 
32 modes of metric and self-dual five-form fluctuations.
This is compatible with the results of sections \ref{sugraspectrum} and \ref{zeromodes}. 
These $n=0$  \opt s are descendents of chiral-primaries (they are in fact 1/4 BPS) and hence we  
expect them to be exact eigenstates of $\cd-\cj$ with $\Delta-J=2$.

We may also build double-trace \opt s with $\Delta_0-J=2$. One can easily recognize two different 
possibilities; a combination of \eqref{singlevac} type \opt s and \eqref{ijBMNstringmode} type or
two \eqref{oneimpurity} type \opt s:
\bea\label{doubletracedeltaj2}
\ct^{J,r}_{ij, n}&=&:\co^{r\cdot J}_{ij, n} \co^{(1-r)\cdot J}: \cr
\ct^{J,r}_{ij}&=&:\co^{r\cdot J}_{i} \co^{(1-r)\cdot J}_j: 
\eea
These \opt s are conjectured to correspond to double-string states. As we will see in 
section \ref{interactingstrings}, once the string coupling is turned on and we have the 
possibility of strings joining and splitting, because of \opt\ mixing effects, there is a mixture of single, double and multi-trace \opt s which correspond to string states diagonalizing the string field theory Hamiltonian. We remind the reader that, as stated in section \ref{tHooft}, string loop diagrams correspond to non-planar graphs in the gauge theory.

Finally we would also like to note that the set of BMN operators we have introduced in this 
subsection is  invariant under the $\mathbb{Z}_2$ action which exchanges the two $SO(4)$ factors.

\subsubsection{BMN \opt s with arbitrary number of impurities}\label{multiimpurity}

The above discussion can  readily be generalized to arbitrary number of impurities to obtain
BMN \opt s with $\Delta_0-J=k$. These states can be constructed by $k$ insertions 
of $\Delta_0-J=1$ operators
or in general $p$ insertions with $\Delta_0-J=2$ and $q$ $\Delta_0-J=1$ insertions where $k=2p+q$. As the previous cases the inserted field can be any of the fields of TABLE \ref{tableone}, bosonic or fermionic. If the 
number of fermionic fields is odd we obtain a BMN \opt\ which corresponds to a fermionic string 
excitation, otherwise the state corresponds to a bosonic string mode.  As before, single-trace
BMN \opt s correspond to higher excitations of single {\it free} string and double-trace ones to
higher excitations of double free string states and so on. As an example we present 
a generic BMN operator with $k$ $\Delta_0-J=1$ insertions. This \opt\ is indexed by two sets of integers, $i_j,~j=1,...,k$ which shows the $SO(4)\times SO(4)$ structure and $n_j,~j=1,...,k$, 
subject to $\sum_{j=1}^k n_j=0$, which  gives the (worldsheet) momentum:
\be
{\co}^{J}_{i_1 i_2\cdots i_k,\ n_1\dots n_k}={\cal N}_{J,n}
\sum_{\stackrel{p_0,\dots ,p_k=0}{p_0+\dots+p_n=J}}^J
\left(\prod_{l=1}^k e^{2\pi i (p_0+\dots+p_k)\frac{n_l}{ J}}\right)
\Tr\left[\phi_{i}Z^{p_0}\prod_{j=0}^{k}(\phi_{i_j}Z^{p_j})\right]\ . 
\label{multiimpurityBMNoperator}
\ee
The function ${\cal N}_{J,k}$ is the normalization factor and 
is chosen such that the {\it planar} two point function of these operators at free field theory 
limit is $\frac{1}{ |x|^{2(J+k+1)}}$. It is easily seen that the $N$ and $g^2_{YM}$ 
dependence of ${\cal N}_{J,k}$ is ${N_0^{-(J+k+1)/2}}$.

We should warn the reader that the \opt s of the 
the form \eqref{multiimpurityBMNoperator} are {\it not} precise BMN \opt s in the sense that 
they are only made out of $\Delta_0-J=1$ insertions. As we see from \eqref{ijBMNstringmode}, 
\eqref{abBMNstringmode} and \eqref{aiBMNstringmode}, generically we require $\Delta_0-J=2$ insertions as well. Obtaining the exact form of $\Delta_0-J=2$ insertions is generally a hard task involving detailed calculations with two point functions of generic $\Delta_0-J=k$ BMN \opt s, which so far has not been performed in the literature. Since we do not find it illuminating
we skip this problem here.
As a generalization of the second part of the \pl/SYM duality, the operators of the type 
(\ref{multiimpurityBMNoperator}) are conjectured to be in one-to-one 
correspondence with the following string states in the plane-wave background:
\be
{\co}_{i_1 i_2\cdots i_k,\ n_1\dots n_k} \llra
\prod_{j'=1}^{k'} {\alpha}^\dagger{}_{i_{j'}\ n_{j'}} \prod_{j=k'+1}^k 
\tilde\alpha^\dagger{}_{i_j,\ n_j}|{\rm v}\rangle\ , 
\ee
subject to $\sum_{j'=1}^{k'} n_{j'}=\sum_{j=k'+1}^{k} n_j$.
Similarly to the two impurity case,  the above correspondence should be 
modified at finite string coupling, due to mixing between single and multi-trace \opt s.

\subsection{Extensions of the  BMN proposal}\label{Extensions}

The \pl /SYM duality, as presented earlier in this section, gives a correspondence between type IIB 
closed strings on the maximally \susyc\ \pl\  and the BMN sector of $\cn=4$ SCFT. This 
duality can be (and in fact has been) extended to several other cases. One of the interesting 
extensions is to 
unoriented open and closed strings on the {\it orientifold} of the maximally \susyc\ \pl\  which 
has been conjectured to be dual to the BMN sector of $Sp(N)$  gauge theory 
\cite{Berenstein:2002zw, Gomis:2003kb}.

As argued in section \ref{penroseguven}, Penrose limits of $AdS_5\times T^{1,1}$ and 
$AdS_5\times S^5/\Gamma$ lead to maximally \susy\ plane-waves or their orbifolds 
\cite{Itzhaki:2002kh, Alishahiha:2002ev}. On 
the other hand type IIB string theory on these two geometries is dual to $\cn = 1$ superconformal 
field theory (SCFT) \cite{Klebanov:1998hh} and $\cn =2$ superconformal  quiver gauge theory 
\cite{Douglas:1996sw, Kachru:1998ys, Oz:1998hr}, 
respectively. 
It is then natural to ask what the ``BMN'' sector of these theories is. In fact,
the BMN sector of many other cases such as where the theory is not conformal and there is an RG 
flow, have been studied, for examples 
see \cite{Gimon:2002sf, Corrado:2002wi, Oz:2002ku, Niarchos:2002fc, Naculich:2002fh, 
Bigazzi:2002gw, Bigazzi:2003jk}. 
Other cases, such as $AdS_5/\Gamma\times S^5$, have also been studied and 
argued that in the BMN limit we the conformal symmetry is restored \cite{Alishahiha:2002bc}.

Here, we very briefly review the $\cn =1$ SCFT case (section \ref{T11}), and the $\cn =2$ 
superconformal case (section \ref{quiver}), in which we show how one may get a description of the 
DLCQ of strings on the \pl\ \bg .

\subsubsection{ BMN sector of $\cn =1$ SCFT equals BMN sector of $\cn =4$ SCFT}\label{T11}

As discussed in \cite{Klebanov:1998hh}, the $\cn =1$ SCFT which arises as the low energy effective 
theory of $N$
D3-branes probing a conifold geometry is an $SU(N)\times SU(N)$ gauge theory with a $U(1)_R$ 
and global $SU(2)\times SU(2)$ symmetries. Noting \eqref{T11penrose}, \eqref{BMNconjecture1} 
should be modified as \cite{Itzhaki:2002kh, Gomis:2002km}
\be
H=\cd-\frac{1}{2}\cj+\cj_3+\cj '_3 \, ,
\ee
where $\cj$ is the R-charge and $\cj_3$ and $\cj'_3$ are the $SU(2)\times SU(2)$ quantum numbers and the BMN sector of the theory is the set of operators with $\Delta_0\to\infty$ while 
\[
H_0=\Delta_0-\frac{1}{2}J+J_3+J'_3
\]
is kept finite. 

In order to work out the BMN-type \opt s we need to know more about the details of 
the matter content of the theory and their $J$ charges. Besides the gauge multiplets of 
$SU(N)\times SU(N)$ groups, which consist of covariant derivatives $D_a$, $\tilde{D}_a$, the 
gauginos $\psi, \tilde{\psi}$ and their complex conjugates, we have four superfields which are
in bi-fundamentals of the $SU(N)\times SU(N)$. The bosonic scalars of these superfields will be 
denoted by $A_i$ and $B_i$, $i=1,2$, and the corresponding fermionic fields by $\chi_{A_i}$ and
$\chi_{B_i}$ (and their complex conjugates). The gauge multiplets are singlets of the $SU(2)\times SU(2)$ global symmetry while the $A_i$ and $B_i$ matter fields are in $({\bf {2}, 1})$ and  $({\bf 1, {2}})$, respectively. The engineering dimension, $\Delta_0$, of $A_i$
and $B_i$ fields are $\frac{3}{4}$ while for their fermionic counterparts, $\chi_{A_i}$ and 
$\chi_{B_i}$ have $\Delta_0=\frac{5}{4}$. The covariant derivatives, as in the $\cn =4$ case, 
have $\Delta_0=1$ and the gauginos have $\Delta_0=\frac{3}{2}$.\footnote{To see these, we note that the superpotential of the theory is of the form \cite{Klebanov:1998hh}
\[W=\epsilon_{ij}\epsilon_{kl} A_i B_j A_k B_l\] 
and the fact that the square of the derivative of the superpotential, being a term in the Lagrangian, should have dimension four. Also note that this superpotential should have R-charge equal to two.}
As for the R-charge, $J$, in the usual $\cn=1$ conventions $A_i$ and $B_i$ have $J=\frac{1}{2}$ while their fermionic counterparts $J=-\frac{1}{2}$ and gauginos have $J=1$ \cite{Klebanov:1998hh}.
Therefore for these fields and their complex conjugates the value of $H_0$ are 
\cite{Itzhaki:2002kh}
\bea\label{N=1}
A_2,\ B_2:\ H_0=0, &&\ \ \  \cr
A^\dagger_1,\ B^\dagger_1,\ {\overline{\chi}_{A_1}},\ {\overline{\chi}_{B_1}}:\ H_0=\frac{1}{2}\ , 
\ \ A^\dagger_2 , &B^\dagger_2&,\ {\overline{\chi}_{A_2}},\ {\overline{\chi}_{B_2}}:\ 
H_0=\frac{3}{2}\ 
,\\  
D_a,\ {\tilde{D_a}},\ A_1,\ B_1,\ \chi_{A_2},\ \chi_{B_2},\psi,\ \tilde{\psi}:\ H_0=1,
&&  
{\overline\psi},\ {\overline{\tilde{\psi}}},\ \chi_{A_1},\ \chi_{B_1}:\ H_0=2. \nonumber
\eea

As the next step we need to figure out what corresponds to the $Z$-field. The obvious choice is of course a combination of $A_2$ and $B_2$ which have $H_0=0$, however,  noting that 
$A_2$ and $B_2$ are bi-fundamentals of $SU(N)\times SU(N)$, the right combination is $A_2B_2$ (or $B_2A_2$) which is in the adjoint of the first (or second) $SU(N)$. So, the string theory vacuum should correspond to $\Tr (A_2B_2)^J$.

Next we need to identify $8$ bosonic and $8$ fermionic fields at $H_0=1$ level. Again noting \eqref{N=1} this can be done by insertions 
of two $H_0=\frac{1}{2}$ fields or one $H_0=1$ field. The only complication compared to the $\cn=4$ case is that the $A$ and $\chi$ 
fields are bi-fundamentals while one is only allowed to insert adjoints into the trace. Taking this into account, the form of the BMN-type $H_0=1$ \opt s is
\bea\label{N=1bosonBMN} 
{\rm Bosons:} && \ 
\Tr(A_1 B_2 (A_2B_2)^J)\ ,\ \ \Tr(A_2 B_1 (A_2B_2)^J) , \cr
&&  
\Tr(A_2 A^\dagger_1 (A_2B_2)^J)\ ,\ \ \Tr(B_2 B^\dagger_1 (B_2A_2)^J) , \\
&&\Tr\left[(D_aA_2)B_2(A_2B_2)^J+ A_2 ({\tilde D_a} B_2) (A_2B_2)^J\right]. \nonumber
\eea
\bea\label{N=1fermionBMN} 
{\rm Fermions:} && \ 
\Tr\left[(\chi_{A_2} B_2+A_2(\chi_{B_2})(A_2B_2)^J\right] , \cr
&&  
\Tr\left[{\overline{\chi}_{A_1}}B^\dagger_1 (B_2A_2)^J\right]\ ,
\Tr\left[{\overline{\chi}_{B_1}}A^\dagger_1 (A_2B_2)^J\right], \\
&&\Tr\left[\psi (A_2B_2)^J+ {\tilde{\psi}}(B_2A_2)^J\right]. \nonumber
\eea
In the above, traces can be over $N\times N$ matrices of either of the $SU(N)$ factors.
In the same spirit, using \eqref{N=1}, one may build the $H_0=2$ BMN \opt s which we will not 
present here, leaving it to the reader.

\subsubsection{BMN sector of $\cn=2$ superconformal quiver theory}\label{quiver}

The gauge theory dual to string theory on $AdS_5\times S^5/Z_K$ orbifold is an $SU(N)^K$ $\cn=2$ superconformal quiver gauge theory with bi-fundamental hypermultiplets. This theory has an $SU(2)\times U(1)$ R-symmetry 
\cite{Kachru:1998ys}. 
The 't Hooft coupling for all the $SU(N)$ factors are equal to $\lambda_K$
\be\label{lambdak}
\lambda_K=KNg^2_{YM} \, .
\ee 
The $\cn =2$ gauge multiplet in the $\cn=1$ notation is composed of a vector-multiplet and a 
complex chiral-multiplet 
$\varphi_i,\ i=1,\cdots ,K$, where $\varphi_i$ is in the adjoint \rep\ of the $i^{{\rm th}}$ 
$SU(N)$ factor of $SU(N)^K$.
(We may use $\varphi_i$ for the whole chiral-multiplet or its complex scalar component.) As for the bi-fundamental hypermultiplets we have $(Q^\alpha_i, {\tilde{Q}}^\alpha_i)$ with $\alpha=1,2$, and under a generic $SU(N)^K$ gauge transformation $Q^\alpha_i\to U_iQ^\alpha_i U^{-1}_{i+1}$ and $\tilde{Q}^\alpha_i\to U_{i+1}\tilde{Q}^\alpha_i U^{-1}_{i}$, where $U_i$ belongs to the $i^{{\rm th}}$ $SU(N)$ factor. 

As discussed in section \ref{penroseorbifold}, depending on the choice of the light-like geodesic, there are two different Penrose 
limits that can be taken. Therefore one expects to find two ``BMN'' sectors of the above quiver theory. Of course 
the difference between the two 
BMN sectors lies in the choice of the R-charge, which parallels the choice of the geodesic in the 
Penrose limit. First we consider the  
case which leads to the orbifold of the \pl\ and then the one leading to the maximally \susyc\ 
\pl .

\vskip .3cm

{\it i)} Gauge theory description of strings on the \pl\ orbifold:
\vskip .3cm

In the case of the $Z_K$ orbifold, on the string theory side we have one untwisted vacuum and 
$K-1$ twisted vacua. Therefore, we need $K$ ``$Z$-fields'' and the proper choice of the $Z$-fields 
are the $\varphi_i$ which have $\Delta_0-J=0$.  Explicitly if $\co^J(i)=\Tr(\varphi_i^J)$
\be
{\rm Untwisted\ vacuum:}\  \sum_{i=1}^k \co^J(i)\ ,\ \ \
{\rm Twisted\ vacua:}\  \co^J(i)-\co^J(i+1)\ ,i=1,\cdots, K-1 .
\ee

At $\Delta_0-J=1$ level, we have $D^i_a$ and $Q^\alpha_i,\ \tilde{Q}^\alpha_i$ fields. The 
{\it gauge invariant} BMN \opt s, however, can only be made through insertion of covariant 
derivative $D^i_a$ into $\co^J(i)$.

At $\Delta_0-J=2$ level, corresponding to the single free closed string states on the orbifold of the \pl\ (see discussions of section \ref{penroseorbifold}), we can place two derivative or two $Q$ insertions into $\co^J(i)$. $D^i_a$ insertions are 
quite similar to \eqref{abBMNstringmode}, while the $Q$ insertions are more involved and there are some number of different possibilities:
\bea
{\cal O}^J_{1, n}(i)=\sum_{p=0}^J{\Tr}\left(\varphi_i^p\,Q^{\mu}_p\,
\varphi_{i+1}^{J-p}\,{\bar Q}^{\nu}_i\right)e^{\frac{2\pi i np}{KJ}} &,&  \ 
{\cal O}^J_{2,n}(i)=\sum_{p=0}^J{\Tr}\left(\varphi_i^p\,Q^{\mu}_i
\,\varphi_{i+1}^{J-p}\,{\tilde Q}^{\nu}_i\right)e^{\frac{2\pi i np}{KJ}},
\cr &&\cr
{\cal O}^J_{3, n}(i)=\sum_{p=0}^J{\Tr}\left(\varphi_i^p\,
{\bar {\tilde Q}}^{\mu}_i\,
\varphi_{i+1}^{J-p}\,{\tilde Q}^{\nu}_i\right)\;e^{\frac{2\pi i np}{KJ}} &,& \ 
{\cal O}^J_{4, n}(i)=\sum_{p=0}^J{\Tr}\left(\varphi_i^p\,
{\bar {\tilde Q}}^{\mu}_i\,\varphi_{i+1}^{J-p}\,{\bar Q}^{\nu}_i\right)\;e^{\frac{2\pi i np}{KJ}}.
\nonumber\label{DELJ2}
\eea
Note the fact that we have $K$ in the denominator of the phase factors, which guarantees the 
correct ``twisted'' string modes.

In the same way one may construct higher $\Delta_0-J$ states which we do not present \cite{Alishahiha:2002ev, Kim:2002fp, Takayanagi:2002hv, Oh:2002sv}.
\vskip .3cm

{\it ii)} Gauge theory description of DLCQ of strings on  \pl s:
\vskip .3cm
 
In \ref{penroseorbifold} we showed that in taking the Penrose limit of $AdS_5\times S^5/{Z_K}$
the orbifolding may disappear and we may end up with the maximally \susyc\ \pl . However, as we 
mentioned, in a specific large $K$ limit the compactification radius of the light-like direction $x^-$ becomes finite. This in particular, as we studied in \ref{bosonicstrings} (\cf\ 
\eqref{momentumX-}), leads to Discrete 
Light-Cone Quantization (DLCQ) of strings. It is therefore quite plausible to expect that the DLCQ of strings on the \pl\  should somehow be described by the BMN sector of $\cn=2$ $SU(N)^K$ quiver theory in the large $K$, large $N$ limit. 

The  light-like compactification radius (\cf\ \eqref{R-}, $R_-$, is 
proportional to $1/K$ and \eqref{lambdak}
\be\label{R-gauge}
R_-=\sqrt{\frac{KNg^2_{YM}}{K^2}}=g_{YM}\sqrt{\frac{N}{K}}\ .
\ee
Therefore, for fixed $g_{YM}$, if $K\sim N\to\infty$, $R_-$ remains finite.

In this case we can safely keep $J$ finite. In fact, it is now $KJ$ that specifies the BMN sector, which should scale like $(KN)^{1/2}$, and $J-1$ plays the role of the winding number of strings along the light-like direction \cite{Mukhi:2002ck, Alishahiha:2002jj}.
Let us focus on the $J=1$ case and define 
$Z_i=Q^1_i+iQ^2_i$. The string vacuum state in the sector with zero 
light-like winding corresponds to
\be
\co_{vac}=\Tr (Z_1Z_2\cdots Z_K).
\ee
This state has $H_0=0, w=0$.  Higher winding vacuum states are of the form 
$\Tr ({\mathbb{Z}}^{w+1})$ where ${\mathbb{Z}}\equiv Z_1Z_2\cdots Z_K$. Other stringy 
excitations can be obtained through insertions
of $\varphi_i$, $D^i_a$ or $\tilde{Q}_i$'s. For a more detailed discussion on these \opt s
the reader is referred to \cite{Mukhi:2002ck, Alishahiha:2002jj}.
The  BMN gauge theory duals of other $AdS$ orbifolds can also be found in 
\cite{Bertolini:2002nr}.

%% file: noninter.tex

%
%
%


In this section we focus only on planar results in the $\neqf$ gauge theory, which according to the
BMN correspondence, should connect with the string theory side at zero string coupling. Higher
genus corrections will be postponed until section \ref{interactingstrings}, where a new
complication arising from the need to re-diagonalize the basis of BMN operators, at each order in
the genus expansion, will be discussed. We start this section by studying the two-point functions
of BMN operators with their conjugates, in the free field theory limit, and use the results to set
the normalization of these operators. We then move on to discuss the quantum corrections to the
scaling dimensions, i.e., the anomalous dimensions. We first present a very brief but general
overview of the scaling behaviour of correlation functions, and the appearance of anomalous
dimensions through the renormalization group equation. While this discussion provides the physical
context in which anomalous dimensions are normally encountered in quantum field theory, the main
point of this section is the actual calculation of anomalous dimensions in the interacting theory
at planar level, first at one-loop, and then using superspace techniques, deriving the
result to all orders in perturbation theory. An important concept in the renormalization of
composite operators, operator mixing, appears when loop corrections are taken account of. Operator
mixing, together with the requirement that BMN operators have a well-defined scaling dimension, are
used to motivated the choice of the BMN operators. As a stringent test of the BMN correspondence,
we compare the calculations of the corrected scaling dimensions to the masses on the string theory
side, and find agreement.

Another key point of this section is the appearance of the new modified 't Hooft coupling 
$\lambda^\prime$ \eqref{lambda'}, and will first be seen when taking the BMN limit of the one-loop 
anomalous dimension.

Whereas most of this section in devoted to the study of two-point functions, the question of the 
relevance of three and higher point functions to the correspondence must also be dealt with. We 
take a preliminary look at this issue via the operator product expansion (OPE) of the BMN 
operators, demonstrating a very important property, which is the closure of the OPE for the set of 
BMN operators. This property will serve as yet another argument in favor of the choice of BMN 
operators. The OPE will turn out to also play a practical role, providing us with a tool to study 
two-point functions of multiple trace operators, but such actual applications will be deferred to 
section \ref{interactingstrings}.

\subsection{Normalization of BMN operators}
\label{norm:bmn}

The propagator for the scalars in the $\neqf$ supermultiplet, which transform in the adjoint of 
$U(N)$, are
\be \label{phi:propagator}
  \Big\langle \phi_i^{ab} (x) \phi_j^{cd} (0) \Big\rangle_0 \: = \:
  \frac{g_{YM}^2 \delta_{ij}}{8 \pi^2 |x|^2}
  \delta^{ad} \delta^{bc}
\ee
where we explicitly display the matrix indices on the fields. We denote correlation functions in 
the {\it free} theory with a subscript $0$, as above. With the convention \eqref{define-Z} for the 
fields carrying the $U(1)_J$ charge, the propagator for them is
\be \label{z:propagator}
  \Big\langle Z^{ab} (x) (Z^\dagger)^{cd} (0) \Big\rangle_0 \: = \:
  \frac{g_{YM}^2 \delta_{ij}}{8 \pi^2 |x|^2}
  \delta^{ad} \delta^{bc}
\ee
Using these propagators, we can demonstrate a set of rules which facilitate the evaluation of 
correlation functions involving traces over algebra valued fields (which we denote by $\Tr$). We assume that the composite operators we work with are normal-ordered, so no contractions between fields in the same operator (i.e. at the same spacetime point) will appear. Such contractions would lead to infinite renormalizations of the operator.
We start with the simplest such structures, evaluated in the free theory.
We have the following fission rules
\be
  \Tr [ : \phi_i \cA : : \phi_j \cB  : ] \: \sim \:
  \delta_{ij} \: : \Tr [ \cA ] :  \: : \Tr [ \cB ] :
  \: \: \: \: \:  \: \: \: \: , \: \: \: \: \: \:  \: \: \: \: \:
  \Tr [ : \phi_i : : \phi_j \cA : ] \: \sim \:
  \delta_{ij} \: N \: : Tr [ \cA ] :
\ee
where for clarity we have dropped some obvious prefactors arising from the propagators, remembering that the rank of $U(N)$ is $N$.
Clearly the second identity is a special case of the first (with one of the operators taken to be the identity matrix in the space of color indices).
We have explicitly kept the normal-ordering symbols here for clarity.
Caution must be used when applying these rules not to allow contractions between fields at the same 
spacetime point (appearing in the same normal-ordering).
In the second identity, we can take $\cA = 1$, which gives $\Tr[\phi_i \phi_j] \sim \delta_{ij} N^2$.
We have also the fusion rule
\be
  : \Tr [ \phi_i \cA ] : : \Tr [ \phi_j \cB ] : \: \sim \:
  \delta_{ij} \Tr [ : \cA : : \cB : ]
\ee
In the future, we will drop the normal-ordering symbol, but all calculations are implicitly 
assumed to account for their presence.

Consider now the normalization of the operator \eqref{Op} in the free theory and at planar level. 
We assume that the vacuum of the theory leaves the $SU(4)$ R-symmetry unbroken, as is the case for 
the superconformal points in the moduli space of $\neqf$ SYM. The correlation function of any set 
of operators then vanishes if they do not form an $SU(4)$ singlet.

Keeping the planar contributions amounts, as is usual with 't Hooft expansions, to keeping the 
leading order contribution in $1/N^2$. The normalization of the operator $\widetilde{\co}$ 
\eqref{Op} is fixed by requiring 
$|x|^{2(J+1)}\langle \widetilde{\co}_p(x) \bar{\widetilde{\co}}_p(0) \rangle_0 = 1$ at planar 
level.
The two-point functions provide a natural notion of an inner product on the space of BMN 
operators, and in the BMN correspondence are the analogue of the inner product between string 
states
(\cf\ discussions of section \ref{deltaj2}).

We work with ($g_{YM}=0$) free theory and use Wick contractions to write the correlation function as sums of products of scalar propagators. We first write out the traces explicitly
\bea
  \langle \widetilde{\co}_p (x) \bar{\widetilde{\co}}_q (0) \rangle_0 &=&
  \langle  
    \Tr ( Z^p \phi_j Z^{J-p} \phi_i )(x) \Tr ( \phi_i \bar{Z}^{J-q} \phi_j \bar{Z}^q )(0)
  \rangle \cr
  &=&
  \Big\langle
    \left(Z^p_{ab} (\phi_j)_{bc} Z^{J-p}_{cd} (\phi_i)_{da}\right) (x)
    \left((\phi_i)_{ef} \bar{Z}^{J-q}_{fg} (\phi_j)_{gh} \bar{Z}^q_{he} )\right)(0)
  \Big\rangle_0  
\eea
having used the cyclicity of the trace, and defining $\bar{Z} \equiv Z^\dagger$. A sum over repeated $U(N)$ color indices $a ... h$ is implied.
The normal-ordering symbols can be safely dropped in this correlation function if we assume that $i 
\ne j$ (since then $\phi$'s at the same point can't be contracted, as is also the case for the $Z$'s and $Z^\dagger$'s). We will make this assumption since it also simplifies some of the combinatorics.
Repeatedly taking Wick contractions on the $\phi$'s and $Z$'s that are nearest to each other using \eqref{phi:propagator} and \eqref{z:propagator}, we arrive at
\be
  \Big\langle \widetilde{\co}_p (x) \bar{\widetilde{\co}}_q (0) \Big\rangle_0 =
  \left( \frac{g_{YM}^2 N}{8 \pi^2 |x|^2} \right)^{J+2} \delta_{p,q}
\ee
The requirement that these operators to be normalized as
$|x|^{2(J+1)}\langle \widetilde{\co}_p(x) \bar{\widetilde{\co}}_q(0) \rangle=\delta_{p,q}$ can 
be satisfied by taking
$\widetilde{\co}_p \rightarrow \left(\frac{8 \pi^2}{g_{YM}^2 N} \right)^{\frac{J+2}{2}} 
\widetilde{\co}_p$.
Similar reasoning gives the normalization of the other BMN operators.
For example, the normalization of the BMN operator with $\Delta_0 - J=2$ in 
\eqref{ijBMNstringmode}, is fixed by the normalization we have just considered, but an extra 
factor of $\frac{1}{\sqrt{J+1}}$ enters from the $J+1$ terms appearing in the sum.


\subsection{Anomalous dimensions}
\label{amonalous}

In a conformal field theory such as $\neqf$ super-Yang-Mills, the content of the theory can be 
extracted via the correlation functions of gauge invariant operators, and is embodied in their 
scaling dimensions, how they mix amongst each other under renormalization and the coefficients in 
their operator product expansions (OPE). We will now present a brief overview of the first two 
topics, leaving the discussion of the OPE for a later section. A discussion of these points in 
general QFT can be found in \cite{Peskin:1995ev,Zinn-Justin:1989mi}.

Denote a bare correlation function built of $n$ bare fields $\phi_b$
and the renormalized correlation function, built in the same way, but using renormalized fields as
\footnote{Generically, a 
number of different types of fields may enter into a correlation function; however, we are most 
concerned with the scaling behavior of such correlators, and for the $\neqf$ SYM theory of 
interest to us, supersymmetry implies that all fields in a supermultiplet receive the same 
anomalous dimensions. We therefore simplify our notation and write only one type of field.}
\be \label{bare-correlation-function}
  \Gamma^{(bare)}_n ( \{ x_i \} , \lambda^{(bare)} , \Lambda ) =
  \Big\langle \phi^{(bare)} (x_1) ... \phi^{(bare)} (x_n) \Big\rangle
  \: \: \: \: , \: \: \: \:
  \Gamma^{(ren)}_n ( \{ x_i \} , \lambda^{(ren)} , \mu ) =
  \Big\langle \phi^{(ren)} (x_1) ... \phi^{(ren)} (x_n) \Big\rangle
  \: .
\ee
The bare correlation functions depend implicitly on a set of bare parameters defined at the cut-off scale $\Lambda$ of the theory, while the renormalized ones depend on the renormalized parameters defined at the renormalization scale $\mu$.
The renormalized fields are proportional to the bare fields, via the wave-function 
renormalization, $\phi^{(ren)} (x) = Z_{\phi}^{-1/2} (\mu) \phi^{(bare)} (x)$.
The dependence of the field strength of the renormalized field on the renormalization scale $\mu$ 
is the source of the anomalous dimension.

A simple consequence of \eqref{bare-correlation-function}
is that the bare and renormalized $n$-point functions are related by powers of the wave-function 
renormalization
\be \label{bare-to-renormalized}
  \Gamma^{(ren)}_n ( \{ x_i \} , \lambda^{(ren)} , \mu ) =
  Z_{\phi}^{-n/2} (\mu)
  \Gamma^{(bare)}_n ( \{ x_i \} , \lambda^{(bare)} , \Lambda )
\ee
The renormalization scale dependence enters the renormalized $n$-point function via the wave-function renormalization $Z_\phi$ and the renormalized parameters $\lambda^{(ren)}$ of the theory, which are defined at that scale, but not the bare $n$-point functions, hence
\be
  \frac{\partial}{\partial \ln \mu}
  \Gamma^{(bare)}_n ( \{ x_i \} , \lambda^{(bare)} , \Lambda ) = 0 \: .
\ee
The chain rule then gives
\be \label{rg-equation}
  \left( \mu \frac{\partial}{\partial \mu} +
  \beta ( \lambda^{(ren)} )
  \frac{\partial}{\partial \lambda^{(ren)}} + n \: \gamma (\lambda^{(ren)}) \right)
  \Gamma^{(ren)}_n (\lambda^{(ren)},\mu) = 0 \: .
\ee
For a single coupling massless theory (like $\neqf$ SYM),
we have written this relation in terms of the dimensionless functions $\beta$ and $\gamma$,
which take account of shifts in the field strength and coupling constants that compensate for 
changes in the renormalization scale to keep the bare correlation functions constant. They are 
defined as\footnote{The $\beta$-function and anomalous dimension $\gamma$ are universal in the 
sense that they are the same for all correlation functions in a given renormalizable theory.}
\be
  \beta ( \lambda^{(ren)} ) = \mu \frac{\partial \lambda^{(ren)} (\mu)}{\partial \mu}
  \Biggr|_{\lambda^{(bare)}}
  \: \: \: \: \: \: \: \: \: \: , \: \: \: \: \: \: \: \: \: \:
  \gamma ( \lambda^{(ren)} ) = \mu \frac{\partial \ln Z_{\phi} (\mu) }{\partial \mu}
  \Biggr|_{\lambda^{(bare)}}
\ee

For a small change in the renormalization scale $\mu \rightarrow \mu + \delta \mu$,
as a result of which the coupling and fields change as $\lambda \rightarrow \lambda + \delta 
\lambda$ and $\phi \rightarrow (1 + \delta \eta) \phi$,
the change in the field strength is related to the anomalous dimension via $\delta \eta = (\delta 
\mu / 2 \mu) \gamma$.


The renormalization group equation
\eqref{rg-equation} is a highly non-trivial statement about the behaviour of correlation functions 
in a quantum field theory, with deep implications (for example the running of  couplings and masses).
The scale dependence introduced into the renormalized theory in the guise of the renormalization
scale $\mu$ generically breaks any classical scale invariance which might be present in a massless
theory with dimensionless couplings. However, there may exist fixed points of the renormalization
group (special value of the parameters $\lambda_*$) at which the $\beta$-function 
vanishes.\footnote{There is of course always the trivial fixed point for which the couplings 
vanish, and hence so do the anomalous dimensions. For $\neqf$ SYM, there is in fact a line of 
fixed points, 
and the $\beta$-function vanishes at all values of the coupling.}
At these fixed points, the classical scale invariance of the renormalized theory is restored.
However, the classical scaling of the fields and the correlation functions might be modified by 
the presence of anomalous scaling dimensions, with the scaling dimension of the field becoming 
$\Delta =  \Delta_0 + \gamma$. Unlike the classical scaling dimensions, the anomalous dimensions 
may take on a continuum of values, which however are constrained by the conformal 
algebra.\footnote{For example, for unitary representations, the dimensions are 
bounded from below, while the anomalous dimensions can be positive or negative 
\cite{Minwalla:1998ka}. Also, 
as a result of supersymmetry, in $\neqf$ SYM, all 
fields in the same $\neqf$ multiplet receive the same anomalous dimension.}
At a fixed point,
the behaviour of the correlation functions reflects the dependence on the non-trivial scaling
\be
  \Gamma^{(ren)}_n (s x_i,\lambda_*,s^{-1} \mu) =
  s^{- n \Delta} \: \Gamma^{(ren)}_n (x_i,\lambda_*,\mu)
\ee

We will encounter composite operators which are local monomial products of fields.
The process of renormalization of a given composite operator might generate new divergences which 
are proportional to other composite operators, requiring their introduction as counterterms, 
leading to a mixing of operators under renormalization. In general a composite operator may mix 
under renormalization with any operator of equal or lower dimension which carry the same quantum 
numbers. For a massless theory with no dimensionful parameters, only operators of the same 
classical 
dimension mix. If we choose as a basis for these local gauge invariant operators a set, which we 
will label $\{\co_i\}$, then multiplicative renormalization occurs in the form of matrix multiplication
\be \label{mult-renorm}
  \co^{(bare)}_i (x) = \sum_j Z_{ij} \co^{(ren)}_j (x)
\ee
The statement regarding the operator only mixing with those of lower or equal classical dimensions
implies that the matrix $Z_{ij}$ can be cast in triangular form when the basis is arranged in 
order of dimensions of the operators.
Correlation functions with insertions of composite operators also satisfy a renormalization group 
equation, generalizing \eqref{rg-equation}, with a new anomalous dimension matrix
\be
  \gamma_{ij} ( \lambda^{(ren)} ) = \mu \frac{\partial \ln Z_{ij} (\mu)}{\partial \mu}
  \Biggr|_{\lambda^{(bare)}}
  \: .
\ee




We make a few final comments about general properties of conformal field theories, which clarify 
some of the points we shall encounter in later sections. 
It is believed that unitary interacting scale invariant quantum field theories generally exhibit a 
larger symmetry containing scale invariance, the group of conformal transformations. Conformal 
invariance turns out to be restrictive enough to completely fix the dependence of two and three 
point functions on the spacetime coordinates (in a suitable basis); those of higher point 
functions, while not completely fixed, are restricted by the requirement that they depend on 
certain special combinations of the coordinates (the conformal ratios) \cite{DiFrancesco:1997nk}.
In a unitary conformally invariant quantum field theory, we can choose a basis of operators with 
definite scaling dimensions (eigenstates of the dilatation operator). These are the quasi-primary 
operators. In each multiplet of the conformal (or super-conformal) algebra, the operators of 
lowest dimension\footnote{These are the operators which are annihilated by the generator of 
special conformal transformations (or the super-conformal supercharges).} are the conformal (or 
super-conformal) primaries.
Two quasi-primary operators are correlated if and only if they have the same scaling dimensions, 
and the two-point correlation function takes the form (dropping normalization factors)
\be \label{2-point:functions:cft}
  \Big\langle \mathcal{O}_i(x_1) \mathcal{O}_j(x_2) \Big\rangle
  \: = \: \frac{\delta_{\Delta_i,\Delta_j}}{|x_{12}|^{2\Delta_i}},
\ee
with $x_{12} \equiv x_1 - x_2$. $\Delta_i$ is the full (engineering plus anomalous) scaling 
dimension of operator $\co_i$. The three-point functions are similarly constrained and satisfy
\cite{DiFrancesco:1997nk}
\be \label{3-point:functions:cft}
  \Big\langle \mathcal{O}_i(x_1) \mathcal{O}_j(x_2) \mathcal{O}_k(x_3) \Big\rangle \: = \:
  \frac{C_{\Delta_i,\Delta_j,\Delta_k (g_{YM}^2,N)}}{
    |x_{12}|^{\Delta_i+\Delta_j-\Delta_k}
    |x_{13}|^{\Delta_i+\Delta_k-\Delta_j}
    |x_{23}|^{\Delta_j+\Delta_k-\Delta_i}
  }.
\ee
For the two-point functions, quantum corrections can enter only through anomalous dimensions for 
the operator, while for three-point functions there is the more general possibility that the 
coefficient $C_{\Delta_i,\Delta_j,\Delta_k} (g_{YM}^2,N)$ may also receive corrections at higher 
loops.
When computing the anomalous dimension of an operator in perturbation theory, we have a power 
series expansion $\gamma = \gamma_1 + \gamma_2 + \ldots$, and $\gamma_n$ includes $n^{{\rm th}}$ 
power of the 't Hooft coupling $\lambda^n$.
The dependence of the two-point function on the positions of the operators, when computed in 
perturbation theory, will take the form
\be \label{log-anomalous-dim}
  \frac{1}{|x|^{2 \Delta}} \approx \frac{\mu^{2 \gamma_1}}{|x|^{2 \Delta_0}}
  \left( 1 - \gamma_1 \ln | x \mu |^2 \right)
\ee
to one-loop order, with the renormalization scale entering to keep the argument of the log 
dimensionless. This approximation is valid so long as $\gamma_1 \ll \ln(x \mu)^{-2}$.
While this expression suggests that scale invariance has been broken, the scale $\mu$ will drop 
out when it is re-summed to all orders in perturbation theory to reproduce the left-hand side of 
the expression. The scale $\mu$ is merely an artifact of perturbation theory.

In the next two sections we move on to a practical calculation of the anomalous dimension of 
composite BMN operators, first at one-loop, and then to all orders in perturbation theory.

\subsection{Anomalous dimensions of the BMN operators, first order in $g^2_{YM}$}
\label{anomalous-dimensions-first-order}

The goal of this section is to compute the anomalous dimension of a class of BMN operators to first 
loop order on the gauge theory side, and to compare the result to the appropriate 
computation of the string theory masses. This will provide the first check of the  BMN 
correspondence stated in section \ref{BMNproposal}.
In this section we concentrate on anomalous dimensions only at planar level, and 
revisit the issue at non-planar level in section \ref{Mixingclues}.

Consider a local gauge invariant operator of the form
\be \label{generic-op}
  \widetilde{\co}_p^J (x) = \Tr \left( \phi_i Z^p \phi_j Z^{J-p} (x) \right)
\ee
with engineering dimension $\Delta_0 = J+2$.
Such a generic operator would not remain an eigenstate of the dilatation operator after 
renormalization, as a result of the operator mixing discussed in the previous section, and would 
therefore not have a well-defined scaling dimension.
This means that after computing loop corrections, even at planar level, the two-point function of 
the operators \eqref{generic-op} would not remain diagonal.
Quantum effects induce a mixing with operators in the same $SU(4)$ representation with Dynkin 
labels $(2,J-2,2)$ and the same engineering dimension.
Any operator of the form \eqref{generic-op} for any $0\leq p\leq J$ satisfies the mixing criteria, 
and in general 
any operators with this charge and dimension would take the form \eqref{generic-op} for some $p$.
The interaction term in the Hamiltonian which connects only the scalars takes the form $H_{int} 
\sim :g_{YM}^2 \sum_{ij} \Tr \left([\phi_i,\phi_j][\phi^i,\phi^j]\right):$.\footnote{We work with a 
normal-ordered Hamiltonian, which amounts to discarding all self-contractions in a given insertion 
of the Hamiltonian in perturbation theory. Contractions across different insertion are not removed 
by normal-ordering.}
It therefore contains terms which exchange the order of $U(1)_J$ charged and neutral fields 
$\phi_i$ and $Z$, and also two different $\phi_i$ fields.
We will refer to such exchanges as ``hopping''.
The interactions mix operators $\widetilde{\co}_p^J$ and $\widetilde{\co}_q^J$, both of the form 
\eqref{generic-op}, but with $p \ne q$.
For each additional loop, the mixing would extend to operators with the insertions of the 
impurities shifted by one more position.
For example, at one-loop order, these interactions generate diagrams with no hopping,
and those which have one hop, either forward or backward.

Since the operators \eqref{generic-op} do not have well-defined scaling dimensions, they cannot be
put in a simple correspondence with the string theory side of the BMN conjecture. One of the main
points of this section will be to construct operators with well-defined renormalized scaling
dimensions at planar level, and hence diagonal two-point functions, which can be put in one-to-one 
correspondence with string theory objects.

In the free theory ($\gym = 0$), operators with $p \ne q$ do not mix at planar level, and the 
two-point function remains diagonal. We can write the non-diagonal contributions at higher orders 
as
\be
  \Big\langle \widetilde{\co}_p^J (x) \widetilde{\co}_q^J (y) \Big\rangle \propto
   \sum_{l=0}^{\infty} \lambda^l M_{p,q}^{(l)} (x-y)
   \equiv M_{p,q} (x-y) \: ,
\ee
where we have dropped proportionality constants coming from the normalization of the tree-level 
two-point function. The zeroth order term is simply the identity $M_{p,q}^{(0)} = \delta_{p,q}$.
The matrices $M_{p,q}^{(l)}$ are proportional to $l^{{\rm th}}$ powers of logs of the separation 
$(x-y)$ of the two operators,
$M^{(l)}_{p,q} (x) = [ \ln (x \mu)^2 ]^l \: \mathcal{M}^{(l)}_{p,q}$,
coming from perturbation theory at $l$-loops.
The matrices ${\mathcal{M}}_{p,q}^{(l)}$ are symmetric in $p,q$, because for each insertion of the 
Hamiltonian which generates a hop to the right, there is one generating a hop to the left.
The hopping can be exhibited more explicitly by separating $\mathcal{M}_{p,q}^{(l)}$ into 
``hopping'' matrices $\mathfrak{m}_j^{(l)}$
\be \label{hopping-matrices}
  \mathcal{M}_{p,q}^{(l)} = \sum_{j=-l}^l \delta_{p,q+j} \: \mathfrak{m}_j^{(l)}
\ee
with the interpretation that $\mathfrak{m}_j^{(l)}$ captures all the effects at loop $l$ coming 
from $j$ hops ($j$ can be positive or negative), and  
$\mathfrak{m}_j^{(l)}=\mathfrak{m}_{-j}^{(l)}$ because forward and backward hops are governed by 
essentially the same term in the Hamiltonian.
We were able to extract a $p$ and $q$ independent term $\mathfrak{m}_j^{(l)}$ here because in the 
interaction Hamiltonian, the commutator terms which generate the various hops, all enter with
precisely the same coefficient. \eqref{hopping-matrices} makes it explicit that 
the range of allowed hops is set by the number of loops (or insertions of Hamiltonian) which are 
included, a point we noted earlier.
Using \eqref{log-anomalous-dim}, we can read the $l$-loop anomalous dimensions directly from
$\mathcal{M}_{p,q}^{(l)}$.

The sum (form all $p=0,\ldots,J$) of the operators in \eqref{generic-op} is protected by a BPS 
condition, and this gives the relation among the coefficients
\be \label{BPS-condition-on-M}
  \sum_{p=0}^J \mathcal{M}_{p,q}^{(l)} = 0 \: \: \: \: \: \: \: \: \: \: \forall l, q > 0 \: .
\ee

As mentioned previously, to specify precisely the dictionary translating between the gauge theory 
and string sides of the duality, we need to find a basis of operators with well-defined scaling 
dimensions. Such a basis would contain operators formed as linear combinations of the above
\be \label{linear-comb}
  \co^J_n (x) = \sum_{p=0}^J \mathcal{F}_{np} (J) \:
  \widetilde{\co}_p^J (x)
\ee
for some $\mathcal{F}$ to be determined by the condition that \eqref{linear-comb} have a well 
defined scaling dimension. We can think of $\mathcal{F}$ as a change of basis on the vector space 
of operators $\widetilde{\co}_p^J$.
We also impose an additional constraint on the expansion coefficients, requiring
$\mathcal{F}_{0p}(J)=1$, which is another statement of the BPS condition.

A few comments are in order regarding the range of the summation in \eqref{linear-comb}. The 
endpoints $p=0$ and $p=J$ correspond, for $i \ne j$ in \eqref{generic-op}, to the case where the 
position of $\phi_i$ and $\phi_j$ are reversed. Both orderings must be included since the 
interaction Hamiltonian will generate such exchanges, and in principle these terms can mix with 
each other. In addition, for the BPS condition to hold when $n=0$ in \eqref{linear-comb}, the 
summation must include both arrangements. Lastly, if we drop one of $p=0$ or $p=J$, we will 
compute an anomalous dimension with a finite piece in the BMN limit, and one that scales as 
$\lambda$, and hence diverges in the double scaling limit. The divergent piece is exactly 
canceled when the missing term is included \cite{Constable:2002hw, Kristjansen:2002bb}.

We are now ready to determine the form of the matrix $\mathcal{F}_{np}(J)$, which at each order in
perturbation theory acts on the operators
\eqref{generic-op}, after which the transformed operators are diagonal,
and hence their two-point functions have perturbative expansions in $\lambda$ of the form
\be \label{diag-two-point-expansion}
  \Big\langle \co_m^J (x) \bar{\co}_n^J (y) \Big\rangle =
  \delta_{m,n} \: \sum_{l=0}^{\infty} \lambda^l f_l (x-y)
  \equiv \delta_{m,n} \:  f_m(x-y)
\ee
and $f_m(x-y)$ can be different for each $\co_m^J$.
We have the similarity transformation
\be
  \left( \mathcal{F} M \mathcal{F}^\dagger \right)_{m,n} =
  \delta_{m,n} \: f_m(x-y)
\ee
with $\mathcal{F}$ admitting a power series expansion in $J$. A suitable, though not unique 
choice, for $\mathcal{F}$ is
\be \label{diag-f}
  \mathcal{F}_{np} = e^{2 \pi i  n p / J}
\ee
which diagonalizes the above operators up to order $\mathcal{O} (1/J^2)$ for any order in 
perturbation theory\footnote{That this orthogonalization is good to all orders in the coupling at 
planar level follows from the results of section \ref{anomalous-dimensions-all-orders}.}, where the $\lambda$ dependence appears in $f_m$. In the BMN limit where $J 
\rightarrow \infty$, the correction terms vanish and the diagonalization, at planar level, is exact.
Note that at planar level, the quantum corrections do not induce mixing between operators with 
different numbers of traces.
When we come to consider the non-planar corrections in section \ref{interactingstrings}, this lack of mixing will 
no longer be the case, and the mixing between operators with different numbers of traces will have 
be dealt with also. In fact, even in the free theory, the single-trace BMN operators will mix among 
themselves at non-planar level.
The significance of this second type of mixing and its role in the duality 
will be the central theme of section \ref{interactingstrings}.

The statement that $\co^J_n$ has a well-defined scaling dimension can be translated into the 
requirement that after renormalization, the bare and renormalized quantities are related by an 
overall scaling, and not a matrix that connects it to other operators as in \eqref{mult-renorm}.
Then,
\be \label{comp-op-renorm}
  \co_n^{J \ (bare)} = \mathcal{Z}_n (\lambda,\mu) \co_n^{J \ (ren)}
\ee
with the renormalization constant generically a function of the coupling (going to the identity 
for $\lambda=0$), and the renormalization scale $\mu$, or alternatively $\epsilon=2-D/2$ in 
dimensional regularization.
The re-scaling $\mathcal{Z}$ depends on the composite operator renormalization $Z_{\co}$ of the 
operator $\co^J_n$ in addition to the usual wave-function renormalizations $Z_Z$ and $Z_{\phi}$ 
for the fields $Z$ and $\phi$, and takes the form
\be \label{define-overall-Z}
  \mathcal{Z}_n = Z_{{\co}_n} Z_\phi (Z_Z)^{J/2} ,
\ee
since there are $J$ fields charged under the $U(1)_J$ and $2$ neutral fields.


The anomalous dimension $\gamma_n$ of the operator $\co_n$
can be computed order by order in perturbation theory, and has a power series expansion in the 't 
Hooft coupling $\lambda$
\be
  \gamma_n (\lambda) = \sum_{l=1}^\infty \lambda^l c^{(n)}_l
\ee
where the $l=0$ term vanishes since the anomalous dimension appears as a quantum correction to the 
classical scaling dimension. The coefficients of this expansion can be Fourier transformed
\be \label{Fourier-coeffs}
  c^{(n)}_l = \sum_{h=-l}^l c^{(n)}_{l,h} \: \: e^{- 2 \pi i  n h / J}
\ee
with a natural interpretation that, as we will see below, $c^{(n)}_{l,h}$ represents the 
portion of the anomalous dimension of the operator $\co_n$ which arises at loop $l$, from the sum 
of diagrams with $h$ hops. By convention, positive $h$ will correspond to hops to the right.
We now compute the Fourier coefficients $c^{(n)}_{l,h}$ at one-loop, working at the planar level.

Our goal is to compute the counterterms necessary to absorb the divergences generated by insertion 
of the composite operator \eqref{linear-comb} and the wave-function renormalizations of the $Z$ 
and $\phi$ scalar fields, and use these to derive the anomalous dimension of the composite operator,
via \eqref{define-overall-Z}. Here we will only focus on the BMN \opt s with two {\it 
non-identical} scalar impurities which can be in $({\bf 9},{\bf 1})$
or $({\bf 3^{\pm}},{\bf 1})$ \soff \rep s (the explicit calculations regarding the singlet case 
$({\bf 1},{\bf 1})$ may be found in \cite{Gomis:2003kj}).
All the other BMN \opt s should have the same anomalous 
dimensions, due to the supersymmetry (\cf\ \ref{deltaj2}).
We work in position space and use dimensional regularization. In dimensional regularization, the
anomalous dimension becomes
\be \label{anom-dim-dim-reg}
  \gamma (\lambda) =
  \: {\epsilon} \frac{\lambda}{\mathcal{Z}_n} \frac{\partial \mathcal{Z}_n}{\partial \lambda}
\ee

Consider a two-point function of the operators \eqref{linear-comb} with  
the choice \eqref{diag-f} 
for the diagonalizing matrix. We can expand this correlation function of sums of operators into a 
double sum of correlation functions of individual composite operators. In a generic QFT, a 
correlation function of ordinary operators with an insertion of a single composite operator has 
divergences which can be removed, in addition to the usual counterterms, with a wave-function 
renormalization of the composite operator. Insertions of additional composite operators will in 
general produce additional divergences requiring subtractions. However, for a conformal field 
theory, the form of the two-point function is fixed, as shown in \eqref{2-point:functions:cft}, 
and the wave-function renormalization \eqref{comp-op-renorm} suffices to absorb all divergences 
coming from the composite operators.



\begin{figure}[ht]
\centering
\epsfig{figure=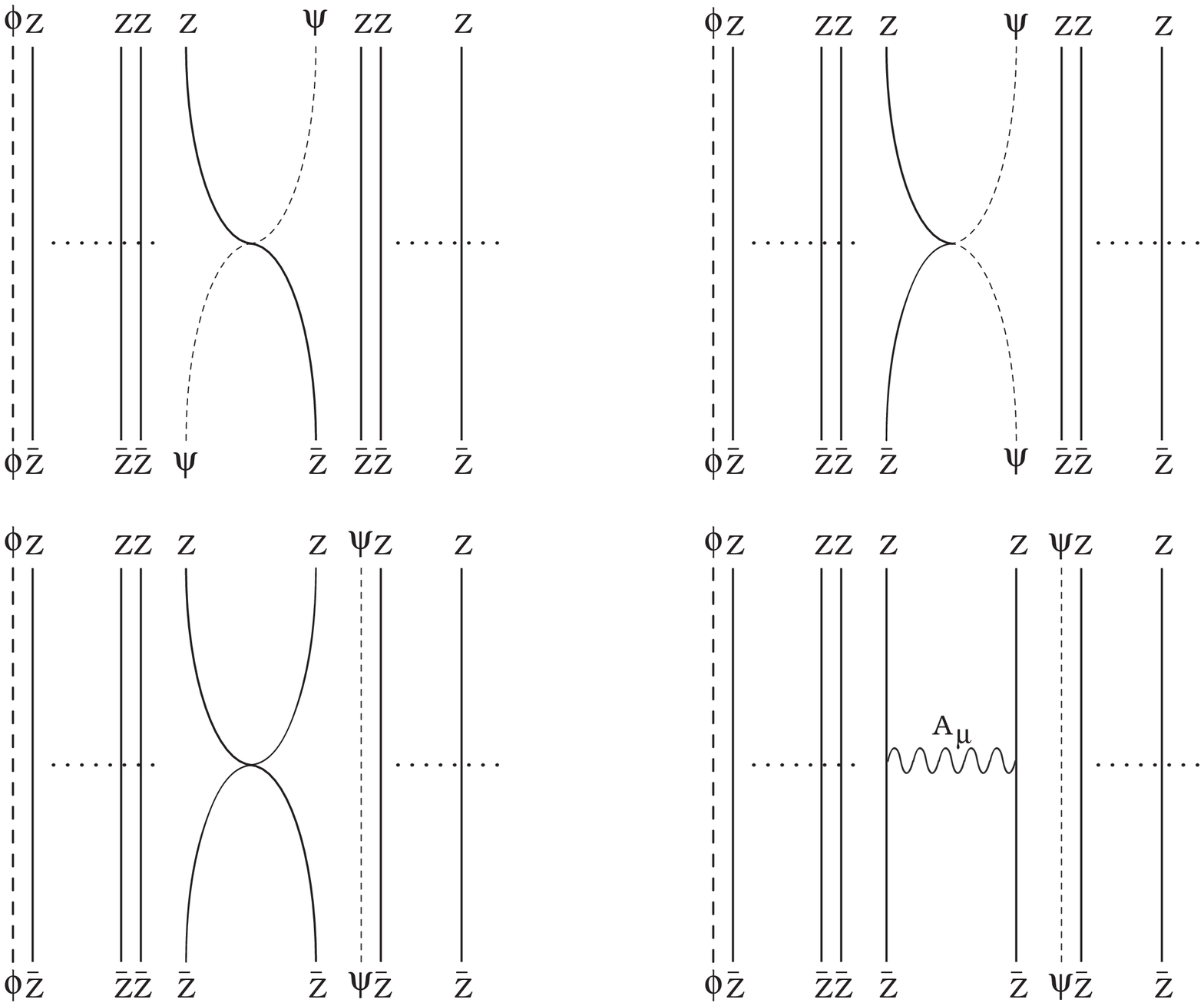,width=165mm,height=123.75mm}
\begin{center}
\caption{Feynman diagrams containing a single insertion of the interaction
Hamiltonian. The two different impurities are labeled $\phi$ and $\psi$.}
\label{feynman1}
\end{center}
\end{figure}

The correlation function will then include the overlap of all operators of the form 
\eqref{generic-op} with the appropriate exponential factors, in other words, the sum of the 
correlators of all pairs of operators \eqref{generic-op}, with some exponential coefficient.
At one-loop, where we have a single insertion of the interaction Hamiltonian, there will in 
general be two classes of diagrams: $(i)$ those in which the correlator receives contributions 
from two-point functions with the same $p$, corresponding to diagrams with no exchange of $\phi$ and $Z$, and $(ii)$ those where one exchange of $\phi$ and $Z$ takes place. The corresponding 
diagrams are presented in FIG.\ref{feynman1}.

The first diagram arises from contractions where one $\phi$ field has ``hopped'' past a $Z$ field, in this case to the left. The exponential factor appearing in front of this term is
$exp(2 \pi i n / J)$, since the amplitude for this term is
\be
  e^{2 \pi i p n / J} \: e^{-2 \pi i (p-1) n / J} \:
  \Big\langle
  \widetilde{\co}_p^J (x) \: \bar{\widetilde{\co}}_{p-1}^J (y)
  \Big\rangle
\ee
There will also be a contribution from a diagram where a $\phi$ field hops to the right, and it 
will be associated with a factor $exp(-2 \pi i n / J)$, with the amplitude otherwise the same.

We will now compute the amplitude for this diagram at planar level, but only keep track  of the 
divergent parts which determine the counterterm structure and eventually the anomalous dimension.
In position space, this diagram consists of $J+2$ fields located at spacetime position $x$,
interacting with $J+2$ fields located at $y$. The divergence arises from the loop at the center of 
the diagram, which corresponds to the integration over all spacetime (i.e. $\int d^4 w$) of one 
insertion of the Hamiltonian and four propagators.
The loop integral will contribute, beyond the tree level result,
\be\label{integral-1}
  \frac{1}{64 \pi^4} \lambda e^{2 \pi i n / J} \int d^Dw \frac{1}{|w-x|^4 |w-y|^4} \sim
  \frac{1}{16 \pi^2 |x-y|^4} \frac{\lambda}{\epsilon} e^{2 \pi i n / J}
\ee
where we have continued to $D=4-2 \epsilon$ dimensions to regulate the ultraviolet divergence 
coming from $x \rightarrow w$ and $y \rightarrow w$, which now appears as a pole in $\epsilon$.
We have the 't Hooft coupling appearing here because a factor of $\gymsq$ combines with a factor 
of $N$ at planar level when the first contraction across the traces are taken. We drop the 
contributions from the part of the diagram outside the interaction, since these do not modify the 
counterterm structure  we are seeking.
We see the appearance  of the combination $\lambda / \epsilon$ appropriate to one-loop.
We ignore the issue with infrared divergences when the external momenta vanish; these do not affect the anomalous dimension.

There are also diagrams in which $\phi$ and $Z$ fields interact, but which nonetheless do not lead to hopping. The hop-less diagrams in which $\phi$ and $Z$ fields interact arise in two ways. The first such diagram is similar to the one we considered above, but with a different ordering of the fields in the interaction term. There is also a diagram in which the interaction between the scalar $\phi$ and $Z$ fields is due to gluon exchange. These two diagrams contain the same divergences in their loops, but with opposite sign, and so their sum is finite. We ignore finite contributions since they do not give rise to anomalous dimensions.

The action \eqref{neqf-action-components} contains an interaction term in which only $Z$ fields 
interact with each other and a term in which $Z$ fields interact with gluons, and clearly lead to no hopping. Such interactions give rise to diagrams in which the four scalar $Z$ fields interact  directly, and diagrams where their interaction is a result of gluon exchange. Both these diagrams contribute equal divergences with the same sign. The divergence part of these is the same as in \eqref{integral-1}, but since there is no hopping, the exponential prefactor is missing. At planar level, there are $J-2$ possible ways the $Z$ fields can interact among each other.

The ultraviolet divergences in these diagrams can be removed by the addition of counterterms to 
the action to absorb the divergences. Computing the correlation function above, to one-loop, with 
an insertion of the composite operator, and including the counterterms appropriate to this order, 
we find the finite renormalized $Z_{\co}$, whose value is
\be
  Z_{{\co}_n} =
 1 -
 \frac{\lambda}{8 \pi^2 \epsilon}
 \left( 2 e^{2 \pi i n / J} + 2 e^{- 2 \pi i n / J} + (J-2) \right)
\ee
where the first two terms absorb the divergences from the diagrams with one hop to the left or 
right respectively, with a factor of two multiplying the exponential due to the hopping,
since we are considering composite operators with two impurities.
The last term absorbs the divergences from the two diagrams which do not result in a hop and come from the interactions of $Z$ fields alone, and contribute $J-2$ such counterterms.

\begin{figure}[ht]
\centering
\epsfig{figure=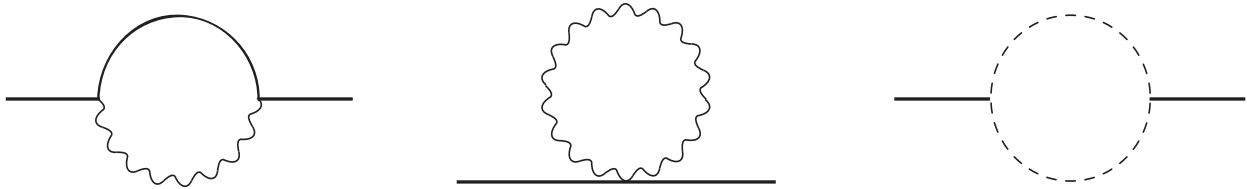,width=165mm,height=24.75mm}
\begin{center}
\caption{Diagrams contributing to the scalar wavefunction renormalization at one loop. The first is a scalar tadpole, the second a gauge boson loop, and the third a fermion loop.}
\label{wf-ren}
\end{center}
\end{figure}

We are now almost ready to compute the anomalous dimension of the operator $\co_n$. The only 
remaining piece left to compute is the wave-function renormalizations of the individual fields 
which enter into the correlation function, as seen in \eqref{define-overall-Z}.
There are three types of diagrams which modify the scalar propagators at one-loop.
The wave-function renormalizations, which are the only kind of renormalization to the bare $\neqf$ propagators, are generated by diagrams in which a closed loop is constructed as in FIG. \ref{wf-ren}, and arise from gauge boson, fermion and scalar loops. Their computation is straightforward, and the resultant one-loop wave-function renormalization is
\be
  Z_\phi = Z_Z = 1 + \frac{1}{4 \pi^2} \frac{\lambda}{\epsilon}
\ee

The factor of $\lambda$ arise because there are two interaction vertices, each contributing 
$\gym$, and a factor of $N$ enters due to the traces over color indices from the closed loop.
Putting these together we have to first loop order
\be \label{final_Z}
  \mathcal{Z}_n = 1 - \frac{\lambda}{4 \pi^2 \epsilon}
  \left( e^{2 \pi i n / J} + e^{- 2 \pi i n / J} - 2 \right)
\ee
which yields the anomalous dimension
\be \label{final-anomalous-dim}
  \gamma_n = 
  - \: \frac{\lambda}{4 \pi^2}
  \left( e^{2 \pi i n / J} + e^{- 2 \pi i n / J} - 2 \right)
\ee
We will meet this general from again in \ref{anomalous-dimensions-all-orders}.
The $2$ is a direct result of supersymmetry, since for $n=0$, the operator is BPS and hence 
protected against receiving quantum corrections.
This is a manifestation of the BPS condition in the form \eqref{BPS-condition-on-M}.
Incidentally, we can decompose this result into the $c_{l,h}^{(n)}$ we met in
\eqref{Fourier-coeffs}, whence  $c_{1,1}^{(n)}=c_{1,-1}^{(n)}=1$ and
$c_{1,0}^{(k)}=-2$.

We mention also that, had we separated the interaction Lagrangian into F and D-terms, at first 
loop we would have found that only the F-terms contribute, and the sum of all the diagrams with 
insertions of D-terms vanish, for two and three-point functions \cite{Constable:2002hw, 
Kristjansen:2002bb}.

The anomalous dimension in \eqref{final-anomalous-dim} has been computed for finite $J$. In the 
BMN limit, when $J$ is taken large, the anomalous dimension becomes
\be \label{large-J-anomalous-dim}
  \gamma_n =
  n^2 \: \lambda^\prime
\ee
and we see explicitly the appearance of the new effective coupling $\lambda^\prime=\lambda/J^2$ because the 't Hooft coupling $\lambda=\gymsq N$ has combined with a $1/J^2$ from the expansion of the exponentials. 
We see that in the BMN limit the anomalous dimensions of BMN operators are finite, since $\gymsq$ 
is held fixed while $N$ and $J$ are scaled such that $\lambda^\prime$ remains finite.
Contrast this with a normal 't Hooft expansion, in which the expansion parameter is $\lambda$, and this diverges in the BMN limit.
This is a key result, since it tells us that in the double scaling limit, BMN operators will have 
finite, and hence well-defined, scaling dimensions, which can be compared to the string side of 
the duality.
Recall that the exponentials entered as the diagonalizing matrix transforming the original 
basis of operators \eqref{generic-op} to one with well-defined scaling dimension, which we took to 
define one set of BMN operators. In turn, the precise structure of this matrix originated in the 
hopping behaviour embodied in the interaction Hamiltonian.

We can compare this result for anomalous dimensions of single-trace operators with two impurities 
to the string theory calculation of the mass spectrum for single-string states \eqref{omega}, with 
excitations of the left and right moving oscillators at level $n$ in the plane-wave 
background.
As we discussed in section \ref{BMNconjecture}, the BMN correspondence states a relationship 
between the effective coupling in the gauge theory and in the BMN limit, and string theory 
parameters on the plane-wave background, which was stated in \eqref{lambda'}, 
\eqref{BMNgenuscounting} and \eqref{stringcoupling}. We noted earlier in section \ref{isometry} 
that $p^+$ is a central charge of the supersymmetry algebra of the plane-wave background, since its 
generator commutes with all the other generators of the algebra. As such, its value specifies a 
sector of the string theory, unmixed by actions of the isometry or string interactions.
This is in distinct contrast to flat space, where the light-cone boosts can change $p^+$.
Therefore it makes sense to think of $\alpha^\prime \mu p^+$ (or 
equally $\mu$ in a sector of fixed $p^+$) as an expansion parameter on the string side.
The effective gauge theory expansion parameter $\lambda^\prime$ is related to the light-cone 
momentum, which is held fixed in the BMN double scaling limit, via \eqref{lambda'}. When the gauge 
theory is weakly coupled and $\lambda^\prime$ is small, the light-cone momentum $\alpha^\prime \mu 
p^+$ is large.
This implies that the tension term in the light-cone string theory action \eqref{LCbosonicaction} 
dominates the gradient terms (since we've taken $\mu$ large), and the quantum mechanics of the 
string becomes that of a collection of massive particles. This has motivated the string bit model
\cite{Vaman:2002ka,Verlinde:2002ig}.
Under these conditions, the mass spectrum \eqref{omega} can be expanded to first order, with the 
result that
\be
  \omega_n \approx \alpha^\prime \mu p^+ \left( 2 + \frac{n^2}{(\alpha^\prime \mu p^+)^2} \right).
\ee
We use the relation ${\cal D}- {\cal J}=\Delta_0+\gamma-J$, which for the operator we have 
considered gives
${\cal D}- {\cal J}=2+\gamma$. We then have
$\frac{2\omega_n}{\alpha^\prime \mu p^+}={\cal D}- {\cal J}$, after using 
\eqref{large-J-anomalous-dim} and \eqref{lambda'}.
The comparison is valid so long as $\lambda^\prime \ll 1/n^2$, or equivalently, when
$|\alpha^\prime \mu p^+| \gg n$. For any finite $n$, we are free to choose $\lambda^\prime$ or 
$\alpha^\prime \mu p^+$ so that these conditions hold.
This is our first direct test of the BMN conjecture, and it has passed with flying colors.

In this section, we computed anomalous dimensions of a class of BMN operators to first order in 
$\gymsq$, and found that it reproduces the string theory calculation, giving a first test of the 
BMN conjecture.
We may wonder whether this result extends to higher loops. The investigation of this question will 
be the focus of the next section, and we will show that the result indeed holds to all orders in 
perturbation theory.

\subsection{Anomalous dimension of the BMN operators, the planar result to all orders in 
$\lambda'$}
\label{anomalous-dimensions-all-orders}


In this subsection we establish the anomalous dimensions of BMN operators in the BMN limit 
\eqref{BMNlimit1} and \eqref{BMNlimit2}, to all orders in perturbation theory, and demonstrate its 
finiteness. We follow 
\cite{Santambrogio:2002sb}, 
relying heavily on superspace techniques and general results from conformal field theory. 
The two-point correlation function was first computed to order $g^4$, using $\mathcal{N}=1$ 
superspace techniques in \cite{Penati:1999ba, Penati:2000zv}.
This analysis was later extended by \cite{Santambrogio:2002sb}, in the planar limit (genus zero),
perturbatively to all orders in the 't Hooft coupling $\lambda^\prime$.
The analysis relies heavily on restrictions on the form of quantum corrections to correlation 
functions in $\neqf$ super-Yang-Mills, arising from supersymmetry and conformal invariance.

The relevant $\neqf$ action, written in $\mathcal{N}=1$ language, together with the relevant 
superspace conventions, is presented in appendix \ref{ConventionD=4}. The $U(1)_J$ 
subgroup we are 
interested in is the one which rotates one of the chiral superfields, which we choose to be 
$Z=\Phi^3$, corresponding to the real scalars in \eqref{define-Z}, and carrying one units of 
positive R-charge.
The propagator for chiral superfields can be derived from 
\eqref{neqf-action1} by expanding the exponentials in the first term, and then inverting the 
quadratic operator connecting $\Phi, \bar{\Phi}$, with the result
\be
\label{super-propagators}
\begin{split}
  \langle \Phi_{ab}^i (z) \bar{\Phi}_{cd}^j (z^\prime) \rangle
  =
  \frac{\delta^{ij} \delta_{ad} \delta_{bc} g_{YM}^2}{8 \pi^2} \bar{D}^2 D^{\prime 2}
  \frac{\delta^4 (\theta - \theta^\prime )}{ | x - x^\prime |^2 } \\
  \langle \bar{\Phi}_{ad}^i (z) \Phi_{bc}^j (z^\prime) \rangle
  =
  \frac{\delta^{ij} \delta_{ad} \delta_{bc}g_{YM}^2}{8 \pi^2} D^2 {\bar{D}^{\prime 2}}
  \frac{\delta^4 (\theta - \theta^\prime )}{ | x - x^\prime |^2 }, 
\end{split}
\ee
with the $U(N)$ adjoint indices explicitly indicated.
The appearance (and the relative order) of the superspace differential operators in these 
propagators can be understood as follows. The fields appearing in the action \eqref{neqf-action1} 
are to be interpreted as chiral superfields, and so the sum over these fields in the partition 
function must be constrained in an appropriate way. We may enforce such a constraint by 
introducing unconstrained potential superfields $\mathcal{U}$ (in a way analogous to Maxwell 
theory), and writing the chiral superfields as \cite{Weinberg:2000}
\be \label{potential-superfield}
  \Phi=\bar{D}^2 \mathcal{U} ,
\ee
whereby $\Phi$ is automatically chiral, satisfying $\bar{D}_{\dot{\alpha}} \Phi=0$. We then 
rewrite the action in terms of the potential superfields, with the partition function measure 
summing over all such field configurations, using \eqref{potential-superfield} in place of the 
chiral fields in the action. When computing correlation functions of chiral superfields, we use 
the new action with the insertions of chiral superfields replaced again according to 
\eqref{potential-superfield}.
A careful treatment then leads us to the form of the propagators in \eqref{super-propagators} 
\cite{Weinberg:2000}.
In particular, the order of the two superderivatives in the two propagators will turn out to be 
important for our purposes, since they do not commute.


Now consider a general operator consisting of $h$ chiral fields and $\bar{h}$ anti-chiral fields. 
The propagators for these are given in \eqref{super-propagators}. We assume the free propagators 
are normal-ordered, and drop all singular terms arising from self contractions of such propagators 
(this is the limit $z \rightarrow z^\prime$).
We'll assume for simplicity that these fields are in the same $\mathcal{N}=1$ multiplet, and 
ignore the index $i$. We will suppress group indices below for notational clarity and
denote such a composite operator as $\cw_{h,\bar{h}}$. We are interested in computing the two-point 
correlator of $\cw$ and its conjugate, $\langle \cw_{h,\bar{h}} (z) \bar{\cw}_{h,\bar{h}} 
(z^\prime) \rangle$ in the free theory\footnote{By $\bar{\cw}_{h,\bar{h}}$ we mean
the conjugate of $\cw_{h,\bar{h}}$. The conjugate actually contains $h$ anti-chiral fields and 
$\bar{h}$ chiral fields.}. As usual, this can be computed by taking Wick 
contractions to write the result as sums of products of the free propagators 
\eqref{super-propagators}. We do not trace over the group indices here, although for gauge 
invariant BMN operators such traces are in place.
Taking traces changes some of the dependence on $N$ in the calculations below, but does not alter 
the anomalous dimension we arrive at in the end.

Some identities which will prove useful are listed in \eqref{useful-identities}.
The simplest case to consider, is of course, that of $h=1,\bar{h}=0$. We define the total 
dimension of the operator $\cw_{h,\bar{h}}$ to be $\Delta=h+\bar{h}$, and the chiral weight to be 
$\omega=h-\bar{h}$. For this simple case, we have $\Delta=\omega=1$.
This example gives rise to only a single propagator, and establishing the result amounts to 
applications of some simple superspace identities.
We would like to rewrite the result in a form that makes clear the differences arising from the 
two propagators in the \eqref{super-propagators}, the source of which is the particular ordering 
of the chiral and anti-chiral derivatives.
We also note that the term $\delta^4 ( \theta - \theta^\prime ) / | x - x^\prime |^2$ is symmetric 
in the primed and unprimed arguments, and so we are free to drop the prime on the derivatives. For 
both ordering of derivatives, $\bar{D}^2 D^2$ and $D^2 \bar{D}^2$, we use commutation relations to 
push one $D^\alpha$ to the left or one $D_\alpha$ to the right, such that both terms generate a 
factor of $D^\alpha \bar{D}^2 D_\alpha / 2$ plus a term proportional to $D^\alpha 
\bar{D}^{\dot{\alpha}}$, but with a relative sign difference depending on the original ordering. 
This sign difference will appear through the chiral weight, which counts the difference between 
numbers of the two types of propagators, and hence the two ordering of the superderivatives. For 
the simplest case we are considering, the result is simple and can be written as
\be
  \langle \cw_{1,0} (z) \bar{\cw}_{1,0} (z^\prime) \rangle = C_{1,0} g_{YM}^2
  \left(
    \frac{1}{2} D^\alpha \bar{D}^2 D_\alpha + \frac{i}{4}
    [ D^\alpha , \bar{D}^{\dot{\alpha}} ] \sigma^\mu_{\alpha \dot{\alpha}} \partial_\mu
  \right)
  \frac{\delta^4 ( \theta - \theta^\prime )}{| x - x^\prime |^2}\ .
\ee
Care must be taken in deriving this expression to drop all singular terms which go like delta 
functions, arising from self contractions of the single propagator (i.e., dropping terms which 
vanish when $x \ne x^\prime$). They arise for example in the above case when $\Delta=1$ through
$\Box |x-x^\prime|^{-2}$. These are removed by the normal-ordering prescription.
The analysis of the case where $h=0,\bar{h}=1$, yields a similar result, with the difference that 
the commutator appears with opposite ordering. There is an overall normalization which we have 
absorbed into the constant $C_{1,0}$.

The next simplest case to consider is with $\Delta=\omega=2$, in which we simply include an 
additional chiral field. We need to evaluate the expression
$\left( \bar{D}^2 D^2 (\delta^4(\theta - \theta^\prime) |x-x^\prime|^{-2}) \right)^2$, which is 
straightforward but tedious. We first expand the superderivatives using the Leibniz rule, then 
square the expression. This generates a large number of terms; however, many
can be dropped by noting that they multiply together delta functions of Grassmann coordinates, or 
such delta functions and Grassmann coordinates; some of these terms will also involve products of 
superderivatives. Most of these terms then vanish because of the Grassmann nature of the delta 
functions and coordinates. Terms of the form
$\delta^4(\theta-\theta^\prime) F(D_\alpha,\bar{D}_{\dot{\alpha}}) \delta^4(\theta-\theta^\prime)$
vanish unless all the $\theta$'s in the second delta function are removed by the combination of 
chiral and anti-chiral derivatives in $F$, and this implies that the only non-vanishing terms of 
this form are those for which $F$ contains at least two chiral and two anti-chiral derivatives; 
for the case with two pairs of such derivatives, the only terms which contribute are those without 
the partial derivative pieces in the superderivatives, since the partial derivatives are always 
paired with anticommuting coordinates.
It also simplifies the calculation to always keep the remaining Grassmann delta function 
explicitly, without applying the Grassmann derivatives to it.
Using the relation
$\Box |x-x^\prime|^{-2 \Delta}=4 \Delta (\Delta - 1) |x-x^\prime|^{-2(\Delta+1)}$
and the identities in \eqref{useful-identities}, we can write the result as
\be
  \langle \cw_{2,0} (z) \bar{\cw}_{2,0} (z^\prime) \rangle = C_{2,0} g_{YM}^4
  \left(
    D^\alpha \bar{D}^2 D_\alpha + \frac{i}{2}
    [ D^\alpha , \bar{D}^{\dot{\alpha}} ] \sigma^\mu_{\alpha \dot{\alpha}} \partial_\mu +
    \Box
  \right)
  \frac{\delta^4 ( \theta - \theta^\prime )}{| x - x^\prime |^4}.
\ee

Had we considered adding an anti-chiral field instead of a chiral field, so that $h=\bar{h}=1$, we 
would have found that both the commutator term (because we would have generated two commutators, 
but with opposite sign), and the Laplacian term would have vanished.
The form of terms with successively more fields can be deduced in the same way, once the form of 
the previous one in the sequence is known, and this suggests an inductive derivation of the 
general result. Given the form for a correlator with an arbitrary number $h$ chiral and $\bar{h}$ 
anti-chiral fields, multiplying by another propagator for a chiral or anti-chiral field and 
performing superspace algebra as above, generates the form of the term with $h+1$ chiral or 
$\bar{h}+1$ anti-chiral fields.

If we assume that the result for $h$ chiral fields and no anti-chiral fields is proportional to
\be \label{results-1}
  g_{YM}^{2h}
  \left( \frac{h}{2} D^\alpha \bar{D}^2 D_\alpha +
  \frac{i}{4} [ D^\alpha , \bar{D}^{\dot{\alpha}} ] \sigma^\mu_{\alpha \dot{\alpha}} \partial_\mu +
  \frac{h}{2} \Box
  \right)
  \left( \frac{\delta^4 ( \theta - \theta^\prime )}{| x - x^\prime |^{2h}} \right), 
\ee
then we can find the result for $h+1$ chiral fields by noting
\be \label{results-2}
  g_{YM}^{2(h+1)}
  \left( \frac{(h+a)}{2} D^\alpha \bar{D}^2 D_\alpha +
  \frac{i(h+b)}{4} [ D^\alpha , \bar{D}^{\dot{\alpha}} ] \sigma^\mu_{\alpha \dot{\alpha}}
    \partial_\mu +
  \frac{(h+c)}{2} \Box
  \right) \bar{D}^2 D^2
  \left( \frac{\delta^4 ( \theta - \theta^\prime )}{| x - x^\prime |^{2h}}
         \frac{\delta^4 ( \theta - \theta^\prime )}{| x - x^\prime |^{2}} \right)
  = 0, 
\ee
which vanishes because of the two delta functions.
For $\bar{h}$ anti-chiral fields and no chiral fields, \eqref{results-1} would be by replaced an 
equation in which $h \rightarrow \bar{h}$, and with the sign of the second term reversed.
Judicious use of the Leibniz rule and dropping all terms which vanish because of the presence of too many Grassmann coordinates will generate a result of the form \eqref{results-1} for which $a=b=c=1$.
Had we reversed the order of $\bar{D}^2 D^2$ (to $D^2 \bar{D}^2$) in \eqref{results-2}, to add one more anti-chiral field, we would have arrived at $a=-b=c=1$.
By induction, we arrive at the general form of the correlation function for arbitrary values of $h$ and $\bar{h}$ (still in the free theory),
\be \label{general-correlator-free}
  \langle \cw_{h,\bar{h}} (z) \bar{\cw}_{h,\bar{h}} (z^\prime) \rangle = C_{h,\bar{h}}
  g_{YM}^{2(h+\bar{h})}
  \left(
    (h+\bar{h}) D^\alpha \bar{D}^2 D_\alpha + 
    \frac{i(h-\bar{h})}{2}
    [ D^\alpha , \bar{D}^{\dot{\alpha}} ] \sigma^\mu_{\alpha \dot{\alpha}} \partial_\mu +
    \frac{h(h-1)+\bar{h}(\bar{h}-1)}{2(h+\bar{h}-1)} \Box
  \right)
  \frac{\delta^4 ( \theta - \theta^\prime )}{| x - x^\prime |^{h+\bar{h}}}, 
\ee
for some overall constant $C_{h,\bar{h}}$ depending on the number of chiral and anti-chiral fields.
In addition, the Laplacian term gives zero when $h=1,\bar{h}=0$ or $h=0,\bar{h}=1$.

We are now interested in computing the value of such correlation functions in the interacting 
theory, taking quantum corrections into account. We take advantage of the conformal invariance of 
the theory, which is preserved (by virtue of the $\neqf$ supersymmetry), in the quantum theory. As 
we have discussed in \ref{amonalous}, conformal invariance fixes the form of the 
two and three point correlation functions. For two-point functions, the only modifications appear 
in corrected scaling dimensions, and the overall normalization of the correlation functions, which 
takes account of the allowed composite operator renormalizations.\footnote{Supersymmetry restricts 
the form of all renormalizations to be in the form of wave-function renormalizations for fields, 
and overall renormalizations for composite operators.} The scaling dimension of the operators 
differ from their classical dimensions through the introduction of anomalous dimensions, which 
vanish at zero coupling. The chiral weight is not renormalized because the chiral and anti-chiral 
fields receive the same anomalous dimensions, as CPT commutes with the scaling operator in the 
superalgebra. Therefore, the two-point function \eqref{general-correlator-free} in the 
full interacting theory, written in terms of the scaling dimension $\Delta$ and chiral weight $\omega$, 
becomes
\be \label{general-correlator-interacting}
  \langle \cw_{h,\bar{h}} (z) \bar{\cw}_{h,\bar{h}} (z^\prime) \rangle =
  C_{h,\bar{h}} (g_{YM}^2 N ) g_{YM}^{2 \Delta_0}
  \left(
    \Delta D^\alpha \bar{D}^2 D_\alpha + 
    \frac{i\omega}{2}
    [ D^\alpha , \bar{D}^{\dot{\alpha}} ] \sigma^\mu_{\alpha \dot{\alpha}} \partial_\mu +
    \frac{\Delta^2 + \omega^2 - 2 \Delta}{2 \Delta - 1} \Box
  \right)
  \frac{\delta^4 ( \theta - \theta^\prime )}{| x - x^\prime |^{2 \Delta}}, 
\ee
with the full scaling dimension $\Delta=\Delta_0 + \gamma$ now the sum of the classical scaling 
and anomalous dimension. So far we have been considering a general $U(N)$ gauge theory at 
arbitrary $N$; however, we are interested in the BMN limit of such operators.
The coefficients $C_{h,\bar{h}}$ are universal in the sense that they depend only on $h$ and 
$\bar{h}$, and not on the particular layout of the chiral and anti-chiral fields in the operator
$\cw_{h,\bar{h}}$.

At genus zero $C_{h,\bar{h}} (g_{YM}^2 , N )$ depends on $g_{YM}^2$ and $N$ through the 't Hooft 
coupling $\lambda = g_{YM}^2 N$. This is in fact where the assumption of planarity appears.
Its dependence on the 't Hooft coupling can be expanded in a power series,
$C_{h,\bar{h}} (\lambda) \propto 1 + \sum_{n=1}^\infty \lambda^n d^n_{h,\bar{h}}$, with an overall 
proportionality factor coming from the normalization of the propagators \eqref{super-propagators}.

We would now like to specialize our discussion, so far in the general $\neqf$ framework, to the 
case of BMN operators, such as those in \eqref{ijBMNstringmode}.
The $\neqf$ multiplet, when written in $\mathcal{N}=1$ language, consists of three chiral superfields.
We single out one of these chiral superfields, which we take to be $Z=\Phi^3$.
It carries unit charge under the $U(1)$ subgroup of the
$SU(4)$ R-symmetry of the superalgebra, rotating the scalars $\phi^5,\phi^6$ into
each other. The total R-charge counts the number of $Z$ fields appearing in the
correlation function.
The remaining chiral superfields $\Phi^1,\Phi^2$ are neutral under this $U(1)$.

For definiteness, we consider operators of the form
\be \label{building-blocks}
\begin{split}
  \mathcal{V}^J_n & = \sum_{p=0}^J e^{\frac{2 \pi i n p}{J}} Z^p \Phi^1 Z^{J-p} \\
  \mathcal{W}^J_n & = \sum_{p=0}^J e^{\frac{2 \pi i n p}{J}} Z^p \bar{\Phi}^2 Z^{J-p}, 
\end{split}
\ee
which can be used as building blocks for operators having more impurities.  
As we discussed in section \ref{BMNproposal}, $n$ is the excitation level on the string side of the 
duality.
The operators in \eqref{building-blocks} are related to each other by supersymmetry (they sit in 
the 
same supermultiplet, up to charge conjugation), with the important consequence that they receive the same quantum corrections, and hence the same anomalous dimensions.
The equations of motion governing the fields in \eqref{building-blocks} can be used to relate the two set of operators in \eqref{building-blocks}.
Interactions can be read off from the action \eqref{neqf-action1}, and connect the three chiral superfields via terms proportional to $\int d^4 x \int d^2 \theta \Tr \left( \Phi^1 [ \Phi^2, Z ] \right) + h.c.$, where the trace is taken with respect to the $U(N)$ indices.
The equation of motion for $\Phi^2$ is derived after rewriting the chiral integrals as chiral derivatives,
\be \label{supereom}
  D^2 \Phi^2 = -i \sqrt{2} \left( \bar{\Phi}^1 \bar{Z} - \bar{Z} \bar{\Phi}^1 \right), 
\ee
together with its conjugate. The trace in the interaction has disappeared; $\Phi^2$ on the 
left-hand side carries the same $U(N)$ matrix indices as the right-hand side, where $\Phi^1$ and 
$Z$ are multiplied in the matrix sense.
\eqref{supereom} allows us to relate $\mathcal{V}^J_n$ and 
$ \mathcal{W}^J_n$. From the equations of motion we can show
\be \label{trade}
  \bar{D}^2 \mathcal{W}^J_n = i \sqrt{2}
    \left( 1 - e^{\frac{- 2 \pi i n}{J}} \right)
    \mathcal{V}^J_n\ .
\ee
Using \eqref{trade} we can rewrite the two point function of two $\bar{D}^2 \mathcal{W}^J_n$ as
\be \label{trade-2}
  \langle \bar{D}^2 \mathcal{W}^J_n (z) D^2 \mathcal{\bar{W}}^J_n (z^\prime) \rangle =
  2 \: \left( 2 - 2\: cos (\frac{2 \pi n}{J})  \right)
  \langle \mathcal{V}^{J+1}_n (z) \mathcal{\bar{V}}^{J+1}_n (z^\prime) \rangle.
\ee
The correlation function on the right differs from that of $\mathcal{V}^J_n$ through the 
insertion of one $Z$. In the limit we are considering, any single impurity interacts with $Z$ 
fields within a finite number of places from the location of the impurity,
implying that the renormalization of $\mathcal{V}^J_n$ and $\mathcal{V}^{J+1}_n$
at any order in the perturbative expansion are the same, and hence they receive the same anomalous 
dimension. Alternatively, we can argue that in the large $J$ limit these operators tend to 
each other, and hence receive the same corrections.
Parenthetically, if we rewrite the prefactor in parentheses on the right-hand side above as
$2-e^{i 2 \pi n / J}-e^{-i 2 \pi n / J}$, we see the similarity to 
\eqref{final-anomalous-dim}, with the $2$ being a consequence of supersymmetry, as we argued in 
\ref{anomalous-dimensions-first-order}.

We point out that these relations are derived using the equations of motion, and hence hold only 
on-shell. The important point to note is that it is the on-shell operators (built from fields 
satisfying the equations of motion) which appear in the correspondence; the duality relates these 
to on-shell string states, satisfying the Virassoro physical state constraints.

The number of chiral and anti-chiral fields appearing in $\mathcal{V}^J_n$ is $h=J+1,\bar{h}=0$, 
while for $\mathcal{W}^J_n$, $h=J,\bar{h}=1$, giving their classical scaling dimension and chiral 
weight as $\Delta=J+1,\ \omega=J=+1$ and $\Delta=J+1,\ \omega=J-1$ respectively.
We now make use of the relation \eqref{general-correlator-interacting} for the two-point function 
of composite operators assembled from chiral or anti-chiral superfields, using it  to replace 
both sides of \eqref{trade-2}. 
Using manipulations similar to those above to maneuver the $\bar{D}^2$ and $D^2$ into proper 
place, we then have, introducing $f(\lambda,\Delta,\omega)$, from which we have extracted the tree 
level normalization,
\be \label{relation}
  N_0^{J} f(\lambda,J+1,J-1)
  \left( \gamma^2 + 2 \gamma \right)
  = 4 \: \left( 1 - cos (\frac{2 \pi n}{J}) \right)
  \: N_0^{J+1} f(\lambda,J+2,J+2), 
\ee
with $N_0$ defined in \eqref{singlevac}. Both sides are multiplying
$\bar{D}^2 D^2 \delta^4 ( \theta - \theta^\prime) / |x-x^\prime|^{2(J+2+\gamma)}$, which we have 
stripped. $f(\lambda,\Delta,\omega)$ tracks the overall quantum corrections to the operators,
and may be expanded in a perturbative expansion in $\lambda$, for any $\Delta,\omega$,
$f(\lambda,\Delta,\omega) = 1 + \sum_{n=1}^\infty \lambda^n f_n(\Delta,\omega)$.
In the large $J$ limit, the factors of $f$ on both sides of \eqref{relation} tend to the same 
function (because their arguments approach each other), with subleading corrections in $J$ which 
we will drop.
The factors of $N$ appearing in \eqref{relation} arise here because we are computing correlation 
functions of fields carrying matrix indices, and the fields in the composite operators
$\mathcal{V}^J_n$ and $\mathcal{W}^J_n$ are multiplied in the matrix sense.
The powers of $N$ appearing are those appropriate to the planar contractions. As a result, the 
anomalous dimension $\gamma$ is determined by solving a quadratic equation, with the solution
\be
  \gamma = -1 + \sqrt{1 - 4 \: N_0 \: \left( cos(\frac{2 \pi n}{J}) - 1 \right)}\ .
\ee
The other solution to the quadratic equation would yield a non-zero anomalous dimension in the free theory limit, and must be discarded.
In the large $J$ limit, the anomalous dimension becomes
\be
  \gamma = -1 + \sqrt{1 + \lambda^\prime n^2}\ , 
\ee
making evident the explicit dependence on the modified 't Hooft coupling $\lambda^\prime$.

This result holds to all orders in perturbation theory,
but only at the planar level, confirming the proposal of
\cite{Berenstein:2002jq} for the mass spectrum of the corresponding string states, and thus
provides a non-trivial check of the duality.

The above result can be generalized for BMN operator with more impurities, for example 
\eqref{multiimpurityBMNoperator}. The technique used to calculate the anomalous dimension remains
the same, and revolves around the key equation \eqref{trade}. Given any operator 
$\mathcal{V}^{J+1}_n$ with more impurities, we can find a corresponding operator
$\mathcal{W}^{J+1}_n$ to which it can be related via an equation analogous to \eqref{trade} 
(which, however, is in general more complicated). The end result is a relation similar to 
\eqref{relation}, which can then be solved for the anomalous dimension. The result is 
\cite{Gross:2002mh, Gomis:2003kj}
\be
\gamma_{n_1\dots n_m}=\sum_{i=1}^m\left( -1 + \sqrt{1 +  n_i^2 \lambda'} \right) \ ,
\ee
with the understanding that $\sum_i n_i \ll J$.

\subsection{Operator Product Expansions in the BMN subsector}
\label{OPE}

In this section, we first present a brief review of the operator product expansion (OPE), then 
move on to discuss its relevance in the context of the BMN correspondence. The most salient point 
will be a demonstration of the closure of the BMN subsector of operators in the $\neqf$ 
super Yang-Mills theory, which can serve as a motivation for the selection of this class of 
operators.
On the more practical side, the OPE will allow us to rewrite certain correlators involving 
multi-trace operators in terms of operator product expansions of single-trace operators in 
prescribed pinching limits. We will use this technique is section \ref{n-point-functions} to 
discuss three and higher point functions of BMN operators.


In quantum field theory, the product of operators is in general divergent if the location of any 
of the operators coincide, and require renormalization.
For free fields, the divergence can be removed by normal-ordering the operator
product, which amounts to subtracting the vacuum expectation value. However, in a general 
interacting theory, the product remains divergent even after normal-ordering.
In the short distance (high momentum) limit, the operator product expansion (OPE) allows one to 
express the singular behaviour as
\be \label{ope-equation}
  O_i (x) \: O_j (y) \: \sim  \: \sum_k \: C_{ij}^k (x-y) \: O_k (y) \: + \:
  \text{non-singular terms}
\ee
with the coefficients $C_{ij}^k(x-y)$ singular in the limit $x \rightarrow y$,
and the other terms regular in this limit.  The $O_k$ are assumed to form a linearly
independent basis of local
operators for the theory under consideration, which commute or anticommute among themselves.
In a unitary conformal field theory, this basis can be taken to be orthonormal 
\cite{DiFrancesco:1997nk}. The sum on the 
right-hand side of (\ref{ope-equation})
receives contributions from a finite number of terms in the limit $x \rightarrow y$.
The OPE (\ref{ope-equation}) is to be understood as an operator relation, i.e., it holds as a 
matrix element between any sets of states, or equivalently, as an insertion into any expression of 
the form
\be
  \Big\langle \cdots O_i (x) \: O_j (y) \cdots \Big\rangle \: \sim  \: \sum_k \:
  C_{ij}^k (x-y) \: \Big \langle \cdots O_k (y) \cdots \Big\rangle , 
\ee
with $\cdots$ denoting other operators which lie a distance to $y$ is greater than $|x-y|$.
In a general quantum field theory, the OPE is an asymptotic expansion and hence not convergent. In 
the special case of a conformal field theory, the OPE can be shown to converge, with radius of 
convergence given by the distance to the nearest operator other than those which coincide.
The proof of the operator product expansion, under some restrictive assumptions, can be found in
\cite{Zimmerman:1970}.
A general discussion of the operator product expansion in quantum field theory can be found in
\cite{Weinberg:1996kr}, while a discussion applicable to two-dimensional conformal field theory, 
for example string theory, can be found in \cite{Polchinski:1998rq,Polchinski:1998rr}.

We have presented a discussion of two and three point functions in unitary conformal field 
theories in section \ref{amonalous}.
The arguments of that section can be applied also to the OPE coefficients, and a renormalization 
group equation for them can be derived. More importantly, the OPE of quasi-primary operators 
simplifies, after choosing an orthonormal basis of operators, into
\be \label{ope-quasi-primary}
  \Big\langle \mathcal{O}_i(x_1) \mathcal{O}_j(x_2) \Big\rangle
  \: \sim \: \sum_k \frac{C_{ij}^k(g_{ym}^2,N)}{|x_{12}|^{\Delta_i+\Delta_j-\Delta_k}}
  \mathcal{O}_k(x_2), 
\ee
where again, the form of quantum corrections is limited to anomalous dimensions and corrections to 
the OPE coefficients. For a single-trace BMN operator $\co_k^J$, the scaling dimension is
$\Delta_k=J_k+\mathcal{I}_k+\gamma_k$, with $\mathcal{I}_k$ the engineering dimension of the 
impurities and $\gamma_k$ the anomalous dimension of the operator.
We now wish to demonstrate the important result \cite{Chu:2002qj} that the operator product 
expansion of BMN operators is closed. Closure here is to be interpreted as follows: the OPE of BMN 
operators has an expansion where only BMN operators appear and the expansion coefficients are 
finite in the BMN limit. The OPE of a set of BMN operators with non-BMN operators has an expansion 
where the OPE coefficients vanish in the BMN limit. This is suggestive that in the double scaling 
limit, the operators of interest to us are in fact the BMN operators. This result is closely 
related to fact that the anomalous dimensions of non-BMN operators, which have some $\lambda$ 
dependence, which we take to $\infty$ in the double scaling limit, generically diverge in the 
BMN limit.
In the operator product expansion \eqref{ope-quasi-primary}, we consider the case of two BMN 
operators on the left-hand side. The total $R$-charge of the two sides must match, as well as the 
total scaling dimensions (which is already included in \eqref{ope-quasi-primary}).
We separate the dependence on the scaling dimension of the operators as follows
\be \label{ope-BMN}
  \Big\langle \mathcal{O}_i(x_1) \mathcal{O}_j(x_2) \Big\rangle
  \: \sim \: \frac{1}{|x_{12}|^{\Delta_i + \Delta_j - J}}
  \sum_k C_{ij}^k(g_{ym}^2,N) |x_{12}|^{\Delta_k - J}
  \mathcal{O}_k(x_2)\ , 
\ee
which is suggested by the R-charge. For BMN operators with a finite number of impurities,
$\Delta_i + \Delta_j - J = \mathcal{I}_i + \mathcal{I}_j$ which is finite, where
$\mathcal{I}_i$ is the dimension of the impurities in the BMN operator $\mathcal{O}_i$,
and $J$ is taken to be the magnitude of the total R-charge $J=|J_i| + |J_j|$.
Now, $\Delta_k - J=\mathcal{I}_k + \gamma_k$, using R-charge conservation, and it is positive by 
virtue of the BPS bound. This is either finite or infinite. Since $J$ is taken large, $\Delta_k - 
J$ finite implies $\mathcal{O}_k$ is a BMN operator. Otherwise $\Delta_k - J$ is infinite and 
$\mathcal{O}_k$ is non-BMN, but then $|x_{12}|^{\Delta_k - J}$ vanishes faster than $|x_{12}|^{J - 
\Delta_i - \Delta_j}$, and therefore the contributions of the non-BMN operators drop out of the OPE.
This behaviour is a direct result of the finiteness of the anomalous dimensions of BMN operators, 
a feature of the double scaling limit we noted earlier, and the divergence of the anomalous 
dimensions of non-BMN operators in the same limit. This is another facet of the requirement that 
BMN operators have well-defined total scaling dimensions in the BMN limit. To summarize, 
only the set of BMN \opt s contribute to the sum  in \eqref{ope-BMN}. 

One can also show, using the results of \cite{Lee:1998bx,Mann:2003qp}, that in the large $J$ 
limit, the OPE of a set of BMN operators with non-BMN operators has an expansion where the OPE 
coefficients vanish.


%% file: inter.tex

Having carefully considered the planar structure of BMN operators, we are now ready to move on and 
examine non-planar corrections to quantities we have been studying in section 
\ref{noninteractingstrings}, first considering higher genus corrections to two-point correlation 
functions of chiral-primary operators (these receive no loop, i.e. $\lambda'$, corrections). 
The  BMN limit of these correlators is examined, showing that in the double scaling limit, certain 
higher genus contributions survive. This result distinguishes the BMN limit from the standard 't 
Hooft limit, wherein all contributions from higher genus diagrams are seen to vanish. This 
consideration will demonstrate explicitly the appearance of the genus counting parameter in the 
BMN limit. We next look at correlators of BMN (near-BPS) operators, first in the free field theory 
limit but with first non-planar contributions, and then after turning on interactions, computing 
the first non-trivial contributions in both the genus counting parameter and the modified 't Hooft 
coupling $\lambda'$. Mixing between single and multiple trace BMN operators, and the requisite 
re-diagonalization of the basis, leading the so called ``improved BMN operators'', will play a 
central role in the precise formulation of the correspondence between gauge theory operators and 
string states. We collect the above results in an elegant form suggested by 
\cite{Gomis:2002wi,Constable:2002vq}.
Up to this point, our focus has been on the calculation of two-point functions of BMN operators; 
three and even higher point functions are introduced and some pathology in their behaviour noted. 

\subsection{Non-planar contributions to correlators of chiral-primary 
operators}\label{NP-chiral-primary}

We review the expansion {\it to all genus} of the 
two-point functions of chiral-primary operators, which are protected against quantum corrections
by virtue of being BPS. Hence, the results we present can be calculated in the free theory, but 
extend to all values of the coupling.

\begin{figure}[ht]
\centering
\epsfig{figure=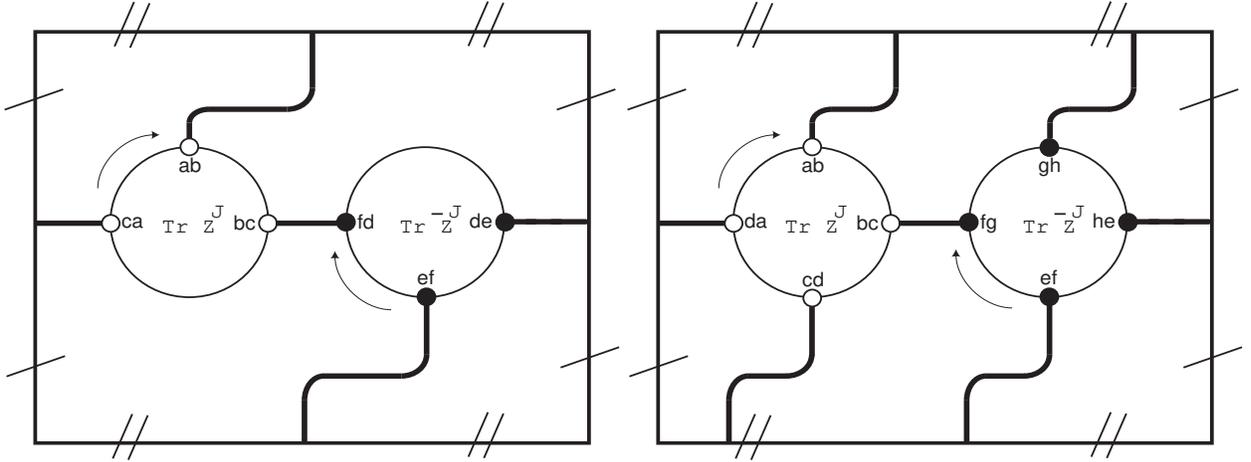,width=165mm,height=61.29mm}
\begin{center}
\caption{Irreducible toroidal diagrams contributing to $\langle \Tr Z^J \Tr \bar{Z}^J \rangle$. 
The arrows indicate the direction in which traces are taken.}
\label{torus-figures}
\end{center}
\end{figure}

To gain some insight into the genus expansion, consider the simplest correlation function that 
receives contributions from higher genus diagrams, the two-point function of chiral-primary 
operators \eqref{singlevac} with $J=3$ (the case of $J=2$ only receives planar corrections)
\be
  \Big\langle \co^J(x) \bar{\co}^J(0) \Big\rangle_{J=3}=
  \frac{1}{3 N_0^3} \langle Z_{ab} Z_{bc} Z_{ca} \bar{Z}_{de} \bar{Z}_{ef} \bar{Z}_{fd} \rangle
  \ .
\ee
There are six possible ways of applying Wick contractions. Three of these lead to a factor of 
$N^3$ from contractions (leaving aside for now the prefactor coming from the normalization). 
These correspond to the planar diagrams. Planar diagrams always generate the highest power of $N$, 
and hence are the ones that dominate a large $N$ expansion (for finite $J$).
Planar diagrams are those which can be drawn on a sphere (a one-point compactification of the 
plane) without any lines crossing. There are also three (that this number equals $J$ is a 
coincidence) non-planar diagrams, of genus one. These are diagrams which can not be drawn on a 
sphere without crossing, but can be placed on a torus without crossing. They contribute a single 
power of $N$.
One can see the structure more clearly by the following trick \cite{Kristjansen:2002bb}.
Imagine that each trace corresponds to a loop on which we place beads corresponding to the individual fields $Z$ and $\bar{Z}$, white beads depicting $Z$'s and black ones for the conjugate fields $\bar{Z}$. The beads are free to move on the loop, but can't be pushed past each other (their order is significant).
Changing the ordering of two nearby beads corresponds to crossing or uncrossing the lines connecting them. For the case $J=3$, reversing the order of the beads on one of the loops while keeping the other loop's ordering fixed exchanges planar and non-planar diagrams, showing how the ordering of the beads is relevant.
The cyclicity of the trace is reflected in the fact that rotating the beads around the loop results 
in an identical loop. One of the possible non-planar contractions is depicted in FIG. 
\ref{torus-figures} (the left figure). 

For $J=3$, the maximum genus contributing is the torus.
This trick can be generalized to higher $J$ and genus. First we need the notion of an irreducible 
diagram. Replace all lines in a diagram that are topologically parallel (call these reducible) 
with a single line (irreducible). The resulting diagram built only from irreducible lines is 
itself irreducible.
Diagrams can be grouped into equivalence classes, where the equivalence is defined as follows: two 
diagrams are considered equivalent if they both collapse to the same irreducible diagram.
For $J=4$ there are diagrams which reduce to the one we have already considered for $J=3$, and new 
ones which reduce to the one depicted in FIG. \ref{torus-figures}, the right 
figure.
For higher $J$, all toroidal diagrams can be reduced to the two already considered. More 
generally, at genus $h$, the set of irreducible diagrams consists of those where the number of 
irreducible lines $l$ ranges between $l=2h+1$ and $l=4h$, which for genus one gives $l=3,4$ and 
for genus two the range is $l=5...8$ \cite{Kristjansen:2002bb}.

At genus one, for arbitrary $J \ge 3$, there are
$
J! / ( (J-3)! \: 3! )
$
ways of grouping the beads into three sets (the three irreducible lines in FIG. \ref{torus-figures}) 
while maintaining the order associated with the operator, and for for $J \ge 4$ the number of such 
groupings into sets of four is
$
J! / ( (J-4)! \: 4! )
$.
We denote the number of inequivalent irreducible diagrams with $l$ irreducible lines at genus $h$ 
by $n_{h,l}$.
The calculation of this number is the trickiest part of working out the combinatorics.
For the cases we have already considered $n_{1,3}=1$ and $n_{1,4}=1$, while $n_{1,j}=0$ for $j>4$. 
However, for higher genus, there exist $n_{h,k}$ greater than one.
The total number of diagrams in an equivalence class with $l$ irreducible lines for fixed $J$ can 
be found as follows: given a set of $J$ elements, place the elements into $l$ ordered distinct 
sets, maintaining the same overall cyclic ordering among all the elements. The number of possible 
ways of doing this is 
$
J! / ( (J-l)! \: l! )
$.
The total number of diagrams at genus $h$ with $l$ irreducible lines for fixed $J$ is
$
n_{h,l} \:
J! / ( (J-l)! \: l! )
$.
At fixed genus, to arrive at the total number of graphs we must sum up the contribution from 
graphs in all equivalence class for all allowed $l$. For the torus, this gives
\be
n_{1,3}
\left(
\begin{matrix}
J \\ 3
\end{matrix}
\right) \: + \:
n_{1,4}
\left(
\begin{matrix}
J \\ 4
\end{matrix}
\right) \: \approx \: \frac{J^4}{4!} \, ,
\ee
where in the last step we have shown the scaling in the large $J$ limit.
Notice that sums of this form are always $N$ independent. The $N$ dependence in the combinatorics 
arise from traces over indices of Kronecker deltas appearing in the propagators 
\eqref{phi:propagator} and \eqref{z:propagator} after all the Wick contractions are applied (of 
course keeping only diagrams at a fixed genus), and this dependence defines the genus order, via 
the standard 't Hooft argument, where the suppression factor at any genus relative to the next 
lower genus goes like $1/N^2$ (or $1/N^{2h}$ relative to planar diagrams). As a result, in the BMN 
double scaling limit \eqref{BMNlimit2} (as opposed to the usual 't Hooft limit), diagrams at all 
genera contribute to correlation functions, giving rise to a new effective expansion parameter 
$g_2^2=(J^2/N)^2$, which is fixed at an arbitrary but finite value and measures the relative 
contribution of each genus in perturbation theory. Contributions from diagrams at genus $h$ scale 
as $g_2^{2h}$.
For the planar diagrams, there is an overall suppression by a factor of $J$ due to the 
normalization 
of the operators in \eqref{singlevac}, but a compensating enhancement by the same factor arising 
from the cyclicity of the trace (which amounts to the rotation of the beads on one of the loops 
relative to the other one).
Putting together these observations, we arrive at the planar plus toroidal contribution to the 
two-point function of chiral-primary operators
\be \label{two-pt-chiral-first-order}
  \Big\langle \co^J(x) \bar{\co}^J(0) \Big\rangle =
  \frac{1}{|x|^{2J}}
  \left(
  1 + \frac{g_2^2}{4 !}+{\cal O}(g^4_2)
  \right) \, .
\ee
The normalization of the operator \eqref{singlevac} is chosen to remove the overall dependence
of the two-point function above on $N$ as well as the coupling $\gymsq$ and factors of $8 \pi^2$.
Here we see the appearance of the parameter $g_2^2$ which organizes the expansion by
genus. The planar diagrams contribute at order $g_2^0$ and the toroidal diagrams at order $g_2^2$.
The new observation for the BMN double scaling limit is that the operators considered receive 
contributions from a number of diagrams which grow as $J^{4h}$ at genus $h$,
but these are suppressed by $1/N^{2h}$, and the $J$ and $N$ dependence combine into the new 
effective expansion parameter $g_2^2$, appearing at genus $h$ as $g_2^{2h}$.

We will now describe a method for establishing the all orders (in $g_2^2$) result.
We earlier mentioned two dimensional $QCD$ as a realization of 't Hooft's idea, and its exact
solution via a matrix model \cite{Kostov:1997bs,Kostov:1998bn}.
It turns out that many of the correlation functions we are interested in can be reduced to 
correlation functions in this matrix theory.
Higher genus correlation functions in the complex matrix model, using loop equations, have been 
computed in \cite{Ambjorn:1992xu}. An alternative method for evaluating statistical ensembles of 
complex (or real) matrices can be found in \cite{Ginibre,Mehta}.
We will need only the most rudimentary results from matrix theory, which we collect here.
Consider $N \times N$ complex matrices $Z_{ij}$, with $i,j$ running from $1$ to $N$, and define 
the measure $dZ d\bar{Z}$ as
\be
  dZ d\bar{Z} \: = \: \prod_{ij} \frac{1}{\pi} d(Re \: Z_{ij})d(Im \: Z_{ij}) \, .
\ee
The partition function over these matrices is defined as the above measure weighted by a Gaussian 
function
\be
  \mathcal{Z} \: = \: \int dZ d\bar{Z} e^{-Tr(Z \bar{Z})} \, .
\ee
The measure and the weight (and hence the partition function) are $U(N) \times U(N)$ invariant, 
representing independent multiplications on the left and the right.
Correlation functions in this matrix model are defined as usual in QFT
\be
  \Big\langle \mathcal{O}(Z,\bar{Z}) \Big\rangle_{MM} \: = \:
  \int dZ d\bar{Z} e^{-Tr(Z \bar{Z})} \mathcal{O}(Z,\bar{Z}) \, .
\ee
The normalization of the measure is chosen so that $\langle 1 \rangle = 1$.

The correlation functions we study are not invariant under the full symmetry, but only under
those generated by the diagonal subgroup, acting in the adjoint representation.
For correlators built out of traces which do not mix $Z$ and $\bar{Z}$, the solution can be given 
by using character expansion techniques, expanding the correlation function in terms of group 
characters. These characters are orthogonal, with a proportionality constant that can be evaluated 
from group theory. The expansion coefficients are similarly computed from Young diagram 
considerations. We summarize the relevant result for two-point functions
\begin{widetext}
\be \label{mm:two:point:function}
  \Big\langle
  \Tr Z^J \Tr\bar{Z}^J
  \Big\rangle_{MM} \: = \:
  \sum_{k=1}^J \prod_{i=1}^k ( N - 1 + i ) \prod_{m=1}^{J-k} ( N - m )
  =
  \frac{1}{J+1} \left(
  \frac{\Gamma(N+J+1)}{\Gamma(N)} -
  \frac{\Gamma(N+1)}{\Gamma(N-J)} \right) \, ,
\ee
\end{widetext}
where have assumed $0 < J < N$ in the last step.
In the above, $N$ is the rank of the group $U(N)$ (we have kept $N$ finite thus far).
Up to this point the results are exact.

Let us now return to the correlation function \eqref{two-pt-chiral-first-order}. As we discussed in
section \ref{amonalous} the spacetime dependence of this two-point function is completely fixed by
the conformal invariance. Moreover, being chiral-primary the scaling dimension is also fixed by
supersymmetry to the free field theory engineering dimension. These have already been made manifest
in \eqref{two-pt-chiral-first-order}.
The remaining problem in computing \eqref{two-pt-chiral-first-order} is that of 
computing the dependence on 
factors of $J$ and $N$ arising from the combinatorics of all the Wick contractions.
Separating out the spacetime dependence, and also the numerical and coupling constant factors in 
the scalar field propagators, the correlation function can be rewritten in terms of a correlation 
function in the matrix model we have described, which captures the combinatorics from evaluating 
all the traces over $U(N)$ color indices (producing both planar and non-planar contributions), as 
well as the combinatoric dependences on $J$,
\be \label{cp-two-point}
  \Big\langle \Tr (Z^J(x)) \: \Tr (\bar{Z}^J(0)) \Big\rangle =
  \left( \frac{g_{YM}^2}{8 \pi^2 |x|^2} \right)^J
  \Big\langle \Tr (Z^J) \Tr (\bar{Z}^J) \Big\rangle_{MM} \, ,
\ee
making use of the matrix model result \eqref{mm:two:point:function}.
We are interested in the large $J$ limit of \eqref{cp-two-point}, and hence that of
\eqref{mm:two:point:function}. We can expand it as
\be \label{BMN-limit-MM-result}
  \Big\langle
  \Tr Z^J \Tr\bar{Z}^J
  \Big\rangle_{MM} \: = \:
  J \: N^J \Bigg[
  1 \: + \: \sum_{h=1}^\infty \: \sum_{k=2h+1}^{4h}
  \left(
  \begin{matrix}
  J \\ k
  \end{matrix}
  \right)   
  \frac{n_{h,k}}{N^{2h}}
  \Bigg]
  \approx
  J \: N^J \Bigg[
  1 \: + \: \sum_{h=1}^\infty
  \frac{n_{h,4h}}{(4h)!} \: \left( \frac{J^4}{N^2} \right)^{h} \: + \: \cdots
  \Bigg] \, ,
\ee
where in the last expression we have taken the large $J$ limit, and $\cdots$ denotes terms
which vanish in the large $J$ and $N$ limit if we scale $J \sim \sqrt{N}$.
We see that the genus counting parameter $g_2^2=J^4/N^2$ make a natural appearance in this limit. We will see in the next section when we come to consider non-BPS operators that this continues to be the case. In fact, this is another way to view the BMN limit: the limit is chosen precisely to ensure that the terms involving $n_{h,4h}$ in this limit remain finite and so we receive
contributions from all genera.
We can explicitly evaluate \eqref{cp-two-point} using \eqref{BMN-limit-MM-result} in the
BMN limit, giving for the chiral-primary operators
\be
  \Big\langle \co^J (x) \: \bar{\co}^J (0) \Big\rangle =
  \frac{1}{|x|^{2J}}\cdot \frac{\sinh \left( \frac{g_2}{2} \right)}{\frac{g_2}{2}} \, .
\ee
Expanding this to first order in $g_2^2$ reproduces \eqref{two-pt-chiral-first-order}.

\subsection{Non-planar contributions to BMN correlators}

In this section we move onto the non-BPS (``almost-BPS'') BMN \opt s and compute the $g_2^2$ order
non-planar contributions to their two-point functions, first at free field theory and then at first
order in $\lambda'$.

\subsubsection{Correlators of BMN operators in free gauge theory to first non-trivial order in $g_2$}
\label{BMN-ops-first-order-in-both}

Having studied the two-point function of chiral-primary operators to all orders, we are now ready 
to discuss the inclusion of phases in the more general BMN operators. We will concentrate on 
operators of the form \eqref{ijBMNstringmode} for $i \ne j$, and choose the notation $\phi_i=\phi$ 
and $\phi_j=\psi$. In this section we study the correlator in the free theory,
postponing consideration of interactions to the next section. 
The correlator we are interested in is
\be \label{two-point-torus-free}
  \Big\langle \co^J_{ij,m} (x) \bar{\co}^J_{ij,n} (0) \Big\rangle_0 \, .
\ee
The calculation of the torus level contribution to the two-point function of BMN operators in the 
free gauge theory has been carried out along two different lines, using matrix model technology in 
\cite{Kristjansen:2002bb}, and via direct computation taking account of the combinatorics in 
\cite{Constable:2002hw}. We will see that the scaling with $N$ and $J$, in the BMN limit, is the 
same as for the chiral-primary operators, and $g_2^2=(J^2/N)^2$ will appear again as the genus 
counting parameter. We follow closely the presentation in \cite{Constable:2002hw}.

To count the number of Feynman diagrams that contribute to a two-point function at genus $h$, we 
draw a polygon with $4h$ sides, then place one operator at the center, and divide the other 
operator among the $4h$ vertices. We then pairwise identify all the sides and identify the 
vertices. All allowed diagrams are then generated by connecting the two operators via propagators,
but without allowing the diagram to be collapsed to lower genus by shrinking homology cycles where 
no propagators have been placed. At genus $h$, the irreducible diagrams are those with $2h+1$ to 
$4h$ groups of lines.
The number of ways of dividing $J$ lines into $4h$ sets is
\be
  \left(
  \begin{matrix}
  J \\ 4h
  \end{matrix}
  \right) =
  \frac{J !}{(J - 4h)! (4h)!} \approx
  \frac{J^{4h}}{(4h)!} \, ,
\ee
where the last expression gives the behaviour at large $J$. A similar counting applies to the 
diagrams where we group the lines into $4h-1$ sets and so on, down to $2h+1$, but the number of 
such groupings is suppressed relative to the $4h$ case. For example, the case $4h-1$ yields
\be
  \left(
  \begin{matrix}
  J \\ 4h-1
  \end{matrix}
  \right) =
  \frac{4h}{J-4h+1}
  \left(
  \begin{matrix}
  J \\ 4h
  \end{matrix}
  \right)    
\ee
number of ways of distributing $J$ lines into $4h-1$ sets, and their contributions relative to the $4h$ groupings vanish in the BMN limit.
This is the same behaviour we saw in the previous section at genus one, and it generalizes to 
arbitrary genus and for any finite number of impurities.

\begin{figure}[ht]
\centering
\epsfig{figure=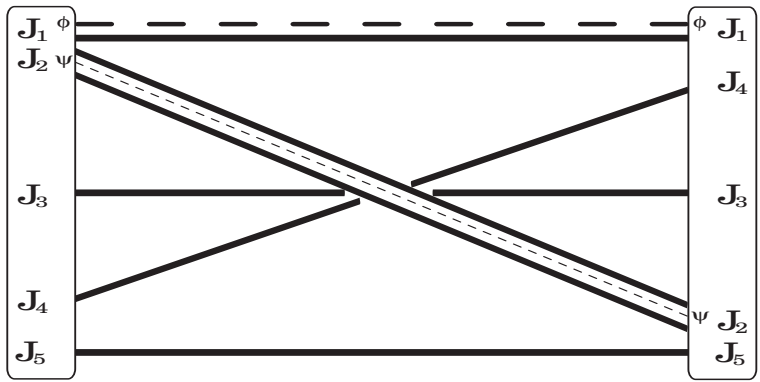,width=100mm,height=50mm}
\begin{center}
\caption{Diagram depicting the phase shift in a torus diagram with no interactions.
The solid lines represent an arbitrary number of $Z$ fields and the dashed lines represent the 
contraction between two $\phi$'s or $\psi$'s.}
\label{phases}
\end{center}
\end{figure}

We can open up FIG. \ref{torus-figures} for the torus diagrams with four groups of lines 
consisting of $J$ scalar fields $Z$ charged under $U(1)_J$ and two different scalar impurities we will label $\phi$ and $\psi$.
Using the cyclicity of the trace, we can always place one of the impurities, say $\phi$, as the 
first field in each operator before applying contractions. This simplifies the counting since the position of the $\phi$ field is fixed, and we only have to worry about placing the $\psi$ field. The diagram can then be drawn as in FIG. \ref{phases}.

Now there are five groups of fields, where the first one begins with the $\phi$ field. Let $J_i$
denote the number of fields, with $i=1,\ldots,5$ (with no $\psi$ field yet). We can place the 
$\psi$ field into any of these groups, and there are $J_i$ ways of doing so for the $i^{\rm th}$
group.
Let us consider first the case where $m=n$ in \eqref{two-point-torus-free}, so the two operators
have similar phase structures.
The two impurities may appear in the same group, in which case when we contract the fields in the two operators, the relative positions of $\phi$ and $\psi$ will remain fixed, and these diagrams will not contribute a phase factor. If the impurities are placed in different groups, then their relative positions in the two operators can in principle change, and the contractions will then be associated with a phase. For example, if $\psi$ is placed in the second group, then it will contract with a field in the conjugate operator where its relative position to the other conjugate scalar will have shifted by $J_3+J_4$ places, and this introduces a relative phase of
$exp(2 \pi i n (J_3+J_4)/J)$. Summing over all ways of placing $\psi$, we have for the two-point 
function the following expression
\be
  \Big\langle \co^J_{ij,m} (x) \bar{\co}^J_{ij,m} (0) \Big\rangle_0^{torus}
  = \frac{1}{J N^{J+2}}
  \left( \frac{1}{|x|^2} \right)^{J+2} N^J
  \sum_{J_1+\ldots+J_5=J+1} \ \
  \sum_{k=1}^5 J_k \ e^{2 \pi i m \theta_k / J} \, ,
\ee
with the phases defined as $\theta_1=\theta_5=0, \theta_2=J_3+J_4, \theta_3=J_4-J_2$ and
$\theta_4=J_2+J_3$.
In performing the sum, we must impose the condition that $\sum_{i=1}^5 J_i=J+1$.
The first term on the right hand side is due to the normalization of the operators in 
\eqref{ijBMNstringmode}. The next term arises from the propagators, with the normalization of the
operators and propagators conspiring to remove the coupling and numerical factors.
The last term comes from all the color index contractions at torus level.
This expression is awkward, but can be turned into an integral representation in the large $J$ 
limit, with a delta function imposing the constraint, which can be evaluated explicitly. To see 
this, define $J_i=J \cdot j_i$. Then in the large $J$ limit, we can rewrite the two-point functions as
\be
  \Big\langle \co^J_{ij,m} (x) \bar{\co}^J_{ij,m} (0) \Big\rangle_0^{torus} =
  \left( \frac{1}{|x|^2} \right)^{J+2} \left(\frac{J^2}{N}\right)^2
  \prod_{k=1}^5 \int_0^1 \: dj_k \: j_k \: e^{2 \pi i m \theta_k / J}
  \delta (1 - \sum_{l=1}^5 j_l) \, ,
\ee
and we again see the appearance of $J^2/N \equiv g_2$ which is held fixed in the BMN limit.
The integral can be evaluated in a straightforward way.
The construction when $m \ne n$ follows along the same lines, with the added complication that the position at which $\psi$ is inserted in each group becomes relevant, since the two operators have different phase structures. This more complicated situation has been considered in
\cite{Constable:2002hw,Kristjansen:2002bb}, and we present only the result.
The final expression for the torus two-point function of BMN operators of the type we have been
considering is
\be \label{final-two-pt-BMN-free}
  \Big\langle \co^J_{ij,m} (x) \, \bar{\co}^J_{ij,n} (0) \Big\rangle_0 \: = \:
  \left( \frac{1}{|x|^2} \right)^{J+2}
  \left( \delta_{mn} \, + \, g_2^2 \,
   M^1_{mn} \right) \, ,
\ee
for $i \ne j$ and where $g_2=J^2/N$, and with the matrix $M^1_{mn}$ is symmetric, i.e. 
$M^1_{mn}=M^1_{nm}$, and  is defined as
\be \label{define-Mmn}
  M^1_{mn} =
  \begin{cases}
    0 \ , & m=0,n \ne 0 \ \ \ \text{or} \ \ \ m \ne 0, n = 0 ; \\
    \frac{1}{24} \ , & m = n = 0 ; \\
    \frac{1}{60} - \frac{1}{24 \pi^2 m^2} + \frac{7}{16 \pi^4 m^4} \ , & m = n \ne 0 ; \\
    \frac{1}{48 \pi^2 m^2} + \frac{35}{128 \pi^4 m^4 } \ , & m = - n \ne 0 ; \\
    \frac{1}{4 \pi^4 (m-n)^2} \left( \frac{\pi^2}{3} + \frac{1}{m^2} + \frac{2}{n^2}
    - \frac{3}{2 m n} - \frac{1}{2 (m-n)^2} \right) \ , & \text{all other cases.}
  \end{cases}
\ee

The case of $m=0$ or $n=0$ corresponds to a two-point function with one of the composite operators 
being BPS; the $m=n=0$ gives the two-point function of BPS operators, while if $m \ne n$, but one 
of $m$ or $n$ zero, we see that the single-trace BPS operators do not mix with the non-BPS ones, 
at torus level. 
We expect the two-point function for $m=n=0$ to be exact to all orders in $\gymsq$, since these 
operators are protected against receiving any anomalous dimensions. The non-BPS cases will receive 
$\gymsq$ corrections, and we will discuss these corrections in section \ref{Mixingclues}.
The other cases show explicitly that in the free theory, single-trace non-BPS operators 
generically mix with each other, and this mixing begins at order $g_2^2$, where $g_2^2$ is the 
genus counting parameter.
The discussion above can be generalized in an obvious way to higher genus diagrams in the free 
theory, with the genus $h$ contributions coming in at order $(g_2^2)^h$.

\subsubsection{Correlators of BMN operators to first order in $\lambda^\prime$ 
and ${J^2}/{N}$}
\label{Mixingclues}


We have already computed the planar anomalous dimension to order $\lambda^\prime$ is section
\ref{anomalous-dimensions-first-order}, and to all orders in section 
\ref{anomalous-dimensions-all-orders}. We are now ready to move beyond planar level, but will work 
only to first order in $\lambda^\prime$. The duality would then put the result in
correspondence with the string theory masses with loop corrections, giving a highly
non-trivial test of the correspondence, and a step beyond what has been possible in the
standard AdS/CFT correspondence.

The result will be proportional to $\lambda^\prime g_2^2$, showing that $g_2^2$ will continue
to play the role of the genus counting parameter even with interactions switched on, and the
role of the effective quantum loop counting parameter is still played by $\lambda^\prime$, in the 
BMN limit.
The computation mirrors that of the previous section, but now taking account of the insertions
of interaction terms. We only present an overview of the calculations; technical details can
be found in \cite{Constable:2002hw}. At this order, only flavor changing interactions
contribute, and therefore the only interactions of relevance are the so called F-terms which
appear as the square of the commutator of scalars in different $\mathcal{N}=1$ chiral
multiplets. We are considering two scalar impurity operators which are in $({\bf 1, 9}$) 
of $SO(4)\times SO(4)$, they are symmetric in $i,j$ indices. Therefore the F-term which involves 
the commutator of the two impurities, being
antisymmetric, does not contribute. The two impurities can therefore be considered
separately, since they do not simultaneously enter into interactions, and only enter into
interactions which are quadratic in the charged fields $Z$. These observations greatly reduce the 
number of possible diagrams which must be considered at this order.

There are three classes of Feynman diagrams to consider, involving nearest neighbor, semi-nearest 
neighbor and non-nearest neighbor.\footnote{These can be classified according to the
possible combinations of contractible and non-contractible homology cycles on a torus, corresponding to the two propagator loops connecting to the interaction vertex. The diagrams
with two non-contractible cycles on the torus do not enter at this order in $\gymsq$ because
they involve interactions other than F-terms.}
Nearest neighbor diagrams are the ones where two lines alongside each other are connected through
an interaction term. One of these lines will always be an impurity. There are four possible interaction types coming from squaring the commutator in the interaction, all with equal
weight, with those that switch the order of the impurity and charged field contributing a
minus sign relative to those which do not. We must sum over all ways of building such diagrams by inserting a single interaction into the free diagrams, taking care with the phases from exchanges and the phase of the free diagram. The phase considerations parallel our discussion in the previous section. Summing all nearest neighbor diagrams, we find that the result
\eqref{final-two-pt-BMN-free} of the previous
section is simply modified by a logarithmic correction which merely changes the scaling
dimension we computed at planar level, since the result does not involve $g_2$. The other two
types of diagrams will, however, involve honest toroidal corrections.

The class of semi-nearest neighbor diagrams are those  in which the fields entering an interaction
are nearest neighbors in one of the composite operators, but not the other. These only contribute
to the two-point function when $m \ne n$. For $m = n$ there are cancellations among
semi-nearest neighbor diagrams. The number of such diagrams is suppressed
relative to the nearest neighbor ones by $1/J$, but this is countered by an enhancement by a factor
of $J$ because these diagrams have a different phase structure which in the large $J$ limit is
larger by a factor $J$ relative to the nearest neighbor diagrams.

The non-nearest neighbor contributions do introduce logarithmic corrections whether or not
$m \ne n$. These diagrams are rarer than the nearest neighbor one by a factor of $1/J^2$,
but we again have an enhancement which compensates this, due to the phase structure.
When we sum over all contributions from the above graphs, we must also consider the phase
associated to the diagram from the placement of the second impurity, as we had to when considering
the two-point function in the free theory.

The final result for the two-point function  of the single-trace BMN operators we have been 
considering is
\be \label{BMN-two-pt-first-order-in-both}
  \Big\langle \co^J_{ij,m} (x) \, \bar{\co^J_{ij,n}} (0) \Big\rangle \: = \:
  \left( \frac{1}{|x|^2} \right)^{J+2}
  \Bigg[
  \delta_{mn}
  \left( 1 + \lambda^\prime L \, m^2 \right)
  + \, g_2^2 \,
  \left( M^1_{mn}
  \, + \, \lambda^\prime L \left(
  mn M^1_{mn} \, + \, \frac{\mathcal{D}^1_{mn}}{8 \pi^2} \right) \right) \Bigg] \, ,
\ee
with $L=-\ln(|x|^2 \Lambda^2)$ and the matrix $M^1_{mn}$ given in \eqref{define-Mmn}.
This result holds for $i \ne j$. The matrix $\mathcal{D}^1_{mn}$ is
\be \label{define-Dmn}
  D^1_{mn} =
  \begin{cases}
    0 \ , & m=0 \ \ \ \text{or} \ \ \ n=0 ; \\
    \frac{2}{3} + \frac{5}{\pi^2 n^2} \ , & m=n \ne 0 \ \ \ \text{or} \ \ \ m=-n \ne 0 ; \\
    \frac{2}{3} + \frac{2}{\pi^2 m^2} + \frac{2}{\pi^2 n^2}\ , & \text{all other cases.}
  \end{cases}
\ee

The next question of interest, the significance of which would become clear in the next subsection, 
is the correlation function of a single-trace operator and a double-trace one, and two-point 
functions of double-trace operators. The double-trace operators have been 
defined in \eqref{doublevac}, \eqref{oneimpuritytwotrace} and \eqref{doubletracedeltaj2}. 
The double-trace operators \eqref{doubletracedeltaj2} contain two scalar impurities, and as 
discussed in section \ref{deltaj2} can be in $({\bf 1},{\bf 9})$, $({\bf 1},{\bf 3^\pm})$ or 
$({\bf 1},{\bf 1})$ tensor representations of $SO(4)\times SO(4)$. 
BPS operators do not occur in the antisymmetric representation of
$\ct^{J,r}_{ij, n}$, since $\co^J_{[ij],n}=-\co^J_{[ij],-n}$.
The correlators of non-singlets have been computed \cite{Beisert:2002bb}, with the result
\be \label{double-trace-2pt}
\begin{split}
 \Big\langle \ct^{J,r}_{ij} (x) \, \bar{\ct}^{J,s}_{ij} (0) \Big\rangle & =
  \left( \frac{1}{|x|^2} \right)^{J+2} \,
  \delta_{rs} \, , \\
  \Big\langle \ct^{J,r}_{ij,m} (x) \, \bar{\ct}^{J,s}_{ij,n} (0) \Big\rangle & =
  \left( \frac{1}{|x|^2} \right)^{J+2} \,
  \delta_{rs} \delta_{mn} \left(
  1 + \lambda' L \, \frac{m^2}{r^2} \right) \, , \\
  \Big\langle \ct^{J,r}_{ij,m} (x) \, \bar{\ct}^{J,s}_{ij} (0) \Big\rangle & = 0 \, ,
\end{split}
\ee
$(i \ne j)$, up to order $\lambda^\prime$ and $g_2$.
We see that the double-trace operators are diagonal to order $g_2$. Their mixing, like the single 
trace operators, begins at order $g_2^2$.
We are considering here the $SO(4)$ non-singlet operators, but not making a distinction between 
the {\bf 9} and ${\bf 3^\pm}$ representations. The results for the singlet representations are 
complicated by the inclusion of the $\Tr(Z^\dagger Z^{J+1})$ term in the definition 
\eqref{ijBMNstringmode}.
The {\bf 9}, ${\bf 3^\pm}$ and {\bf 1} \rep s all receive the same anomalous dimensions, and this 
degeneracy is a result of supersymmetry (\cf\ discussions of section \ref{deltaj2}). Starting from 
any one of these \rep s we may reach the 
others by transformations generated by combinations of supercharges, and these combinations 
commute with the dilatation operator. 

The single and double-trace operators mix at order $g_2$, with the overlaps, at first order in 
$\lambda'$ being
\be \label{double-single-trace-overlap}
\begin{split}
  \Big\langle \ct^{J,r}_{ij,m} (x) \, \bar{\co}^J_{ij,n} (0) \Big\rangle & =
  \left( \frac{1}{|x|^2} \right)^{J+2}
  \frac{g_2}{\sqrt{J}} \frac{r^{3/2} \sqrt{1-r} \sin^2(\pi n r)}{\pi^2 (m-n r)^2}
  \left( 1 + \frac{\lambda^\prime L \, (m^2-mn r+n^2r^2)}{r^2} \right) \, , \\
  \Big\langle \ct^{J,r}_{ij} (x) \, \bar{\co}^J_{ij,n} (0) \Big\rangle & =
  \left( \frac{1}{|x|^2} \right)^{J+2}
  \frac{g_2}{\sqrt{J}} \left( \delta_{n,0} r -\frac{\sin^2(\pi n r)}{\pi^2 \, n^2} \right)
  \left( 1 + \lambda^\prime L n^2 \right) \, .
\end{split}
\ee
Don't be alarmed by the appearance of $1/\sqrt{J}$ in these expressions. When we come to 
rediagonalize the single-trace operators in the next section, we will see that the $1/J$ terms are 
compensated by sums (over $r$), and the two-point functions of the rediagonalized single-trace 
operators will receive contributions from such terms.

Extracting an overall power of $g_2$ in expression like \eqref{double-single-trace-overlap},
the remaining terms can be arranged into an expansion in powers of $g_2^2$, i.e. in terms of 
planar and non-planar diagrams.

At this order in $g_2$, there are non-zero overlaps between double and triple trace operators, and 
at order $g_2^2$ even overlaps between single-trace and triple-trace operators.
More generally, the overlap of a single-trace operator with any $t$-trace operator begins at order 
$g_2^{t-1}$. We have ignored these corrections since they do not affect the anomalous dimensions 
of single-trace operators at order $g_2^2$.

\subsection{Operator mixings and improved BMN conjecture}\label{mixing}

The results of the previous two sections have been computed and presented in the BMN basis.
These results are to be compared to those on the string theory side of the duality according to
the identification \eqref{BMNconjecture1}, in which we are instructed to compare the eigenvalue
spectrum of the string field theory Hamiltonian to the spectrum of the dilatation operator minus
the R-charge in gauge theory. This is a basis independent comparison.
Alternatively, we may compare the matrix elements of the operators on the two
sides of the duality.
On the other hand, the two sides of the duality involve different Hilbert spaces, and
the mapping between the bases of these distinct Hilbert spaces
is part of the statement of the duality.
Denote the basis on the gauge theory side by $\{ | \, \mathfrak{a} \, \rangle{}_{{}_{gauge}} \}$ 
and on the string theory side by $\{ | \, \tilde{\mathfrak{a}} \, \rangle{}_{{}_{string}} \}$, with
$\mathfrak{a}$ labeling gauge theory states, and
$\tilde{\mathfrak{a}}$ the labels on the string side.
We need an isomorphism between the states of the two theories
\be
  \{ | \, \mathfrak{a} \, \rangle{}_{{}_{gauge}} \} \leftrightarrow
  \{ | \, \tilde{\mathfrak{a}} \, \rangle {}_{{}_{string}} \} \, ,
\ee
under the condition that the inner products on both sides agree
\be\label{Hilbertmetric}
  {}_{{}_{gauge}}\langle \, \mathfrak{a} \, | \, \mathfrak{b} \, \rangle{}_{{}_{gauge}} \: = \:
  {}_{{}_{string}}\langle \, \tilde{\mathfrak{a}}
    \, | \, \tilde{\mathfrak{b}} \, \rangle{}_{{}_{string}} \, .
\ee
The duality, in the proper basis, holds between these matrix elements
\be
  {}_{{}_{gauge}}\langle \, \mathfrak{a} \, |
  \, \left( \mathcal{D} - \mathcal{J} \right) \,
  | \, \mathfrak{b} \,
  \rangle{}_{{}_{gauge}} \: = \:
  {}_{{}_{string}}\langle \, \tilde{\mathfrak{a}} \, |
  \, \frac{H}{\mu} \,
  | \, \tilde{\mathfrak{b}} \,
  \rangle{}_{{}_{string}} \, .
\ee

In section \ref{deltaj2} the text around \eqref{BMNconjecture2} (the second part of the 
SYM/plane-wave duality), we introduced a specific mapping between the Hilbert spaces on the either 
sides of the duality, however, we warned the reader that \eqref{Hilbertmetric} does not hold for 
the identification \eqref{BMNconjecture2} (more precisely it only holds at $g_2^0$ level).	
In this section we intend to refine the dictionary between the gauge theory and string theory 
Hilbert spaces taking account of higher $g_2$ orders.

On the string theory side, there is a natural basis, the one which diagonalizes the free string 
theory Hamiltonian. We will refer to this basis as the free-string basis.
In this basis, $m$-string states are orthogonal to $n$-string states
for $m \ne n$, and in fact this basis is orthonormal (\cf\ discussions of section 
\ref{physicalstringspectrum}).
The interactions induce mixings between these states; this basis does not diagonalize the full 
string field theory Hamiltonian. For example, at order $g_s$, the cubic string field theory 
Hamiltonian will cause transitions between one and two string states.

On the gauge theory side we start with the BMN basis, but if we are interested in the full
scaling dimensions, including the anomalous dimensions, then we should choose a basis
which diagonalizes the dilatation operator. This basis is referred to as the
$\Delta$-BMN basis \cite{Georgiou:2003kt}. Incidentally, in this basis the operators are conformal 
primaries, and this is the basis in which the two and three-point functions
take the forms \eqref{2-point:functions:cft} and \eqref{2-point:functions:cft} required by 
conformal invariance.
This basis would correspond to the one on the string theory side
that diagonalizes the full string field theory Hamiltonian, and is not the free-string basis we
defined above.
The basis of BMN operators we have been working with above are neither of these.
They have well-defined scaling dimensions at planar level, but non-planar corrections
induce non-diagonal mixings between the single-trace operators and between single and multi-trace 
operators in general. For example, at toroidal level, the classical ($\lambda^\prime = 0$) scaling 
dimensions are no longer well-defined because of order $g_2^2$ mixings.
This is seen easily by noting that single and double-trace BMN operators overlap at
order $g_2$ \eqref{double-single-trace-overlap},
and shows up at order $g_2^2$ in two-point functions of single-trace operators
\eqref{BMN-two-pt-first-order-in-both}.

The results of section \ref{BMN-ops-first-order-in-both} can be cast in the form
\be
  |x|^{2 \Delta_0} \: \Big\langle \co_\mathfrak{a} (x) \bar{\co}_\mathfrak{b} (0) \Big\rangle =
  G_{\mathfrak{a} \mathfrak{b}} - \lambda^\prime
  \Gamma_{\mathfrak{a} \mathfrak{b}} \ln (|x|^2 \Lambda^2) \: + \: {\cal O}(\lambda'^2) \, ,
\ee
written in the BMN basis.
We have introduced a notation whereby the indices $\mathfrak{a}$ range over single, double 
and in general $n$-trace operators, and the operators within each such class.
This expression is written up to first order in $\lambda^\prime$, with the remaining
terms of higher order in $\lambda^\prime$. $\Delta_0$ is the classical (non-anomalous) scaling 
dimension.
The matrix $G_{\mathfrak{a} \mathfrak{b}}$ is the inner product on the Hilbert space of states
created by the BMN operators, and $\Gamma_{\mathfrak{a} \mathfrak{b}}$ is the
matrix of anomalous dimensions.

The free-string basis can be constructed on the gauge theory side by taking linear
combinations of the original BMN operators
\be
  | \, \mathfrak{a} \, \rangle{}_{{}_{gauge}} \: = \:
  \mathcal{U}_{\mathfrak{a} \mathfrak{b}} \co_{\mathfrak{b}}(0) | \, 0 \, \rangle{}_{{}_{gauge}}
  \, ,
\ee
with the BMN operator $\co_{\mathfrak{a}}$ acting on the gauge theory vacuum.
When $g_2$ vanishes, this basis coincides with the original BMN basis.
Therefore, at order $g_2^0$ the change of basis matrix $\mathcal{U}$ is simply the
identity.

Perturbative corrections in powers of $g_2$ results in a mixing between BMN operators with 
different numbers of traces, and we must rediagonalize this set of operators at each order in $g_2$
to maintain orthonormality of the inner product $G_{\mathfrak{a} \mathfrak{b}}$, to
preserve the isomorphism with the free-string basis.
The change of basis is chosen such that
\be \label{change-of-basis-matrix}
  \mathcal{U} G \mathcal{U}^\dagger = 1 \, ,
\ee
leading to
\be \label{gauge-matrix-free-string-basis}
  {}_{{}_{gauge}}\langle \, \mathfrak{a} \, |
  \, \left( \mathcal{D} - \mathcal{J} \right) \,
  | \, \mathfrak{b} \,
  \rangle{}_{{}_{gauge}} \: = \:
  \left( \mathcal{U} (\Delta_0 - J) G \mathcal{U}^\dagger
  + \mathcal{U} \Gamma \mathcal{U}^\dagger \right)_{\mathfrak{a} \mathfrak{b}} \: = \:
  n \delta_{\mathfrak{a} \mathfrak{b}} + \tilde{\Gamma}_{\mathfrak{a} \mathfrak{b}} \, ,
\ee
with $n$ counting the number of impurities in the operator $\co_{\mathfrak{a}}(0)$ creating
the state $| \, \mathfrak{a} \, \rangle{}_{{}_{gauge}}$, and $\tilde{\Gamma}$ the
anomalous dimension matrix in the free-string basis.
In the basis where the inner product $G$ is diagonal, the anomalous dimension
matrix is symmetric.
The matrix elements \eqref{gauge-matrix-free-string-basis} are to be compared to the matrix 
elements of the string field theory Hamiltonian in the free-string basis. We will return to
this in section \ref{SFT}.

The change of basis implemented by $\mathcal{U}$
is not unique, but all such choices are related by orthogonal transformations.
We may make a unique choice, with one subtlety involving BPS operators which we mention 
momentarily, by requiring that the matrix $\mathcal{U}$ implementing the change of basis 
\eqref{change-of-basis-matrix} be a real symmetric matrix. This turns out to be the choice for which the matrix elements
of the rediagonalized operators can be matched to the matrix elements on the string side
in the free-string basis.
As we have already pointed out, in the BMN basis, single-trace BPS operators do not mix with single 
trace non-BPS operators, and likewise for pairs of double-trace operators, but they may mix with each other. However, this mixing does not involve $\lambda^\prime$ corrections since both operators are BPS. This mixing will not affect the anomalous dimensions. A similar pattern occurs in the string field theory, where the sums over intermediate BPS states do not alter the string masses.
The dictionary translating between the string and gauge theory sides of the duality
then seems to contain ambiguities for the BPS operators and their corresponding
string states \cite{Beisert:2002bb};
for example,
we are unable to distinguish between single string and double string vacuum states, as well
as single and double graviton states. We will comment on this point briefly
in section \ref{conclusion}.
However, mixing between BPS and non-BPS operators can be dealt with by choosing a
basis in which BPS operators do not mix with non-BPS operators, regardless of the number of traces; however, the degeneracy in the BPS subspace remains.

We expand the diagonalizing matrix, the inner product matrix and the matrix of
anomalous dimensions, to order $g_2$:
\begin{alignat}{3}
\label{expand-in-g2}
  \mathcal{U} & = 1 && + g_2 \mathcal{U}^{(1)} && + \co(g_2^3) \, , \nonumber \\
  G & = 1 && + g_2 G^{(1)} && + \mathcal{O}(g_2^3) \, , \\
  \Gamma & = \Gamma^{(0)} && + g_2 \Gamma^{(1)} && + \co(g_2^3) \, . \nonumber
\end{alignat}
$\mathcal{U}$ and $G$ are the identity at zeroth order in $g_2$ since the BMN operators start 
mixing among each other only at order $g_2$ for single-trace and double-trace overlaps, and at 
order $g_2^2$ for single-trace overlaps with single-trace with double-trace intermediate channels,
while the non-vanishing of $\Gamma^{(0)}$ to this order captures the first order
(in $\lambda^\prime$) anomalous dimensions of the unmixed BMN operators.

Inserting the expansions \eqref{expand-in-g2} in \eqref{change-of-basis-matrix}, we find
that the change of basis matrix $\mathcal{U}$, to order $g_2$ involves the term of the same order
in the expansion of the inner product matrix, and since $\mathcal{U}$ is unitary, we have
\be
  \mathcal{U}^{(1)} \: = \: - \frac{1}{2} G^{(1)} \, .
\ee

We may also solve for $\tilde{\Gamma}$ to first order in $g_2$, using
\eqref{gauge-matrix-free-string-basis} and 
expanding $\tilde{\Gamma}$ as above in \eqref{expand-in-g2}, with
$\tilde{\Gamma}^{(0)}=\Gamma^{(0)}$. This yields
\be
  \tilde\Gamma^{(1)}=\Gamma^{(1)}-\frac{1}{2}\{G^{(1)},\Gamma^{(0)}\},
\ee
We then have for the order $g_2$ rediagonalized matrix of anomalous dimensions
\be \label{gamma-tilde-g2}
  \tilde\Gamma^{(1)}=
  \left(
  \begin{matrix}
  0 & \tilde{\Gamma}^{(1)}_{n,qs} & \tilde{\Gamma}^{(1)}_{n,s} \\
  \tilde{\Gamma}^{(1)}_{pr,m} & 0 & 0 \\
  \tilde{\Gamma}^{(1)}_{r,m} & 0 & 0
  \end{matrix} \right) \, ,
\ee
In writing this matrix, we have chosen to discard the entries corresponding to the BPS operators
$\co_{ij,n=0}^J$ and the combination $\sqrt{r}\ct^{J,r}_{ij, n=0}+\sqrt{1-r}\ct^{J,r}_{ij}$. This 
combination is chosen because it is orthogonal to $\co_{ij,n}^J$, $n\neq 0$, which can be easily 
seen from \eqref{double-single-trace-overlap}.  
The sub-matrix involving these BPS operators can be diagonalized using the
freedom we mentioned in the discussion following \eqref{gauge-matrix-free-string-basis}.
The remaining basis elements are chosen to correspond to the non-BPS single and double
trace BMN operators given in $\co_{ij,n}^J,\ \ct^{J,r}_{ij, n}$ ($n\neq 0$) 
and $\sqrt{1-r}\ct^{J,r}_{ij, n=0}-\sqrt{r}\ct^{J,r}_{ij}$, in order.
The entries of \eqref{gamma-tilde-g2} in this basis can be read off from 
\eqref{double-single-trace-overlap}, and are
\begin{subequations}\label{Gamma1-components}
\begin{align}
  \tilde\Gamma^{(1)}_{n,r} & = \tilde{\Gamma}^{(1)}_{r,n}
   = - \frac{\sin^2(\pi nr)}{\sqrt{J} 2\pi^2} \, \\ 
  \tilde\Gamma^{(1)}_{n,pr} & = \tilde{\Gamma}^{(1)}_{pr,n}
   = \frac{\sqrt{1-r}}{\sqrt{Jr}} \frac{\sin^2(\pi nr)}{2\pi^2} \, . 
\end{align}
\end{subequations}
This procedure can be continued to higher orders in $g_2$ in an obvious way.

To read off the anomalous dimensions, we must choose a basis which diagonalizes the
dilatation operator. This basis would simultaneously diagonalize both the matrices
$G_{\mathfrak{a} \mathfrak{b}}$ and $\Gamma_{\mathfrak{a} \mathfrak{b}}$.
That such a diagonalization is possible (i.e. that these two matrices commute), can be
argued from conformal invariance, since it implies that a basis of operators with definite
scaling dimensions (classical plus anomalous) can be chosen.
This choice of basis has been presented in \cite{Constable:2002vq,Beisert:2002bb}.
Going to the $\Delta$-BMN basis, we find for the scaling dimension of single-trace
BMN operators with two impurities, at order $g_2^2$,
\be\label{masscorrection-first-order-lambda'g2}
\Delta= \Delta_0 + \lambda^\prime
  \left( n^2 + g_2^2 \left(
  \frac{1}{48 \pi^2} + \frac{35}{128 \pi^4 n^2} 
  \right) \right) \, ,
\ee
for $n \ne 0$ and with $\Delta_0 = J + 2$ for two impurities.
For $n = 0$, the classical scaling dimension is protected against quantum corrections by virtue of 
supersymmetry. \eqref{masscorrection-first-order-lambda'g2} is the main (basis independent) result 
of this section, to be directly compared with the corresponding string field theory results of
section \ref{SFT}.

\subsection{$n$-point functions of BMN operators}\label{n-point-functions}

Up to this point we have dealt with two-point functions of BMN operators, taking into account 
both non-planar corrections and interactions.
We may wonder what role higher $n$-point functions play in the plane-wave/SYM duality.
We address this issue in the context of three and four-point functions below.

\subsubsection{Three point functions of BMN \opt s}  

As discussed in section \ref{amonalous}, the spacetime dependence of three-point functions in a 
conformal field theory is completely fixed and once a basis of quasi-primary \opt s is chosen, two 
and three point functions take the form of \eqref{2-point:functions:cft} and 
\eqref{3-point:functions:cft}. Moreover, in such basis the operator product expansion takes a 
particularly simple form of \eqref{ope-quasi-primary}.
The remaining task is then to find
$C_{\Delta_i,\Delta_j,\Delta_k} (g_{YM}^2,N)$.
Taking the pinching limit of the three-point function \eqref{3-point:functions:cft}
(e.g. $x_1 \rightarrow x_2$) and using the operator product expansion together with the
the special form for two-point functions of quasi-primary operators,
we can compute $C_{\Delta_i,\Delta_j,\Delta_k} (g_{YM}^2,N,J)$ as
a sum over $C_{ij}^k(g_{YM}^2,N,J)$, where the sum runs over the non-singular
constant terms in the OPE and with the dimension of the OPE coefficient equal to that of
the third operator away from the pinching.
We see that {\it three-point functions of BMN \opt s carry no extra information beyond those of 
two-point functions}.
As an aside, one may use the fact that three-point functions have a simple
form in a basis of quasi-primary operators \eqref{3-point:functions:cft} as a check of gauge 
theory computations.

Here, however, we present the three-point function of chiral-primary \opt s. These correlators are 
protected against quantum corrections \cite{Lee:1998bx,D'Hoker:1998tz,D'Hoker:1999ea}. Furthermore, 
for chiral-primary \opt s the anomalous dimension vanishes, therefore
\be \label{three-pt-chiral-primary}
  \Big\langle \co^{J_1} (x) \co^{J_2} (y) \bar{\co}^{J_3} (0) \Big\rangle \: = \:
  \frac{1}{|x|^{2 J_1} |y|^{2 J_2}} \Big\langle \Tr Z^{J_1} \Tr Z^{J_2} \Tr\bar{Z}^{J_1 + 
J_2}
  \Big\rangle_{MM},
\ee
where $J_1+J_2=J_3$, and vanishing otherwise. The correlator on the right-hand side is a correlator 
in the Matrix model introduced in section \ref{NP-chiral-primary} and using the Matrix theory 
techniques we can evaluate them to all orders in the genus expansion \cite{Kristjansen:2002bb}
\begin{widetext}
\be \label{three-point-mm}
\begin{split}
  \Big\langle
  \Tr Z^{J_1} \Tr Z^{J_2} & \Tr\bar{Z}^{J_1 + J_2}
  \Big\rangle_{MM} =
  \left(\sum_{k=J_2+1}^{J_1+J_2}- \sum_{k=1}^{J_1}\right) 
  \prod_{i=1}^k ( N - 1 + i ) \prod_{m=1}^{J_1+J_2-k} ( N - m ) \\
  & = \frac{1}{J_1 + J_2 + 1}\biggl( \frac{\Gamma (N+1)}{\Gamma (N-J_1-J_2)} 
  + \frac{\Gamma (N+J_1+J_2+1)}{\Gamma (N)} -
  \frac{\Gamma (N+J_1+1)}{\Gamma (N-J_1 )} -
  \frac{\Gamma (N+J_1 +1)}{\Gamma (N-J_1)} \biggr) \\
  & \simeq \sqrt{\frac{J_1 + J_2}{J_1 J_2}} \frac{\sinh \left( \frac{J_1 (J_1 + J_2)}{2 N}
  \right)
  \sinh \left( \frac{J_2 (J_1 + J_2)}{2 N} \right)}
  {\frac{\left( J_1 + J_2 \right)^2}{2 N}} \, ,
\end{split}
\ee
\end{widetext}
where the first equality is obtained assuming (without loss of generality) that $0<J_1, J_2 <N$
and in the final step we have taken the BMN limit $J_1, J_2 \sim \sqrt{N} \to \infty$ 
and $J_1/J_2={\rm fixed}$. The explicit expressions for three-point functions of generic BMN 
operators may be found in \cite{Huang:2002wf, Huang:2002yt}.

\subsubsection{Higher point functions of BMN \opt s and pinching limits}  

As previously discussed, conformal invariance constrains (in a suitable basis)
the dependence of two and three-point correlation functions on spacetime coordinates.
Higher point functions can, however, pick up an arbitrary dependence on certain conformally 
invariant functions of the spacetime coordinates, the conformal ratios. (It is not 
possible to construct such invariants from only two or three coordinates, which is why the two and 
three point functions are so highly constrained.)
The conformal ratios are functions only of the differences of the spacetime points, and so
all higher point functions remain translationally invariant, as required by conformal
symmetry.

A well known result in $\neqf$ SYM is that two and three-point functions of BPS
operators are protected against any quantum corrections, and are hence independent of the
coupling. This allows one to establish results at weak coupling which then
extend by virtue of the protected nature of the quantity to all values of the coupling.
Such non-renormalization theorems have also been demonstrated for certain extremal
and next-to-extremal\footnote{A correlator of $n$ operators is extremal if the scaling dimension of one of the operators
is the sum of the remaining $n-1$ operators and is next-to-extremal if the dimension of one of the 
operators is equal to the sum of the dimensions of the other $n-1$ \opt s plus two.}
higher point correlation functions of chiral primaries 
\cite{Lee:1998bx,D'Hoker:1998tz,D'Hoker:1999ea}. A more extensive list of references can
be found in \cite{D'Hoker:2002aw}.
It is assumed that these results, established in the context of the AdS/CFT correspondence,
still hold when the number of fields in the correlators are taken large.

We are interested in the connected four-point function of chiral-primary operators 
\eqref{singlevac} \cite{Beisert:2002bb},
considering the non-extremal case
\be \label{four-point-function}
  \mathcal{G}^{J_1,J_2,J_1^\prime,J_2^\prime} (x_1,x_2,x_1^\prime,x_2^\prime) =
  \Big\langle \co^{J_1} (x_1) \co^{J_2} (x_2)
  \bar{\co}^{J_1^\prime} (x_1^\prime) \bar{\co}^{J_2^\prime} (x_2^\prime)
  \Big\rangle^{conn} \, .
\ee
(Note that $\co^{J_i}$ are chiral primaries.) Charge conservation requires 
$J_1+J_2=J_1^\prime+J_2^\prime$.

The fact that the spacetime dependence of the four-point function \eqref{four-point-function}
is not completely fixed by conformal invariance appears even in its form in the free theory at 
planar level.
Summing all the connected diagrams \cite{Beisert:2002bb}, the result, to leading order
in $\gymsq$ and $N$ and still for finite $J \equiv J_1+J_2=J_1^\prime+J_2^\prime$, is
\be \label{four-point-free}
  \mathcal{G}^{J_1,J_2,J_1^\prime,J_2^\prime} (x_1,x_2,x_1^\prime,x_2^\prime) =
  \mathcal{D}^{J_1,J_2,J_1^\prime,J_2^\prime} (x_1,x_2,x_1^\prime,x_2^\prime)
  \frac{\sqrt{J_1 J_2 J_1^\prime J_2^\prime}}{N^2}
  \Bigg[ (J_2 - J_1^\prime) + q^{J_1} (J_2 - J_2^\prime) - \frac{q - q^{J_1}}{q - 1} \Bigg] \, ,
\ee
separating the spacetime dependence into
\be
  \mathcal{D}^{J_1,J_2,J_1^\prime,J_2^\prime} (x_1,x_2,x_1^\prime,x_2^\prime) =
  \left(
  |x_1 - x_2^\prime|^{J_1} \: \:
  |x_2 - x_1^\prime|^{J_1^\prime} \: \: |x_2 - x_2^\prime|^{J_2 - J_1^\prime}
  \right)^{-2} \, ,
\ee
and the conformal ratio
\be \label{conformal-ratio}
  q \equiv \frac{|x_1 - x_2^\prime|^2 |x_2 - x_1^\prime|^2}
  {|x_1 - x_1^\prime|^2 |x_2 - x_2^\prime|^2} \, ,
\ee
which depends on the spacetime coordinates in a continuous fashion.
Now take $J_1$ and $J_2$ simultaneously large. The result depends on whether
$q>1$, $q<1$, or $q=1$. The three cases are
\be
  \mathcal{G}^{J_1,J_2,J_1^\prime,J_2^\prime} (x_1,x_2,x_1^\prime,x_2^\prime) =
  \mathcal{D}^{J_1,J_2,J_1^\prime,J_2^\prime} (x_1,x_2,x_1^\prime,x_2^\prime)
  \frac{\sqrt{J_1 J_2 J_1^\prime J_2^\prime}}{N^2}
  \begin{cases}
    (J_2 - J_1^\prime) \ , & q < 1 \, ; \\
    J_2 \, & q = 1 \, ; \\
    (J_2 - J_2^\prime) \: q^{J_1} \, & q > 1 \, .
  \end{cases}
\ee
Now consider letting $x_1^\prime=x_2^\prime+\epsilon$. Holding $x_1$ and $x_2$ fixed, we may
let $\epsilon$ range from a small positive number to a small negative one, continuously passing
through zero.
Even though the conformal ratio $q$ changes continuously, the correlation
function develops a discontinuity at $\epsilon=0$, corresponding to the pinching limit
$x_1^\prime \rightarrow x_2^\prime$. Such behavior is expected to be present also for BMN 
operators with impurities.
Note the order of limits: we have first taken the BMN double scaling limit, and then analyzed the behaviour of the correlation functions when varying its arguments.

Let us now move beyond the free theory and consider interactions, but still at planar level.
We present the main points of the result. The reader can find the details of the computation
in \cite{Beisert:2002bb}. The first order (in $\lambda^\prime$) correction to the
correlation function \eqref{four-point-function} is
\be
  \delta
  \mathcal{G}^{J_1,J_2,J_1^\prime,J_2^\prime} (x_1,x_2,x_1^\prime,x_2^\prime) =
  \mathcal{D}^{J_1,J_2,J_1^\prime,J_2^\prime} (x_1,x_2,x_1^\prime,x_2^\prime)
  \frac{\lambda^\prime J^2}{8 \pi^2} \frac{q-q^{J_1}}{q-1}
  f(r_1,r_2)
  \frac{\lambda^\prime J^2}{8 \pi^2} \frac{q-q^{J_1}}{q-1}
  \mathscr{X}(x_1,x_2,x_1^\prime,x_2^\prime) \, ,
\ee
where the function $\mathscr{X}$ depends only on the spacetime coordinates through
conformal ratios and is non-singular for all values of its arguments, and $f$ is
some constant function of the ratios $r_1=J_1/(J_1+J_2)$ and $r_2=J_2/(J_1+J_2)$.
In the BMN limit
\be
\begin{split}
  \delta
  \mathcal{G}^{J_1,J_2,J_1^\prime,J_2^\prime} (x_1,x_2,x_1^\prime,x_2^\prime) = &
  \mathcal{D}^{J_1,J_2,J_1^\prime,J_2^\prime} (x_1,x_2,x_1^\prime,x_2^\prime)\times \\
  \times & \frac{\sqrt{J_1 J_2 J_1^\prime J_2^\prime}}{N^2}
  \frac{\lambda^\prime J^2}{8 \pi^2} \frac{q-q^{J_1}}{q-1}
  f(r_1,r_2)
  \mathscr{X}(x_1,x_2,x_1^\prime,x_2^\prime) \, ,
  \begin{cases}
    \frac{1}{r-1} \ , & q < 1 \, ; \\
    - \: J_1 \ , & q = 1 \, ; \\
    \frac{1}{1-q} q^{J_1}  \ , & q > 1 \, .
  \end{cases}
\end{split}
\ee
The quantum correction is not even finite in the BMN limit, and the scaling of the
divergence with $J$ changes discontinuously with $q$.
Such behaviour is expected to continue, and in fact become worse, for higher $n$-point
functions.
These pathologies make it unlikely that four or higher point functions can be made 
sense of in a dictionary relating the BMN subsector of the gauge theory to strings
on the plane-wave.

As we mentioned when discussing three-point functions, there are specific 
``pinching limits'' of the four-point functions of the BMN \opt s which are well-behaved in the BMN limit. The function $\mathscr{X}$ vanishes in the limit
$x_1 \rightarrow x_2$ or $x_1^\prime \rightarrow x_2^\prime$, and hence the quantum correction
vanishes in either limit. In this limit we reproduce an extremal three-point function of BPS
operators which, as we discussed, is protected against any quantum
corrections, and is well-defined in the BMN limit. We could also take the pinching limit 
where we reproduce a two-point function of BPS double-trace operators, which would of course be 
protected. In this double pinching limit, we end up with
a two-point function of double-trace BMN operators.
Although as we discussed above, $n$-point functions of generic BMN \opt s are not well-behaved in the BMN limit, they reduce to well-behaved two-point 
functions after pinching $(n-2)$ points. Note that we are to perform the pinching {\it after} taking the BMN limit.
In general, many different ``hierarchies'' of pinchings might be legal \cite{Chu:2003qd}, and the end result will of course depend on how the pinching pairs are formed.

The lesson to take away is that ultimately, {\it the two-point functions carry all
the information relevant to the duality.}





%% file: SFT.tex
As a theory which is described by a two dimensional $\sigma$-model plus vertex \opt s, string theory is a first quantized theory \cite{Polchinski:1998rq} in the sense that all its 
states are always on-shell states and can only be found as external ``particles'' of an S-matrix. However, one may ask if we can have a theory allowing (describing) off-shell string propagation. Such a theory, which is necessarily a field theory (as opposed to first quantized Quantum mechanics), is called string field theory (SFT). The on-shell part of ``Hilbert space'' of SFT should then, by definition, match with the spectrum of string theory. There have been many attempts to formulate a superstring field 
theory, see for example \cite{Witten:1986qs, Witten:1986cc, Berkovits:2000hf} and for a review 
\cite{Siegel:1988wk}, however, the final formulation has not been achieved yet. One 
of the major places where a string field theory description becomes useful and necessary  is when the vacuum (or background) about which we are expanding our string theory is not a true, stable vacuum. Such cases generally have the pathology of having tachyonic modes. This line of research has attracted a lot of attention \cite{Berkovits:2000hf}. In this section, we 
study a simpler question, 
string field theory after fixing the light-cone gauge, the \lc SFT, in 
the \pl\ \bg .  Being a \lc field theory,  \lc SFT in the zero coupling limit only describes 
on-shell particles. Therefore the ``Hilbert space'' of \lc SFT, where the corresponding 
\opt s act, is exactly the same as the one discussed in section \ref{physicalstringspectrum}.
The \lc SFT in flat space for bosonic closed and open strings was developed even 
before two dimensional conformal field theory techniques were available 
\cite{Mandelstam:1974hk, Arfaei:1975bk, Arfaei:1976xt} and 
then generalized to supersymmetric open \cite{Green:1983tc} and closed  
\cite{Green:1983hw} strings. 

Here we first very briefly review the basic tools and concepts needed to develop \lc closed 
superstring field theory and then focus on  the \pl\ \bg . Using the symmetries, including \susy , we fix the form of the cubic string 
vertices and then in section \ref{SFTcontact} study second order terms (in string coupling) in the \lc SFT Hamiltonian. 

\subsection{General discussion of the \lc String Field 
Theory}\label{SFTgeneral}
 
The fundamental object in \lc SFT is the string field \opt \ $\Phi$ which creates or 
destroys complete strings, i.e.
\be
\Phi :\ {\mathbb{H}}_m \longrightarrow {\mathbb{H}}_{m\pm 1}\ ,
\ee
and ${\mathbb{H}}_{m}$ is the $m$-string Hilbert space (\cf\  section 
\ref{physicalstringspectrum}).
In the \lc SFT $\Phi$ is a function of $x^+,\ p^+$ (\lc time and momentum), as well as string 
worldsheet fields $X^I(\sigma),\ \theta_{\alpha\beta}(\sigma)$ and 
$\theta_{\dot\alpha\dot\beta}(\sigma)$,
where $X^I(\sigma)=X^I(\sigma, \tau=0)$ and likewise for the other fields. Of course it is also possible to consider the ``momentum'' space 
\rep , in which $\Phi$ is a function of $P^I(\sigma), \ \lambda_{\alpha\beta}(\sigma)$ and 
$\lambda_{\dot\alpha\dot\beta}(\sigma)$
, with $\lambda$ equal to $-i$ times the momentum conjugate to $\theta$, i.e.
\be
\lambda_{\alpha\beta}=\frac{1}{2\pi\alpha'}\theta^\dagger_{\alpha\beta}\ ,\ \ 
\lambda_{\dot\alpha\dot\beta}=\frac{1}{2\pi\alpha'}\theta^\dagger_{\dot\alpha\dot\beta}\ .
\ee
Here we mainly consider the momentum space \rep . Noting the commutation relations \eqref{[XP]}
and \eqref{thetalambda} we find that
\be\label{Xtheta}
X^I(\sigma)=i\frac{\delta}{\delta P_I(\sigma)}\ ,\ \ \ \
\theta_{\alpha\beta}(\sigma)=i\frac{\delta}{\delta \lambda^{\alpha\beta}(\sigma)}\ .
\ee

As in any {\it light-cone} field theory, the \lc  dynamics of $\Phi$ is governed by the 
non-relativistic 
Schrodinger equation
\be\label{HPhi}
{\cal H}_{SFT}\ \Phi =i\frac{\partial}{\partial x^+}\ \Phi\ ,
\ee
where ${\cal H}_{SFT}$ is the \lc string field theory Hamiltonian. In principle, in order to study the dynamics of the theory we should know the Hamiltonian, and obtaining the Hamiltonian is the main goal of this section. As usual we assume that ${\cal H}_{SFT}$ has an expansion in powers of string coupling and at free string theory limit it should be equal to the Hamiltonian coming from the string theory $\sigma$-model, in our case this is ${\cal H}^{(2)}_{l.c.}$ ({\it cf.} \eqref{H2}):
\be\label{Hexpansion}
{\cal H}_{SFT}={\cal H}^{(2)}_{l.c.}+g_s {\cal H}^{(3)}+g^2_s {\cal H}^{(4)}+\cdots
\ee
Our guiding principle for obtaining $g_s$ corrections to the Hamiltonian is using all
the symmetries of the theory, bosonic and fermionic, to restrict the form of such corrections. In the 
case of flat space these symmetries are so restrictive that they completely fix the form of ${\cal 
H}^{(3)}$ and all the higher order corrections \cite{Green:1983tc, Green:1983hw}.
In the \pl \ case, 
as we discussed in section \ref{maxsusy} the number of symmetry generators is less than 
flat space. Nevertheless, as we will see, the number of symmetry generators is
nevertheless large enough to determine ${\cal H}^{(3)}$ up to an overall $p^+$ dependent factor.

Let us now come back to equation \eqref{HPhi} and try to solve it for free strings. This will 
give some idea of what the free string fields $\Phi$ look like. Let us first consider the bosonic 
strings with the Hamiltonian \eqref{bosoniclcH}. We will work in the momentum basis. 

{\it Conventions:} Hereafter we will set $\alpha'=2$; instead of $p^+$ we will use 
$\alpha\equiv \alpha' p^+$ and $e(x)\equiv sign(x)=\frac{|x|}{x}$. If necessary, powers of 
$\alpha'$ can be recovered on dimensional grounds.

Since in the Hamiltonian there are $\partial_\sigma X$ terms it is more convenient to use Fourier 
modes of $X^I(\sigma)$ and $P^I(\sigma)$, i.e. we use \eqref{bosonicmode} at $\tau=0$, however, in order to match our conventions with that of the literature \cite{Spradlin:2002ar, Pankiewicz:2003kj, 
Spradlin:2003xy}
we need to redefine the $\alpha_n$ and $\tilde\alpha_n$ modes:\footnote{For fermions we have a similar 
expansion
\be\label{thetalambdaFourier}
\theta(\sigma)={\theta}_0+\frac{1}{\sqrt{2}}\sum_{n\neq 0} 
\left(\theta_{|n|}-i e(n) \theta_{-|n|}\right)\ 
e^{in\sigma/\alpha},\ \ 
\lambda(\sigma)=\frac{1}{2\pi\alpha}\left[\lambda_0+\frac{1}{\sqrt{2}}\sum_{n\neq 
0} \left(\lambda_{|n|}-i e(n) 
\lambda_{-|n|}\right)\ e^{in\sigma/\alpha}\right] ,
\ee
for both $\theta^{\dot\alpha\dot\beta}$ and $\theta^{\dot\alpha\dot\beta}$ modes.
These modes and our fermionic modes in section \ref{stringbg} are related by
\newline
$\theta_{n}-i 
\theta_{-n}=\frac{1}{\sqrt{p^+}}(c_{-n}(1+\rho_{-n})\beta_n^\dagger+
c_{n}(1-\rho_{n})\tilde\beta_n),\ 
\lambda_{n}-i\lambda_{-n}=\sqrt{p^+}(c_{-n}(1-\rho_{-n})\beta_n^{\dagger}+c_{n}(1+\rho_{n})\tilde\beta_n
), \ n>0 $ 
and $\theta_{0\alpha\beta}=\frac{1}{\sqrt{p^+}}\beta_{0\alpha\beta},\   
\lambda_{0\alpha\beta}={\sqrt{p^+}}\beta^\dagger_{0\alpha\beta},\ 
\theta_{0\dot\alpha\dot\beta}=\frac{1}{\sqrt{p^+}}\beta^\dagger_{0\dot\alpha\dot\beta},\   
\lambda_{0\dot\alpha\dot\beta}=\frac{1}{\sqrt{p^+}}\beta_{0\dot\alpha\dot\beta}$.}
\begin{equation}\label{XPFourier}
X^I(\sigma)=x^I_0+\frac{1}{\sqrt{2}}\sum_{n\neq 0} \left(x^I_{|n|}-i e(n) x^I_{-|n|}\right) 
e^{in\sigma/\alpha},\
P^I(\sigma)=\frac{1}{2\pi\alpha}\left[p^I_0+\frac{1}{\sqrt{2}}\sum_{n\neq 0} \left(p^I_{|n|}-i e(n) 
p^I_{-|n|}\right)e^{in\sigma/\alpha}\right],
\end{equation}
where 
$x^I_{n}-i x^I_{-n}=\sqrt{\frac{2}{\omega_n}}(\tilde\alpha_n+\alpha_n^\dagger), \ 
ip^I_{n}+ p^I_{-n}=\sqrt{\frac{\omega_n}{2}}(\tilde\alpha_n-\alpha_n^\dagger),  \ n>0
$.
Using these $x_n$ and $p_n\ (n\in \mathbb{Z})$ one can introduce another basis for 
creation-annihilation \opt s which is usually used in the \lc SFT 
\cite{Spradlin:2002ar}, and
whose indices range from $-\infty$ to $+\infty$:
\be\label{abasis}
a_n=\frac{1}{\sqrt{2}i} (\alpha_n+\tilde\alpha_n)\ ,\ \ \ \  
a_{-n}=\frac{1}{\sqrt{2}} (\tilde\alpha_n-\alpha_n)\ ,\ \ n>0 \ ,
\ee   
and $a_0=\alpha_0$ and likewise for fermions
\be\label{bbasis}
b_n=\frac{1}{\sqrt{2}i} (\beta_n+\tilde\beta_n)\ ,\ \ \ \  
b_{-n}=\frac{1}{\sqrt{2}} (\tilde\beta_n-\beta_n)\ ,\ \ n>0 \ ,
\ee   
and $b_0=\beta_0$. It is readily seen that
\[
[a_n, a^\dagger_m]=\delta_{mn}\ ,\ \ \ \
\{b_n, b^\dagger_m\}=\delta_{mn}\ ,\ \ n\in\mathbb{Z}\ ,
\]
where all the bosonic and fermionic indices have been suppressed. The \lc Hamiltonian \eqref{H2} in 
this basis is ${\cal H}^{(2)}_{l.c.}=\sum_{n\in \mathbb{Z}} \omega_n(a^\dagger_n a_n+b^\dagger_n b_n)$.

Since $[x_n^I, p_m^J]=i\delta^{IJ}\delta_{mn}$ or equivalently $x_n^I=i\frac{\delta}{\delta 
p_n^I}$, 
the Hamiltonian \eqref{bosoniclcH} written in terms of these Fourier modes 
becomes
\[
{\cal H}^{(2)}=\frac{1}{\alpha}\sum_{n=-\infty}^{+\infty} \left[ p_n^2 +\frac{1}{4}\omega_n^2 
x_n^2\right]=
\frac{1}{\alpha}\sum_{n=-\infty}^{+\infty}\left[ p_n^2 -\frac{1}{4}\omega_n^2 
(\frac{\delta}{\delta 
p_n^I})^2\right]\ ,
\] 
and hence the eigenfunctions of the Schrodinger equation \eqref{HPhi} are products of (an infinite 
number of) momentum eigenfunctions 
$\psi_{N_n}(p_n)$, where ${N_n}$ is the excitation number of the $n^{\rm th}$ oscillator with 
frequency 
${\omega_n/\alpha}$. Being a momentum 
eigenstate, $\frac{\sqrt{\omega_n}}{2}(a_n^\dagger+a_n)\psi(p_n)=p_n\psi(p_n)$, implies that
\be\label{psi(pn)}
\psi(p_n)=\left(\frac{2}{\pi\omega_n}\right)^{1/4}\ {\rm exp}\left[-\frac{1}{\omega_n}p^2_n+
\frac{2}{\sqrt{\omega_n}} p_n a_n^\dagger -\frac{1}{2} a_n^\dagger a_n^\dagger \right].
\ee
The string field $\Phi$ is a linear combination of these 
modes, i.e.
\be\label{Phi0}
\Phi[p_n]=\sum_{\{N_n\}} \phi_{\{N_n\}}\ \prod_{n=-\infty}^{+\infty} \psi_{\{N_n\}}(p_n)\ .
\ee

To quantize the string field theory, as we do in any field theory, we promote $\phi_{\{N_n\}}$ to \opt s acting on the string Fock space where it destroys or creates a complete string with excitation number $\{N_n\}$ at  $\tau=0$. Explicitly $\phi_{\{N_n\}}: {\mathbb{H}}_m \longrightarrow {\mathbb{H}}_{m\pm 1}$ and $\phi_{\{N_n\}}|{\rm vacuum}\rangle=|\{N_n\}\rangle$. Next we promote all the superalgebra generators to \opt s acting on the SFT Hilbert space. We will generically use hatted letters to distinguish SFT \rep s from that of first quantized string theory. As for the generators in the \pl\ superalgebra, as discussed in section \ref{stringsusy}, the kinematical ones depend only on 
the zero modes of strings and dynamical ones are quadratic in string creation-annihilation \opt s. Therefore, at free string theory 
limit (zeroth order in $g_s$), the dynamical $PSU(2|2)\times PSU(2|2)\times U(1)_-$ superalgebra generators, 
$\widehat {J}_{ij},\ \widehat {J}_{ab},\ \widehat {Q}^{(0)}_{\alpha\dot\beta},\ \widehat 
{Q}^{(0)}_{\dot\alpha\beta}$ and 
${\widehat {\cal H}}^{(2)}$, should be quadratic in the string field $\Phi$, for example
\be\label{hatH2}
{\widehat {\cal H}}^{(2)}=\frac{1}{2}\int \alpha d\alpha D^8p(\sigma) D^8\lambda(\sigma)
\Phi^\dagger {\cal H}^{(2)}_{l.c.} \Phi\ ,
\ee
with $D^8p(\sigma)=\prod_{n=-\infty}^{\infty} dp_n$ and 
$D^8\lambda(\sigma)=\prod_{n=-\infty}^{\infty} d\lambda_n^{\alpha\beta}d\lambda_n^{\dot\alpha\dot\beta}
d\lambda_n^{\dagger\alpha\beta}\ d\lambda_n^{\dagger\dot\alpha\dot\beta}$.
Note that all these \opt s preserve the string number; i.e. they map ${\mathbb{H}}_m$ onto $
{\mathbb{H}}_m$. 
 
Now let us use the \susy\ algebra (\cf\ sections \ref{isometry} and 
\ref{planewavesusy}) to restrict and obtain the corrections 
to \susy\ generators once string interactions are turned on.
The kinematical sector of the superalgebra as well as $P^+$ are not corrected by the 
string interactions, because they only depend on the 
zero modes  (or center of mass modes) of the strings and do not have the chance to mix 
with other string modes. Among the dynamical 
generators, $\widehat {J}_{ij},\ \widehat {J}_{ab}$, being generators of a compact \soff 
group, cannot receive corrections, because
their eigenvalues are quantized and cannot vary continuously (with $g_s)$. Therefore, 
only $\widehat {Q}$ and $\widehat {\cal H}$
can receive $g_s$ corrections. We have parametrized the corrections to 
${\widehat {\cal H}}$ as in \eqref{Hexpansion} and similarly 
$\widehat {Q}$'s can be expanded as
\be\label{Qexpansion}
{\widehat Q}_{\alpha\dot\beta}={{\widehat Q}_{\alpha\dot\beta}}^{(0)}+g_s {{\widehat 
Q}_{\alpha\dot\beta}}^{(3)}+
g^2_s {{\widehat Q}_{\alpha\dot\beta}}^{(4)}+\cdots
\ee
where the superscript $(3)$ and $(4)$ in \eqref{Hexpansion} and \eqref{Qexpansion} show that they 
are 
cubic and quartic in the string field $\Phi$; more precisely
\begin{subequations}\label{HmHmpmm}
\begin{align}
{\widehat {\cal H}}^{(3)},\ 
{\widehat Q}_{\alpha\dot\beta}^{(3)},\ {\widehat Q}_{\dot\alpha\beta}^{(3)}&: {\mathbb{H}}_m 
\ 
\to\ {\mathbb{H}}_{m\pm 1}, \\
{\widehat {\cal H}}^{(4)},\ 
{\widehat Q}_{\alpha\dot\beta}^{(4)},\ {\widehat Q}_{\dot\alpha\beta}^{(4)}&: {\mathbb{H}}_m 
\ 
\to\ {\mathbb{H}}_{m} \cup {\mathbb{H}}_{m\pm 2}\ . 
\end{align}
\end{subequations}

These $g_s$ corrections, however, should be such that ${\widehat {\cal H}}$ and ${\widehat 
Q}$ still satisfy the superalgebra. This, as we will show 
momentarily, will impose strong restrictions on the form of these corrections.
{}From \eqref{KILI}, \eqref{HKL}, \eqref{KQ}, \eqref{LQ} and the fact that the algebra 
should hold at any $x^+$, we learn that
\begin{subequations}\label{momentumconserve}
\begin{align}
[{\widehat {\cal H}}^{(n)}, {\widehat X}^{I}]=0\ &,\ \ \ \  
[{\widehat Q}^{(n)}, {\widehat X}^{I}]=0\ , \\ 
[{\widehat {\cal H}}^{(n)}, {\widehat P}^{I}]=0\ &,\ \ \ \ 
[{\widehat Q}^{(n)}, {\widehat P}^{I}]=0\ , \ \ \ n>2 . 
\end{align}
\end{subequations}
Note in particular that (\ref{momentumconserve}b) means that the interaction parts of 
${\widehat {\cal H}}$ and $\widehat Q$ are translationally invariant, while the quadratic 
part of ${\widehat {\cal H}}$ and $\widehat Q$  do not have this symmetry (\cf\ 
\eqref{HKL} and \eqref{KQ}). Similarly 
\eqref{Hq} and \eqref{qQ} imply that
\begin{subequations}\label{fermionicmomentumconserve}
\begin{align}
[{\widehat {\cal H}}^{(n)}, {\widehat q}_{\alpha\beta}] & =[{\widehat {\cal H}}^{(n)}, {\widehat
q}_{\dot\alpha\dot\beta}]=0 \ , \\ 
[{\widehat Q}^{(n)}, {\widehat q}_{\alpha\beta}] & =[{\widehat Q}^{(n)}, {\widehat 
q}_{\dot\alpha\dot\beta}]=0\ , \ \ \ n>2 ,
\end{align}
\end{subequations}
and finally since ${\widehat P}^+$ commutes with all generators:
\be\label{HQP+}
[{\widehat {\cal H}}^{(n)}, {\widehat P}^+]=0\ ,\ \ \ \ 
[{\widehat {Q}}^{(n)}, {\widehat P}^+]=0\ \ \ n>2.
\ee

\subsection{Three string vertices in the \pl\ \lc SFT}
\label{cubicSFT}

Let us now focus on 3-string vertex. Here we will be working in the sector with light-cone 
momentum $p^+\neq 0$. Hereafter we will relax the positivity condition on $p^+$ (\cf\ 
\eqref{lightconeX+}) and take the incoming states to have $p^+ >0$ and the outgoing states
$p^+ < 0$ \cite{Spradlin:2002ar}. Without loss of generality we can assume 
that string one
and string two are incoming and string three is outgoing. The physical quantities, such as 
$P^I$ and $\lambda^{\alpha\beta}$ of the $r^{\rm th}$ string ($r=1,2,3$) will be denoted by
$P_{(r)}^I$ and $\lambda_{(r)}^{\alpha\beta}$.
In order to guarantee (\ref{momentumconserve}b), \eqref{fermionicmomentumconserve} and 
\eqref{HQP+}, which are nothing but the {\it local} momentum conservations of bosonic and 
fermionic fields, ${\widehat {\cal H}}^{(3)},{\widehat Q}^{(3)}$ must be proportional to
\[
\Delta^8\left[\sum_{r=1}^3 P^I_{(r)}(\sigma)\right] 
\Delta^8\left[\sum_{r=1}^3 \lambda^{\alpha\beta}_{(r)}(\sigma)\right] 
\Delta^8\left[\sum_{r=1}^3 \lambda^{\dot\alpha\dot\beta}_{(r)}(\sigma)\right] 
\delta(\sum_{r=1}^3 \alpha_{(r)})
\]
where $\Delta$-functionals are products of (infinite number of) $\delta$-functions of the 
corresponding 
argument at different values of $\sigma$. 
In sum, so far we have shown that
\be\label{calH3Q3}
{\widehat {\cal H}}^{(3)}=\int d\mu_3\ {H}_3\ \Phi(1)\Phi(2)\Phi(3)\ ,\ \ \ \ 
{\widehat {Q}}^{(3)}=\int d\mu_3\ {Q}_3\ \Phi(1)\Phi(2)\Phi(3)
\ee
where $\Phi(r)$ is the string field of the $r^{\rm th}$ string, ${H}_3, Q_3={H}_3, Q_3 (
\alpha_{(r)}, P_{(r)}, X_{(r)}, \theta_{(r)}, \lambda_{(r)})$ are to be determined later using the 
dynamical part of the superalgebra
and
\be\label{measurethree}
d\mu_3=
\left(\prod_{r=1}^3 d\alpha_{(r)}D^8P_{(r)}(\sigma)D^8\lambda_{(r)}(\sigma)\right)
\Delta^8\left[\sum_{r=1}^3 P^I_{(r)}(\sigma)\right] 
\Delta^8\left[\sum_{r=1}^3 \lambda^{\alpha\beta}_{(r)}(\sigma)\right] 
\Delta^8\left[\sum_{r=1}^3 \lambda^{\dot\alpha\dot\beta}_{(r)}(\sigma)\right] 
\delta(\sum_{r=1}^3 \alpha_{(r)})\ .
\ee
We would like to note that (\ref{momentumconserve}a) and the fermionic counterpart of that (which is 
a combination 
of (\ref{fermionicmomentumconserve}a) and (\ref{fermionicmomentumconserve}b)) should still be 
imposed on ${\widehat {\cal H}}^{(3)}$ and ${\widehat {Q}}^{(3)}$.
Since \eqref{momentumconserve}, \eqref{fermionicmomentumconserve} and \eqref{HQP+} are
exactly  the same as their flat space counterparts \cite{Green:1983hw, Green:1983tc},
much of the analysis of \cite{Green:1983hw, Green:1983tc} carries over 
to our case. 

\subsubsection{Number \opt\ basis}

Since in the string scattering processes we generally start and end up with states which are 
eigenstates of number \opt\ $N_n$ (i.e. they have definite excitation number) rather than the 
momentum eigenstates, it is more convenient to rewrite \eqref{calH3Q3} in the number \opt\ basis; 
in 
fact this is what is usually done in the \lc SFT literature (e.g. see \cite{Green:1987se} chapter 
11). 

Since $H_3$ and $Q_3$ do not depend on the string field, for the purpose of converting the basis to 
number \opt\ basis we can simply ignore them and focus on the measure $d\mu_3$ and $\Phi(r)$. For 
this change of basis we need to explicitly write down $\psi_{\{N_n\}}(p_n)$'s (\cf\ \eqref{Phi0}) 
and perform the momentum integral. To identify 
${\widehat{\cal H}}^{(3)}$ and ${\widehat Q}^{(3)}$  it is enough to find their matrix elements 
between {\it two}
incoming strings and {\it one} outgoing string (\cf\ \eqref{HmHmpmm}),  however, it is more 
convenient to work with $|H^{(3)}\rangle, |Q^{(3)}\rangle \in {\mathbb{H}}_3$ where
\be\label{|H3>}
\langle 1| \otimes \langle 2|  {\cal H}^{(3)} |3\rangle  \equiv
\langle 1| \otimes\langle 2|\otimes \langle 3'| H^{(3)}\rangle 
\ee
and similarly for $|Q^{(3)}\rangle$. In the above $\langle 3'|$ and $|3\rangle$ are related 
by worldsheet time-reversal, in other terms  $\langle 3'|=\langle {\rm v}|\Phi(3)^\dagger$ while
$\Phi(3)|{\rm v}\rangle=|3\rangle$ (for more details see \cite{Green:1987se}). Then, defining 
$|V_3\rangle$ as
\be\label{V3}
|V_3\rangle=\left[\int d\mu_3\ \prod_{r=1}^3\prod_{n=-\infty}^{\infty} \psi(p_n)\right]|{\rm 
v}\rangle_3 
\ee
($|{\rm v}\rangle_3$ is three-string vacuum)
$|H^{(3)}\rangle$ and $|Q^{(3)}\rangle$ take the form
\be\label{H3Q3}
|H^{(3)}\rangle=H_3|V_3\rangle\ , \ \ \ |Q^{(3)}\rangle=Q_3|V_3\rangle.
\ee
$H_3$ and $Q_3$ are \opt s acting on three-string Hilbert space ${\mathbb{H}}_3$ and as we will 
state 
in the next subsection $Q_3$ is linear 
and $H_3$ is quadratic in {\it bosonic} string {\it creation} \opt s.
$|V_3\rangle$ itself maybe decomposed into a bosonic part $|E_a\rangle$ and a fermionic part
$|E_b\rangle$ 
\cite{Spradlin:2002ar, Green:1987se}
\be
|V_3\rangle=|E_a\rangle \otimes |E_b\rangle \ \delta(\sum_r \alpha_r)
\ee
The notation $a$ and $b$ for bosons and fermions stems from our earlier notation in which the 
bosonic and fermionic creation \opt s where denoted by $a_n^\dagger$ and $b_n^\dagger, \ n\in 
\mathbb{Z}.$

{\it Note:} Here we mainly focus on the bosonic part, for the fermionic part the calculations are 
essentially the same and we only present the results.
Also in this section we will skip the details of calculations which are generally straightforward 
and standard, more details for the flat space case may be found in \cite{Green:1987se} chapter 11, 
and for the plane-wave \bg\ in \cite{Spradlin:2002ar, Pankiewicz:2002tg}.
\begin{figure}[ht]
\begin{center}
\epsfig{figure=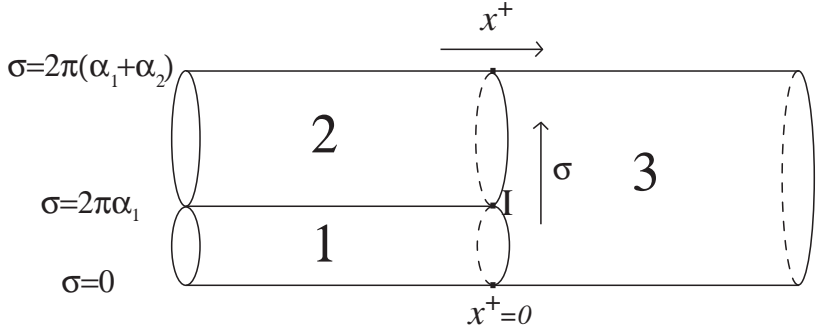, width=0.6\textwidth}
\caption{Three string interaction vertex in the \lc gauge. Note that, due to closed string boundary 
conditions, $\sigma=0$, 
$\sigma=2\pi\alpha_1$  and $\sigma=2\pi(\alpha_1+\alpha_2)$ are identified and $I$ is the 
interaction point.}
\label{stringinteractionvertex}
\end{center}
\end{figure}

To evaluate the integral \eqref{V3} we need to parametrize the interaction vertex from the 
worldsheet point of view. This has been depicted in Fig. \ref{stringinteractionvertex}.
It is convenient to define $\sigma_r$ as
\bea\label{sigmar}
\sigma_{(1)}&=&\sigma\ \ \ \ \ \ \ \ \ \ \ 0\leq \sigma\leq 2\pi\alpha_1\cr
\sigma_{(2)}&=&\sigma-2\pi\alpha_1\ \ \ \ \ \ \ 2\pi\alpha_1\leq \sigma\leq 
2\pi(\alpha_1+\alpha_2)\\
\sigma_{(3)}&=&-\sigma\ \ \ \ \ \ \ \ \ 0\leq \sigma\leq 2\pi(\alpha_1+\alpha_2)\nonumber
\eea
Then in general it should be understood that $P_{(r)}(\sigma)$ is only defined on the domain of 
$\sigma_r$ and otherwise it is zero.
As first step we rewrite the $\Delta$-functionals in terms of the Fourier modes, for that we make 
use of $\Delta[F(\sigma)]=\prod_{n=-\infty}^{\infty}\ \delta\left(\int_0^{2\pi|\alpha_3|}d\sigma\
e^{in\sigma/|\alpha_3|} F(\sigma)\right)$, hence 
\be
\Delta[P_{(r)}(\sigma)]=\prod_{m=-\infty}^{\infty}\ \delta\left(\sum_{r,n} X^r_{mn} p_{n(r)}\right)
\ee
where $X^3_{mn}=\delta_{mn}$ and
\be\label{X1mn}\begin{array}{cc}
X^1_{mn}(\beta)=\frac{2(-1)^{m+n+1}}{\pi} \frac{m\beta\sin \pi m\beta}{n^2-m^2\beta^2}\ ,&\ \ \ \
X^1_{m,0}(\beta)=\frac{\sqrt{2}(-1)^m}{\pi}\frac{\sin \pi m\beta}{m\beta}\cr \ \ \ \ \cr
X^1_{-m,-n}(\beta)=\frac{2(-1)^{m+n+1}}{\pi} \frac{n\sin \pi m\beta}{n^2-m^2\beta^2}\ ,&\ \ \
X^1_{0,0}=1\ ,\ \ \ \ X^1_{m,n}=0 \ \ {\rm otherwise}\ ,
\end{array}
\ee
$\beta=\frac{\alpha_{1}}{\alpha_{3}}$ and in \eqref{X1mn} $m,n>0$. Then, $X^2_{m,n}$ can be 
written in terms of $X^1$ as $X^2_{m,n}(\beta)= (-1)^n X^1_{m,n}(\beta+1)$ for any $m,n\in \mathbb{Z}$. 
Using \eqref{psi(pn)} we can perform the Gaussian momentum integrals of \eqref{V3} to obtain 
the bosonic part of $|V_3\rangle$:
\be\label{Ea}
|E_a\rangle ={\rm exp}\left[ \frac{1}{2}\sum_{r,s=1}^3\sum_{m,n\in\mathbb{Z}} \delta^{IJ} 
a^{I\dagger}_{m(r)} {\overline{N}}^{(rs)}_{mn} a^{J\dagger}_{n(s)}\right]|{\rm v}\rangle_3
\ee
where the Neumann matrices ${\overline{N}}^{(rs)}_{mn}$ are given by
\be\label{Bneumann}
{\overline{N}}^{(rs)}_{mn} =\delta^{rs}\delta_{mn}-2\sqrt{\omega_{m(r)}\omega_{n(s)}}
(X^{(r)}\Gamma^{-1}_a X^{(s)})_{mn}\ ,
\ee
in which 
\[
(\Gamma^{-1}_a)_{mn}=\sum_{r=1}^3\sum_{p\in\mathbb{Z}} \omega_{p(r)}X^{(r)}_{mp}X^{(r)}_{pn}\ 
\]
and $\omega_{n(r)}=\sqrt{n^2+\mu^2\alpha^2_r}$.
Similarly one can work out the fermionic integrals with fermionic Neumann functions 
${\overline{Q}}^{(rs)}_{mn}$ \cite{Pankiewicz:2003kj}
\be\label{Eb}
|E_b\rangle ={\rm exp}\left[ \frac{1}{2}\sum_{r,s=1}^3\sum_{m,n\geq 0} 
(b^{\dagger\alpha\beta}_{-m(r)} b^{\dagger}_{n(s)\alpha\beta}+
b^{\dagger\dot\alpha\dot\beta}_{m(r)} b^{\dagger}_{-n(s) \dot\alpha\dot\beta})
{\overline{Q}}^{(rs)}_{mn} \right]|{\rm v}\rangle_3
\ee
(Explicit formulas for ${\overline{Q}}^{(rs)}_{mn}$ and ${\overline{N}}^{(rs)}_{mn}$   can be found 
in \cite{Pankiewicz:2003kj, He:2002zu}.) 

As mentioned earlier (\ref{momentumconserve}a) and a part of \eqref{fermionicmomentumconserve} 
should still be imposed on $| H^{(3)}\rangle$ and $| Q^{(3)}\rangle$. These are nothing but the 
{\it worldsheet continuity} conditions
\be\label{worldsheetcontin}
\sum_{r=1}^3 e(\alpha_r) X_{(r)}(\sigma)| H^{(3)}\rangle=0,\ \ \
\sum_{r=1}^3 e(\alpha_r) \theta^{\alpha\beta}_{(r)}(\sigma)| H^{(3)}\rangle=0,\ \ \
\sum_{r=1}^3 e(\alpha_r) \theta^{\dot\alpha\dot\beta}_{(r)}(\sigma)| H^{(3)}\rangle=0,
\ee
and similarly for $| Q^{(3)}\rangle$. One can show that $|V_3\rangle$ already satisfies these 
conditions \cite{Spradlin:2002ar}. We would like to comment that the Neumann 
matrices ${\overline{N}}^{(rs)}_{mn}$ and ${\overline{Q}}^{(rs)}_{mn}$ are invariant under CPT (\cf\ 
\eqref{timereversal}) \cite{He:2002zu, Schwarz:2002bc}. 

\subsubsection{Interaction point \opt }

In this part we use the dynamical $PSU(2|2)\times PSU(2|2)\times U(1)_-$ superalgebra to determine 
the ``prefactors'' $H_3$ and $Q_3$ (\cf\ \eqref{calH3Q3} or \eqref{H3Q3}). For that we expand both 
sides 
of \eqref{HQ} and \eqref{QQ} in powers of $g_s$ and note that the equality should hold at any order 
in $g_s$. At first order in $g_s$ we obtain
\begin{subequations}\label{gsfirstorder}
\begin{align}
[{\widehat {\cal H}}^{(3)}, {\widehat Q}^{(0)}_{\alpha\dot\beta}]+[{\widehat {\cal H}}^{(2)}, 
{\widehat Q}^{(3)}_{\alpha\dot\beta}] & =0\ , \\
\{{\widehat {Q}}^{(3)}_{\alpha\dot\beta}, ({\widehat Q}^{(0)})^\dagger_{\rho\dot\lambda}\}+
\{{\widehat {Q}}^{(0)}_{\alpha\dot\beta}, ({\widehat Q}^{(3)})^\dagger_{\rho\dot\lambda}\} & =
2\epsilon_{\alpha\rho}\epsilon_{\dot\beta\dot\lambda}{\widehat {\cal H}}^{(3)}.
\end{align}
\end{subequations}
The equations for ${\widehat Q}_{\dot\alpha\beta}$ is quite similar and hence we do not present 
them here.
In fact, as in the flat space case, one can show $H_3$ and $Q_3$ as a function of worldsheet 
coordinate $\sigma$ should only be non-zero at the interaction point $\sigma=2\pi\alpha_1$ 
\cite{Green:1987se} chapter 11. This and the necessity of these prefactors may be seen by first 
setting $H_3=1$ and 
demanding (\ref{gsfirstorder}a) to hold, 
i.e.
\[
\sum_{r=1}^3 {\widehat {\cal H}}^{(2)}_r |Q^{(3)}_{\alpha\dot\beta}\rangle +
\sum_{r=1}^3 {\widehat {Q}}^{(0)}_{\alpha\dot\beta(r)} |V_3\rangle =0\ .
\]
Then the conservation of energy at each vertex implies that $\sum_{r=1}^3 {\widehat {\cal 
H}}^{(2)}_r=0$ for the physical string states (which are necessarily on-shell), and hence the above 
equation reduces to
\bea\label{interactionpoint}
\sum_{r=1}^3 \int
d\sigma_r &\biggl[& (4\pi P^i_{(r)}-i\mu X^i_{(r)})(\sigma_i)_{\alpha}^{\ 
\dot\rho}\theta^\dagger_{\dot\rho\dot\beta{(r)}}+
(4\pi P^a_{(r)}+i\mu X^a_{(r)})(\sigma_a)_{\dot\beta}^{\ \rho}\theta^\dagger_{\alpha\rho{(r)}}\cr
&+&
i\partial_\sigma X^i_{(r)} (\sigma_i)_{\alpha}^{\ \dot\rho}\theta_{\dot\rho\dot\beta{(r)}}+
i\partial_\sigma X^a_{(r)} (\sigma_a)_{\dot\beta}^{\ \rho}\theta_{\alpha\rho{(r)}}
\biggr]|V_3\rangle =0\ .
\eea
where we have used (\ref{Qstring}a) for ${Q}^{(0)}$'s. In is easy to check that the integrand of 
\eqref{interactionpoint} on $|V_3\rangle$ is generically vanishing and the only non-zero 
contribution comes from the interaction point $\sigma=2\pi\alpha_1$. However, at this point the 
integrand is singular and after integration over $\sigma$ yields a finite result, i.e. 
\eqref{interactionpoint} is not satisfied and hence $H_3\neq 1$.

To work out $H_3$ and $Q_3$ we again use the number \opt \ basis and try to solve 
\eqref{gsfirstorder}. These equations in terms of $H_3$ and $Q_3$ are
\begin{subequations}\label{H3Q3equation}
\begin{align}
\sum_{r=1}^3 \left({\widehat {\cal H}}^{(2)}_r {(Q_3)}_{\alpha\dot\beta} + 
{\widehat {Q}}^{(0)}_{\alpha\dot\beta(r)} H_3 \right)|V_3\rangle  & =0\ , \\
\sum_{r=1}^3 \left( {\widehat {Q}}^{(0)\dagger}_{\alpha\dot\beta(r)} 
{(Q_3)}_{\rho\dot\lambda}+
{\widehat {Q}}^{(0)\dagger}_{\rho\dot\lambda(r)}  {(Q_3)}_{\alpha\dot\beta}\right)|V_3\rangle  & 
=
2\epsilon_{\alpha\rho}\epsilon_{\dot\beta\dot\lambda} H_3 |V_3\rangle\ .
\end{align}
\end{subequations}
These equations, being linear in $Q_3$ and $H_3$, only allow us to determine 
${\widehat{\cal H}}^{(3)}$ and ${\widehat{Q}}^{(3)}$ up to an overall $\mu$ (or more 
precisely $\alpha' \mu p^+$) dependent 
factor. This should be contrasted with the flat space case, where besides the above there is an 
extra condition coming from the boost in the \lc directions (generated by $J^{+-}$ in the notations 
of section \ref{isometry}) \cite{Green:1983hw}. In the \pl\ \bg , however, 
this boost symmetry 
is absent and this overall factor should be fixed in some other way, e.g. by comparing the SFT 
results by their gauge theory correspondents (which are valid for $\alpha'\mu p^+\gg 1$) or by 
the results of \sugra\ on the \pl\ \bg\ (which are trustworthy for $\alpha'\mu p^+\ll 1$).

First we note that, in order to guarantee the continuity conditions 
\eqref{momentumconserve} and \eqref{fermionicmomentumconserve}, the prefactors should
(anti)commute with the kinematical \susy\ generators. Then, one can show that there exist  
{\it linear} combinations of the bosonic and fermionic {\it creation \opt s} which satisfy these 
continuity conditions:
\begin{subequations}\label{basicprefactors}
\begin{align}
{\cal K}^I=\sum_{r=1}^3 \sum_{n\in\mathbb{Z}} K_{n(r)} a^{I\dagger}_{n(r)}\ &,\ \  \
{\widetilde{\cal K}}^I=\sum_{r=1}^3 \sum_{n\in\mathbb{Z}} {\widetilde K}_{n(r)} 
a^{I\dagger}_{n(r)},\ \  {\widetilde K}_{n(r)}=K_{n(r)}^* \\
Y^{\alpha\beta}=\sum_{r=1}^3 \sum_{n\geq 0} {\bar G}_{n(r)} b^{\dagger\alpha\beta}_{n(r)}\ &,\ \ \ 
Z^{\dot\alpha\dot\beta}=\sum_{r=1}^3 \sum_{n\geq 0} {\bar G}_{n(r)} 
b^{\dagger\dot\alpha\dot\beta}_{-n(r)}\ , \ \ \  {\bar G}_{n(r)}={\bar G}_{n(r)}^*.
\end{align}
\end{subequations}
$K_{n(r)},\ {\widetilde K}_{n(r)},\ {\bar G}_{n(r)}$ have complicated expressions and are functions 
of $\mu$ and $p^+$. Since we do not find their  explicit formulas illuminating we do not present 
them here, however, the interested reader may find them in 
\cite{Spradlin:2002rv, Pankiewicz:2002gs, He:2002zu, DiVecchia:2003yp, Pankiewicz:2003kj}.
Therefore, taking prefactors to be functions of 
${\cal K}^I,\ {\widetilde{\cal K}}^I,\ Y^{\alpha\beta}$ and $Z^{\dot\alpha\dot\beta}$ would 
guarantee the continuity conditions.

Equipped with \eqref{basicprefactors} we are ready to solve \eqref{H3Q3equation}. Here we only 
present the results and for more detailed calculations, which are lengthy but straightforward, we 
refer the reader to \cite{Pankiewicz:2003kj}. The main part of the calculation is to 
work out some number 
of relations and identities among $K$'s, $Y$'s and $Z$'s.
\bea
(Q_3)_{\alpha\dot\beta}&=&e^{i\pi/4}|\alpha_3|^{3/2} \sqrt{-\beta(\beta+1)}\left(
S^+_{\dot\rho\dot\beta}(Z)T^+_{\alpha\lambda}(Y^2) {\widetilde{\cal K}}_1^{\dot\rho\lambda}+
iS^+_{\alpha\lambda}(Y)T^-_{\dot\rho\dot\beta}(\tilde{Z}^2) {\widetilde{\cal K}}_2^{\dot\rho\lambda}
\right)\times f(\mu)
\label{fullQ3}\\
(Q_3)_{\dot\alpha\beta}&=&-e^{-i\pi/4}|\alpha_3|^{3/2} \sqrt{-\beta(\beta+1)}\left(
S^-_{\lambda\beta}(Y)T^-_{\dot\alpha\dot\rho}(Z^2){\widetilde{\cal K}}_1^{\lambda\dot\rho}
+ iS^-_{\dot\alpha\dot\rho}(Z)T^+_{\lambda\beta}(\tilde{Y}^2){\widetilde{\cal 
K}}_2^{\lambda\dot\rho}
\right) \times f(\mu) \\
H_3 &=& \biggl[\left({\cal K}^i{\widetilde{\cal 
K}}^j+\frac{\mu\beta(\beta+1)}{2}\alpha_3^3\delta^{ij}\right)V_{ij}-
\left({\cal K}^a{\widetilde{\cal 
K}}^b+\frac{\mu\beta(\beta+1)}{2}\alpha_3^3\delta^{ab}\right)V_{ab}\cr
&-& {\cal K}_1^{\dot\alpha\rho}{\widetilde{\cal K}}_2^{\dot\beta\lambda}
S^+_{\rho\lambda}(Y)S^-_{\dot\alpha\dot\beta}(Z) -{\widetilde{\cal K}}_1^{\dot\alpha\rho}
{\cal K}_2^{\dot\beta\lambda}
S^-_{\rho\lambda}(Y)S^+_{\dot\alpha\dot\beta}(Z)\label{fullH3}\biggr]\times f(\mu)
\eea
where $\beta=\frac{\alpha_1}{\alpha_3},\ |\alpha_3|=\alpha' p^+$ (note that in our conventions 
$\alpha_3<0$ and $-1\leq \beta < 0$) and 
\bea
{\widetilde{\cal K}}_1^{\dot\alpha\rho}\equiv 
{\widetilde{\cal K}}^i {(\sigma_i)}^{\dot\alpha\rho}\ ,\ \ \
{{\cal K}}_1^{\dot\alpha\rho}\equiv 
{{\cal K}}^i {(\sigma_i)}^{\dot\alpha\rho} &,& \ \
{\widetilde{\cal K}}_2^{\dot\alpha\rho}\equiv{\widetilde{\cal K}}^a {(\sigma_a)}^{\dot\alpha\rho}\ 
,\ \ \ {{\cal K}}_2^{\dot\alpha\rho}\equiv 
{{\cal K}}^a {(\sigma_a)}^{\dot\alpha\rho}\ ,\ \ \
\\
S^{\pm}(Y) =Y\pm\frac{i}{3} Y^3\ &,&\ \ \ T^{\pm}(Z^2)=\epsilon \pm{i} Z^2-\frac{1}{6} Z^4\ ,
\\
V_{ij}=\delta_{ij}\left[ 1+\frac{1}{12}(Y^4+Z^4)+\frac{1}{144}Y^4Z^4\right] 
&-&\frac{i}{2}\left[Y^2_{ij}(1+Z^4)-Z^2_{ij}(1+\frac{1}{12}Y^4)\right]+\frac{1}{4} (Y^2Z^2)_{ij}
\\
V_{ab}=\delta_{ab}\left[ 1-\frac{1}{12}(Y^4+Z^4)+\frac{1}{144}Y^4Z^4\right] 
&-&\frac{i}{2}\left[Y^2_{ab}(1-Z^4)-Z^2_{ab}(1-\frac{1}{12}Y^4)\right]+\frac{1}{4} (Y^2Z^2)_{ab}
\eea
In the above,
\bea
Y^2_{\alpha\beta}\equiv Y_{\alpha\rho}Y_{\beta}^{\ \rho} ,\ \ \
{\tilde{Y}}^2_{\alpha\beta}\equiv Y_{\rho\alpha}Y^{\rho} _{\ \beta}\ ,\ \ \ 
Y^4_{\alpha\beta}&\equiv& Y^2_{\alpha\rho}(Y^2)_{\beta}^{\ \rho}=\frac{-1}{2}\epsilon_{\alpha\beta} 
Y^4\ ,\ \ \
\tilde{Y}^4_{\alpha\beta}\equiv \tilde{Y}^2_{\alpha\rho}
(\tilde{Y}^2)_{\beta}^{\ \rho}=\frac{1}{2}\epsilon_{\alpha\beta} Y^4\ ,
\cr
Y^3_{\alpha\beta}\equiv Y_{\alpha\rho}Y^{\lambda\rho}Y_{\lambda\beta},\ \ \
Y^2_{ij}=(\sigma_{ij})^{\alpha\beta} Y^2_{\alpha\beta}&,&\ \ 
Z^2_{ij}=(\sigma_{ij})^{\dot\alpha\dot\beta} Z^2_{\dot\alpha\dot\beta}\ ,\ \ \ 
(Y^2Z^2)_{ij}=Y^2_{k(i}Z^2_{j)k}\ ,\nonumber
\eea
where $Y^4\equiv Y^2_{\alpha\beta}{(Y^2)}^{\alpha\beta}=-
{\tilde Y}^2_{\alpha\beta}{({\tilde Y}^2)}^{\alpha\beta}$. Note that $Y^2$, ${\tilde{Y}}^2$ (and 
similarly 
$Z_{\dot\alpha\dot\beta}^2,\ \tilde{Z}_{\dot\alpha\dot\beta}^2$) are symmetric {\it matrices}, i.e. 
both of their indices belong to only one of $SO(4)$'s; in fact $Y^2$ and $Z^2$ are matrices in the 
first $SO(4)$ and ${\tilde{Y}}^2$ and $\tilde{Z}^2$ in the second one, moreover $V_{ij}$ and 
$V_{ab}$ 
are Hermitian, $V_{ij}^*=V_{ji}$ and $V_{ab}^*=V_{ba}$. The function $f(\mu)$ (or more precisely
$f(\alpha'\mu p^+)$) is an overall factor which is not fixed through the superalgebra requirements 
(\cf\ discussions following \eqref{H3Q3equation}).
 
In order to have a better sense of the above it is instructive to consider the bosonic case. This 
can be done by setting $Y$ and $Z$ equal to zero (and hence 
$V_{ij}=\delta_{ij}$ and $V_{ab}=\delta_{ab}$). This would considerably simplify \eqref{fullH3} 
and 
we obtain
\bea\label{bosonicH3}
|H^{(3)}\rangle &=& f(\mu)({\cal K}_i{\widetilde{\cal K}}_i - {\cal K}_a{\widetilde{\cal K}}_a) 
|E_a\rangle \delta(\sum_{r=1}^3\alpha_r)\cr
&=& \frac{f(\mu)}{4\pi} |\alpha_3|^3 
\beta(\beta+1)\sum_{r=1}^3\sum_{n\in\mathbb{Z}}\frac{\omega_{n(r)}}{\alpha_r}(
a^{i\dagger}_{n(r)}a^{i}_{-n(r)}-a^{a\dagger}_{n(r)}a^{a}_{-n(r)})|E_a\rangle 
\delta(\sum_{r=1}^3\alpha_r)\ ,
\eea
where $\omega_{n(r)}=\sqrt{n^2+\mu^2\alpha_r^2}$.
To obtain the second line, some identities among $K_{n(r)}$ have been employed \cite{Lee:2002vz,
Pearson:2002zs}.
We would like to note  the $\ztwo$ behaviour of $|H^{(3)}\rangle$. This $\ztwo$, as discussed in 
section \ref{isometry} exchanges the two $SO(4)$'s of \soff isometry. {}From \eqref{bosonicH3}
it is evident that ${\cal K}_i{\widetilde{\cal K}}_i - {\cal K}_a{\widetilde{\cal K}}_a$ is odd 
under $\ztwo$. However, as we argued (\cf\ section \ref{physicalstringspectrum}) the vacuum $|{\rm 
v}\rangle$ is odd under $\ztwo$, therefore altogether $|H^{(3)}\rangle$ is $\ztwo$ even. Of course 
with a little bit of work, one can show that this property is also true for the full expression of 
\eqref{fullH3}.

Before closing this subsection we should warn the reader that in the most of the \pl\ SFT 
literature 
(e.g. \cite{Spradlin:2002ar, Spradlin:2002rv}) \soe fermionic \rep s together with an \soe 
invariant vacuum 
$|0\rangle$ or $|\dot{0}\rangle$ (\cf\ \eqref{vacuums}) have been used. In the \soe notation, 
unlike 
our case, this $\ztwo$ symmetry is not manifest. It has been shown that the \soff formulation we 
presented here and the  \soe one are indeed equivalent \cite{Pankiewicz:2003ap}.
In the \soe notation it is very easy to observe that, 
as one would expect, in the $\mu\to 0$ limit goes over to the well-known flat space result; 
this point can be (and in fact have been) used as a cross check for the calculations.

\subsection{Plane-wave light-cone SFT contact terms}\label{SFTcontact}

In this section we will test the \pl /SYM duality at $\co (g_2^2)$ by working out the one-loop 
corrections to single-string spectrum. Explicitly, we run the machinery of quantum mechanical time 
independent perturbation theory with the Hilbert space ${\mathbb{H}}$ and Hamiltonian 
${\widehat{\cal H}}$. One might also try to use time {\it dependent} perturbation theory starting
with string wave-packets to study 
strings scattering processes, the point which will not be studied here and we will only make some 
comments 
about that later on in this section and also in section \ref{conclusion}. 

It is easy to see that at first order in $g_s$ time independent perturbation theory gives a 
vanishing result for energy shifts, i.e. $\langle \psi|{\widehat{\cal H}}^{(3)}|\psi\rangle=0$ for 
any 
$|\psi\rangle\in {\mathbb{H}}_1$ (of course one should consider degenerate perturbation theory, 
nevertheless this result is obviously still true). Therefore we should consider the second order 
corrections. For that, however, we need to work out ${\widehat{\cal H}}^{(4)}$. So, in this section 
first 
we will continue the analysis of section \ref{cubicSFT} and work out the needed parts of 
${\widehat{\cal H}}^{(4)}$. As we will see, to compare the gauge theory results of section 
\ref{interactingstrings} against the string (field) theory side we do not need to have the full 
expression of ${\widehat{\cal H}}^{(4)}$, which considerably
simplifies the calculation. 

\subsubsection{Four string vertices}\label{quarticSFT}

The procedure of finding ${\widehat{\cal H}}^{(4)}$ and ${\widehat{Q}}^{(4)}$ is essentially a 
direct continuation of the lines of previous section; i.e. solving the continuity conditions
\eqref{momentumconserve}, \eqref{fermionicmomentumconserve} and \eqref{HQP+} together with  the 
constraints coming from  the dynamical \susy\ algebra, which are
\begin{subequations}\label{gssecondorder}
\begin{align}
[{\widehat {\cal H}}^{(3)}, {\widehat Q}^{(3)}_{\alpha\dot\beta}]+
[{\widehat {\cal H}}^{(2)}, {\widehat Q}^{(4)}_{\alpha\dot\beta}]+ 
[{\widehat {\cal H}}^{(4)}, {\widehat Q}^{(0)}_{\alpha\dot\beta}] & =0\ , \\
\{{\widehat {Q}}^{(3)}_{\alpha\dot\beta}, ({\widehat Q}^{(3)})^\dagger_{\rho\dot\lambda}\}+
\{{\widehat {Q}}^{(0)}_{\alpha\dot\beta}, ({\widehat Q}^{(4)})^\dagger_{\rho\dot\lambda}\}+ 
\{{\widehat {Q}}^{(4)}_{\alpha\dot\beta}, ({\widehat Q}^{(0)})^\dagger_{\rho\dot\lambda}\} 
& =2\epsilon_{\alpha\rho}\epsilon_{\dot\beta\dot\lambda}{\widehat {\cal H}}^{(4)}.
\end{align}
\end{subequations}

The important point to be noted is that ${\widehat {\cal H}}^{(4)}$ contains two essentially 
different pieces, one is the part which does not change the string number and the other is the 
part which changes string number by two (\cf\ (\ref{HmHmpmm}b)).
In fact in our analysis to find mass corrections to single-string states we need to calculate
$\langle \psi | {\widehat{\cal H}}^{(4)}|\psi\rangle,\ \ |\psi\rangle\in \mathbb{H}_1$. We then
note that ${\widehat{Q}}^{(4)}$ is quartic in string field $\Phi$ and that ${\widehat{Q}}^{(0)}$ maps
$\mathbb{H}_1$ onto $\mathbb{H}_1$. Therefore the terms in   
(\ref{gssecondorder}b) involving ${\widehat{Q}}^{(4)}$ do not contribute to energy shift of 
single-string states at the $g_s^2$ level.
This in particular means that we need not calculate ${\widehat {Q}}^{(4)}$ and therefore we have 
all the necessary ingredients for calculating the one-loop string corrections to the strings mass 
spectrum.

\subsubsection{One-loop corrections to string spectrum}\label{OneloopSFT}
 
In this subsection we compute the mass shift to the string state in $({\bf 9}, {\bf 1})$ \rep\ of 
\soff (\cf\ section \ref{physicalstringspectrum}), i.e.
\be\label{(9,1)}
|(ij), \ n\rangle\equiv \frac{1}{\sqrt{2}}\alpha^{(i\dagger}_n\tilde\alpha^{j)\dagger}_n
|{\rm v}\rangle=\frac{1}{\sqrt{2}}(\alpha^{i\dagger}_n\tilde\alpha^{j\dagger}_n
+\alpha^{j\dagger}_n\tilde\alpha^{i\dagger}_n-\frac{1}{2}\delta^{ij}
\alpha^{k\dagger}_n\tilde\alpha^{k\dagger}_n)|{\rm v}\rangle\ ,
\ee
where it is easy to show that 
\be\label{Tijkl}
\langle (kl),\ m|(ij), \ n\rangle \equiv
\delta_{mn} T^{ijkl}=\delta_{mn} 
(\delta^{ik}\delta^{jl}+\delta^{il}\delta^{jk}-\frac{1}{2}\delta^{ij}\delta^{kl})\ .
\ee
The computation 
for the mass shift should, in principle, be repeated for all the string states in different \soff 
\rep s discussed in \ref{physicalstringspectrum}. However,  we only present that of $({\bf 9}, {\bf 
1})$. Noting that the states with the same excitation number $n$, but with different \soff \rep s 
are generically related by the \pl\ superalgebra (\cf\ section \ref{maxsusy})  and also the fact 
that we have constructed the $g_s$ correction to \lc SFT so that they respect the same \susy\ 
algebra, imply that the mass shifts for these states should only depend on the excitation number 
$n$ and not the details of their \soff \rep . In more precise terms, in the $PSU(2|2)\times 
PSU(2|2) \times U(1)_-$
superalgebra the Hamiltonian commutes with the supercharges \eqref{HQ} implying that all the states
in the same $PSU(2|2)\times PSU(2|2) \times U(1)_-$ supermultiplet should necessarily have the same 
mass for any 
value of string coupling. On the other hand it is easy to show that all the states presented in 
\eqref{bosonicsinglestrings} and \eqref{fermionicsinglestrings} form a (long) multiplet of this 
algebra specified with one single quantum number, $n$. 
(Of course there is another physical number which is not encoded in the superalgebra \rep s, the 
string number or from the BMN gauge theory point of view the number of traces in the BMN \opt s.)
This result have also been checked through 
explicit one-loop SFT calculations for $({\bf 1}, {\bf 1})$ \cite{Gomis:2003kj}, for 
$({\bf 3}^{\pm}, {\bf 1})$ \cite{Roiban:2002xr} and for $({\bf 4}, {\bf 4})$  
\cite{Pankiewicz:2003kj}.

The corrections to the mass at order $g_s^2$ receives contributions from second order perturbation 
theory with ${\widehat {\cal H}}^{(3)}$ and first order perturbation with ${\widehat {\cal 
H}}^{(4)}$:
\be\label{oneloopshift}
\delta E_n^{(2)}=g_s^2\left(\sum_{1,2\in {\mathbb{H}}_2}\frac{1}{2} 
\frac{|\langle 1,2|{\widehat {\cal H}}^{(3)}| (ij),n\rangle|^2}{E^{(0)}_n-E_{1,2}^{(0)}}+
\frac{1}{8}\langle (ij),n|\{{\widehat{Q}}^{(3)}_{\alpha\dot\beta},{\widehat 
Q}^{(3)\dagger\alpha\dot\beta}\}
| (ij),n\rangle\right)\ .
\ee
The extra $\frac{1}{2}$ factor in the first term comes from the fact that this term arises from a 
second order perturbation theory $\left({\rm e}^{S+\delta S}={\rm e}^{S}(1+\delta 
S+\frac{1}{2}(\delta 
S)^2\right)$ 
or in other words it is due to the reflection symmetry of the one-loop \lc string diagram 
\cite{Roiban:2002xr} while $\frac{1}{8}$ factor in the second term is obtained 
noting 
(\ref{gssecondorder}b) after taking the trace over $\alpha\rho$ and $\dot\beta\dot\lambda$ indices.
Note that since the Hamiltonian is a singlet of $SO(4)\times SO(4)\rtimes \ztwo$ and also following 
our superalgebra arguments, we expect states in different {\it irreducible} \soff \rep s not to mix 
and hence we use non-degenerate perturbation theory.

To evaluate the right-hand side of \eqref{oneloopshift} we note that since 
${\widehat{Q}}^{(3)}_{\alpha\dot\beta}$ is cubic in string fields the second term can be written as
\[
\frac{1}{8}\langle (ij),n|\{{\widehat{Q}}^{(3)}_{\alpha\dot\beta},{\widehat 
Q}^{(3)\dagger\alpha\dot\beta}\}
| (ij),n\rangle=
\frac{1}{4} 
\sum_{1,2\in {\mathbb{H}}_2}
\langle (ij),n|
{\widehat{Q}}^{(3)}_{\alpha\dot\beta}| 1,2\rangle \langle 1,2|{\widehat 
Q}^{(3)\dagger\alpha\dot\beta}| (ij),n\rangle .
\]
Next we note that the \lc momentum $p^+$ ($\alpha_3$ in the notation of section \ref{cubicSFT})
can be distributed among the ``internal'' states $| 1,2\rangle$ as 
$\beta=\frac{\alpha_1}{\alpha_3}$ and $-(\beta+1)=\frac{\alpha_2}{\alpha_3}$, and then the sum over 
$1,2\in\mathbb{H}_2$ would reduce to a sum over only string excitation modes together with an 
integral over $\beta$, explicitly
\be\label{oneloopbetaintegral}
\delta E_n^{(2)}=-g_s^2 \int_{-1}^0 
\frac{d\beta}{\beta(\beta+1)}\sum_{\begin{array}{cc}&{\rm stringy}\\ &{\rm modes}\end{array}}
\left(\frac{1}{2}\frac{|\langle 1,2|{\widehat {\cal 
H}}^{(3)}|(ij),n\rangle|^2}{E^{(0)}_n-E_{1,2}^{(0)}}+
\frac{1}{4} |\langle (ij),n|{\widehat{Q}}^{(3)}_{\alpha\dot\beta}| 1,2\rangle |^2\right)\ .
\ee
The $\frac{1}{\beta}$ and $\frac{1}{\beta+1}$ factors may be understood as the ``propagator'' of 
the $1,2$ states in the \lc or equivalently they arise from the normalization of $1,2$ 
states indicating the length conservation in the $\sigma$ direction (this may be put in other 
words: the \lc Hamiltonian 
$P^-$ and the contribution of the transverse momenta to energy differ by a factor of 
$\frac{1}{p^+}$ \cite{Bak:2002ku}).

Let us first spell out the steps of performing the calculations and work out the necessary 
ingredients:

$\bullet$ {\it 0)} Since in the plane-wave \bg\ $P^+$ commutes with all the other \susy\ generators
we can always restrict ourselves to a sector with a given $p^+$ (this is to be contrasted with the 
flat space case where $J^{+-}$ and $J^{-I}$, which are absent in the \pl\ case, can change $p^+$). 
Therefore, instead of the dimensionless parameter $\alpha' \mu p^+$ in our 
calculations  without any ambiguity  we will simply use $\mu$. For example $\lambda'$ expansion in
the gauge theory side would correspond to  large $\mu$ expansion (\cf\ \eqref{lambda'}) on the SFT 
side.

$\bullet$ {\it i)} In the gauge theory calculations of section \ref{interactingstrings}
we diagonalized the dilatation \opt\ only in a subspace of the BMN \opt s, the sector which had the 
same number of impurities, i.e. the {\it impurity conserving} sector. Although this truncation was 
not physically strongly justified, for the matter of comparison, in the SFT calculations 
only the same subsector must be included. Explicitly, in the sum over the two string states in 
\eqref{oneloopbetaintegral} only the string states with two stringy excitations (one left and 
one right mover) should be included. We will have more discussion on this point later on in this 
section.

$\bullet$ {\it ii)} Noting that $| (ij),n\rangle$, defined in \eqref{(9,1)}, is in $({\bf 9},{\bf 
1})$ and ${\widehat{\cal H}}^{(3)}$ a singlet of $SO(4)\times SO(4)$, in the first term of 
\eqref{oneloopbetaintegral} only $|1\rangle\otimes |2\rangle$ states which are in $({\bf 9},{\bf
1})$ \rep\ would contribute and as discussed in section \ref{genericsinglestring} this state should 
necessarily be in the ``NSNS'' sector. This in particular implies that fermionic modes (i.e. ``RR'' 
modes) do not contribute to the first term of \eqref{oneloopbetaintegral} and hence to evaluate 
this term we can simply use \eqref{bosonicH3}.

$\bullet$ {\it iii)} Since ${\widehat{Q}}^{(3)}$ is in 
$({\bf (2,1)}, {\bf (1, 2)})$ of \soff only the $|1\rangle\otimes |2\rangle$ states in
$({\bf 4}, {\bf 1})\otimes ({\bf (1,2)}, {\bf (1, 2)})$ \rep\ (which is of course an ``RNS'' state, 
\cf\ section \ref{genericsinglestring}) would contribute.

$\bullet$ {\it iv)} The gauge theory results of section \ref{interactingstrings} are at first order
in $\lambda'$ and this expansion is valid for $\lambda'\ll 1$. Therefore the SFT results to be 
compared with, should be calculated in the corresponding limit, the large $\mu$ limit
(\cf\ \eqref{lambda'}). The large $\mu$ expansion of the bosonic Neumann functions which would 
appear in our calculations are \cite{Spradlin:2002rv}
\be\label{largemuneumann}
{\overline{N}}^{(r3)}_{mn}(\beta)=-\sqrt{-\frac{\alpha_r}{\alpha_3}} X^{r}_{nm}(\beta) + \co 
(\mu^{-2}),\ \ 
r\in\{1,2\},\ m,n\in\mathbb{Z}\ ,
\ee
where $X^{r}_{nm}$ is given in \eqref{X1mn}. 

$\bullet$ {\it v)} As stated above, $|1,2\rangle$ states contributing to the first term of the sum 
\eqref{oneloopbetaintegral} in the impurity conserving sector should be of the form 
$|(kl), \ m\rangle_{\alpha_2}\otimes |{\rm v}\rangle_{\alpha_1}$ or
$\alpha^{(k\dagger}_{0(r)}|{\rm v}\rangle_{\alpha_1}\otimes \alpha^{l)\dagger}|{\rm 
v}\rangle_{\alpha_2}$ where the index $\alpha_r$ 
indicates the portion of the light-cone momentum carried by that state.\footnote{ As discussed 
in section \ref{physicalstringspectrum} for the states created by $\alpha^{k\dagger}_{0}$ it is 
not necessary to have right and left movers, because it already satisfies 
\eqref{totallevelmatching}.} 
Noting the form of $|H^{(3)}\rangle$ given in \eqref{bosonicH3} we observe that the non-zero 
contribution to
$_{\alpha_3}\langle (ij), n|\otimes _{\alpha_2}\langle (kl), m|\otimes  _{\alpha_1}\langle{\rm v}|  
|H^{(3)}\rangle$ should be proportional to $({\overline{N}}^{(r3)})^2$ because
$_{\alpha_3}\langle (ij), n|\otimes _{\alpha_2}\langle (kl), m|\otimes  _{\alpha_1}\langle{\rm v}|$
contains four stringy annihilation \opt s and the prefactor of $|H^{(3)}\rangle$ has another one, 
hence we need to pick the term which contains five creation \opt s. This implies that the 
exponential of \eqref{Ea}  should be expanded to second order. Performing the calculations and 
using the identities \cite{Roiban:2002xr}
\[
\sum_{p\geq 0} {\overline{N}}^{(r3)}_{\pm m,\pm p}{\overline{N}}^{(r3)}_{\pm n,\pm p}=
-\frac{(-1)^{m+n}\delta_{r,1}+\delta_{r,2}}{\pi}\left[
\frac{\sin(\pi(n-m)\beta)}{n-m}\pm \frac{\sin(\pi(n+m)\beta)}{n+m}\right],\ \ \ \ m,n>0
\]
we find that \cite{Spradlin:2002rv, Roiban:2002xr}
\begin{subequations}\label{H3matrixelement}
\begin{align}
\langle {\rm v}| \alpha^{(i\dagger}_{n(3)}\tilde\alpha^{j)\dagger}_{n(3)} 
\alpha^{(k\dagger}_{0(r)}\tilde\alpha^{l)\dagger}_{0(s)}|H^{(3)}\rangle
&=\frac{g_s 
f(\mu)}{2\pi^2\mu\alpha_3^2}(\delta^{rs}+\frac{\sqrt{\alpha_r\alpha_s}}{\alpha_3})\sin^2n\pi\beta\ 
T^{ijkl}+\co(\mu^{-2})\ \\
\langle {\rm v}| \alpha^{(i\dagger}_{n(3)}\tilde\alpha^{j)\dagger}_{n(3)} 
\alpha^{(k\dagger}_{m(r)}\tilde\alpha^{l)\dagger}_{m(r)}|H^{(3)}\rangle
&=\frac{g_s f(\mu)}{2\pi^2\mu\alpha_3\alpha_r}\beta(\beta+1)\sin^2n\pi\beta\ 
T^{ijkl}+\co(\mu^{-2})\ .
\end{align}
\end{subequations}
These two matrix elements should be compared to their gauge theory correspondents given
\eqref{Gamma1-components}, in which $r\to -\beta$. As we see there is a perfect match if 
we remember that \lc string states are normalized to their light-cone momentum, 
explicitly \eqref{H3matrixelement} equals to those in \eqref{Gamma1-components} multiplied with 
$\sqrt{Jr(1-r)}$. 
Inserting \eqref{H3matrixelement} into \eqref{oneloopbetaintegral} and performing the $\beta$ 
integral as well as the sum over $m$ the first term of \eqref{oneloopbetaintegral} is obtained to be
\be\label{deltaone}
\delta E^{(2,1)}_n= \mu \frac{\lambda' g_2^2}{4\pi^2} \left(\frac{f(\mu)}{2\pi\mu^2 \alpha^2_3}\right) 
\frac{15}{16\pi^2n^2}\ ,
\ee
where $g_2$ is defined in \eqref{BMNgenuscounting}. 

$\bullet$ {\it vi)} Similarly one can work out the contributions from the ${\widehat{Q}}^{(3)}$ 
term. Here we only present the result and for more details the reader is referred to 
\cite{Roiban:2002xr, Pankiewicz:2003kj}
\be\label{deltatwo}
\delta E^{(2,2)}_n= \mu \frac{\lambda' g_2^2}{4\pi^2} 
\left(\frac{f(\mu)}{2\pi\mu^2\alpha_3^2}\right) 
(\frac{1}{24}+\frac{5}{64\pi^2n^2})\ .
\ee

Now we can put all the above together. The one-loop contribution to single-string mass spectrum is 
the sum of \eqref{deltaone} and \eqref{deltatwo}: 
\be\label{deltaE}
\delta E^{(2)}_n= \mu \frac{\lambda' g_2^2}{4\pi^2} 
\left(\frac{f(\mu)}{2\pi\mu^2 \alpha^2_3}\right)(\frac{1}{12}+\frac{35}{32\pi^2n^2}) .
\ee
Choosing $f(\mu)=2\pi \mu^2\alpha^2_3$ for large $\mu$, this result is in precise agreement with 
the 
gauge theory result of \eqref{masscorrection-first-order-lambda'g2}.
In fact it is possible to absorb $f(\mu)$ into $g_s$, the SFT expansion 
parameter, i.e. the effective string coupling is
\be
g_s^{eff}=g_s f(\alpha'\mu p^+)\sim g_s(\alpha'\mu p^+)^2 =g_2^2
\ee
where $\sim$ in the above shows the large $\mu$ limit. 

\subsubsection{Discussion of the SFT one-loop result}

Here we would like to briefly discuss some of the issues regarding the large $\mu$ expansion and 
the SFT one-loop result \eqref{deltaE}. As we discussed \eqref{deltaE} has been obtained by only 
allowing the ``impurity conserving'' intermediate string states in the sums 
\eqref{oneloopbetaintegral}. However, at the same order one can have contributions from string 
states which change impurity by two. For the impurity non-conserving channel the matrix elements of 
the first term of \eqref{oneloopbetaintegral} are of order $\mu^2$ while they are of order $\mu$ in 
the impurity conserving channel \cite{Spradlin:2002rv, Spradlin:2003xy}. Moreover, the energy 
difference denominator 
in the impurity changing channel is of order $\mu$ while it is of order $\mu^{-1}$ in the impurity 
conserving channel. Therefore, altogether the contributions of the impurity conserving and 
impurity non-conserving channels are of the same order and from the string theory side it is quite 
natural to consider both of them. However, the available gauge theory calculations are only 
in the impurity conserving channel; this remains an open problem to tackle.

The other point which should be taken with a grain of salt is the large $\mu$ expansion. In fact as 
we see in \eqref{oneloopbetaintegral} sums contain energy excitations ranging from zero to 
infinity. On the other hand to obtain the large $\mu$ expansion generically it is assumed that
$\omega_n=\sqrt{n^2+(\alpha' \mu p^+)^2}$ can be expanded as $\alpha'\mu p^++\frac{n^2}{\alpha' \mu 
p^+}+\cdots$; this expansion is obviously problematic when $n$ is very large. In other words the 
large $\mu$ expansion and the sum over $n$ do not commute. In fact it has been shown that if we 
do the large $\mu$ expansion first, we will get contributions which are linearly divergent (they 
grow like $\mu$) \cite{Roiban:2002xr}, leading to energy corrections of the order $\mu 
g^2_2\sqrt{\lambda'}$. However, if we did the sum first and then perform the large $\mu$ expansion, 
we would get a finite result for any finite value of $\mu$. This is expected if the results are 
going to 
reproduce the flat space results in the $\mu\to 0$ limit.
This divergent  result from the gauge theory point of view, being proportional to 
$\sqrt{\lambda'}$, seems like a non-perturbative effect \cite{Klebanov:2002mp, Spradlin:2003xy}.


%% file: conclusion.tex

In this review we presented a new version of the string/gauge theory correspondence,
the \pl/SYM duality, and spelled out the correspondence between various parameters and quantities 
on the two sides. As evidence for this duality we reviewed the gauge theory calculations leading 
to the spectrum of free strings on the \pl\ as well as one-loop corrections to this spectrum, 
showing strong support for the duality. There have been many other related directions pursued in 
the literature, and although being interesting in their own turn, they went beyond the scope of a 
pedagogical review. However, we would like to mention some of these topics:

\begin{itemize}
\item {\it Holography in the \pl\ \bg\ OR  what is the gauge theory whose 't Hooft strings are 
type IIB strings on the plane-wave \bg ?}

 As a descendent of the AdS/CFT duality, one may wonder if the \pl/SYM duality is also holographic. 
The first and (following AdS/CFT logic) natural guess for the dual theory is a gauge 
theoryresiding on the boundary of the \pl\ \bg, which is a one dimensional light-like direction 
(\cf\ section \ref{Penrosediagram}). This implies that the holographic dual of strings on the \pl\ 
\bg \ is a quantum mechanical model. Although there have been many attempts in this direction 
(e.g.see \cite{Das:2002cw, Kiritsis:2002kz, Leigh:2002pt, Dobashi:2002ar, Siopsis:2002vw, 
Yoneya:2003mu}), a widely accepted holographic model is still lacking.

\item {\it Plane-wave/SYM for open strings}

The \pl/SYM duality we discussed in this review was constructed for (type IIB) closed strings. The 
extension of the duality to the case of open strings has been studied and may be found for e.g. in 
\cite{Berenstein:2002zw, Gomis:2003kb, Imamura:2002wz, Lee:2002cu, Stefanski:2003zc, 
Skenderis:2002wx}.

\item {\it Flat space limit}

In the \pl/SYM correspondence, the perturbative gauge theory calculations are only 
possible at large $\mu$ while the \sugra\ description of the string theory side can only 
be trusted for small $\mu$, where the gauge theory is strongly coupled. 
Since in the $\mu\to 0$ limit the \pl\ \bg\ reduces to flat space, one may wonder if it is 
possible to take the same limit on the gauge theory side and finally obtain a gauge theory 
description of strings on flat space. This, of course, amounts to knowing about the 
(non-perturbative)  finite $\mu$ behaviour of the BMN gauge theory. It is conceivable that 
at finite $\mu$ non-perturbative objects such as instantons and D-branes would dominate the 
dynamics (at least in some corners of the moduli space). 
Presumably the $\mu\to 0$ limit is not a smooth one and we lose some of the normalizable states 
of the Hilbert space. This line of questions remains open and should be addressed.

\item {\it String bit model and Quantum Mechanical model for BMN gauge theory} 

In the large $\mu$ limit one can readily observe that in \eqref{LCbosonicaction} we can drop 
the $(\partial_\sigma X)^2$ term against the mass term $\mu^2X^2$. This in particular implies that 
in such a limit strings effectively become a collection of some number of massive particles, the 
string bits. Hence it is quite natural to expect the large $\mu$ dynamics of strings on the \pl\ 
\bg\ to be governed by a string bit model \cite{ Verlinde:2002ig, Vaman:2002ka, Zhou:2002mi} in 
which the effects of string tension and 
interactions are introduced as interaction terms in the string bit Lagrangian. The proposed string 
bit model consists of $J$ string bits of mass $\mu$, with the permutation symmetry and more 
importantly, the $PSU(2|2)\times PSU(2|2)\times U(1)_-$ symmetry built into the model.
The action for the string bit model, besides the kinetic (quadratic) term, has cubic and quartic 
terms, but terminates at the quartic level, as dictated by \susy .
The model has been constructed (or engineered) so that it gives the free-string mass spectrum. 
Remarkably it also reproduces the one-loop results of section \ref{interactingstrings} 
or \ref{SFT}. Based on this model it has been  conjectured that \cite{Pearson:2002zs, Vaman:2002ka}
the ``mixing'' between the two impurity BMN
states, to all orders in $g_2$, is given by $|\tilde\psi\rangle={\rm 
e}^{-\frac{g_2}{2}{\widehat{\Sigma}}}|\psi\rangle$
and where the \opt\ ${\widehat\Sigma}$ is defined as
\[
{\widehat\Sigma}\co_{ij,\ n}^J= \sum_{r, m} M_{nm}^r \ct_{ij,\ m}^{J,r}-\sum_r M_n^r 
\ct_{ij}^{J,r},
\]
where $M_{nm}^r$ and $M_n^r$ are matrix elements of $\mathcal{U}^{(1)}$ defined in section 
\ref{mixing}.
Moreover, one of the basic predictions of the string bit model is that the genus counting 
parameter $g_2$ would always appear through the combination $\lambda' g_2^2$ (\cf\  
\eqref{stringcoupling}). This result, however, has been challenged by yet another quantum 
mechanical model of the BMN gauge theory constructed to capture the dynamics of BMN \opt s. The 
Hamiltonain for this quantum 
mechanical model is the dilatation \opt\ of the $\cn=4$ SYM and its Hilbert space is the BMN 
states with two impurities \cite{Eynard:2002df, Beisert:2002ff, Spradlin:2003bw, Kristjansen:2003uy}. 
Based on this 
model it has been argued that there are $\lambda' g_2^4$ corrections to the 
string mass spectrum at genus two \cite{Beisert:2002ff}, where they also conjectured that 
to all orders, both in $\lambda'$ and $g_2^2$, the string spectrum is given by the eigenvalues of 
a ``full'' Hamiltonian of the form \cite{Spradlin:2003bw, Beisert:2002ff} 
\[
H_{\rm full}= 2\mu\sqrt{1+\lambda' H} \, ,
\] 
and where 
\[
H=H_0+\frac{1}{2} g_2 (V+V^\dagger)+ \frac{1}{8} g_2^2 [{\widehat \Sigma},V-V^\dagger] \, .
\]
Here $V$ is fixed by the requirement of \susy\ and is such that at $g_2=0$ it gives the 
free-string spectrum and reproduces the one-loop result of \eqref{deltaE}. This conjecture has so 
far passed $\co(\lambda')$ (including $\lambda'^2$) tests when compared to the direct gauge 
theory calculations. However, already a mismatch with the exact SFT results 
\cite{He:2002zu} has been reported \cite{Spradlin:2003bw}. It has been speculated that this 
mismatch is due to the basic assumption in this quantum mechanical model, where only the two 
impurity BMN states i.e., the impurity preserving sector (\cf\ discussion of section 
\ref{SFTcontact}) has been considered \cite{Spradlin:2003bw}. 
It would be desirable to directly obtain the plane-wave light-cone SFT from the study of BMN gauge 
theory, some step in this direction has been taken in \cite{deMelloKoch:2002nq, 
deMelloKoch:2003pv}.

\item {\it Spectrum of dilatation operator}

According to the plane-wave/SYM duality the dilatation \opt\ should be identified with the 
light-cone string field theory Hamiltonian. As an alternative way to study and verify this duality 
one may choose to focus only on the representation of the dilatation \opt\ $\widehat{\cal D}$ on 
the space of all possible (gauge invariant)  \opt s of the $\cn=4$ SYM theory in the BMN sector. 
(Note that $\widehat{\cal D}$ is the Hamiltonian of $\cn=4$ SYM theory in the radial 
quantization.) Therefore, working out $\widehat{\cal D}$ and its spectrum (in powers of the genus 
counting parameter) would help to establish the BMN duality, as well as solving the full $\cn=4$ 
SYM theory. This direction of study has recently attracted some attention with a view to solving 
the full gauge theory, see for e.g. \cite{Beisert:2003tq, Beisert:2003jj, Dolan:2003uh, 
Beisert:2003yb, Beisert:2003xu, Belitsky:2003ys}.

\item {\it String interactions and scattering}

In section \ref{SFT}, we worked out in detail the cubic string interaction terms in the \pl\ \lc 
string field theory Hamiltonian. However, we never used it to evaluate amplitudes for any 
string scattering process on the \pl\ \bg. It has been argued that \cite{Bak:2002ku}
generically field theories on \pl\ \bg s admit an S-matrix description. Because of the harmonic 
oscillator potential the fields see in the \pl\ \bg , the only directions along which the 
particles can move off to infinity are the $x^+,x^-$ directions. In this sense the 
S-matrix is essentially an S-matrix for a $1+1$ dimensional (non-local)  theory 
\cite{Bak:2002ku}. (Note that the results of \cite{Bak:2002ku} should be 
taken with a grain of salt and they only hold when the
explicit mass of the particles are non-zero, i.e. the \sugra\ modes of strings cannot
be used as external states in an S-matrix, simply because one cannot form a wave-packet with a 
non-zero group velocity out of them. The problem of defining an S-matrix for supergravity modes 
appears to be a physical one and has a counterpart in the gauge theory: as we discussed briefly in 
section \ref{mixing} there is an arbitrariness in the mixing among the single-trace and 
double-trace chiral-primary \opt s, and likewise for their descendants which are supergravity 
modes, that cannot be fixed through similar arguments to those used for higher stringy 
excitations. We would also like to point out that the fact that for the massless case light-cone 
spectrum becomes $p^+$ independent can be evaded for the case where we have an NSNS three-form 
background, such as the case of parallelizable pp-waves discussed in \cite{Sadri:2003ib}. 
Question of S-matrix for these cases have been addressed in \cite{D'Appollonio:2003dr}.)
So, the immediate question 
which one might ask is what is the gauge theory dual for 
the string scattering amplitudes. This question can only be answered in the strict $J =\infty$ 
limit, because otherwise the only ``space'' direction along which the particles can travel to 
infinity, $x^-$, is essentially compact (\cf\ \eqref{x+-xy}). It has been argued that  due to 
instability of massive string modes, one really needs to use time-dependent perturbation theory 
\cite{Freedman:2003bh, Bonderson:2003xs}. Furthermore, studying the correlators of two BMN  \opt s 
and a generic non-BMN \opt\ it has been argued that the BMN dictionary is very sensitive to 
thefluctuations of the \bg\  \pl\ \cite{Mann:2003qp}. The resolution of the above issues and 
puzzles is not yet available, and calls for thorough study.
 
\item {\it D-branes in plane-wave backgrounds}

Here we have only studied strings on the \pl\ \bg , however, type IIB string theory on this \bg\ 
also has D-brane solutions. Similar to the flat space case, D-branes on the \pl\ \bg\ can be 
studied by introducing open strings in the type II theory and imposing Dirichlet boundary 
conditions on them \cite{Polchinski:1995mt}, or equivalently by giving the closed string 
description through the boundary state formulation \cite{Callan:1996xx}.
Both of the approaches have been pursued for D-branes in \pl \ \bg ; see \cite{Bergman:2002hv, 
Billo:2002ff, Dabholkar:2002zc} for examples.

In general,  D-branes in the \pl\ \bg\ can be classified into two sets, those which are
``parallel'', meaning that they include $x^-$ along their worldvolume, and ``transverse'', in 
which the $x^-$ direction is transverse to the worldvolume. It has been shown that in the \pl\ 
\bg\ we can have (half supersymmetric) ``parallel'' $D_p$-branes for $p=3,5,7$, and where they are 
localized at the origin of the space transverse to the brane \cite{Dabholkar:2002zc}. ``Parallel'' 
$D_p$-branes 
in \pl\ \bg s, other than the maximally \susyc\ one, and their supersymmetric intersections, 
$D$-brane interactions, their worldvolume theory as well as the corresponding \sugra\ solution 
have been under intensive study e.g. see \cite{Skenderis:2002vf, Bain:2002nq, Alishahiha:2002rw, 
Biswas:2002yz, Ganor:2002ju, Michishita:2002jp, Hyun:2002xe, Kim:2003zw, Gaberdiel:2003sb, 
Sadri:2003ib, Kumar:2002ps, Chandrasekhar:2003fq, Ohta:2003rr, Bain:2002tq, Sarkissian:2003jn}.

As for the transverse D-branes, one can in fact show that the only half supersymmetric brane 
solution of the maximally \susyc\ \pl\ \bg\ is a spherical threebrane, which is a giant graviton 
\cite{McGreevy:2000cw}. The role of these giant gravitons in the context of 
\pl/SYM duality has not been explored in detail, however some useful preliminary analysis can be 
found in \cite{Balasubramanian:2002sa, Metsaev:2002sg}.

\item {\it T-duality on \pl\ \bg s}

One of the other interesting directions which has been pursued in the literature is the question 
of extending usual T or S dualities, which are generally studied for the flat space backgrounds, 
to plane-waves. T-duality is closely tied with compactification.
Compactification is possible along directions which have translational symmetry (or along the 
Killing vectors). In the coordinates we have adopted for plane-waves (\cf\ 
\eqref{plane-wave-metric}) such isometries are not manifest. However, as we have extensively 
discussed, there are a pair of eight space-like Killing vectors ($L_I$'s and $K_I$'s \cf\ 
\eqref{KILI}), and hence by a suitable coordinate transformation we can make them manifest. Such a 
coordinate transformation would necessarily involve using a ``rotating frame'' 
\cite{Michelson:2002wa}. (Of course the possibility of light-like compactification along the $x^-$ 
direction always exists.) Upon compactification, in the fermionic sector we need to impose 
non-trivial boundary conditions on the (dynamical) supercharges and we may generically lose some 
supersymmetries. That is, T-duality may change the number of supercharges. One can also study 
T-duality and the Narain lattice at the level of string theory. However, on the plane-wave \bg\ 
the T-duality group is generally smaller than its flat space counterpart; study of 
compactification and T-duality on the plane-wave can be found in, for example, 
\cite{Michelson:2002wa,Sadri:2003ib,Mizoguchi:2003be, Bertolini:2002nr, Alishahiha:2002nf}.

\item {\it ``Semi-Classical'' quantization of strings in the $AdS_5\times S^5$ \bg }

The BMN sector of the $\cn=4$ gauge theory is defined as a sector with large $R$-charge $J$. One may ask whether it is possible to make similar statements about the states with large spin $S$. It has been argued that the string $\sigma$-model on the $AdS_5\times S^5$ \bg\ takes a particularly simple form for strings with large spin and one can quantize them semi-classically \cite{Gubser:2002tv}. This has opened a new line for further explorations of various corners of the AdS/CFT correspondence For some useful references we mention 
\cite{Alishahiha:2002fi, Frolov:2002av, Tseytlin:2002ny, Russo:2002sr, Minahan:2002rc, 
Mandal:2002fs, Arutyunov:2003uj, Beisert:2003xu, Beisert:2003ea}.

\item{\it M(atrix)-theory on Plane-waves}

Another interesting maximally \susyc\ \pl\ \bg\ is the eleven dimensional \pl\ arising as the Penrose limit of $AdS_{4,7}\times S^{7,4}$ \cite{Blau:2002dy}. 
It has been conjectured that DLCQ of M-theory in the sector with $N$ units of \lc\ momentum on 
this \bg\ is described by a Matrix model, the BMN Matrix model \cite{Berenstein:2002jq}. This matrix model, its supersymmetric vacua and spectrum have been worked out \cite{Bak:2002rq, Dasgupta:2002ru, Dasgupta:2002hx, Kim:2002if, Motl:2003rw, Yee:2003ge, 
Kim:2002tj}. The long-standing question of transverse fivebranes in the Matrix model \cite{Maldacena:2002rb} has finally been answered. The transverse five-brane in the BMN Matrix model and its Heterotic version has been studied in \cite{Maldacena:2002rb, Motl:2003rw}.
 
\end{itemize}

%% file: appendix1.tex

There are various formulations of $\mathcal{N}=4$ supersymmetric Yang-Mills
theory. We present the two realizations which are most commonly encountered
in the literature, one based on dimensional reduction of the ten dimensional
component formulation of SYM, the other realized by writing the Lagrangian in terms
of $\mathcal{N}=1$ superspace gauge theory coupled to a set of chiral-multiplets.
In addition, there is also a formulation of $\mathcal{N}=4$ SYM based
on $\mathcal{N}=2$ harmonic superspace, which we will not discuss.

\subsection{$\mathcal{N}=4$ SYM Lagrangian in $\mathcal{N}=1$ superfield language}
\label{appendix1}

In this appendix we fix our conventions for the ${\cal N}=4,\ D=4$  gauge 
theory action in terms of ${\cal N}=1$ gauge theory in superspace.
This formulation is useful when we consider the planar result to all orders in 
$\lambda'$ for the anomalous dimensions of the BMN operators.
An $\neqf$ vector multiplet decomposes into one $\mathcal{N}=1$ vector and three chiral multiplets.

An introduction to $\mathcal{N}=1$ superspace and superfields can be found in
\cite{Wess:1992cp, Buchbinder:1998qv, Gates:1983}.
We follow the conventions of \cite{Gates:1983} in our superspace notation. 
We coordinatize $\mathcal{N}=1$ superspace as $z=(x,\theta)$.

The generators of supertranslation on superspace, written as
chiral and anti-chiral superderivatives, are
\be
\label{superderivatives}
\begin{split}
  D_\alpha = \frac{\partial}{\partial \theta^\alpha} +
  \frac{i}{2} \bar{\theta}^{\dot{\alpha}} \sigma^\mu_{\alpha \dot{\alpha}} \partial_\mu \, , \\
  \bar{D}_{\dot{\alpha}} = \frac{\partial}{\partial \bar{\theta}^{\dot{\alpha}}} +
  \frac{i}{2} \theta^\alpha \sigma^\mu_{\alpha \dot{\alpha}} \partial_\mu \, .
\end{split}
\ee
Squares of fields and derivatives are defined with a customary factor of $1/2$, and with the index conventions as in
\be
  D^2 = \frac{1}{2} D^{\alpha} D_{\alpha} \: , \: \: \: \: \:
  \bar{D}^2 = \frac{1}{2} \bar{D}_{\dot{\alpha}} \bar{D}^{\dot{\alpha}} \, ,
\ee
and likewise for the fields. The superderivatives satisfy the $\mathcal{N}=1$ anticommutation relations
\be
  \{ D_\alpha , D_\beta \} = 0 \: , \: \: \: \: \:
  \{ D_\alpha , \bar{D}_{\dot{\alpha}} \} = i \sigma^\mu_{\alpha \dot{\alpha}} \partial_\mu \: .
\ee
Grassmann Delta functions are given by
\be
  \delta^4 ( \theta - \theta^\prime ) =
  ( \theta - \theta )^2 \: ( \bar{\theta} - \bar{\theta}^\prime)^2
\ee
Some useful identities are
\be \label{useful-identities}
  [ D_\alpha , \bar{D}^2 ] = i \sigma^\mu_{\alpha \dot{\alpha}} \partial_\mu \bar{D}^{\dot{\alpha}}
  \: , \: \: \: \: \: \: \:
  D^2 \theta^2 = \bar{D}^2 \bar{\theta}^2 = -1
  \: , \: \: \: \: \: \: \:
  D^2 \bar{D}^2 D^2 = \Box D^2
  \: , \: \: \: \: \: \: \:
  [ \bar{D}^{\dot{\alpha}} , D^\alpha ] \sigma^\mu_{\alpha \dot{\alpha}} \partial_{\mu} =
  \frac{1}{2} \bar{D}^{\dot{\alpha}} D^\alpha \sigma^\mu_{\alpha \dot{\alpha}} \partial_{\mu} +
  i \Box \: .
\ee

The non-Abelian $\mathcal{N}=4$ supersymmetric Yang-Mills action
cast in $\mathcal{N}=1$ superfield form is
\be\label{neqf-action1}
\begin{split}
  S = \frac{2}{g_{YM}^2} \Tr \Big(
  \int d^8 z \: e^{-V} \bar{\Phi}^i e^{V} \Phi_i
  \: + \: \frac{1}{2} \int d^6 z \: W^\alpha W_\alpha \: + \
  \frac{1}{2} \int d^6 \bar{z} \:
  \bar{W}_{\dot{\alpha}} \bar{W}^{\dot{\alpha}} \\
  + \frac{i \sqrt{2}}{3!}
  \int d^6 z \: \epsilon^{ijk}
  \left[ \Phi_i,\Phi_j \right] \Phi_k
  \: + \: \frac{i \sqrt{2}}{3!} \int d^6 \: \bar{z} \epsilon^{ijk}
  \left[ \bar{\Phi}_i,\bar{\Phi}_j \right] \bar{\Phi}_k \Big) \, ,
\end{split}
\ee
with the field strength given by
\be
  W_\alpha \: = \: i \: \bar{D}^2 \left( e^{-V} D_\alpha e^{V} \right) \: .
\ee
Here, $\Phi_i$ ($i=1,2,3$) are chiral superfields and all superfields take values in the Lie 
algebra whose generators obey
\be
  \left[ t^A , t^B \right] \: = \: i \: f^{A B C} \: t^C \, .
\ee
The superspace measures are defined as $d^8 z=d^4 z \: d^2 \theta \: d^2 \bar{\theta}$, $d^6 z=d^4 
x \: d^2 \theta$, and $d^6 \bar{z}=d^4 x \: d^2 \bar{\theta}$.

\subsection{$\mathcal{N}=4$ SYM Lagrangian from dimensional reduction}

The component formulation is more useful when actually computing Feynman diagrams and studying 
the combinatorics which lead to the double expansion characteristic of the double scaling limit 
proposed by BMN.

We use the mostly minus metric convention, $g_{\mu \nu}=diag(+,-,-,-)$.
The Lagrangian (and field content) of the $\neqf$ super-Yang-Mills theory can be deduced by 
dimensionally reducing the ten-dimensional $\mathcal{N}=1$ SYM theory (with $16$ supercharges) on $T^6$ (which preserves all supersymmetries). There is a single vector, four Weyl fermions
and six real scalars, all in the adjoint representation of the gauge group.
The reduced Lagrangian, in component form, is
\be \label{neqf-action-components}
\begin{split}
  \mathcal{L} =
  \frac{1}{g_{YM}^2}
  \Tr \Big(
  &- \frac{1}{2} F_{\mu \nu} F^{\mu \nu}
  + \frac{\theta _I}{16 \pi ^2} F_{\mu \nu} \tilde F^{\mu \nu}
  + \sum_{i=1}^6 D_\mu \phi^i D^\mu \phi^i
  + \sum_{A=1}^4 i \bar{\Psi}^A \Gamma^\mu D_\mu \Psi_A \\
  &+ \frac{1}{2} \sum _{i,j=1}^6 [\phi^i , \phi^j]^2
  + \sum_{i=1}^6 \bar{\Psi}^A \Gamma^i [ \phi^i , \Psi_A ] \, .
  \: \Big)
\end{split}
\ee
Decomposing the ten dimensional Dirac matrices yields four ($\Gamma^\mu$) and six 
($\Gamma^i$) dimensional ones.
This Lagrangian is manifestly invariant under a $U(N)$ gauge symmetry.
The generators of $U(N)$ are chosen with the (non-standard) normalization
\[
  \Tr ( t^A t^B ) \: = \: \delta^{AB} \, ,
\]
($A,B=1,...,N^2)$, and satisfy the appropriate completeness relation
\[
  \delta_{AB} (t^A)^a_b (t^B)^c_d \: = \:
  \delta^a_d \delta^c_b \, ,
\]
$a,b=1,...,N$, since these are the generators in the adjoint representation.
The fields take values in the $U(N)$ algebra
\[
  \chi(x) \: = \: \chi^A(x) t^A \, ,
\]
with $\chi$ any of the fields in the $\neqf$ multiplet.
The sums above are taken over the $N^2-1$ generators of $SU(N)$ and the single generator of the 
$U(1)$ factor in $U(N)$.
The covariant derivative is defined as $D_\mu \chi=\partial_\mu-i[A_\mu,\chi]$.
When diagrams are computed, Feynman gauge is chosen to simplify calculations, taking advantage of the similarity between scalar and vector propagators in this gauge.
There is also a global $SU(4) \sim SO(6)$ R-symmetry,
under which the scalars $\phi^i$ transform in the fundamental of $SO(6)$, and the fermions 
$\Psi_A$ in the fundamental of $SU(4)=Spin(6)$. The vectors are singlets of the 
R-symmetry. The $\theta$ term counts contributions from non-trivial instanton backgrounds, which is 
ignored when one assumes the trivial vacuum.

%% file: appendix2.tex

We briefly review our conventions for the representations of Dirac
matrices in ten dimensions. We use the mostly plus metric.
As for the ten dimensional indices, mainly used in section \ref{stringbg}, we use 
Greek indices $\mu,\nu,...$ to range over the curved (target-space) indices, while 
hatted Latin indices
$\hat{a},\hat{b},...$ denote tangent space indices and $ I,J=1,2,\cdots , 8$ label coordinates 
on the space transverse to the light-cone directions. In the plane-wave background,
it is more convenient to decompose $I,J$ indices into $i,j$ and $a,b$, each ranging from one to 
four. In this review, unless explicitly stated otherwise, the $a,b$ indices will denote these four 
directions. Then the curved space Gamma matrices are defined via contraction with vierbeins as 
usual, $\Gamma^\mu = e^\mu_{\hat a} \Gamma^{\hat a}$. 

We may rewrite the two Majorana-Weyl spinors in ten dimensional type IIA and IIB theories
as a pair of Majorana spinors $\chi^\alpha, \alpha=1,2$,
subject to the chirality conditions appropriate to the theory,
\be \label{chirality}
\Gamma^{11} \: \chi^1 \: = \: + \: \chi^1 \ , \ \ \ \ \ \ \ \ \ \
\Gamma^{11} \: \chi^2 \: = \: \pm \: \chi^2 \ ,
\ee
where for the second spinor we choose $-$ for non-chiral type IIA
and $+$ for chiral type IIB theories, and treat the index $\alpha$ labeling the
spinor as an SL(2,$\mathbb{R}$) index. 
Type II string theories contain two Majorana-Weyl gravitinos
$\psi_\mu^\alpha$, and two dilatinos $\lambda^\alpha$, $\alpha=1,2$, which are of the same 
(opposite) chirality in IIB 
(IIA).

\subsection{Ten dimensional Fermions in \soe \rep s}\label{SO(8)fermions}

The Dirac matrices in ten dimensions obey
\be
  \{ \Gamma^\mu , \Gamma^\nu \} \: = \: 2 \: g^{\mu \nu}
\ee
A convenient choice of basis for
$32 \times 32$ Dirac matrices, which we denote by $\Gamma^\mu$, 
can be written in terms of $16 \times 16$
matrices $\gamma^\mu$ such that
\be
\label{gammabasis}
  \Gamma^{+} \: = \:
  i
  \left(
  \begin{matrix}
     0 & \sqrt{2}  \\
     0 & 0
  \end{matrix}
  \right) ,
  \ \ \ \
  \Gamma^{-} \: = \:
  i
  \left(
  \begin{matrix}
     0 & 0 \\
     \sqrt{2} & 0
  \end{matrix}
  \right) ,
  \ \ \ \
  \Gamma^{I} \: = \:
  \left(
  \begin{matrix}
     \gamma^I & 0 \\
     0 & - \gamma^I
  \end{matrix}
  \right) ,
  \ \ \ \
  \Gamma^{11} \: = \:
  \left(
  \begin{matrix}
     \gamma^{(8)} & 0 \\
     0 & -\gamma^{(8)}
  \end{matrix}
  \right) \ ,
\ee
and the $\gamma^I$ satisfy $\{\gamma^I,\gamma^J\}=2\delta_{IJ}$ with $\delta_{IJ}$
the metric on the transverse space.
Choosing a chiral basis for the $\gamma$'s, we have
$\gamma^{(8)}=diag({\bf 1_{8}},-{\bf 1 _{8}})$. 
The above matrices satisfy
\be
\label{properties}
\begin{split}
(\Gamma^+)^\dagger \: &= \: - \Gamma^- , \ \ \ \ \
(\Gamma^-)^\dagger \: = \: - \Gamma^+ , \ \ \ \ \
(\Gamma^+)^2 = (\Gamma^-)^2 = 0 , \\
\{ \Gamma^{11} &, \Gamma^\pm \} \: = \: 0 , \ \ \ \ \
\{ \Gamma^{11} , \Gamma^{I} \} \: = \: 0 , \ \ \ \ \
\left[ \Gamma^\pm , \Gamma^{IJ} \right] \: = 0 \ ,
\end{split}
\ee
and
$\Gamma^\pm \Gamma^I ... \Gamma^J \Gamma^\pm = 0$ if the same signs appear
on both sides.

We define light-cone coordinates
$x^{\pm} = (x^0 \pm x^9)/\sqrt{2}$ and likewise for the
light-like Gamma matrices $\Gamma^\pm=(\Gamma^0 \pm \Gamma^9) / \sqrt{2}$,
and also define antisymmetric products of $\gamma$ matrices with weight one,
$\gamma^{IJ ... KL} \equiv \gamma^{[I} \gamma^J ... \gamma^K \gamma^{L]}$.

We may choose our ten dimensional, 32 component Majorana fermions $\psi$ to satisfy
\be\label{psipm}
\Gamma^+ \psi^+=0\ ,\ \ \ \ \
\Gamma^- \psi^-=0.
\ee
Noting \eqref{gammabasis}, it is easily seen that 
\be\label{psipmsol}
  \psi^+  = 
  \left(
  \begin{matrix}
     \psi^+_{\alpha} \\
     0 
  \end{matrix}
  \right) \ ,\ \ \ \ \ 
  \psi^-  = 
  \left(
  \begin{matrix}
     0\\
  \psi^-_{\alpha} 
  \end{matrix}
  \right) \ ,\ \ \ \ \alpha=1,2,\cdots, 16\ ,
\ee
where $\psi^{\pm}_\alpha$ can be thought of as \soe Majorana fermions, and the real $\gamma^I$ 
matrices as $16\times 16$ \soe Majorana gamma matrices.
Moreover, we have
\be\label{teneightchirality}
\Gamma^{11}\psi^{+}=   \left(
  \begin{matrix}
    \gamma^{(8)} \psi^+_{\alpha} \\
     0 
  \end{matrix}
  \right) \ ,\ \ \ \ \ 
\Gamma^{11}\psi^{-}=   \left(
  \begin{matrix}
     0\\
 -\gamma^{(8)} \psi^-_{\alpha} 
  \end{matrix}
  \right) \ ,
\ee
i.e. the ten dimensional chirality is related to eight dimensional \soe chirality as indicated
in \eqref{teneightchirality}. 

Now let us focus on the type IIB theory where the maximally \susyc\ \pl\ is defined.
In this case we start with fermions of the same ten dimensional chirality. Then, as stated 
in \eqref{teneightchirality}, $\psi^\pm_\alpha$ should have $\pm$ \soe chirality. Explicitly,
we have
\be\label{eightchirality}
\left(\gamma^{(8)} \psi^{\pm}\right)_\alpha=\pm \psi^\pm_{\alpha}\ .
\ee
Therefore, in the type IIB theory $+/-$ can also be understood as \soe chirality. The above 
equation, however, can easily be solved with the choice $\gamma^{(8)}=diag({\bf 1_{8}},-{\bf 1 _{8}})$, 
where
\[
  \psi^+_\alpha  = 
  \left(
  \begin{matrix}
     \psi^+_{a} \\
     0 
  \end{matrix}
  \right) \ ,\ \ \ \ \ 
  \psi^-_\alpha  = 
  \left(
  \begin{matrix}
     0\\
  \psi^-_{{\dot a}} 
  \end{matrix}
  \right) \ ,\ \ \ \ a,{\dot a}=1,2,\cdots, 8.
\]
$\psi^+_{{a}}$ and $\psi^-_{{\dot a}}$ are then Majorana-Weyl \soe fermions, usually denoted by 
\eights\ and \eightc\ respectively \cite{Green:1987fi}. The gamma matrices can also be 
reduced to 
$8\times 8$ \rep s,
$\gamma^I_{a{\dot a}}$ and $\gamma^I_{{\dot a}a}$, where	the $16 \times 16$ $\gamma^I$ matrices are
\[
     \gamma^I= \left( 
  \begin{matrix}
      0 & \gamma^I_{a{\dot a}}\\
      \gamma^I_{{\dot a}a} & 0
  \end{matrix}
  \right) , \ I=1,2,\ldots,8,\ \ \ \ \ a,{\dot a}=1,2,\ldots,8.
\]

The fermionic coordinates of the IIB superspace consist of
two same chirality ten dimensional Majorana-Weyl fermions, 
$\theta^1$ and $\theta^2$, and after fixing the light-cone gauge
\[
\Gamma^+ \theta^{1,2}= 0,
\]
and as explained above, we end up with two \soe Majorana-Weyl fermions both in the \eights\
\rep , $\theta^1_a$ and $\theta^2_a$, $a=1,2,\cdots, 8$. We may then combine these two real 
eight-component fermions into a single {\it complex} eight-component fermion
\be\label{complexfermions}
\theta_a=\frac{1}{\sqrt{2}} (\theta^1_a+i\theta^2_a)\ ,\ \ \ \
\theta_a^\dagger=\frac{1}{\sqrt{2}} (\theta^1_a-i\theta^2_a).
\ee

As for the 32 supercharges, the 16 kinematical supersymmetries are in the {\it complex}
\eights\ \rep\ while the 16 dynamical ones are in the {\it complex} \eightc\ \rep .
Note that this statement is true both in flat space and in  the \pl\ \bg\ we are interested in.

\subsection{Ten dimensional Fermions in \soff \rep s}\label{SO(4)fermions}

In the \pl\ \bg , due to the presence of RR five-form flux, the \soe symmetry is broken to 
$SO(4)\times SO(4)$. Therefore for the purpose of this review it is more convenient to make this $SO(4)\times SO(4)$, which is already manifest in the bosonic sector, explicit in the fermionic sector by choosing \soff \rep s instead of {\it complex} \soe \eights\ and \eightc\ fermions. 

{\it Note: Unless explicitly stated otherwise, we will use this \soff notation for fermions and 
gamma matrices.}

First, we note that an \sof Dirac fermion $\lambda$ can be decomposed into two Weyl 
fermions $\lambda_\alpha$ and $\lambda_{\dot\alpha}$, $\alpha, {\dot\alpha}=1,2$.
As usual for the $SU(2)$ fermions, these Weyl indices are lowered and raised using the $\epsilon$ tensor 
\be
\lambda_\alpha=\epsilon_{\alpha\beta}\lambda^\beta .
\ee
We have defined $\theta^\dagger_{\alpha \beta}=(\theta_{\alpha \beta})^*$.
Therefore the \soff fermions are labeled by two \sof Weyl indices, i.e.
$\lambda_{\alpha\beta'}$,  
$\lambda_{\alpha{\dot \beta}'}$, 
$\lambda_{{\dot \alpha}\beta'}$, 
$\lambda_{{\dot \alpha}\beta'}$ and 
$\lambda_{{\dot \alpha}{\dot \beta}'}$,
where the ``primed'' indices, such as $\beta'$ and ${\dot 
\beta}'$ correspond to the second $SO(4)$. We may drop this prime whenever there is no confusion and then simply use, e.g. $\lambda_{\alpha\beta}$ where the first (second) Weyl index corresponds 
to the first (second) $SO(4)$ factor. In fact, as  explained in the main text in section 
\ref{isometry}, there is a 
$\mathbb{Z}_2$ symmetry which exchanges these \sof factors and hence the theory  should 
be symmetric under the exchange of  the first and second Weyl indices.

To relate these \soff fermions to those of $SO(8)$ ({\it complex} \eights\ and \eightc ),
we note that in our conventions
\eights\ (\eightc ) has positive (negative) \soe chirality. On the other hand if we denote the two 
\sof
``$\gamma^{(5)}$'''s by $\Pi$ and $\Pi'$, i.e.
\be\label{gammafives}
\Pi=\gamma^{1234}\ ,\ \ \
\Pi'=\gamma^{5678}\ ,
\ee
then it is evident that
\be
\gamma^{(8)}=\Pi \Pi' \ .
\ee
Therefore for \eights\ fermions, the two $SO(4)$'s should have the same chirality while for 
\eightc\ they should have opposite chirality. Explicitly
\bea\label{eighttofour} 
\psi_a &\to& \psi_{\alpha\beta'} \ \ {\rm and}\ \ \psi_{{\dot \alpha}{\dot \beta}'}\cr
\psi_{\dot a} &\to& \psi_{{\alpha}\dot \beta'} \ \ {\rm and}\ \ \psi_{{\dot\alpha}{\beta}'}\ .
\eea
We would like to emphasize that by \eights\ and \eightc\ we mean the {\it complex} \soe 
fermions defined in \eqref{complexfermions}.

Noting that $SO(4)\simeq SU(2)\times SU(2)$, a Weyl \sof fermion can be represented as 
$({\bf 2, 1})$ for $\lambda_{\alpha}$ and  $({\bf 1, 2})$ for $\lambda_{{\dot \alpha}}$ 
and hence an \soff fermion $\lambda_{\alpha\beta'}$ may be expressed as $\left(({\bf 2, 1}),
({\bf 2, 1})\right)$,
and similarly for the others. In this notation, \eqref{eighttofour} can be written as
\be\begin{split}
{\bf 8}_s &\to \left(({\bf 2, 1}),({\bf 2, 1})\right) \oplus \left(({\bf 1, 2}),({\bf 1, 
2})\right) \, , \\
{\bf 8}_c &\to \left(({\bf 2, 1}),({\bf 1, 2})\right) \oplus \left(({\bf 1, 2}),({\bf 2, 
1})\right) \, . 
\end{split}
\ee
As the last step we need to choose a proper \soff basis for the $\gamma^I_{a{\dot a}}$ matrices.
Following the notation we have adopted in the review (e.g. see section \ref{ppwave}), we denote the first four $SO(4)$ directions by $i,j$ and the other four by $a,b$:
\[
\gamma^I_{a{\dot a}}=(\gamma^i_{a{\dot a}},\gamma^a_{a{\dot a}}) \, ,
\]
where
\be\label{gammai}
\gamma^i_{a{\dot a}}\: = \:
\left(
  \begin{matrix}
     0 & {(\sigma^i)}_{\alpha{\dot \beta}}\ \delta_{\alpha'}^{\ \beta'}\\
     {(\sigma^i)}^{{\dot \alpha}\beta}\ \delta_{{\dot \alpha'}}^{\ {\dot \beta'}} & 0
  \end{matrix}
  \right) ,
  \ \ \ \
\gamma^i_{{\dot a}a}\: = \:
\left(
  \begin{matrix}
  0 & {(\sigma^i)}_{{\alpha}{\dot \beta}}\ \delta_{{\dot \alpha'}}^{\ {\dot \beta'}}\\ 
  {(\sigma^i)}^{{\dot \alpha}{\beta}}\ \delta_{\alpha'}^{\ \beta'} & 0\\
  \end{matrix}
  \right) ,
\ee
and 
\be\label{gammaa}
\gamma^a_{a{\dot a}}\: = \:
\left(
  \begin{matrix}
      -\delta_{\alpha}^{\ \beta}\ {(\sigma^a)}_{\alpha'{\dot \beta}'} & 0\\
    0 & \delta_{{\dot \alpha}}^{\ {\dot \beta}}\ {(\sigma^a)}^{{\dot \alpha}'\beta'}
  \end{matrix}
  \right) ,
  \ \ 
\gamma^a_{{\dot a}a}\: = \:
\left(
  \begin{matrix}
      -\delta_{\alpha}^{\ \beta}\ {(\sigma^a)}^{{\dot \alpha}'{\beta}'} & 0\\
    0 & \delta_{{\dot \alpha}}^{\ {\dot \beta}}\ {(\sigma^a)}_{{\dot \alpha}'{\beta'}}
  \end{matrix}
  \right) ,
  \ee
with 
\be\label{sigmai}
(\sigma^i)_{\alpha{\dot \alpha}}=({\bf{1}},\ \sigma^1,\ \sigma^2,\ \sigma^3)_{\alpha{\dot 
\alpha}} \, ,
\ee
and similarly for $\sigma^a$, where ($\sigma^1$, $\sigma^2$, $\sigma^3$) are the Pauli matrices.
In the above 
\be\label{sigmaup}
(\sigma^i)_{\alpha{\dot \alpha}}=
\epsilon_{\alpha\beta}\epsilon_{{\dot \alpha}{\dot \beta}}\ (\sigma^i)^{{\dot \beta}{\beta}}\ 
.
\ee

In this basis, $\Pi$ ({\it cf.}  \eqref{gammafives}), is given by
\be\label{Pi}
\Pi_{ab}=
\left(
  \begin{matrix}
     \bf{1} & 0\\
     0 & -\bf{1} 
  \end{matrix}
  \right)= diag (\bf{1}_4,\ -\bf{1}_4)\ .
\ee
As usual one can show that
\bea
(\sigma^i)_{\alpha{\dot \beta}}\ (\sigma^j)^{{\dot \beta}{\gamma}}+
(\sigma^j)_{\alpha{\dot \beta}}\ (\sigma^i)^{{\dot \beta}{\gamma}} &=& 
2\delta^{ij}\delta_\alpha^\gamma \ , \cr
(\sigma^i)^{{\dot \alpha}{\beta}}\ (\sigma^j)_{\beta{\dot \gamma}}+
(\sigma^j)^{{\dot \alpha}{\beta}}\ (\sigma^i)_{\beta{\dot \gamma}} &=& 
2\delta^{ij}\delta_{{\dot\alpha}}^{{\dot\gamma}} \ .
\eea
The generators of \sof rotations, $\gamma^{ij}=\frac{1}{2}[\gamma^i,\gamma^j]$, can be easily  
worked out in terms of $\sigma^{ij}$. They are
\be\label{so4generator}
(\gamma^{ij})_{ab}=
\left(
  \begin{matrix}
{(\sigma^{ij})}_{{\alpha}{\beta}}\ \delta_{\alpha'}^{\ \beta'} & 0, \\
0 & {(\sigma^{ij})}^{{\dot \alpha}{\dot \beta}}\ \delta_{\dot \alpha'}^{\ {\dot\beta'}} \\
  \end{matrix}
  \right) ,
\ee
where 
\bea
(\sigma^{ij})_{\alpha\beta} 
&=& \frac{1}{2}[(\sigma^i)_{\alpha}^{\dot \gamma}\ 
(\sigma^j)_{{\beta}{\dot\gamma}}- (\sigma^j)_{\alpha}^{\dot \gamma}\ 
(\sigma^i)_{{\beta}{\dot\gamma}}]=(\sigma^{ij})_{\beta\alpha} \ , \cr
(\sigma^{ij})^{{\dot\alpha}{\dot\beta}} &=& 
\frac{1}{2}[(\sigma^i)^{\dot\alpha}_{\gamma}\ (\sigma^j)^{{\dot \beta}{\gamma}}- 
(\sigma^j)^{\dot\alpha}_{\gamma}\ (\sigma^i)^{{\dot \beta}{\gamma}}]=
(\sigma^{ij})^{{\dot\beta}{\dot\alpha}} \ . 
\eea

Finally we gather some other useful identifies regarding $\sigma^i$'s which are used mainly
in the calculations of section \ref{SFT}:
\bea
(\sigma^{i})_{\alpha{\dot\beta}}(\sigma^{j})^{{\dot\beta}}_{\gamma}=
\delta^{ij}\ \epsilon_{\alpha\beta}+
(\sigma^{ij})_{\alpha\gamma}\ &,& \ \ 
(\sigma^{i})^{{\dot\alpha}{\beta}}(\sigma^{j})_{\beta}^{{\dot\gamma}}=
\delta^{ij}\ \epsilon^{{\dot\alpha}{\dot\beta}}+
(\sigma^{ij})^{{\dot\alpha}{\dot\gamma}}\ ,\\
(\sigma^{i})_{\alpha{\dot\alpha}}(\sigma^{i})_{\beta{\dot\beta}}=2
\epsilon_{\alpha\beta}\epsilon_{{\dot\alpha}{\dot\beta}}\ \ &,&\ \ 
(\sigma^{i})_{\alpha{\dot\alpha}}(\sigma^{i})^{{\dot\beta}{\beta}}=2
\delta_{\alpha}^{\beta}\delta_{{\dot\alpha}}^{{\dot\beta}}\ ,
\eea
\be
(\sigma^{ij})_{\alpha\beta} (\sigma^{ij})^{{\dot\alpha}{\dot\beta}}= 0\ ,\ \  
(\sigma^{ij})_{\alpha\beta} (\sigma^{ij})_{\rho\lambda}= 4(
\epsilon_{\alpha\rho}\epsilon_{\beta\lambda}+\epsilon_{\alpha\lambda}
\epsilon_{\beta\rho}),
\ee
\be
(\sigma^{i})_{\alpha{\dot\alpha}}(\sigma^{j})_{\beta{\dot\beta}}=\frac{1}{2}\left[
\delta^{ij}\epsilon_{\alpha\beta}\epsilon_{{\dot\alpha}{\dot\beta}}+
(\sigma^{ij})_{\alpha\beta} \epsilon_{{\dot\alpha}{\dot\beta}}+
\epsilon_{\alpha\beta} (\sigma^{ij})_{{\dot\alpha}{\dot\beta}} 
-(\sigma^{ik})_{{\alpha\beta}}(\sigma^{jk})_{{\dot\alpha}{\dot\beta}}-
(\sigma^{jk})_{{\alpha\beta}}(\sigma^{ik})_{{\dot\alpha}{\dot\beta}}\right] .
\ee

\subsection{SO(6) and SO(4,2)  fermions}\label{sixdimfermion}

Here we briefly present the $spin(6)$ and $spin(4,2)$ fermion conventions used in section
\ref{fermionicpenrosecontraction}. Let us first consider  the $spin(6)$ spinors, i.e.
six dimensional Euclidean fermions (more details may be found in \cite{Polchinski:1998rr}).
In six dimensions we deal with $2^{6/2}=8$ component Dirac fermions. The $so(6)$ $8\times 8$ Dirac matrices satisfy
\[
\{ \Gamma^{\hat{A}},\Gamma^{\hat{B}}\}=2\delta^{\hat{A}\hat{B}}\ ,\ \ \ \hat{A},\hat{B}=1,2,\cdots, 
6.
\]
As usual (and by definition), the commutator of these $\Gamma$ matrices, which is denoted by 
$\Gamma^{\hat{A}\hat{B}}=\frac{1}{2} [\Gamma^{\hat{A}},\Gamma^{\hat{B}}]$, form an $8\times 8$ 
\rep \ 
of $so(6)$. The eight component $so(6)$ Dirac fermions, 
however, may be decomposed into two four component (complex) Weyl spinors. Explicitly, $\psi_A$, where $A=1,\ldots,8$, can be decomposed into $\psi_{I}$ and $\psi_{\dot{I}}$ where $I,\dot{I}=1,2,3,4$ can be thought of as fundamental (anti-fundamental) $su(4)$ indices. The Dirac matrices 
$\Gamma^{\hat{A}}$, similarly to \eqref{gammabasis}, can be decomposed into $\Gamma^\pm$ and 
$\gamma^{I}$, where now $\gamma$'s are $4\times 4$ matrices and act on the Weyl spinors. 
Each of these $so(6)$ Weyl spinors in their own turn can be decomposed into two four dimensional 
(i.e. $so(4)$) Weyl spinors, though with opposite chiralities, i.e. 
\[
\psi_{I}  \to (\psi_\alpha,\ \psi_{\dot{\alpha}}) \, ,
\]
where $\alpha, {\dot\alpha}=1,2$. Since the arguments closely parallel those of appendix 
\ref{SO(8)fermions} (where we explained how to 
reduce $SO(9,1)$ fermions into the $SO(8)$ fermions), we do not repeat them 
here. In fact a similar result is also true for $so(4,2)$ fermions, and a Weyl $so(4,2)$ fermion can be decomposed into two $so(4)$ Weyl fermions of opposite chirality; if we denote the $so(4,2)$ 
Weyl index by $\hat{I}$ ($I=1,2,3,4$), this means
\[
\psi_{\hat{I}}  \to (\psi_\alpha,\ \psi_{\dot{\alpha}}) \, .
\]

The $SO(4,2)\times SO(6)$ fermions naturally carry spinorial indices of both of the groups. 
Therefore in general we can have four different fermions depending on the chirality of the fermions under either of the groups. In our case the spinors that we deal with (those appearing in the $AdS_5\times S^5$ superalgebra), should have the  same chirality under both groups. This 
comes from the fact that we are working with type IIB theory where both of the fermions 
have the same ten  dimensional chirality. So a general $AdS_5\times S^5$ fermion would carry 
two indices, which are fundamentals of $su(2,2)$ and $su(4)$, e.g. $\psi_{\hat{I}J}$ or 
$\psi_{\hat{\dot{I}}\dot{J}}$. (The choice of $\psi_{\hat{I}J}$ or $\psi_{\hat{\dot{I}}\dot{J}}$
fermions is related to the sign of the self-dual fiveform flux on the $S^5$ of the \ads geometry. 
Here we have chosen the positive case and hence we are dealing with $\psi_{\hat{I}J}$ fermions.)
Note that 
since these are complex fermions this spinor has $32$ degrees of freedom.
This fermion can be decomposed as an $SO(4)\times SO(4)$ fermion using the above decompositions:
\be\label{psiIJ}
\psi_{\hat{I}J}\to (\psi_{\alpha\beta},\ \psi_{\alpha\dot\beta},\ \psi_{\dot\alpha\beta},\ 
\psi_{\dot\alpha\dot\beta}).
\ee